\documentclass[aip,amsmath,amssymb,preprint]{revtex4-1}

\usepackage{graphicx}
\usepackage{dcolumn}
\usepackage{bm}

\usepackage[utf8]{inputenc}
\usepackage[T1]{fontenc}
\usepackage{mathptmx}
\usepackage{etoolbox}
\usepackage{booktabs}
\usepackage{multirow}
\usepackage{tikz}
\usepackage[normalem]{ulem}
\usepackage{amsmath}

\usepackage{hyperref}

\hypersetup{
    colorlinks=true,
    linkcolor=black,
    urlcolor=blue,
    citecolor=black,
    pdfpagemode=FullScreen,
    }
\useunder{\uline}{\ul}{}
\usetikzlibrary{shapes.geometric, arrows}
\tikzstyle{spatio-temporal_explanation_node} = [rectangle, rounded corners, minimum width=1cm, minimum height=1cm,text centered, draw=black, fill=red!30]
\tikzstyle{arrow} = [thick,->,>=stealth]

\newcommand{\Rey}[0]{\mathrm{Re}}
\newcommand{\figref}[1]{Figure \ref{#1}}
\newcommand{\equref}[1]{Equation \ref{#1}}
\newcommand{\tabref}[1]{Table \ref{#1}}
\newcommand{\appref}[1]{Appendix \ref{#1}}
\newcommand{\diffd}{\mathrm{d}}
\newcommand{\secref}[1]{Section \ref{#1}}
\newcommand{\spatialfigwidth}{0.225}
\newcommand{\spatiotemporalfigwidth}{0.25}
\newcommand{\multilinecell}[2][c]{\begin{tabular}[#1]{@{}c@{}} #2 \end{tabular}}

\DeclareMathOperator*{\argmin}{arg\,min}

\graphicspath{images/}

\makeatletter
\def\@email#1#2{%
 \endgroup
 \patchcmd{\titleblock@produce}
  {\frontmatter@RRAPformat}
  {\frontmatter@RRAPformat{\produce@RRAP{*#1\href{mailto:#2}{#2}}}\frontmatter@RRAPformat} 

  {}{}
}%
\makeatother
\begin{document}

\preprint{AIP/123-QED}

\title{Deep learning fluid flow reconstruction around arbitrary two-dimensional objects from sparse sensors using conformal mappings}
\author{Ali Girayhan \"Ozbay}
\email{aligirayhan.ozbay14@imperial.ac.uk}
\author{Sylvain Laizet}%
\affiliation{ 
Turbulence Simulation Group, Department of Aeronautics, \\Imperial College London, SW7 2AZ London, UK
}%

\date{\today}

\begin{abstract}

The usage of neural networks (NNs) for flow reconstruction (FR) tasks from a limited number of sensors is attracting strong research interest, owing to NNs' ability to replicate high dimensional relationships. Trained on a single flow case for a given Reynolds number or over a reduced range of Reynolds numbers, these models are unfortunately not able to handle flows around different objects without re-training. We propose a new framework called Spatial Multi-Geometry FR (SMGFR) task, capable of reconstructing fluid flows around different two-dimensional objects without re-training, mapping the computational domain as an annulus. Different NNs for different sensor setups (where information about the flow is collected) are trained with high-fidelity simulation data for a Reynolds number equal to approximately $300$ for 64 objects randomly generated using Bezier curves. The performance of the models and sensor setups are then assessed for the flow around 16 unseen objects. It is shown that our mapping approach improves percentage errors by up to 15\% in SMGFR when compared to a more conventional approach where the models are trained on a Cartesian grid, and achieves errors under 3\%, 10\% and 30\% for pressure, velocity and vorticity fields predictions, respectively. Finally, SMGFR is extended to predictions of snapshots in the future, introducing the Spatio-temporal MGFR (STMGFR) task. A novel approach is developed for STMGFR involving splitting DNNs into a spatial and a temporal component. We demonstrate that this approach is able to reproduce, in time and in space, the main features of flows around arbitrary objects.

\end{abstract}

\maketitle

\section{Introduction}
\label{sec:introduction}
Most fluid dynamics experiments have access to only sparse measurements, due to the intrusive (i.e.\ flow-altering) nature of pitot tubes and pressure probes used for measurements. Although non-invasive methods to obtain full flow fields in experiments such as particle imaging velocimetry\cite{piv_review} (PIV), magnetic resonance velocimetry\cite{mrv_review} (MRV) and laser doppler flowmetry\cite{ldf_review} (LDF) exist, their usage can be limited by practicality, cost or safety constraints; for instance PIV systems often require `Class IV' lasers that can gravely harm human eyes and cost thousands of dollars. Despite these practical limitations in experiments, knowledge of the full flow fields is often critical to understanding the dynamics of many complex fluid flows. Flow reconstruction (FR) methodologies can offer reliable estimation of a full flow field from only sparse measurements.

Callaham et al.\cite{flow_reconstruction_2} describe the FR task in terms of a high- and a low-dimensional state vector $x \in \mathbb{R}^m, \; s \in \mathbb{R}^p, \; m>>p$, where $x$ represents the `full' flow field and $s$ are sparse sensor measurements. The two state vectors are linked through measurement and reconstruction operators $H:\mathbb{R}^m \rightarrow \mathbb{R}^p$ and $P:\mathbb{R}^p \rightarrow \mathbb{R}^m$ such that 
\begin{align}
    s = H(x) \\
    x = P(s)
\end{align}
The goal of the FR task is to find an approximation mapping $R:\mathbb{R}^p \rightarrow \mathbb{R}^m$ such that some measure of error, typically $L_2$, between $\hat{x}=R(s)$ and $x=P(s)$ is minimized. In practice, $F$ is a statistical or deep learning algortihm with a parameter set $w$ optimized to fit some dataset; i.e. $R(s) = R(s,w)$. Deviating from the formulation by Callaham et al.\cite{flow_reconstruction_2} to use some generic objective function $L$ instead of the $L_2$ norm, the flow reconstruction task can be expressed as a minimization problem of the following form
\begin{equation}
    \label{eq:fr_objective}
    \argmin_w \, L(x,R(s,w))
\end{equation}

Historically, methods such as Gappy PODs\cite{gappy_pod,gappy_pod_ex1} and Linear Stochastic Estimation\cite{fr_history_lse} are some of the main methodologies investigated for FR, but they are often unsuitable for multi-geometry FR (the reconstruction of the flow field generated past an arbitrary geometry), as detailed in \secref{sec:relatedwork}. Research in this field has recently intensified, and a new wave of studies -- the vast majority focused on using neural networks (NNs) -- have been published, see \cite{flow_reconstruction_2,flow_reconstruction_3,flow_reconstruction_4,flow_reconstruction_5} to name a few. As a starting point for neural network based flow reconstruction, these publications largely focus on obtaining models that work on a single fluid flow case, typically predicting vorticity fields for incompressible flow past a circular cylinder at a single (or over a narrow range) of Reynolds numbers. As a result, such approaches do not have the potential yet to be used in wind tunnel testing driven shape optimization, as training such NNs first requires collecting large datasets of the flow past specific objects.

Training NNs for multi-geometry FR from sparse sensors is not straightforward. Within the aforementioned setting of vorticity field reconstruction on Cartesian grids for two-dimensional (2D) incompressible flows past arbitrary objects, naively augmenting the dataset with multiple objects results in models that fail to reproduce key flow features; concentrations of vorticity in boundary layers and near stagnation points often disappear, and the objects themselves are engulfed by amorphous blobs of non-physical vorticity concentrations. The root cause of these issues is that, within these settings, the models lack information regarding the shape of the object they are making predictions on, as the need for such information is obviated by the single-shape nature of the datasets. 

Overcoming these issues requires a representation of the flow field in a way that removes the necessity for the model to predict the shape of the object immersed in the fluid. An effective tool for this is a mapping, whereby all possible geometries are mapped to a single shape. For 2D cases, this can be achieved via the Schwarz-Christoffel (S-C) conformal mapping, which can be used to map any $l$-connected domain to a disc with $l$ holes\cite{scmap_book}. Thus, the fluid domain in any bluff body flow with a single object can be mapped to an annulus.

In this work, the Spatial Multi-Geometry Flow Reconstruction (SMGFR) task is introduced, with the objective of reconstructing pressure, velocity and vorticity fields surrounding randomly generated objects immersed in a fluid flow from sparse sensor measurements. S-C mappings are utilized to map the fluid computational domains surrounding the said randomly generated bluff bodies to annuli, and a dense field sampling approach based on grids uniformly spaced in angular and radial directions in the annular domains is developed. Over a comprehensive set of 24 experiments (different models and sensor arrangements), the mapping approach is compared to reconstruction based on uniformly spaced Cartesian grids, and the performance of different sensor setups and NN architectures (feed forward, U-Net\cite{unet} and Fourier Neural Operator \cite{fourier_neural_operator}) in SMGFR are investigated. 

As a further step towards spatio-temporal reconstruction, a modified version of the SMGFR is proposed whereby the model is expected to construct snapshots at future times given sensor readings at a present time. In this task, dubbed Spatio-temporal Multi-Geometry Flow Reconstruction (STMGFR), a model obtained from the spatial-only SMGFR task (called the spatial model) is coupled with a second neural network model. This second model, called the temporal model, accepts the reconstructed dense field as its input and predicts the state of the full flow field $k$ time-steps in the future. The resulting system, composed of the spatial and the temporal models, is thus able to reconstruct the full flow fields $k$ time-steps in the future given current sensor measurements.


The work is organized as follows: first, a brief overview of recent and historical approaches to the FR task is provided in \secref{sec:relatedwork}. Subsequently, an introduction to the S-C mapping and details of its novel application to the FR task are presented in \secref{sec:scmappings}. The dataset constructed to take advantage of the S-C mapping is described in \secref{sec:dataset}, and the models chosen to fit this dataset as well as their training procedures are described in \secref{sec:model}. Finally, the results are showcased in \secref{sec:results} and a summary of this work and plans for future investigations are provided in \secref{sec:conclusion}.

\section{Related work}
\label{sec:relatedwork}
The FR task, as introduced in \secref{sec:introduction}, falls under the broad umbrella of inverse problems\cite{de2005learning}. Commonly encountered in a wide range of scientific and engineering fields including fluid dynamics, inverse problems often lack well-defined, unique solutions and rely on minimization of some objective such as the $L_2$ norm, as encountered e.g. in the Moore-Penrose pseudo-inverse for linear least-squares problems. Dubois et. al.\cite{flow_reconstruction_3} divide the types of approaches to FR into three categories:
\begin{enumerate}
    \item \textbf{Direct reconstruction:} A set of parameters is optimized to learn an approximator $R$ to the inverse operator $G$, as summarized in \equref{eq:fr_objective}.
    \item \textbf{Regressive reconstruction:} The unsupervised learning counterpart of direct reconstruction methods, these methods attempt to fit a series of modes\cite{pod_review1} to the available flow data. Methods relying on the principal orthogonal decomposition (POD) fall within this category.
    \item \textbf{Data assimilation:} Dynamical systems based approaches such as Kalman filtering \cite{kalman_filter} are used to evolve high-dimensional systems in time based on sensor measurements
\end{enumerate}

Historically, some of the earliest works in FR originate from meteorology, dating back to the 1980s. Falling mostly within the third category of this classification, Gustafsson et al.\cite{meteorology_fr_variational_3} provide a comprehensive survey of these methods. Framing the FR task in a variational setting\cite{meteorology_fr_variational, meteorology_fr_variational_2, fr_history_variational}, they aimed to predict the time evolution of weather systems given past states and point observations from satellites and ground stations. Despite widespread adoption in weather prediction throughout the world, such approaches are unsuitable for the specific setting explored in this work (reconstruction of current and future vorticity fields from a single instantaneous measurement), due to the necessity of supplying high-quality initial guesses and the time evolution of sensor measurements over a long period of time, which are assumed to be unavailable.

Interest in FR outside the meteorological community emerged in the late 1980s, with methods that fall within the second category of the above classification. Linear stochastic estimation (LSE)\cite{fr_history_lse} is a prominent tool with roots in this era, which reconstructs flow fields based on the computation of a correlation matrix between the sensor inputs and the full flow fields. Recently, it has been applied to reconstruction of PIV-obtained fields in flows over flat plates from microphone measurements\cite{fr_history_lse_4}, reconstruction of velocity fields in internal combustion engines from sensor measurements to identify locations of vortical structures\cite{fr_history_lse_5} or estimation of all components of the dense velocity field in wall turbulence from hot-wire measurements of the horizontal velocity only\cite{fr_history_lse_3}.

A further early statistical technique for estimating dense flow fields given sensor measurements is the Gappy POD\cite{gappy_pod} method, originally developed for reconstructing images of human faces. Examples of the application of the Gappy POD method to FR have included predicting missing information in Direct Numerical Simulation (DNS) data\cite{gappy_pod_ex1}, estimation of the lift coefficient of a NACA0012 airfoil under plunging motion in a Mach 0.6 flow by reconstructing the simulated dense velocity and pressure fields from pressure sensors on its surface\cite{gappy_pod_ex2}, and filling in missing data points in PIV snapshots from gas turbine combustors\cite{gappy_pod_ex4}. 

The spectrum of statistical methods applied to FR contains several further less-investigated avenues. Notably, the sparse representation\cite{sparse_representation} technique, with origins in facial recognition, has been recently applied for FR of free-shear and mixing layer flows\cite{fr_history_sparse_reconstruction}. New methods, such as the Sparse Fourier Divergence Free method\cite{fr_history_sfdf} dedicated specifically to the investigation of incompressible flows, are being developed. Unfortunately, the linear unsupervised nature of the methods in this category lends itself to making predictions for flow datasets incorporating a single geometry only, and we are not aware of any previous works using such methods for multi-geometry datasets. The universal approximation capabilities of deep neural networks with nonlinear activation functions\cite{nn_universal_approx}, belonging to the first category in the above classification, are therefore a better fit for multi-geometry FR as evidenced by their state-of-the-art performance in many generative Machine Learning (ML) tasks such as text\cite{gpt3} and image\cite{stylegan2} generation. The strength of NNs is evident in FR research as well, with a rapid rise in the number of works investigating the application of neural networks to FR. 


Still, most NN-based FR techniques focus on training models that work for a single flow configuration only (without retraining), some notable examples of which are provided: Erichson et al.\cite{flow_reconstruction_2} reconstructed vorticity fields in the wake of a circular cylinder from measurements on its surface using a `shallow' NN (i.e. feedforward NN with few layers and units), while Kumar et al.\cite{flow_reconstruction_9} improved upon this technique with a recurrent autoencoder based architecture which can reduce the error by over an order of magnitude when extremely sparse sensor setups involving only a single sensor are used. Dubois et al.\cite{flow_reconstruction_3} investigated linear and nonlinear autoencoder NNs with and without variational training, and found that non-linearities in NNs enable identification of dominant flow modes, while variational training leads to higher robustness to noise at the cost of higher error. Fukami et al.\cite{flow_reconstruction_4} compared the performance of feedforward NNs with three non-neural ML techniques on reconstructing the wake behind a circular cylinder and a flapped airfoil based on noisy sparse measurements from the object surfaces, finding that although neural models are not necessarily the best performing at low noise, they constitute the most robust option under high noise situations. Sun and Wang\cite{flow_reconstruction_5} investigated the usage of physics-constrained Bayesian NNs for flow reconstruction from sparse sensors in stenotic vessels and T-shaped geometries, which permit significantly higher noise robustness compared to standard NNs. Usage of NNs in this context have also been extended to experimental as opposed to simulated data, for example by Carter et al.\cite{flow_reconstruction_7} who used feedforward NNs for the reconstruction of experimental PIV-obtained velocity fields above the suction surface of a NACA0012 airfoil, achieving superior results compared to non-neural methods. 

In comparison, the body of works investigating deep learning based FR for the reconstruction of the flow past arbitrary objects without re-training is small. One notable recent work in this area is by Chen et al.\cite{2d_flowpred_random_1}, using graph convolutional neural networks (GCNNs) for reconstructing steady flow fields around random objects. Training a GCNN to predict the velocity and pressure fields around 1600 random objects generated via Bezier curves, they applied the model to predict the pressure and velocity fields around 400 test geometries, which permitted the estimation of drag and lift coefficients with very small mean percentage error levels. The present work differs from Chen et al.\cite{2d_flowpred_random_1} as it focuses on the usage of a novel mapping approach to achieve the geometry invariance as opposed to graph convolutions which permits the usage of traditional NN architectures. Here the aim is to reconstruct instantaneous snapshots as opposed to steady fields, and to explore predicting future instantaneous snapshots from current measurements. Additionally, this work is conducted at a substantially higher $\Rey=300$ as opposed to $\Rey=10$ in Chen et al.\cite{2d_flowpred_random_1} which leads to the emergence of unsteady rotational flows. To assess the performance of the models, the focus is on reconstructing the vorticity fields (which are difficult to predict from pressure and velocity sensors, as shown in \secref{sec:pressure_vel_results}). Reconstructed data for the pressure and velocity fields are also briefly presented for completeness.

\section{Schwarz-Christoffel mappings}
\label{sec:scmappings}

Conformal transformations have been used extensively in fluid dynamics, especially the well-known Joukowsky and Kármán-Trefftz\cite{karmantrefftz} (K-T) transformations which map the unit circle to airfoil shapes. However, the usefulness of these two transformations can be limited due to their inability to generalize to arbitrary shapes. A more flexible alternative is the Schwarz-Christoffel (S-C) mapping, which is a conformal transformation historically used to map polygonal simply connected domains to the unit disc. The S-C mapping has been extended in the recent decades to multiply connected domains. Although the \textit{existence} of a conformal mapping between any given two $l$-connected domains is guaranteed\cite{scmap_book}, the practical computation of such an S-C mapping typically requires the use of numerical methods to determine a number of parameters in the mapping expression, referred to as the Schwarz-Christoffel parameter problem. Numerically implementing the S-C mapping for doubly (or higher) connected domains is not a trivial undertaking. In this work, the \texttt{DSCPACK} code\cite{dscpack}, which is a Fortran package aimed at computing S-C mappings between doubly connected domains bounded by polygons to annuli, was utilized to solve the parameter problem. 

A rigorous treatment of the methodology used in this package lies beyond the scope of this work, for which the reader is directed to established works in S-C mapping literature\cite{scmap_book, scmap_book2, scmap_background1}, though the general strategy can be summarized as follows: denoting $z$ as the complex coordinates in the original domain and $w$ as the complex coordinates in the annulus domain, \texttt{DSCPACK} uses an expression of the form
\begin{align}
    \label{eq:sc_mapping_forward}
    z = g(w) = g(w_c) + C \int_{w_c}^w W(w) \, \diffd w, \\
    W(w) = \prod_{q=1}^M \left[\sigma \left(\mu, \frac{-w}{\mu w_{0q}}  \right) \right]^{\alpha_{0q}-1} \prod_{r=1}^N \left[\sigma \left(\mu, \frac{-\mu w}{w_{1r}}  \right) \right]^{\alpha_{1r}-1}, \\
    \sigma(\mu,w) = 1 + \sum_{b=1}^{\infty} \mu^{b^2} (w^b + w^{-b}),
\end{align}
as the mapping, where $C$ is some complex valued constant; $\mu$ is the radius of the inner ring of the annulus in the $w$-domain; $M,\alpha_{0q},w_{0q}$ and $N,\alpha_{1r},w_{1r}$ are the number of vertices, the turning angles and prevertices\footnote{vertex coordinates in the $w$-domain} of the outer and the inner polygon, respectively. Of these variables, $C,\mu,w_{0q}$ and $w_{1r}$ are unknowns (`accessory parameters' of the mapping) and must be computed by solving a series of nonlinear integral equations
\begin{align}
    z_{01} - z_{0M} = C\int_{w_{0M}}^{w_{01}} W(w) \, \diffd w ,\\
    |z_{0,q+1} - z_{0q}| = |C\int_{w_{0q}}^{w_{0,q+1}} W(w) \, \diffd w|, \, q \in [1,M) ,\\
    z_{11} - z_{1N} = C\int_{w_{1N}}^{w_{11}} W(w) \, \diffd w ,\\
    |z_{1,r+1} - z_{1r}| = |C\int_{w_{1r}}^{w_{1,r+1}} W(w) \, \diffd w|, \, r \in [1,N) ,\\
    z_{1N} - z_{0M} = C\int_{w_{0M}}^{w_{1N}} W(w) \, \diffd w,
\end{align}
where $z_{0q}$ and $z_{1r}$ denote the complex coordinates of the polygon vertices in the original domain. The \texttt{DSCPACK} code solves this nonlinear system using a Newton iteration scheme. The resulting expression maps the outer ring (with unity radius) of an annulus to the outer polygonal boundary, while the inner ring is mapped to the inner polygon.

Once the forward mapping $g$ is known, the inverse mapping $f$ can be approximated for any (fixed) $z=\bar{z}$ using Newton iteration:
\begin{align}
    \bar{z} - g(w) = 0 \\
    w_{n+1} = w_n + \frac{\bar{z} - g(w_n)}{g'(w_n)} \\
    \label{eq:inverse_mapping_derivative}
    g'(w_n) = CW(w_n),
\end{align}
where \equref{eq:inverse_mapping_derivative} follows from the application of the fundamental theorem of calculus to \equref{eq:sc_mapping_forward}.

\begin{figure}[h!]
    \centering
    \includegraphics[width=\columnwidth]{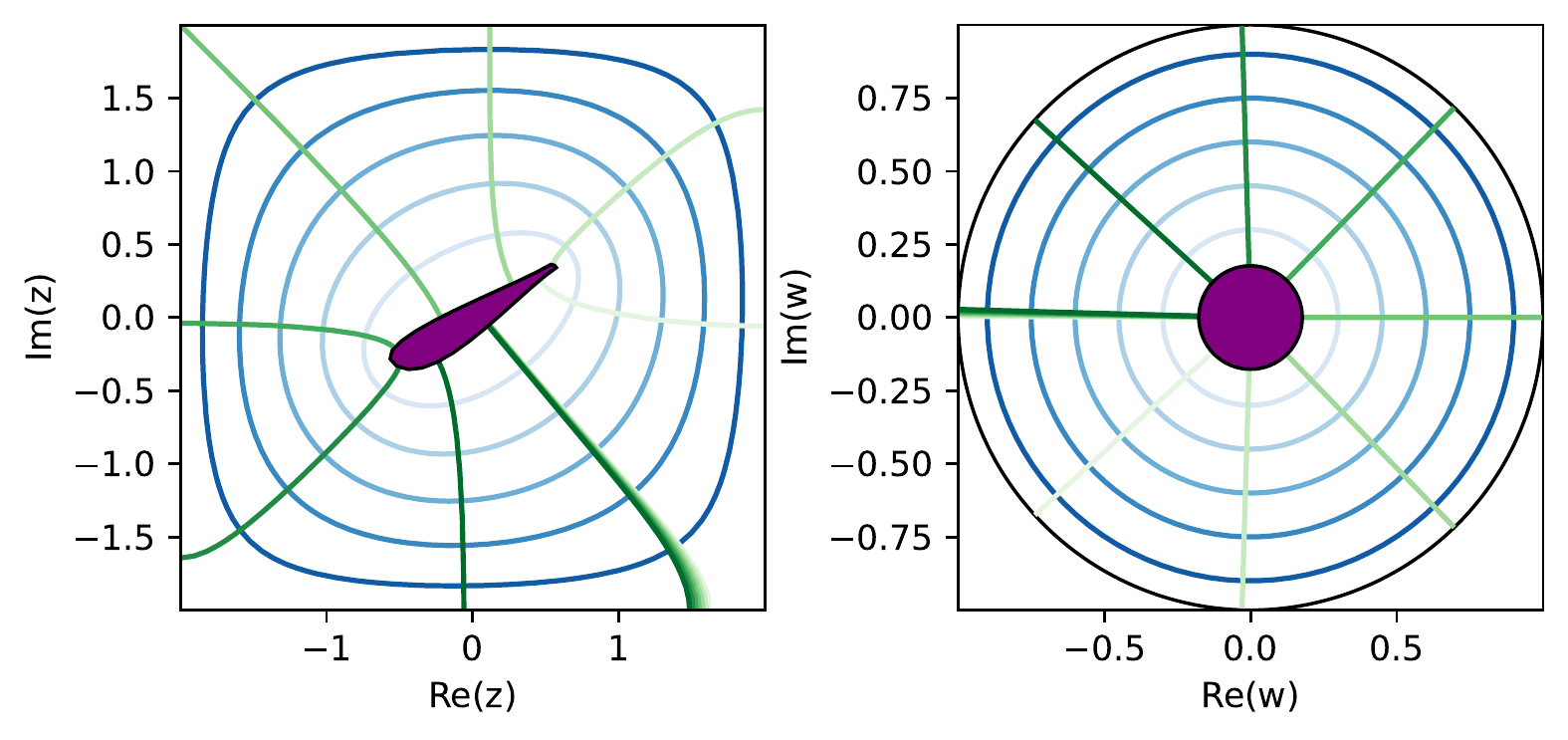}
    \caption{A random geometry (left) and its annular preimage (right). Blue and green contours depict the norm and argument in the $w$ domain, respectively. The outer boundary of the domain is smaller than the ones used in the actual study for illustrative purposes.}
    \label{fig:scmap_example}
\end{figure}

To ensure smooth interoperability of \texttt{DSCPACK} with modern machine learning packages, a set of Python bindings to a modified form of the code were developed, dubbed \texttt{pydscpack}. Furthermore, a number of enhancements to the original code were made to parallelize performance-critical sections with \texttt{OpenMP}. \figref{fig:scmap_example} depicts an S-C mapping computed using \texttt{pydscpack}, for a geometry used in this study.

\section{Dataset}
\label{sec:dataset}

\subsection{Geometries}
\begin{figure}[h!]
    \centering
    \includegraphics[width=\columnwidth]{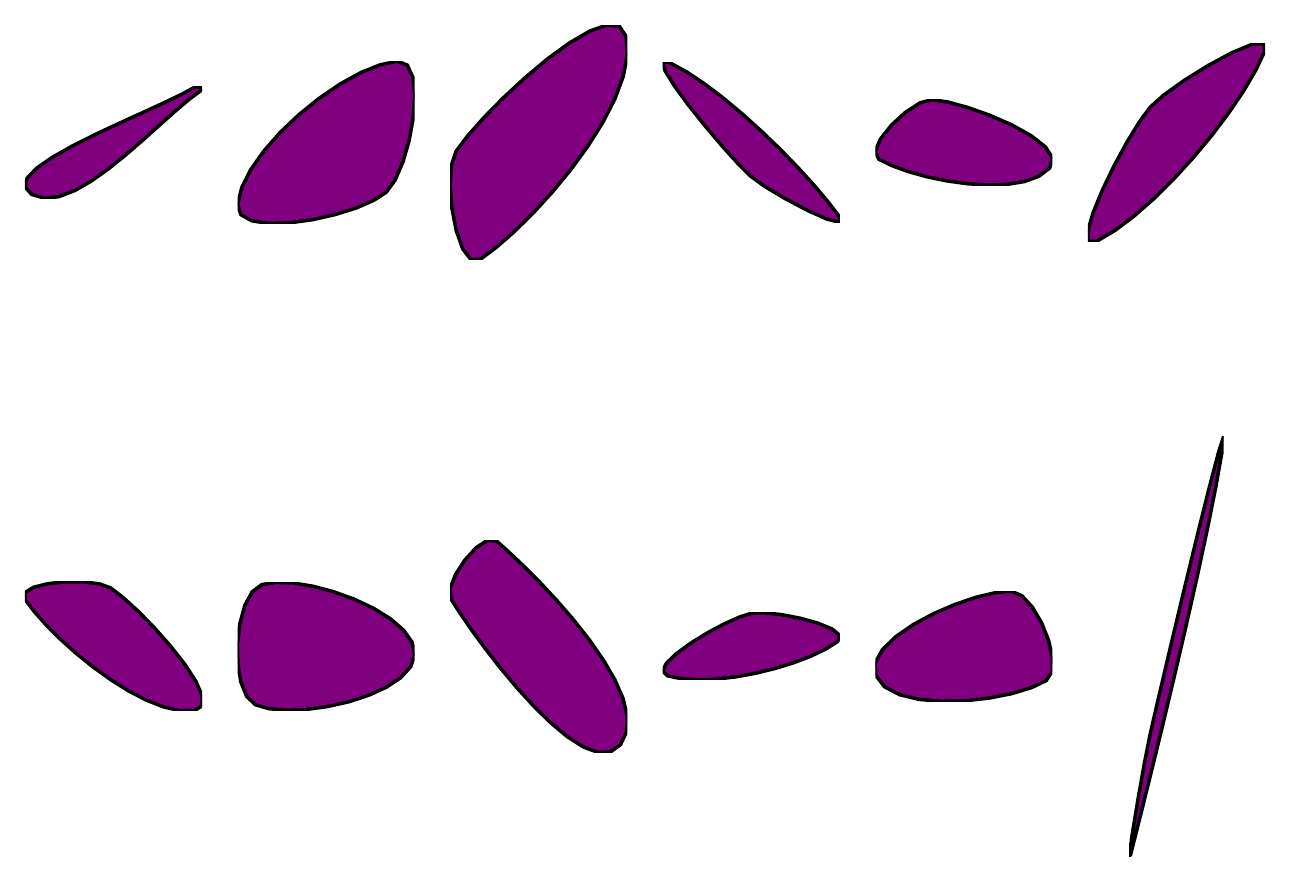}
    \caption{Twelve geometries used in this study, among a total of 80.}
    \label{fig:geometry_examples}
\end{figure}

A collection of 80 geometries $G_i, i \in [0,79]$ were generated using random Bezier curves, via the \texttt{bezier\_shapes} package by Viquerat et al.\cite{bezier_code}. The control points for the Bezier curves were chosen randomly in a square domain with characteristic length $L_m$. Each geometry was placed in the center of a $40L_m/3 \times 40L_m/3$ square domain. The fluid domains were meshed with \texttt{gmsh} using a combination of triangular and tetrahedral elements, with c. 20000 elements per geometry. \figref{fig:geometry_examples} shows 12 of the geometries generated using this method.

\subsection{Ground truth values}
\label{sec:gtvalues}
Using uniform Dirichlet velocity boundary conditions $(u,v) = (1.0,0.0)$ along the external edges of the domain, the flow around each object was computed for $\Rey = uL_m/\nu = 300$ (where $\nu$ is the kinematic viscosity) using the \texttt{PyFR} solver\cite{pyfr}, which is a flux reconstruction\cite{flux_reconstruction} based advection-diffusion equation solver using the artificial compressibility approach to solve the incompressible Navier-Stokes equations. It was chosen for its Python interface and GPU acceleration capabilities. The simulations were performed on two Nvidia V100 GPUs. Normalizing the physical time $\tau$ by the large eddy turnover time to obtain $\tau^*=u\tau/L_m$, $t \in [0..600]$ snapshots (containing the pressure and velocity field components $\mathbf{p}_{t,i}$, $\mathbf{u}_{t,i}$ and $\mathbf{v}_{t,i}$) were recorded per geometry for a total of 48080 snapshots between $\tau^*=3.333$ and $\tau^*=23.333$. 

Following the simulations, referring to the fluid domain around each $G_i$ as $F_i$, the forward and inverse mappings $g_i$ and $f_i$ between $F_i$ and the corresponding annuli $A_i$ were computed using \texttt{pydscpack}. $64 \times 256$ grids uniformly spaced in the radial and angular directions with coordinates $\mathbf{w}_{A,i}$ were generated for each $A_i$. \textcolor{black}{Subsequently, $\mathbf{w}_{A,i}$ were mapped back to the original domains $F_i$ using the computed mappings $g_i$ to obtain the annular grid coordinates in the original domain $\mathbf{z}_{A,i} = g_i(\mathbf{w}_{A,i})$. The velocity, pressure and vorticity fields $\mathbf{u}_{t,i}$, $\mathbf{v}_{t,i}$, $\mathbf{p}_{t,i}$ and $\mathbf{\omega}_{t,i}$ from the high-fidelity simulation data were interpolated to $\mathbf{z}_{A,i}$ to obtain the interpolated fields $\tilde{\mathbf{u}}_{t,i}$, $\tilde{\mathbf{v}}_{t,i}$, $\tilde{\mathbf{p}}_{t,i}$ and $\tilde{\mathbf{\omega}}_{t,i}$, which form the ground truth values of the Annular dataset.}

This sampling strategy and grid resolution provide a high grid density near the object, scaling with a factor of $1/r$ based on distance to the center of the annulus. It ensures that the regions of the flow with high vorticity concentrations have enough grid points for a correct representation of the vortical structures. Additionally, as a baseline case, a further collection of ground truth values sampled naively on a $128 \times 128$ uniformly spaced Cartesian grid was also produced, with the same number of grid point as for the mapping approach. Note that these grids are used solely for the interpolation of flow variables, not to perform the fluid simulations.

\subsection{Inputs}
The inputs of the dataset are vectors of pressure and/or velocity values $s_{t,i}$ at a sparse number of sensor locations, obtained via interpolation of the \texttt{PyFR} solution to the sensor locations. The sensor setup to build the inputs of the dataset is split into two sensor types, chosen to represent a setup that can be practically implemented in a laboratory environment: 
\begin{itemize}
    \item \textbf{Pressure:} Placed on the surface of each $G_i$, with equal angular spacing along the inner ring of each $A_i$.
    \item \textbf{Velocity:} Positioned on a rectangular grid spanning a $2L_m/3 \times 4L_m/3$ region, the left edge of which is $L_m/6$ units behind the rearmost point of each $G_i$ and the centroid of which is vertically level with each $G_i$.
\end{itemize}

\begin{figure}[h!]
    \centering
    \includegraphics[width=0.75\columnwidth]{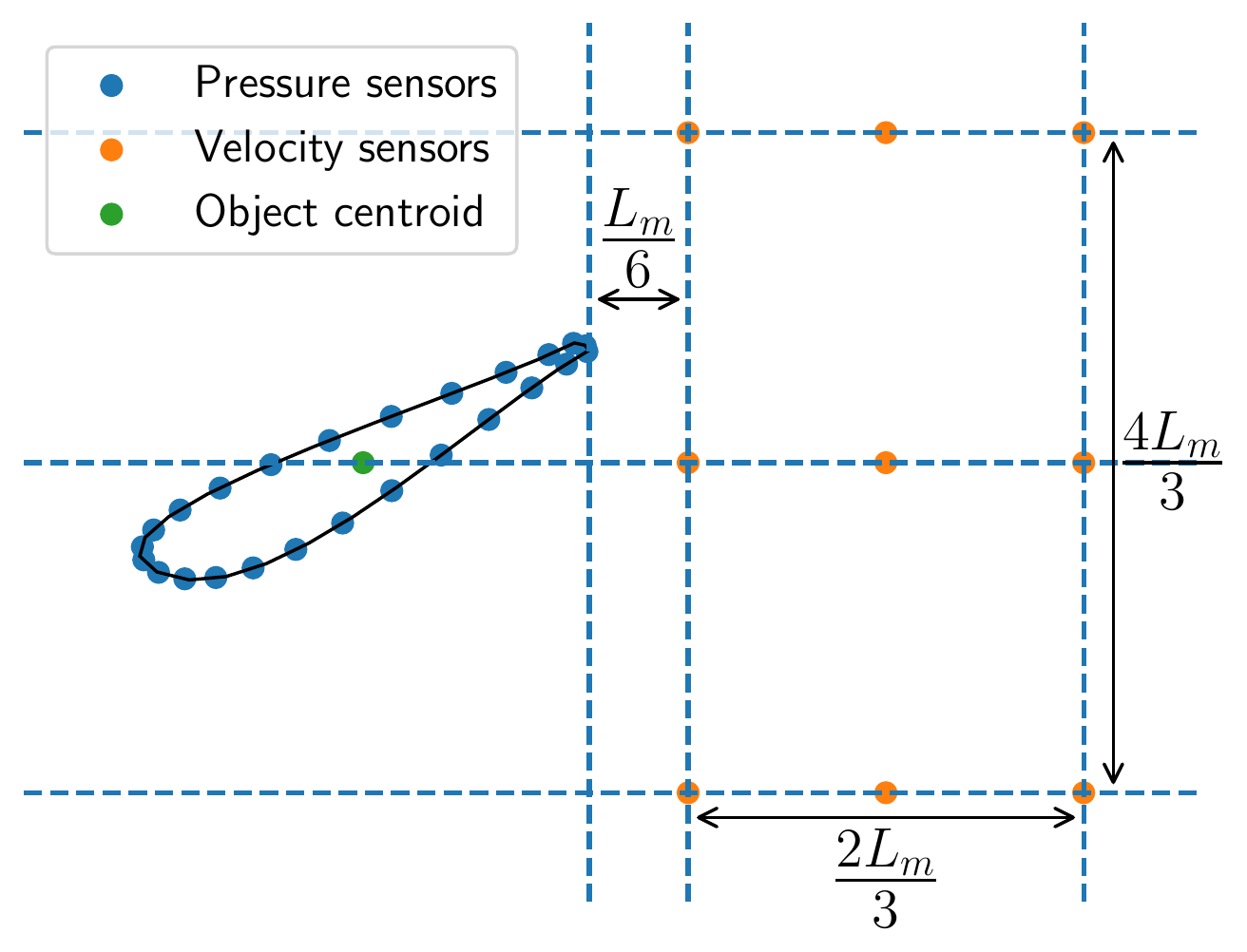}
    \caption{Illustration of the medium sensor setup for one of the geometries used in the study.}
    \label{fig:sensor_setup}
\end{figure}

Based on this general template three setups with varying sensor quantities were considered, summarized in \tabref{tab:sensor_numbers}. \figref{fig:sensor_setup} depicts the medium sensor setup for a sample geometry. 
\begin{table}[h!]
\caption{Number of sensors by type and configuration}
\begin{tabular}{@{}lcc@{}}
\toprule
              & $\#$ of pressure sensors & $\#$ of velocity sensors \\ \midrule
Small         & 12                       & 4                        \\
Medium        & 25                       & 9                        \\
Large         & 50                       & 25                       \\ \bottomrule
\end{tabular}
\label{tab:sensor_numbers}
\end{table}

\subsection{Normalization}
\label{sec:normalization}
Normalizing inputs and/or outputs plays an important role in obtaining good performance from deep learning algorithms, as it permits better conditioning of the gradients within the optimization landscape during training, by keeping the per-layer statistical distribution of the gradients similar\cite{dnn_normalization_general_survey}. A variety of data normalization methodologies, including mean centering, standardization and min-max scaling were tried in a preliminary study. Denoting $X$ as the dataset inputs and $T$ as the target values, $X^+$ and $X^-$ as the maximum and minimum values of $X$, respectively, and $\mu$ and $\sigma$ as the mean and standard deviation, respectively, the three normalization methods can be summarized as follows:
\begin{flalign}
    \label{eq:mean_center}
    \text{Mean centering:} && \hat{X} = X - \mu, \, \hat{T} = T - \mu, \\
    \label{eq:standardize}
    \text{Standardization:} && \hat{X} = (X - \mu)/\sigma, \, \hat{T} = (T - \mu)/\sigma \\
    \label{eq:minmaxscaling}
    \text{Min-max scaling:} && \hat{X} = \frac{X - X^-}{X^+ - X^-} + X^-,  \hat{T} = \frac{T - X^-}{X^+ - X^-} + X^-
\end{flalign}
Mean centering both inputs and ground truth values based on the ground truth mean values was chosen as the data normalization method, as it provides the results with the lowest validation loss levels for models trained using either the Cartesian and Annulus datasets.


\section{Experimental setup}
\label{sec:model} 

Two related tasks have been investigated, with the dataset detailed in \secref{sec:dataset}. The first is identical to spatial flow reconstruction tasks from sparse sensors in previous literature\cite{flow_reconstruction_2,flow_reconstruction_3,flow_reconstruction_7,flow_reconstruction_8}, but with the inclusion of snapshots taken from a multitude of geometries in the training and validation datasets. As a reminder the name Spatial Multi-Geometry Flow Reconstruction (SMGFR) is used to describe the FR task in this specific configuration. The second task is a generalization of SMGFR where target snapshots are in the future relative to the sensor measurements by a fixed amount of time $\Delta \tau^*$, as opposed to SMGFR where the target snapshots are contemporaneous with the measurements, dubbed Spatio-temporal Multi-Geometry Flow Reconstruction (STMGFR).

For both tasks, the dataset from \secref{sec:dataset} was split by randomly choosing the data associated with 64 geometries as the training set; the remaining 16 geometries constituted the validation set. Training in all experiments was conducted using the Adam\cite{adam_optimizer} algorithm using an initial learning rate (LR) of $10^{-3}$, reduced by $90\%$ each time the loss values plateaued.

\begin{table}[ht!]
\caption{Number of parameters per model and sensor setup, broken down by layer type, for SMGFR. Only the first feedforward layer's parameter numbers are affected by the sensor setup.}
\label{tab:param_numbers_spatial}
\footnotesize
\begin{tabular}{@{}lccc|ccc|ccc@{}}
\toprule
         & \multicolumn{3}{c|}{Large sensor setup}     & \multicolumn{3}{c|}{Medium sensor setup}    & \multicolumn{3}{c}{Small sensor setup}      \\
         & Dense        & Conv/FNO        & Total      & Dense        & Conv/FNO        & Total      & Dense        & Conv/FNO        & Total   \\ \midrule
SD       & 677,424      & -               & 677,424    & 675,144      & -               & 675,144    & 674,224      & -               & 674,224    \\
SD-Large & 46,399,488   & -               & 46,399,488 & 46,282,752   & -               & 46,282,752 & 46,235,648   & -               & 46,235,648 \\
SD-UNet  & 34,683,392   & 7,711,301       & 42,394,693 & 34,654,208   & 7,711,301       & 42,365,509 & 34,642,432   & 7,711,301       & 42,353,733 \\
SD-FNO   & 34,683,392   & 2,105,857       & 36,789,249 & 34,654,208   & 2,105,857       & 36,760,065 & 34,642,432   & 2,105,857       & 36,748,289 \\ \bottomrule
\end{tabular}
\end{table}

\subsection{Spatial multi-geometry flow reconstruction (SMGFR)}
\label{sec:model_spatial_reconstruction}
\begin{figure}[hb!]
    \centering
    \includegraphics[width=\textwidth]{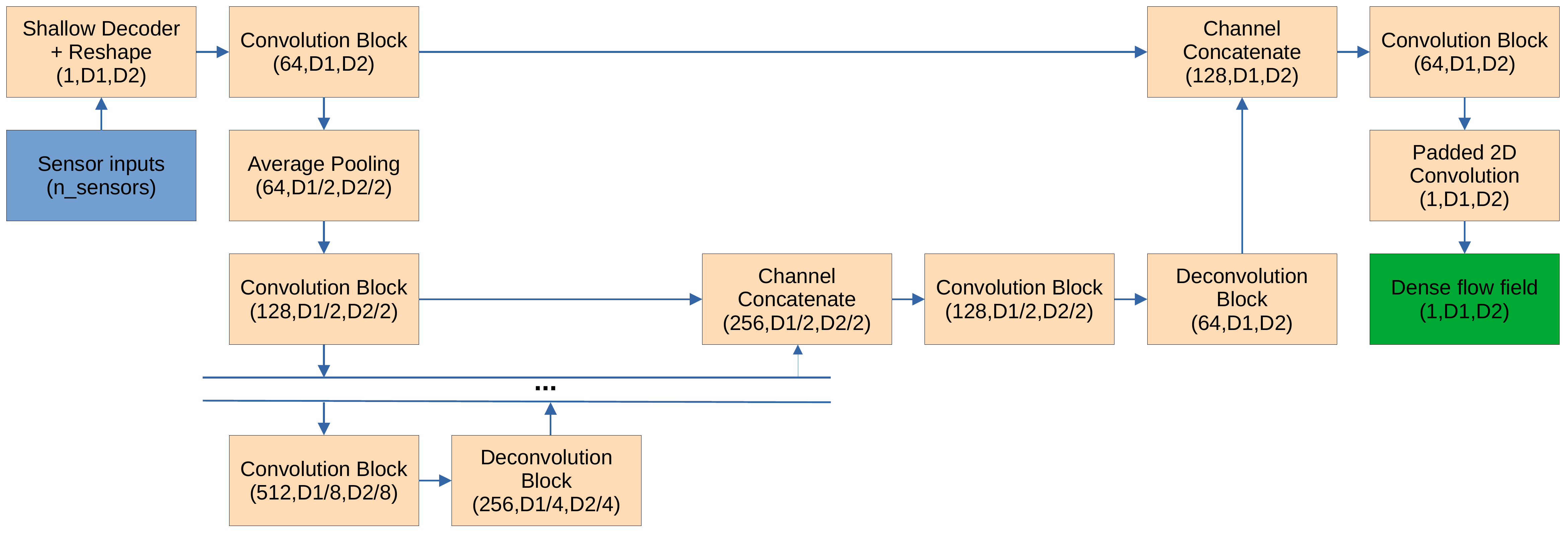}
    \vspace{0.05cm}
    \hrule
    \vspace{0.05cm}
    \includegraphics[width=\textwidth]{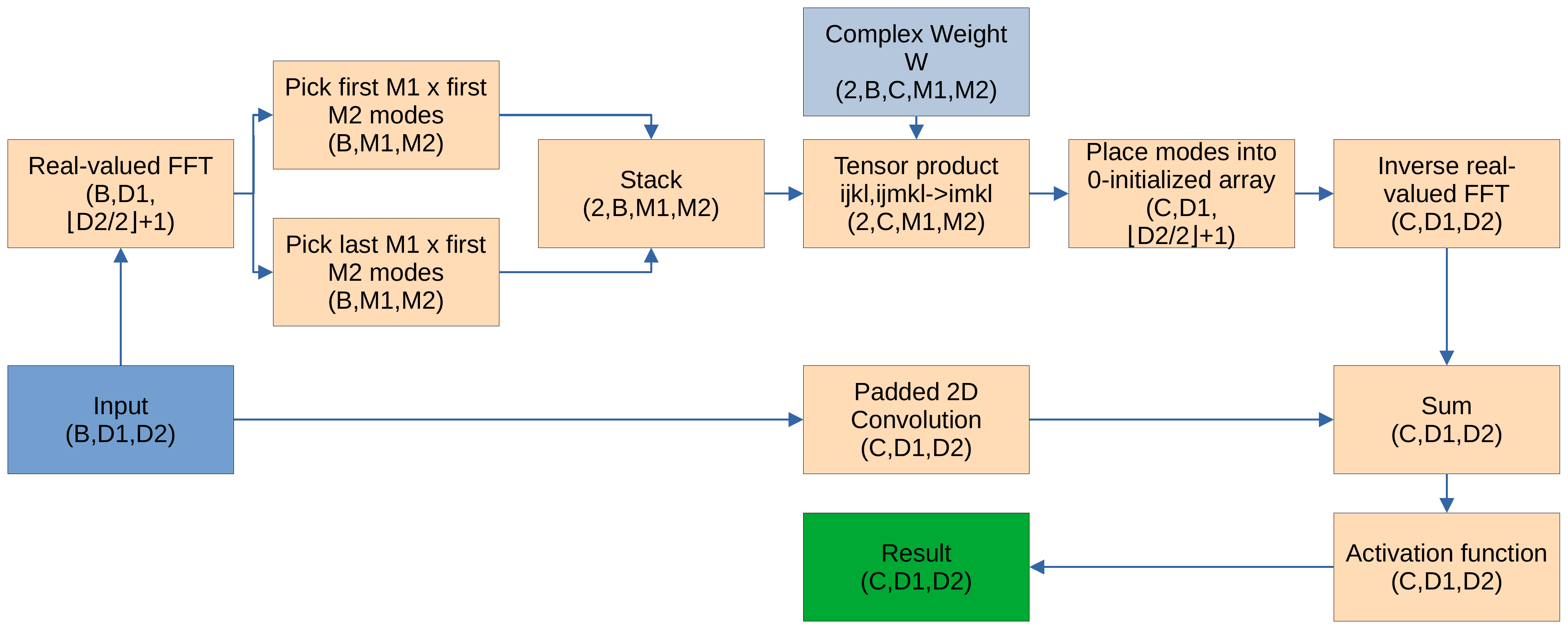}
    \vspace{0.05cm}
    \hrule
    \vspace{0.05cm}
    \includegraphics[width=\textwidth]{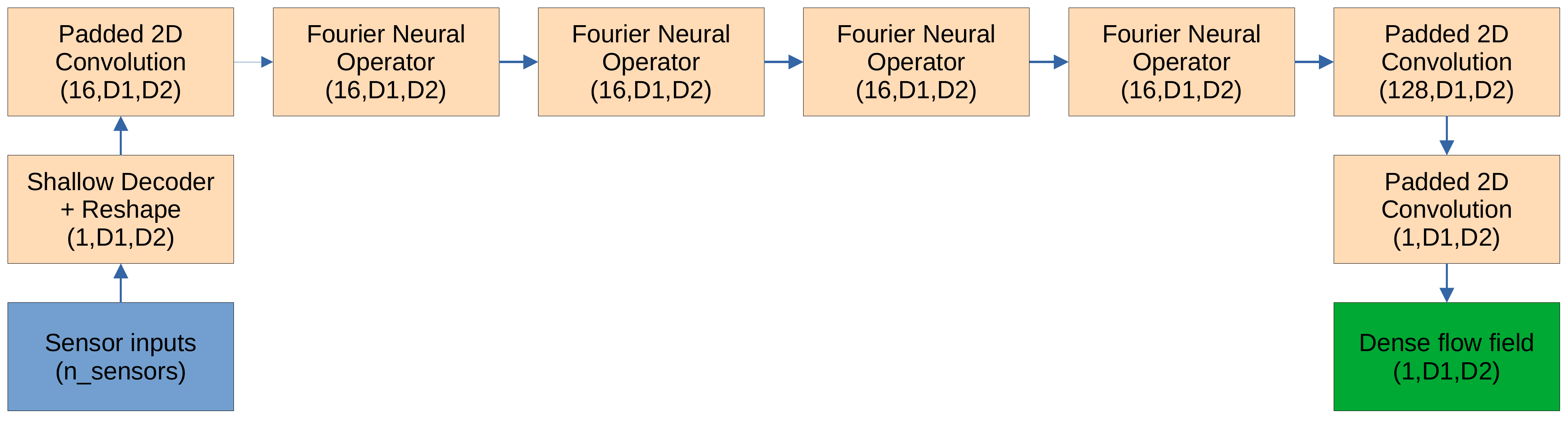}
    \caption{Diagrams of the SD-UNet (top), an FNO layer (middle), and the SD-FNO (bottom). The values in the parentheses indicate the shape of each block's output tensor; D1=64 and D2=256 for the Annulus dataset, and D1=D2=128 for the Cartesian dataset. In the SD-UNet, each convolution block is formed of two convolution layers, each preceded by a batch normalization layer and followed by a dropout layer, each deconvolution block is formed of a batch normalization layer followed by a deconvolution with stride 2.}     
    \label{fig:model_diagrams}
\end{figure}

Using the notation in \secref{sec:dataset}, the SMGFR task can be summarized as predicting $\tilde{\mathbf{p}}_{t,i}$, $\tilde{\mathbf{u}}_{t,i}$, $\tilde{\mathbf{v}}_{t,i}$ or $\tilde{\omega}_{t,i}$ given $s_{t,i}$. The experiments investigate the performance of four different models, all implemented using Tensorflow\cite{tensorflow} v2.5.1, with the parameter counts available in  \tabref{tab:param_numbers_spatial}. The latter two models are described schematically in details in \figref{fig:model_diagrams}. Below is an overview of the models investigated, with some arguments to justify their use in the present study:
\begin{enumerate}
    \item \textbf{Shallow Decoder (SD)\cite{flow_reconstruction_2}:} A 3 layer feedforward neural network with 40 units and ReLU activations in the intermediate layers, identical to the setup in Erichson et al.\cite{flow_reconstruction_2}.
    \item \textbf{SD-Large:} A larger SD with 4 layers and 2048 units in each layer, included to assess whether the SD is sufficiently parametrized. Additionally, this model incorporates leaky ReLU activations and batch normalization\cite{batchnorm}, motivated by the well-known positive impact of rectifier non-linearities \cite{relu_good} and activation normalization\cite{dnn_normalization_general_survey} on DNN performance.
    \item \textbf{SD-UNet:} An SD model with 512 and 2048 units in the intermediate layers, followed by a reshape operation to a 2D grid and a four-level U-Net\cite{unet} model with 64 channels in the base level, leaky ReLU activations, batch normalization and $p=0.25$ dropout. This model is included to assess the performance of a well-studied convolutional image-to-image translation model, as opposed to the fully connected SD architecture.
    \item \textbf{SD-FNO:} An SD model identical to the one present in the SD-UNet, followed by four Fourier Neural Operator (FNO)\cite{fourier_neural_operator} layers. Included due to remarkable performance of the FNO architecture in previous studies related to fluid flows.
\end{enumerate}


\clearpage

\subsection{Spatio-temporal multi-geometry flow reconstruction (STMGFR)} 
\label{sec:model_spatio-temporal_reconstruction}

The spatio-temporal multi-geometry flow reconstruction (STMGFR) task extends the purely spatial SMGFR task presented under \secref{sec:model_spatial_reconstruction}. Whereas SMGFR focuses on obtaining a reconstruction $\hat{x}_{t,i}$ of the (interpolated) target field $\tilde{x}_{t,i}$ (e.g.\ $\tilde{x}_{t,i} = \tilde{\omega}_{t,i}$) given the measurements $s_{t,i}$, STMGFR is a generalization to flow fields at $k$ time-steps in the future given the current measurements; i.e.\ obtaining a reconstruction $\hat{x}_{t+k,i}$ of $\tilde{x}_{t+k,i}$ given $s_{t,i}$. In this work, two values for the temporal interval are investigated: a short interval where $k=20$ (equal to $\Delta \tau^*=0.667$, or 3.33\% of the total simulation time), and a longer interval where $k=80$ (equal to $\Delta \tau^*=2.667$ or approximately 13\% of the total simulation time). This way it is possible to investigate temporal gaps both shorter and longer than the large eddy turnover time $\tau^*=1.0$. 

\begin{figure}[h!]
    \centering
    
    \begin{tikzpicture}[node distance=6cm]
    \centering
    
    \node(sensors)[spatio-temporal_explanation_node] {\begin{tabular}{c} Current sensor values \\ $s_{t,i}$ \end{tabular}};
    \node(fr_spatial)[spatio-temporal_explanation_node, right of = sensors] {\begin{tabular}{c} Current vorticities \\ $\hat{\omega}_{t,i}$ \end{tabular}};
    \node(fr_spatio-temporal)[spatio-temporal_explanation_node, right of = fr_spatial] {\begin{tabular}{c} Future vorticities \\ $\hat{\omega}_{t+k,i}$ \end{tabular}};
    
    \draw [arrow] (sensors) -- node[] {\begin{tabular}{c} Spatial model \\ (SD-UNet) \end{tabular}} (fr_spatial);
    \draw [arrow] (fr_spatial) -- node[] {\begin{tabular}{c} Temporal model \\ (FNO) \end{tabular}} (fr_spatio-temporal);
    \end{tikzpicture}
    
    \caption{Summary of the two-step spatio-temporal reconstruction approach.}
    \label{fig:spatio-temporal_summary}
\end{figure}
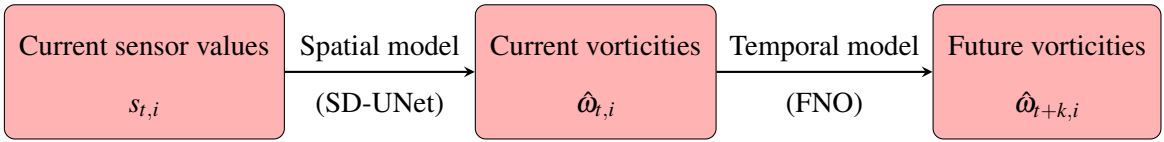

An effective way to tackle this FR problem is first obtaining $\hat{x}_{t,i}$ from $s_{t,i}$ using one of the models detailed in \secref{sec:model_spatial_reconstruction}, dubbed the spatial model. Subsequently, a second model (the temporal model) is used to obtain $\hat{x}_{t+k,i}$ from $\hat{x}_{t,i}$, as summarized in \figref{fig:spatio-temporal_summary}. A stack of six FNO layers was used as the temporal model, which have been demonstrated to be accurate when used to time-march the two-dimensional Navier-Stokes equations \cite{fourier_neural_operator}. Every FNO layer has 64 channels and 32 modes per spatial dimension, while the convolutions use $[1,1]$ kernels, translating to 50,635,313 parameters. The spatial model was chosen as the SD-UNet based on its high performance in the spatial task (as detailed in \secref{sec:results-spatial}, with identical weights), with a configuration identical to the one in \secref{sec:model_spatial_reconstruction}. Likewise, the chosen sampling strategy is annular, due to its superior performance in \secref{sec:results-spatial}.

Training the temporal model is done in a supervised manner. Since the temporal model would be expected to perform well given reconstructed inputs $\hat{x}_{t,i}$ from the spatial model in an inference scenario, in each epoch the input associated with every sample is randomly chosen to be either a reconstructed vorticity field $\hat{x}_{t,i}$ or a ground truth field $\tilde{x}_{t,i}$, with a $50\%$ chance for each. Using separate spatial and temporal models with parameter counts $P_s$ and $P_t$ (versus a larger single model with parameter count $P_s+P_t$ directly predicting $\hat{x}_{t+k,i}$ from $s_{t,i}$) is highly computationally efficient as it permits easy re-training of multiple temporal models for different values of $k$. The results from the spatial model can be easily cached and re-used when training a new temporal model, which translates to computational speedups as well as model accuracy benefits owing to the possibility of using larger batch sizes given a fixed pool of memory. 

\section{Results}
\label{sec:results}

The results of a series of SMGFR and STMGFR experiments are detailed in this section, the setups of which are detailed in the previous chapters. First, to compare the accuracy and quality of our reconstruction methodology to the previous work by Chen et al.\cite{2d_flowpred_random_1}, we briefly present results on reconstructing pressure and velocity fields in \secref{sec:pressure_vel_results}. Additionally, we numerically demonstrate that the reconstruction of vorticity presents a greater challenge than the reconstruction of pressure and velocity from pressure and velocity sensors.

Subsequently, we proceed to detailed comparisons of vorticity reconstruction performance, where the differences between the various setups are visible more acutely. \secref{sec:results-spatial} details the results from the spatial reconstruction task with four models and three sensor setups, as detailed in \secref{sec:dataset} and \secref{sec:model_spatial_reconstruction}. With the best performing configuration for the spatial task identified in this section, \secref{sec:results-spatio-temporal} presents the efforts in combining the spatial-only configuration with a further time-marching model to predict future vorticity fields from current sensor measurements, as explained in \secref{sec:model_spatio-temporal_reconstruction}.  \secref{sec:runtime} compares the wall-clock runtimes of all models considered. 

To focus on the most rotational regions of the flows, the analyses are constrained to a region of the computational domain immediately surrounding and downstream of the objects investigated, encompassing a rectangular region with vertices $[-L_m,L_m], [-L_m,-L_m], [4L_m, L_m]$ and $[4L_m, -L_m]$. 

\subsection{Spatial multi-geometry pressure and velocity reconstruction}
\label{sec:pressure_vel_results}
In order to compare the quality of our reconstruction methodology with the previous multi-geometry reconstruction work by Chen et al. \cite{2d_flowpred_random_1}, we briefly present the results of training the $k=0$ SD-UNet+FNO combination from \secref{sec:model_spatio-temporal_reconstruction} on pressure and velocity data in \tabref{tab:pressure_vel_error}, using the large sensor setup.

Additionally, to demonstrate that a fairly complicated and non-linear reconstruction relationship $P$ is present between the sensor measurements and the vorticity fields investigated in Sections \secref{sec:model_spatial_reconstruction} and \secref{sec:model_spatio-temporal_reconstruction}, we include two difficulty measures $\mathcal{D}$ and $\mathcal{M}$ in \tabref{tab:pressure_vel_error}. These measures are based on the Frobenius norms of the Spearman rank correlation coefficient matrix \cite{spearman_rcc} (SRCC) and Mutual Information\cite{mutual_information} (MI). Defining $\psi_{t,i} = [s_{t,i} \quad x_{t,i}] \in \mathbb{R}^{p+m}$ as a vector concatenating the sensor measurements $s \in \mathbb{R}^p$ and the full field $x \in \mathbb{R}^m$ (pressure, vorticity etc.) for a particular snapshot $i$ at a particular time $t$, we construct a large matrix $\Psi$ containing the entirety of the data in our dataset:
\begin{equation}
   \Psi = \begin{bmatrix} \psi_{0,0} \\ \psi_{0,1} \\ . \\ . \\ .  \end{bmatrix} = \begin{bmatrix} s_{0,0} & x_{0,0} \\ s_{0,1} & x_{0,1} \\ . & . \\ . & . \\ . & .  \end{bmatrix}
\end{equation}

Subsequently, we compute the SRCC and MI matrices $\mathbf{D}, \mathbf{M} \in \mathbb{R}^{(p+m) \times (p+m)}$ based on the columns of $\Psi$. Both are composed of four sub-matrices $\mathbf{D}_{s,s}, \mathbf{M}_{s,s} \in \mathbb{R}^{p \times p}$; $\mathbf{D}_{x,x},\mathbf{M}_{x,x} \in \mathbb{R}^{m \times m}$; $\mathbf{D}_{s,x}, \mathbf{M}_{s,x} \in \mathbb{R}^{p \times m}$; $\mathbf{D}_{x,s}, \mathbf{M}_{x,s} \in \mathbb{R}^{m \times p}$. The first two sub-matrices contain the SRCC values of the sensor measurements and full field values among themselves, while the latter two contain the SRCC values between the sensor measurements and the full field. The difficulty measures $\mathcal{D}$ and $\mathcal{M}$ are defined as the Frobenius norms of $\mathbf{D}_{s,x}$ and $\mathbf{M}_{s,x}$, respectively:

\begin{align}
    \mathbf{D} = \begin{bmatrix} \mathbf{D}_{s,s} & \mathbf{D}_{s,x} \\ \mathbf{D}_{x,s} & \mathbf{D}_{x,x} \end{bmatrix} \qquad \qquad \mathbf{M} = \begin{bmatrix} \mathbf{M}_{s,s} & \mathbf{M}_{s,x} \\ \mathbf{M}_{x,s} & \mathbf{M}_{x,x} \end{bmatrix} \\
    \mathcal{D} = ||\mathbf{D}_{s,x}||_2 = ||\mathbf{D}_{x,s}||_2 \qquad \mathcal{M} = ||\mathbf{M}_{s,x}||_2 = ||\mathbf{M}_{x,s}||_2
\end{align}

\begin{table}[h!]
\caption{Mean absolute error (MAE) and mean absolute percentage error (MAPE) levels from the pressure and velocity reconstruction experiments using the $k=0$ SD-UNet+FNO combination from \secref{sec:model_spatio-temporal_reconstruction}. MI scores were computed using $50 \times 50$ bins via \texttt{sklearn}'s\cite{scikit-learn} \texttt{adjusted\_mutual\_info\_score}. $\omega$ results are identical to $k=0$ results in \secref{sec:results-spatio-temporal} - note that MAE is lower with Cartesian sampling despite higher MAPE, as Annulus sampling concentrates grid points near the boundary of the object where larger vorticity concentrations are typically observed.}
\label{tab:pressure_vel_error}
\begin{tabular}{@{}llcccc@{}}
\toprule
                                                            &      & p & u & v & $\omega$ \\ \midrule
\multilinecell[c]{$\mathcal{D}$ \\ (Higher is easier)}                             &     & 306.22 & 298.99 & 301.85 & 242.82  \\ 
\multilinecell[c]{$\mathcal{M}$ \\ (Higher is easier)}                              &     & 302.72 & 296.37 & 269.81 & 207.75  \\ \midrule
\multirow{2}{*}{Annular}                                    & MAE  & $1.18 \times 10^{-2}$  & $2.64 \times 10^{-2}$  & $1.22 \times 10^{-2}$ & $3.10\times 10^{-1}$ \\
                                                            & MAPE & 2.43\%  & 8.26\% & 9.40\%  & 28.89\% \\
\multirow{2}{*}{Cartesian}                                  & MAE  & $1.33 \times 10^{-2}$ & $3.32 \times 10^{-2}$ & $1.64 \times 10^{-2}$  & $4.95 \times 10^{-2}$ \\
                                                            & MAPE & 3.32\%  & 11.56\% & 15.61\%  & 33.56\% \\ \bottomrule
\end{tabular}
\end{table}

The difficulty metrics clearly show the reason for lower performance when predicting $\omega$ with all setups. The $\mathcal{D}$ scores display greater correlation between the sensor inputs and pressure/velocity fields compared to the vorticity field. This translates to, on average, greater monotonicity in the relations mapping the sensor data to the full field data for the pressure and velocity fields compared to the vorticity field. $\mathcal{M}$, meanwhile, demonstrates that the probability distributions of the sensor inputs and full field values are substantially more alike (i.e. have lower relative entropy). Both of these have profound effects on the accuracy of the neural networks, which manifests in the difference in the MAPE scores when predicting different target fields.

Since the higher difficulty associated with predicting the vorticity field is illustrated, we move forward to comparing the quality of our pressure and velocity results with previous works. Chen et al.\cite{2d_flowpred_random_1} reported reconstruction errors amounting to $7.70 \times 10^-3$ in a similar setup, but at a substantially lower $\Rey$, predicting on flow cases within steady, laminar flow regimes only. Thus, considering the substantially higher $\Rey$ in this work which results in the creation of unsteady vortical structures, the differences in data generation methodologies, and the different objective involving the prediction of instantaneous as opposed to steady fields, the error levels exhibited are in line with previous works. The MAPE levels, under $3\%$ and $10\%$ respectively for pressure and the velocity components, clearly demonstrate that our work is a clear step forward for SMGFR. A gallery of sample velocity and pressure predictions is provided in \appref{sec:app:velocity-pressure-pred}.

\subsection{Spatial multi-geometry vorticity reconstruction}
\label{sec:results-spatial}


Considering the higher difficulty of predicting the vorticity field from pressure/velocity sensors, and to push the boundaries of the neural networks for flow reconstruction tasks, a comprehensive set of 24 experiments for the vorticity SMGFR have been conducted to highlight the differences between the combinations of sensor setups, model architectures and sampling strategies. \tabref{tab:spatial_mape} summarizes the performance of all combinations.

\begin{table}[h!]
\caption{\% errors of the model predictions, averaged over the validation dataset, per sensor setup and sampling strategy. Two different values for the \% errors are presented; the standard MAPE, and the High Vorticity MAPE (HV-MAPE) filtering out the data points with vorticity magnitude below 1\% of the max ground truth vorticity. In both cases, sampling points with percentage errors above 200\% were excluded from the MAPE calculation, due to error values approaching infinity near the ground truth zero vorticity field contours. The best performing sampling strategy-model architecture combinations for each sensor setup are highlighted with italics and underscores for the MAPE and HV-MAPE, respectively.}
\label{tab:spatial_mape}
\begin{tabular}{@{}llcc|cc|cc|cc@{}}
\toprule
\multirow{2}{*}{Sensor setup} & \multirow{2}{*}{Sampling} & \multicolumn{2}{c|}{SD}             & \multicolumn{2}{c|}{SD-Large}       & \multicolumn{2}{c|}{SD-UNet}                       & \multicolumn{2}{c}{SD-FNO}                   \\
                              &                           & MAPE             & HV-MAPE          & MAPE             & HV-MAPE          & MAPE                      & HV-MAPE                & MAPE                      & HV-MAPE          \\ \midrule
\multirow{2}{*}{Large}        & Annular                   & \textbf{44.29\%} & \textbf{34.28\%} & \textbf{43.80\%} & \textbf{31.85\%} & \textit{\textbf{39.92\%}} & {\ul \textbf{31.37\%}} & \textbf{40.83\%}          & \textbf{31.78\%} \\
                              & Cartesian                 & 59.88\%          & 46.14\%          & 57.52\%          & 53.36\%          & 47.64\%                   & 39.88\%                & 46.56\%                   & 39.34\%          \\ \midrule
\multirow{2}{*}{Medium}       & Annular                   & \textbf{46.86\%} & \textbf{34.35\%} & \textbf{45.45\%} & \textbf{33.77\%} & \textit{\textbf{42.61\%}} & {\ul \textbf{32.73\%}} & \textbf{43.66\%}          & \textbf{32.93\%} \\
                              & Cartesian                 & 55.41\%          & 48.99\%          & 58.65\%          & 48.68\%          & 48.49\%                   & 42.99\%                & 48.17\%                   & 42.12\%          \\ \midrule
\multirow{2}{*}{Small}        & Annular                   & \textbf{47.35\%} & \textbf{34.72\%} & \textbf{47.20\%} & \textbf{34.06\%} & \textbf{45.03\%}          & {\ul \textbf{33.17\%}} & \textit{\textbf{44.04\%}} & \textbf{33.77\%} \\
                              & Cartesian                 & 57.40\%          & 50.89\%          & 59.77\%          & 48.59\%          & 51.56\%                   & 44.47\%                & 51.81\%                   & 45.81\%          \\ \bottomrule
\end{tabular}
\end{table}

Two different metrics of percentage error, as seen in \tabref{tab:spatial_mape} are used in this section: the standard MAPE, and the High Vorticity MAPE (HV-MAPE) computed from points for which the vorticity magnitude exceeds 1\% of the maximum absolute vorticity in a snapshot. Comparing the MAPE and HV-MAPE, it can be seen that HV-MAPEs are consistently lower than the MAPEs, which is not surprising given the choice of the MAE as the loss function which assigns a greater penalty to the regions of higher target field magnitude given constant percentage errors. As a consequence, the models implicitly prioritize lowering the percentage errors for areas with high vorticity, which is desirable since these are the most important features when analyzing the dynamics of fluid flows. Overall MAPEs are between 40\% and 47\% for the Annular sampling while they are between 46\% and 60\% for the Cartesian sampling. HV-MAPEs can go as low as 31\% for the Annular sampling and 39\% for the Cartesian sampling. 

Among the three major variables differentiating the experiments -- the sampling method, model architecture and sensor setup -- the factor most consistently leading to superior results is the sampling method. The annular sampling method enables substantially lower MAPE and HV-MAPE with all models and sensor setups tested, lowering MAPE and HV-MAPE by up to 15 and 21 percentage points respectively (translating to 26\% and 40\%), in the case of the SD model with the large sensor setup. The error-reducing effect of the annular sampling is especially prominent in the high vorticity regions of the flow, as evidenced by comparing the differences in HV-MAPE versus differences in MAPE when comparing the two sampling strategies; out of the 12 combinations of models and sensor setups, HV-MAPE has a larger absolute gap than MAPE between the two sampling methodologies in all cases except for the SD model using the Large sensor setup, with the differences in HV-MAPE being larger by about 3 percentage points on average.

Next in the order of importance is the model architecture. The two architectures which more explicitly isolate the various spatial length scales, by the way of pooling-convolution-upsampling branches for the SD-UNet or convolution in Fourier space in the case of the SD-FNO, exhibit superior percentage error metrics for all sensor and sampling setups. This is not surprising given the established superiority of convolutional architectures in various computer vision tasks over fully connected networks\cite{conv_good}. However, the importance of the architecture varies by the sampling method. While MAPEs and HV-MAPEs are at most within a 5\% and 3\% range of each other (respectively) between models for a particular sensor setup in the case of annular sampling, this gap increases to as much as 14\% with Cartesian sampling. Controlling for the number of parameters (as seen in \tabref{tab:param_numbers_spatial}), the impact of the architecture declines even further, as the SD-Large model is consistently ahead of the standard SD model. 


The final variable to discuss is the sensor setup, which does not have a great impact on the accuracy of the models. The largest setup provides five times more information as the smallest, only leading to modest improvements in terms of the MAPE which do not exceed 5\% for any model or sampling strategy. However, especially in the case of the SD and SD-Large models, taking advantage of the extra information is possible only in conjunction with the usage of annular sampling. The merit of having a large sensor setup is that, from a purely computational perspective, increasing the number of sensors is very cheap, as it leads only to an increase in the number of parameters and computational cost in the first layer of the model, opening up an avenue for modest accuracy improvements almost `for free'. In an experimental setting, however, this may be counteracted by the burden of a significantly more labor-intensive physical sensor setup.

\begin{figure}
    \centering
    \includegraphics[width=0.40\textwidth]{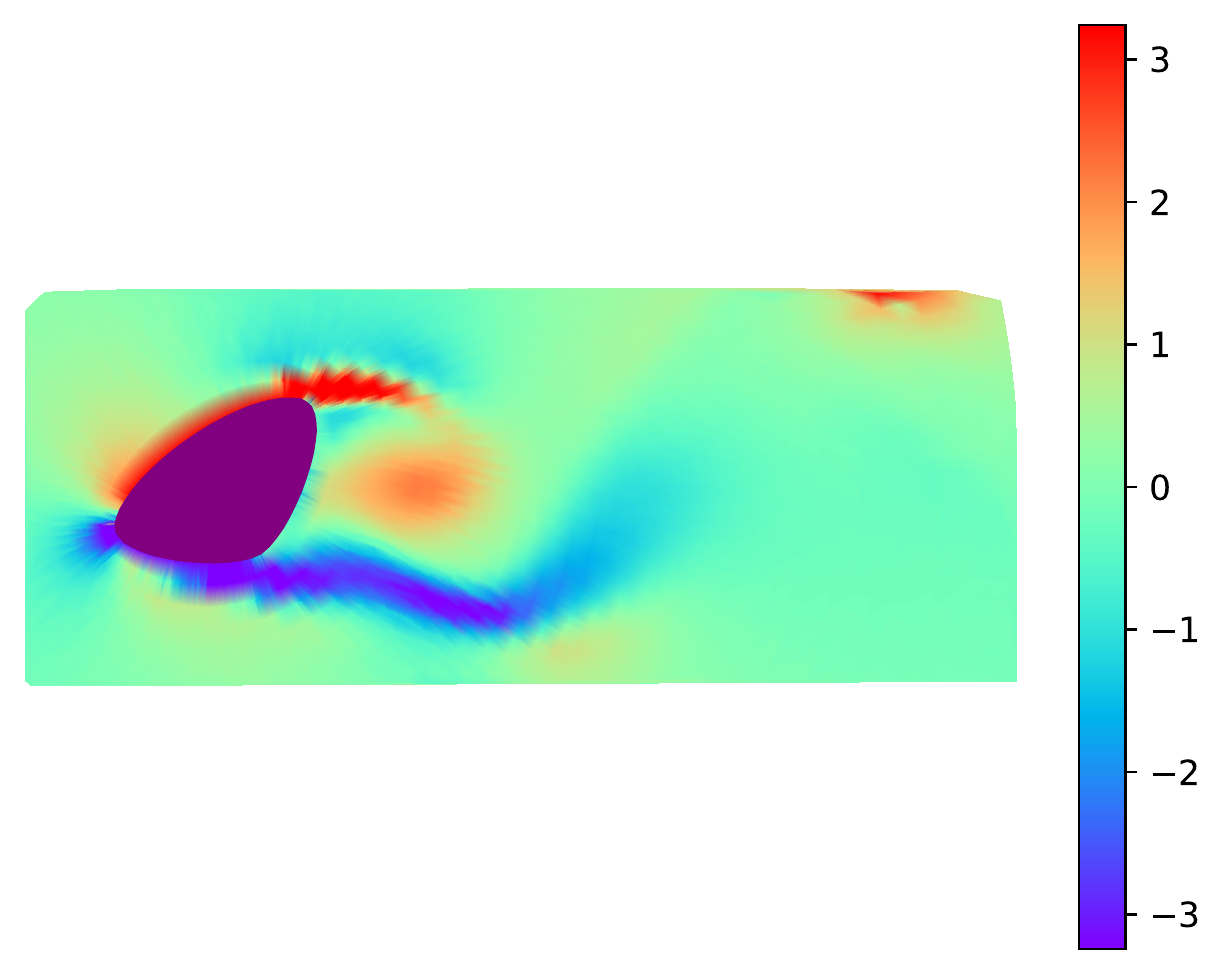}
    \includegraphics[width=0.045\textwidth]{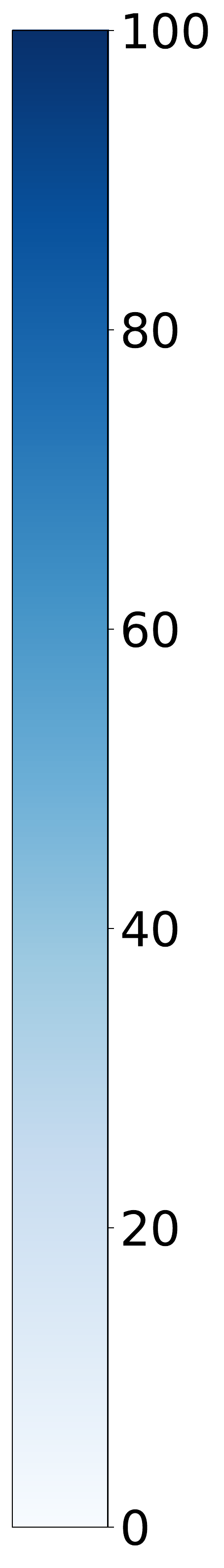}
\begin{tabular}{@{}lcccc@{}}
\toprule
\multirow{2}{*}{Model} & \multicolumn{2}{c|}{Annular Sampling}                        & \multicolumn{2}{c}{Cartesian Sampling}      \\ \cmidrule(l){2-5} 
                       & \multicolumn{1}{c|}{Vorticity} & \multicolumn{1}{c|}{\% Error} & \multicolumn{1}{c|}{Vorticity} & \% Error \\ \midrule
SD                     &    \begin{minipage}{\spatialfigwidth\textwidth}
      \includegraphics[width=\linewidth]{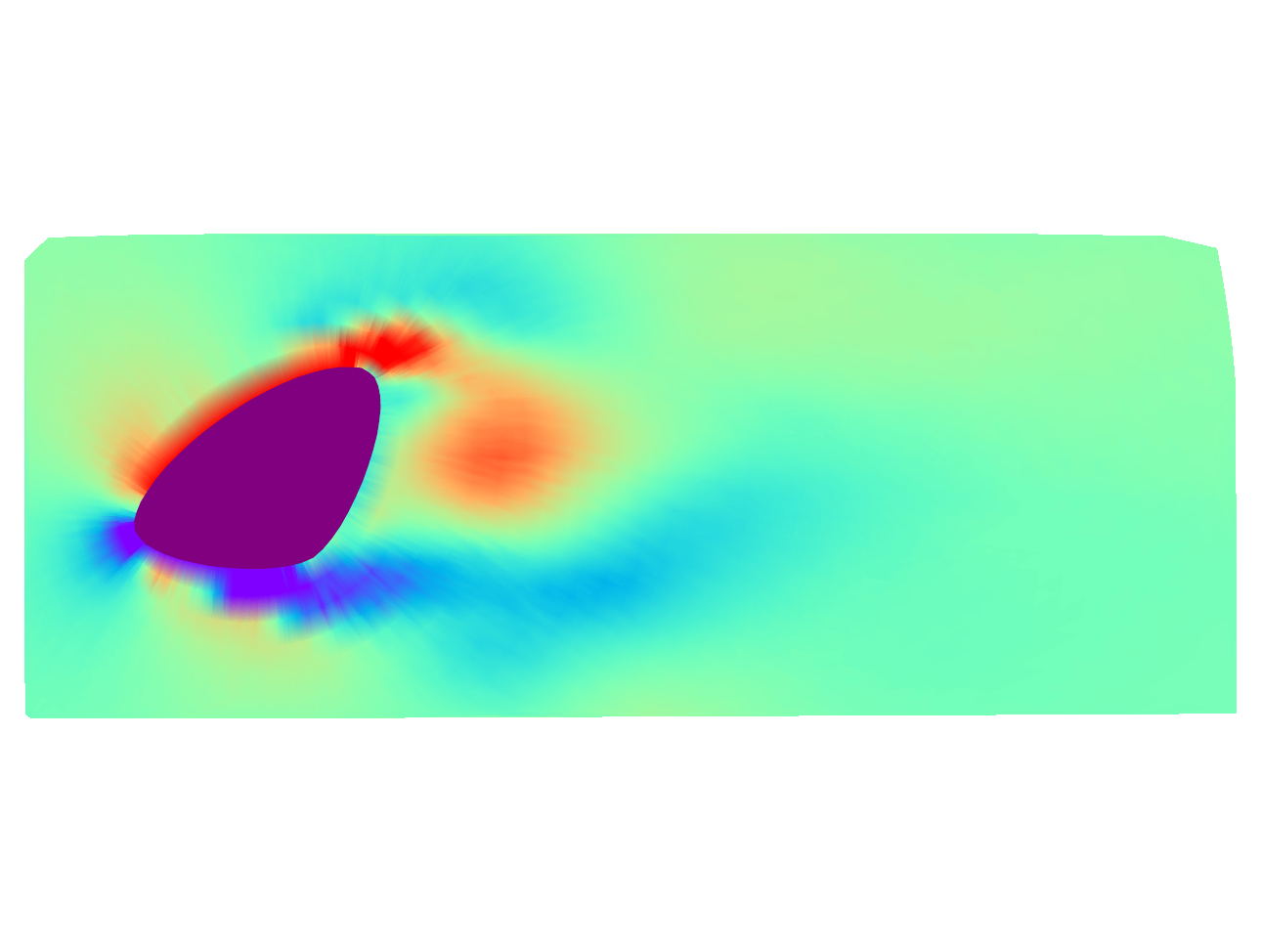}
    \end{minipage}                            &  \begin{minipage}{\spatialfigwidth\textwidth}
      \includegraphics[width=\linewidth]{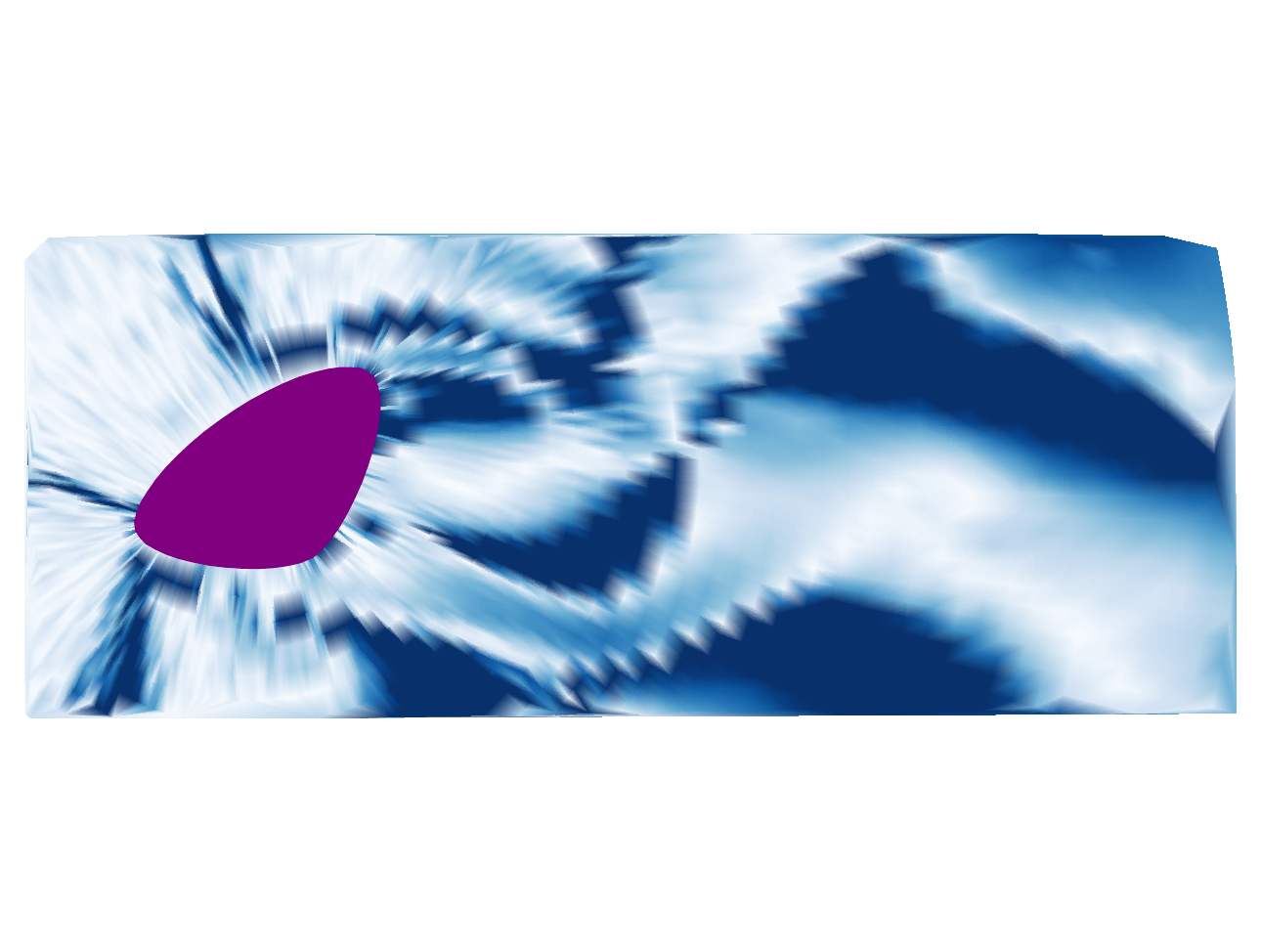}
    \end{minipage}                             &  \begin{minipage}{\spatialfigwidth\textwidth}
      \includegraphics[width=\linewidth]{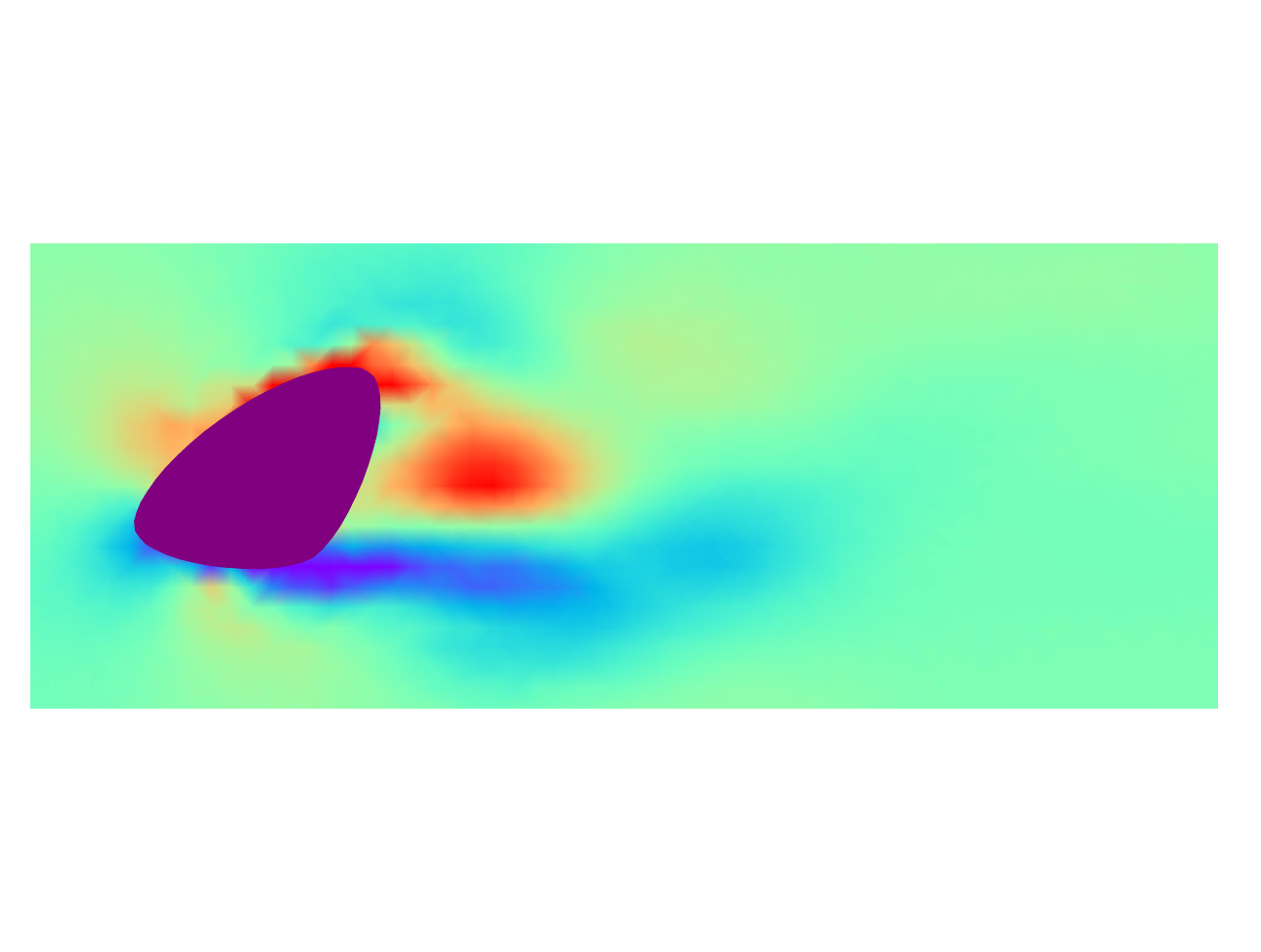}
    \end{minipage}                             & \begin{minipage}{\spatialfigwidth\textwidth}
      \includegraphics[width=\linewidth]{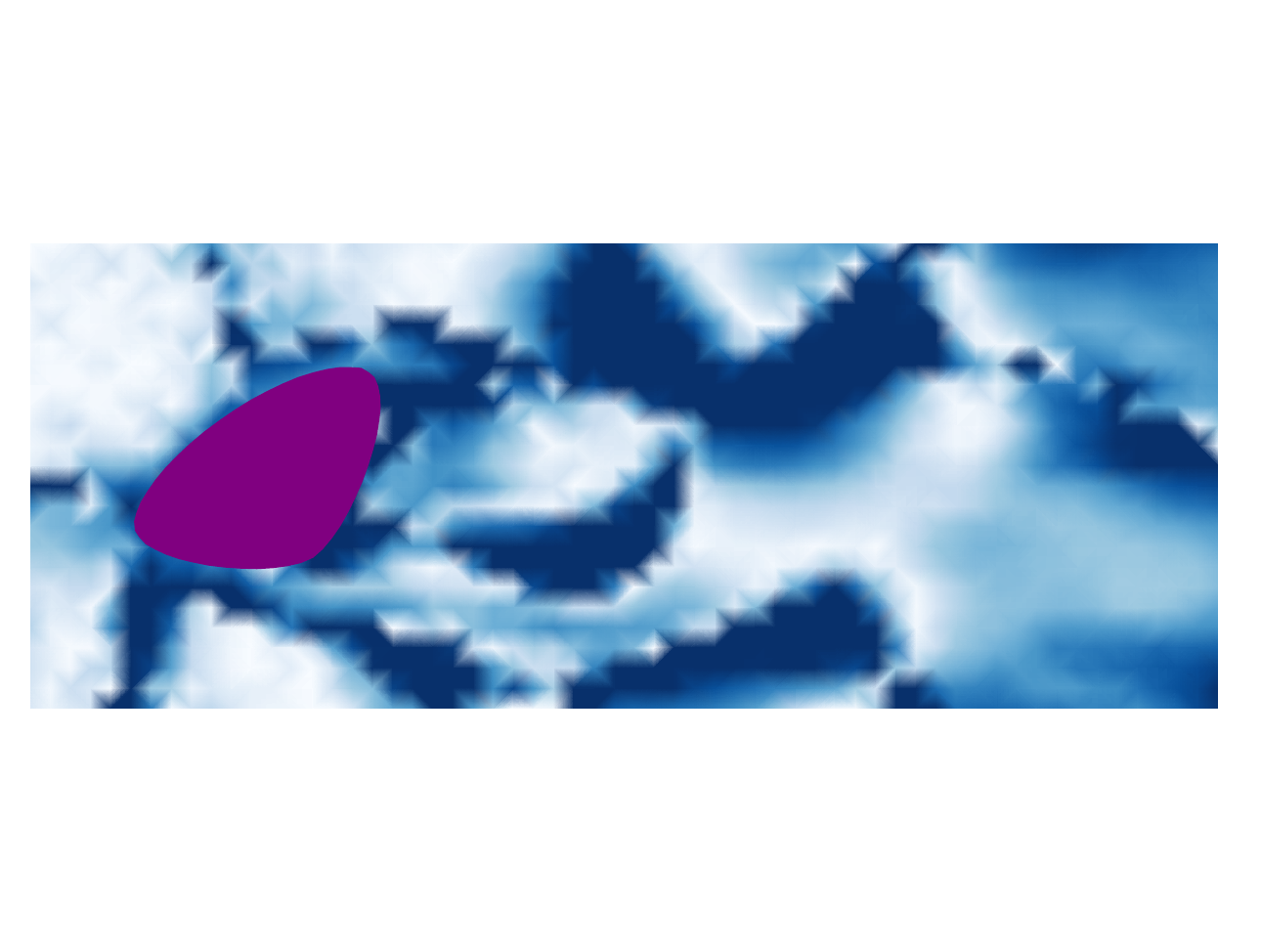}
    \end{minipage}         \\ 
\% Err    & \multicolumn{2}{c}{MAPE \textbf{38.98\%}, HV-MAPE \textbf{33.12\%}} & \multicolumn{2}{c}{MAPE 52.57\%, HV-MAPE 49.25\%}\\ \midrule
SD-Large               &    \begin{minipage}{\spatialfigwidth\textwidth}
      \includegraphics[width=\linewidth]{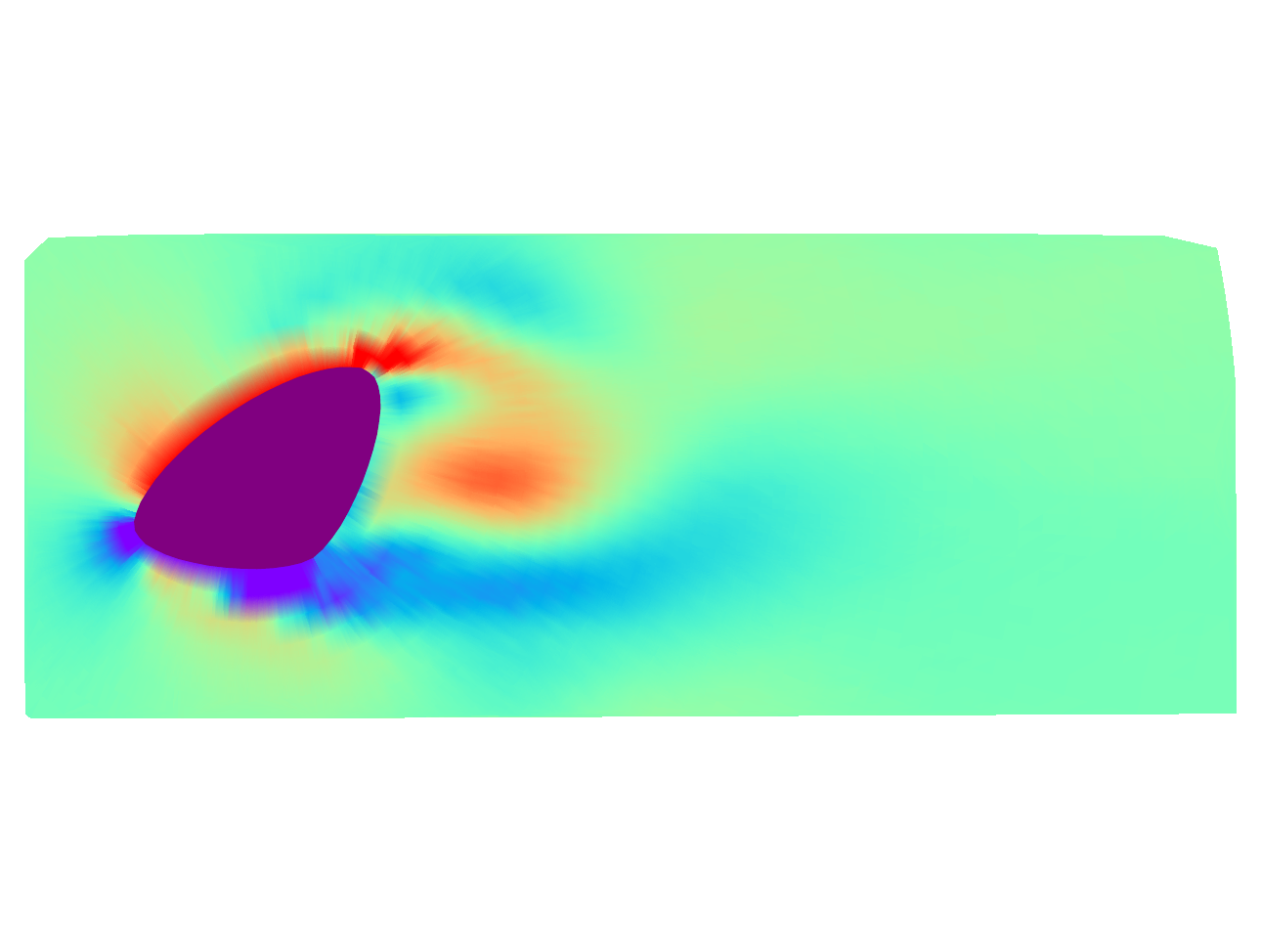}
    \end{minipage}                            &  \begin{minipage}{\spatialfigwidth\textwidth}
      \includegraphics[width=\linewidth]{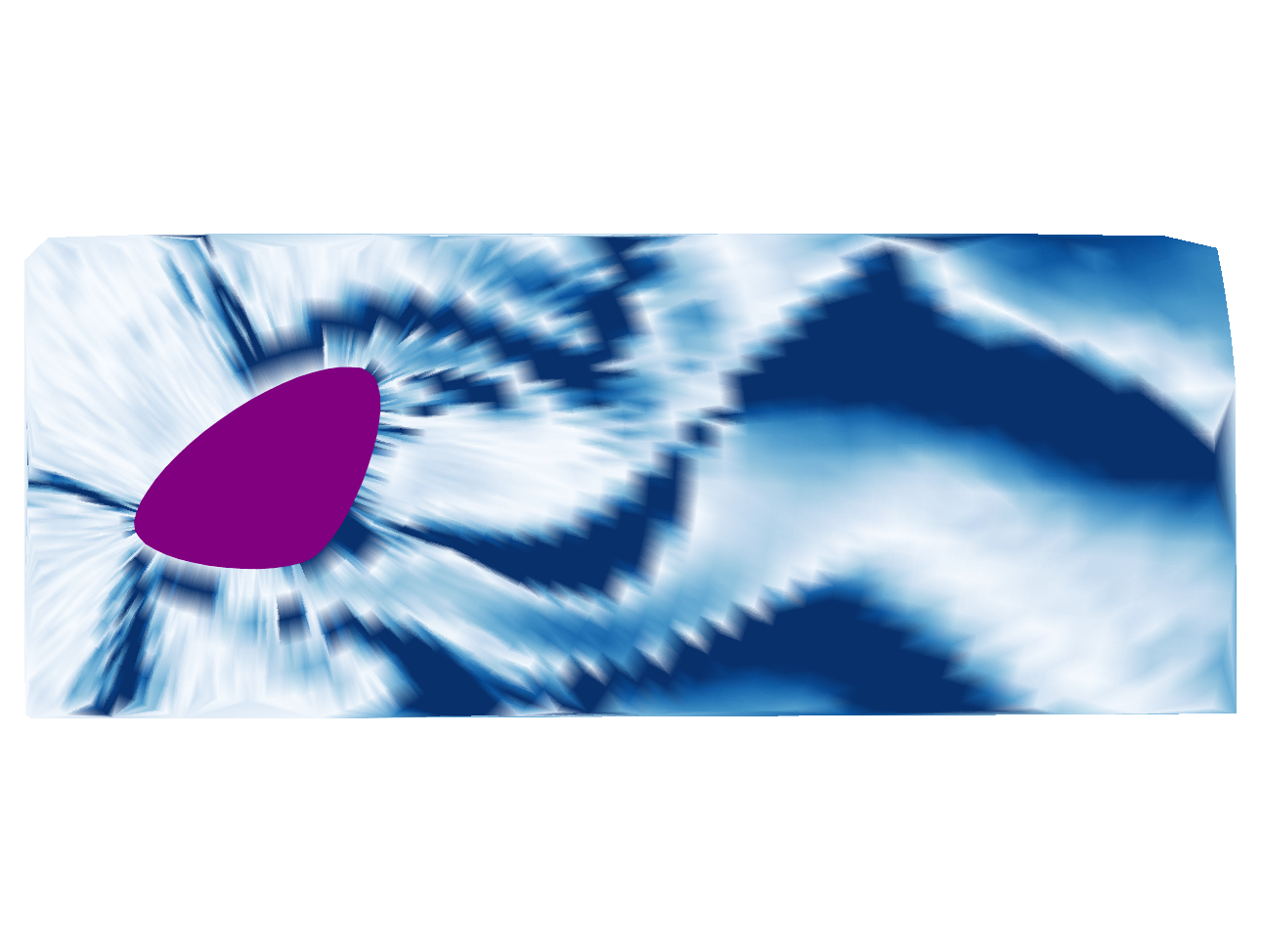}
    \end{minipage}                             &  \begin{minipage}{\spatialfigwidth\textwidth}
      \includegraphics[width=\linewidth]{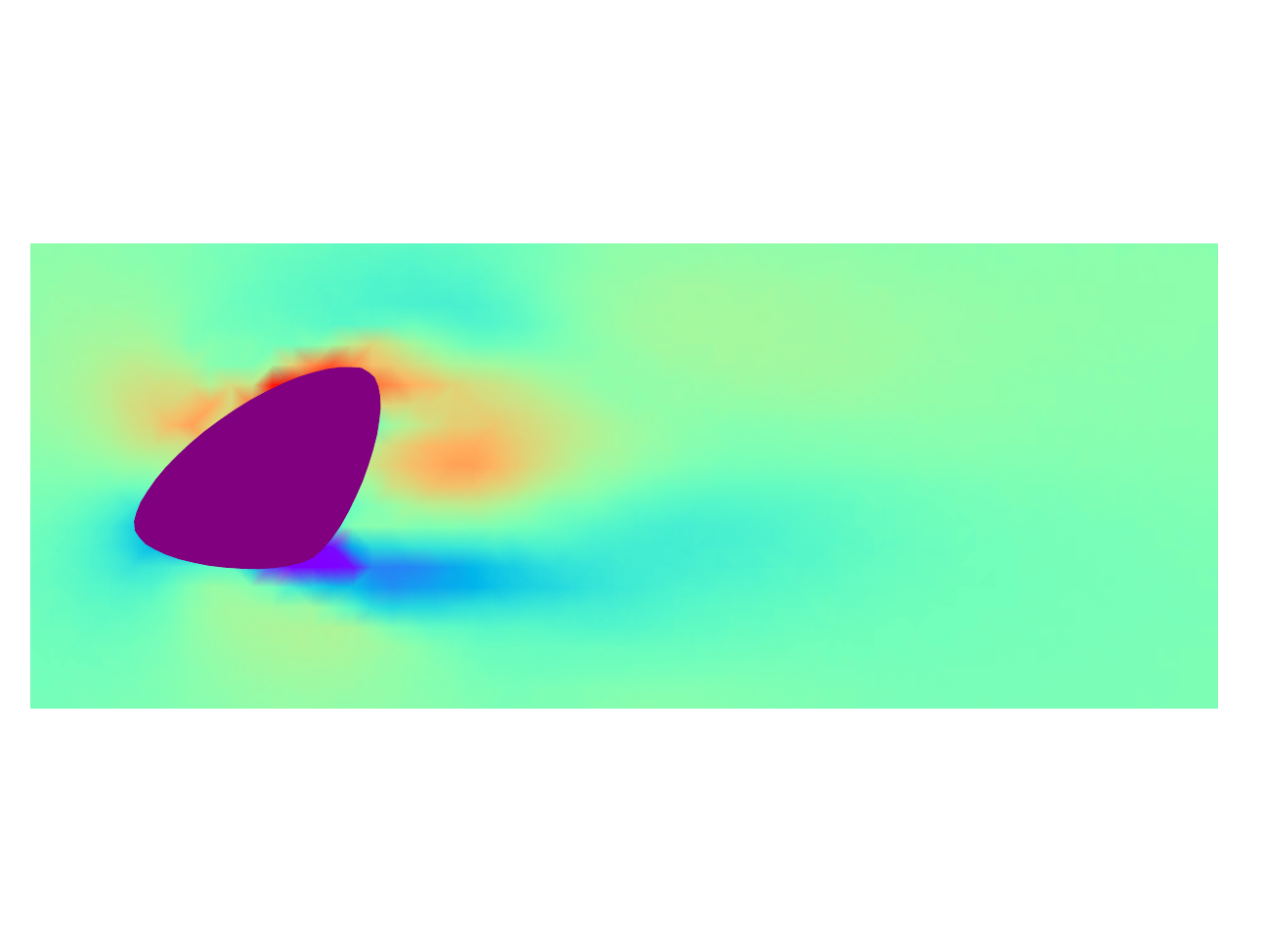}
    \end{minipage}                             & \begin{minipage}{\spatialfigwidth\textwidth}
      \includegraphics[width=\linewidth]{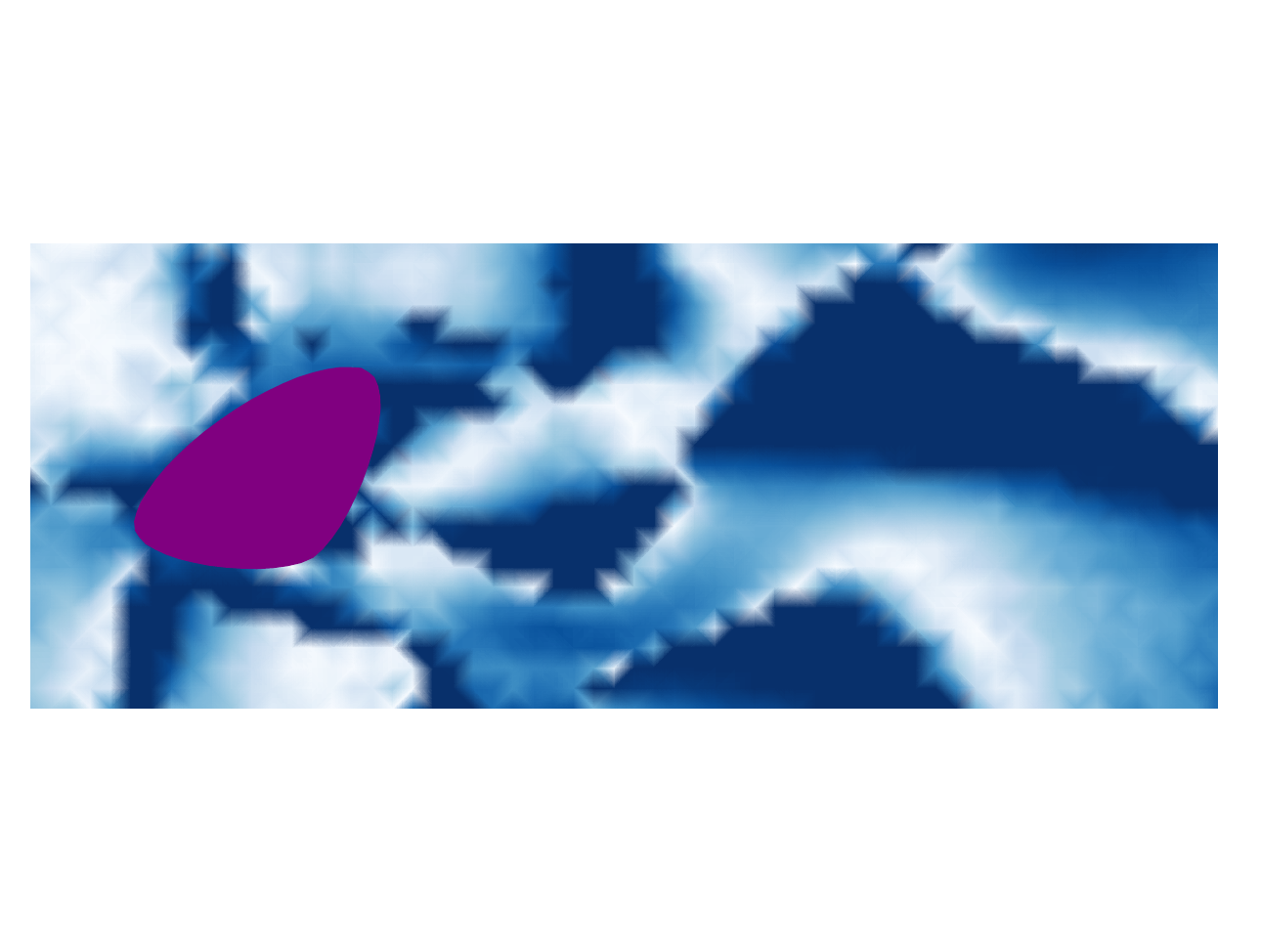}
    \end{minipage}          \\
\% Err    & \multicolumn{2}{c}{MAPE \textbf{39.71\%}, HV-MAPE \textbf{31.79\%}} & \multicolumn{2}{c}{MAPE 61.79\%, HV-MAPE 62.26\%}\\ \midrule
SD-UNet                &    \begin{minipage}{\spatialfigwidth\textwidth}
      \includegraphics[width=\linewidth]{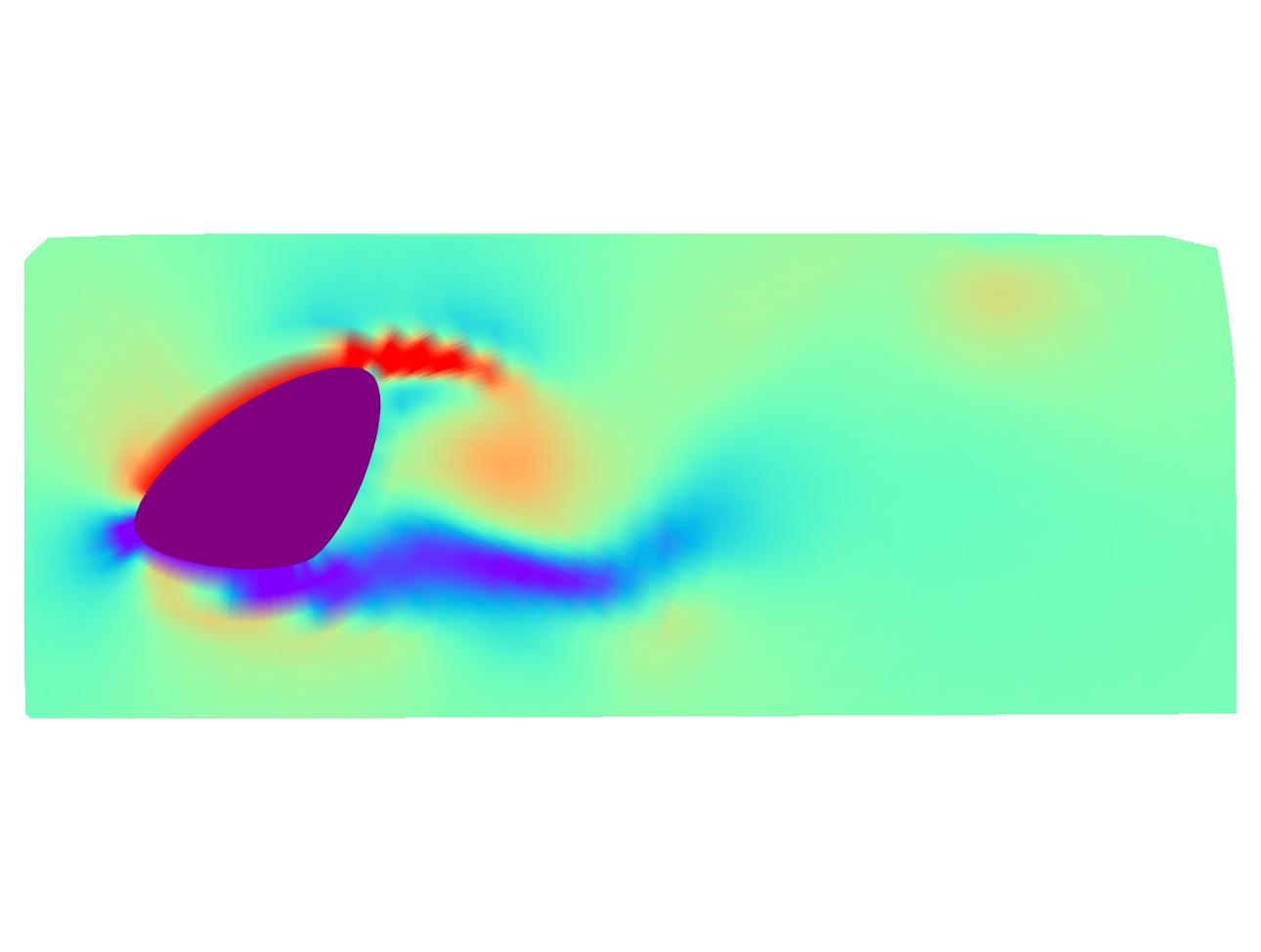}
    \end{minipage}                            &  \begin{minipage}{\spatialfigwidth\textwidth}
      \includegraphics[width=\linewidth]{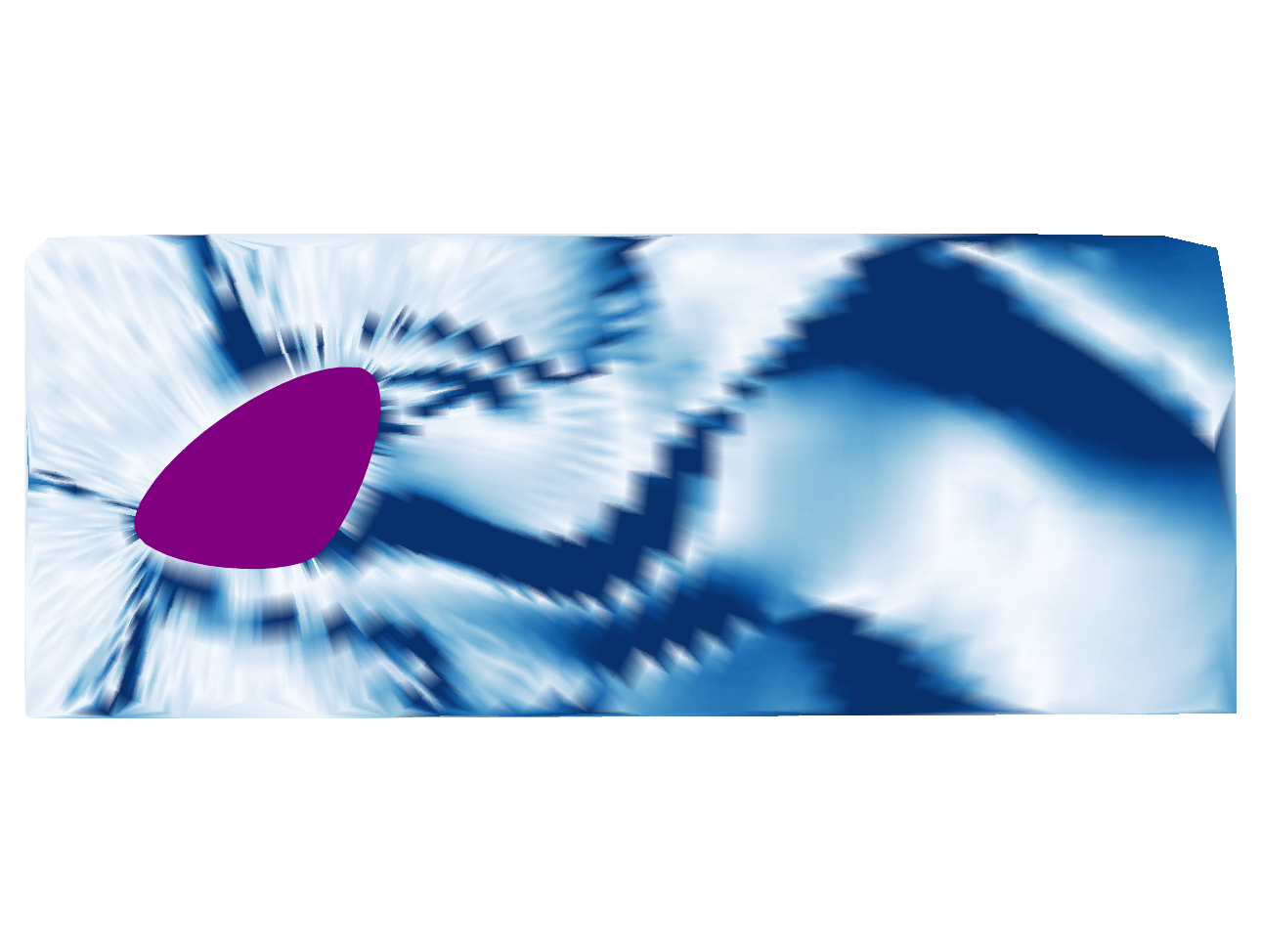}
    \end{minipage}                             &  \begin{minipage}{\spatialfigwidth\textwidth}
      \includegraphics[width=\linewidth]{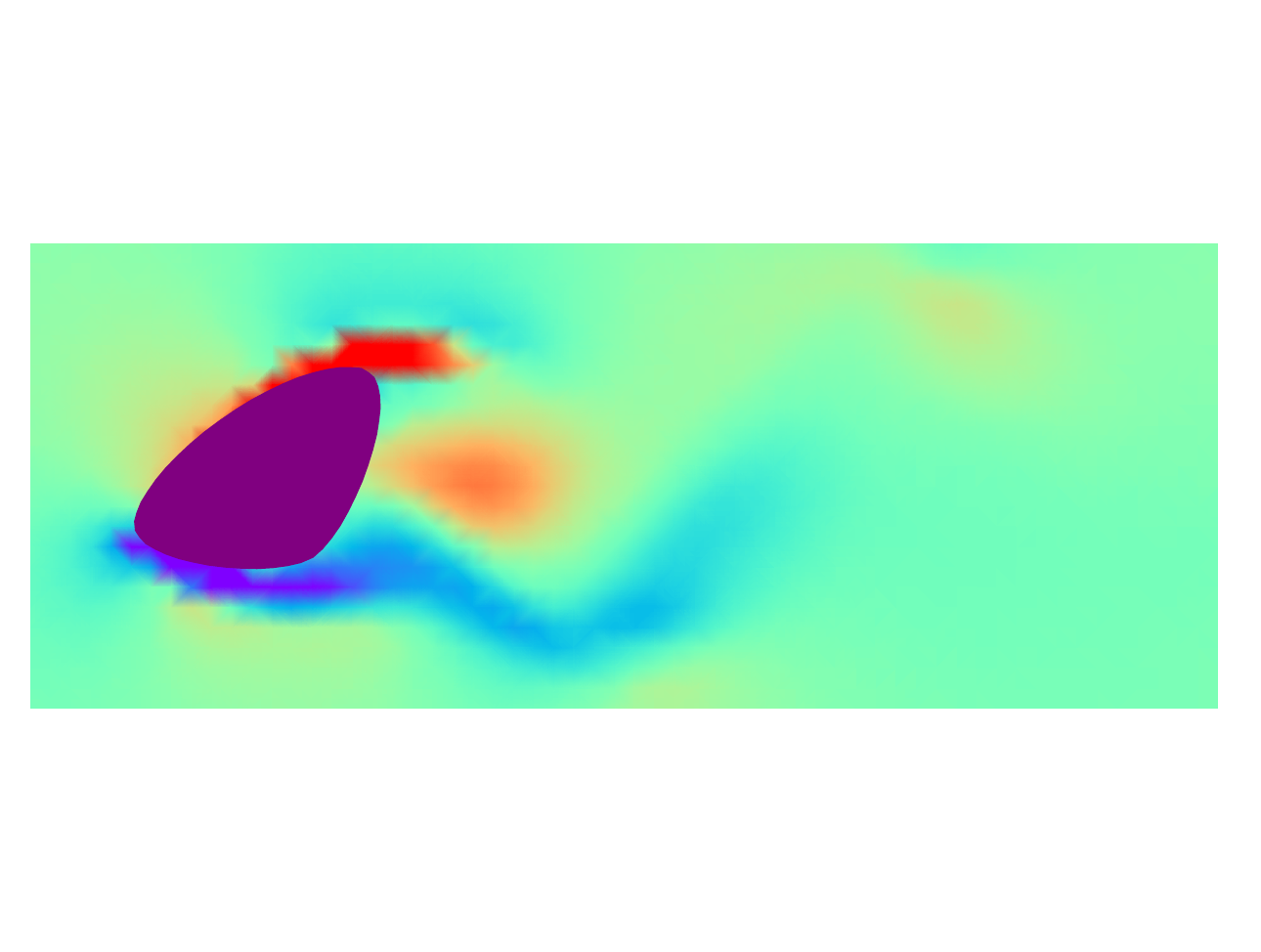}
    \end{minipage}                             & \begin{minipage}{\spatialfigwidth\textwidth}
      \includegraphics[width=\linewidth]{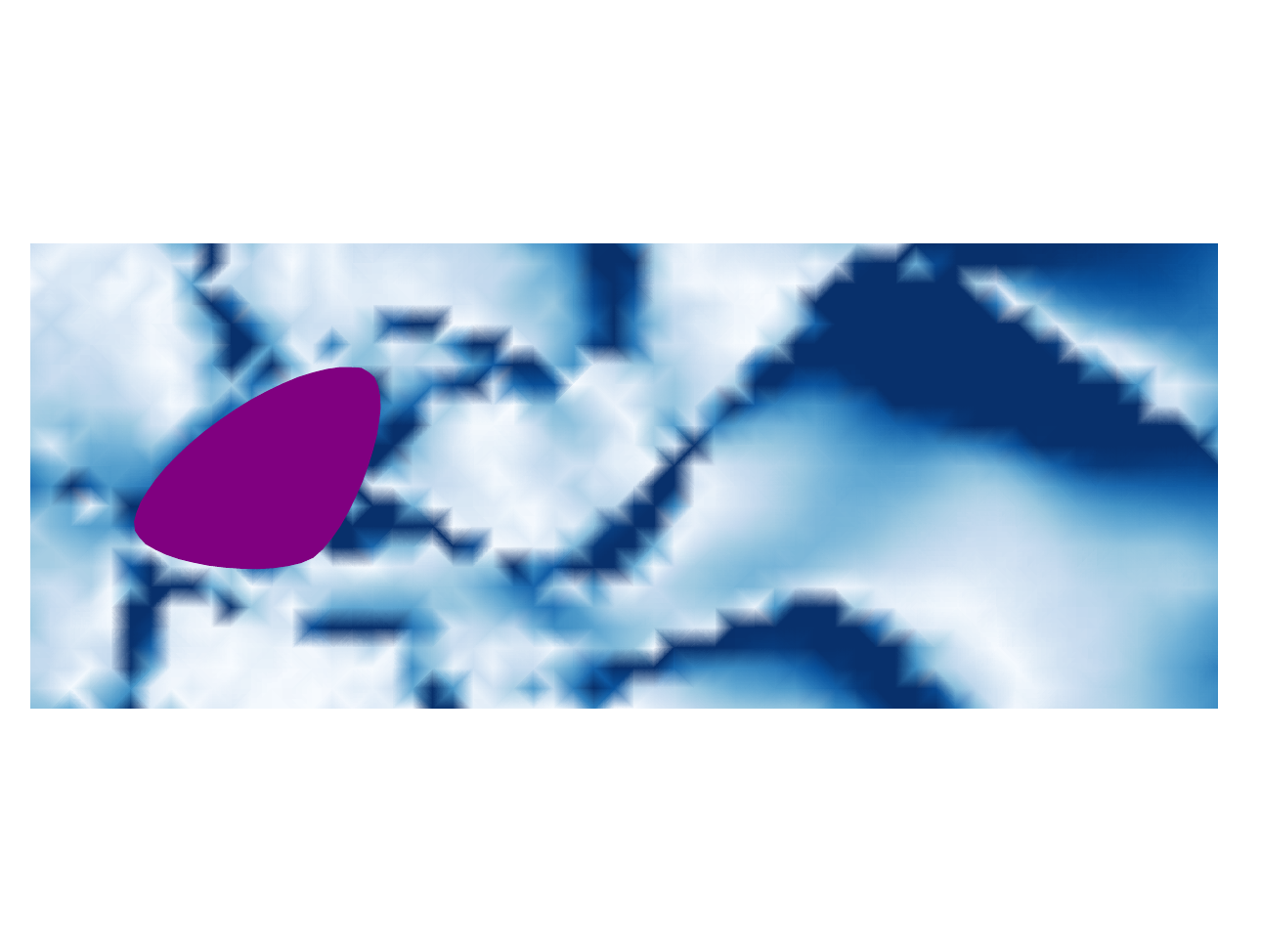}
    \end{minipage}          \\
\% Err    & \multicolumn{2}{c}{MAPE \textbf{36.59\%}, HV-MAPE \textbf{32.13\%}} & \multicolumn{2}{c}{MAPE 46.01\%, HV-MAPE 42.60\%}\\ \midrule
SD-FNO                 &    \begin{minipage}{\spatialfigwidth\textwidth}
      \includegraphics[width=\linewidth]{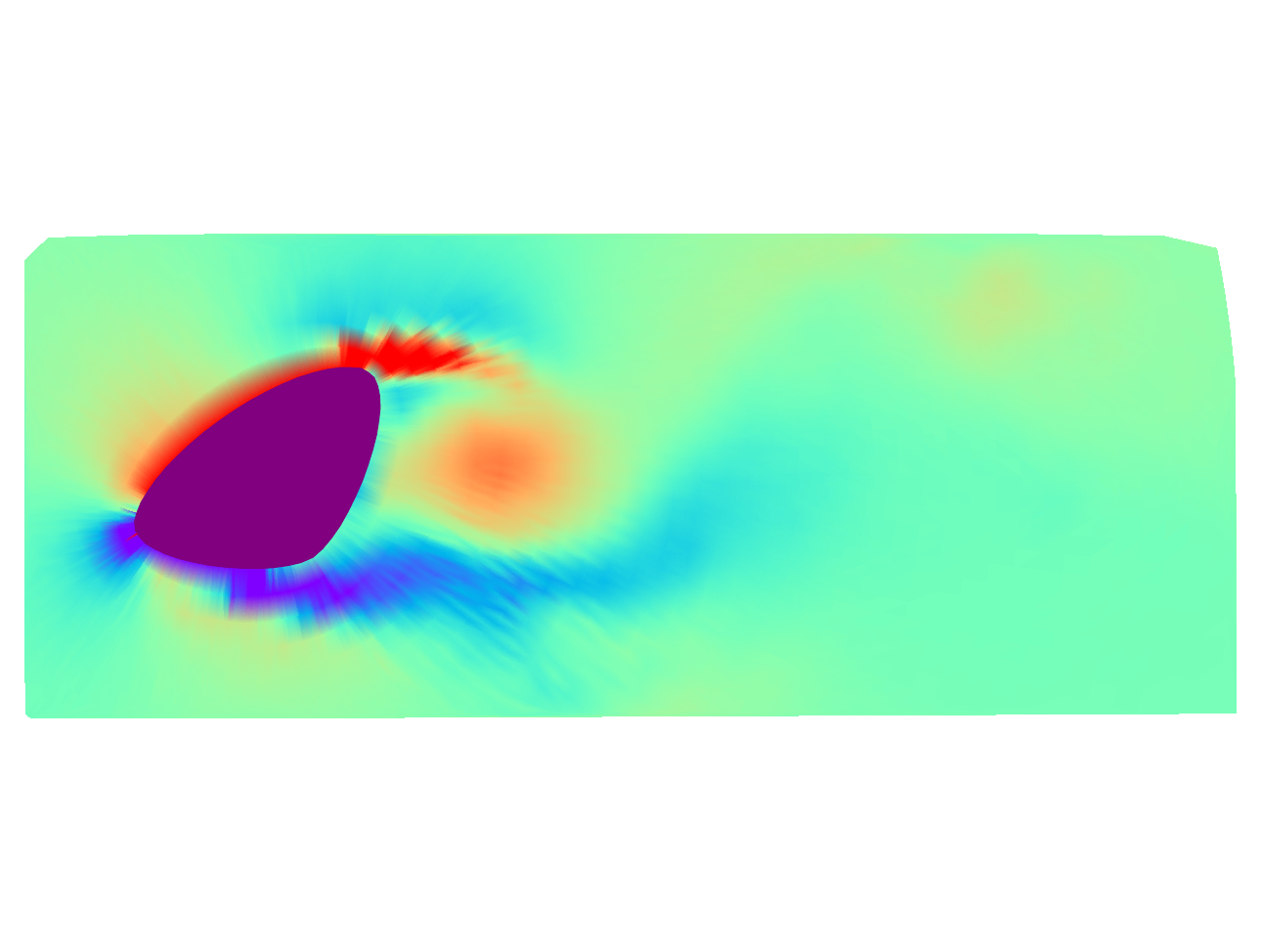}
    \end{minipage}                            &  \begin{minipage}{\spatialfigwidth\textwidth}
      \includegraphics[width=\linewidth]{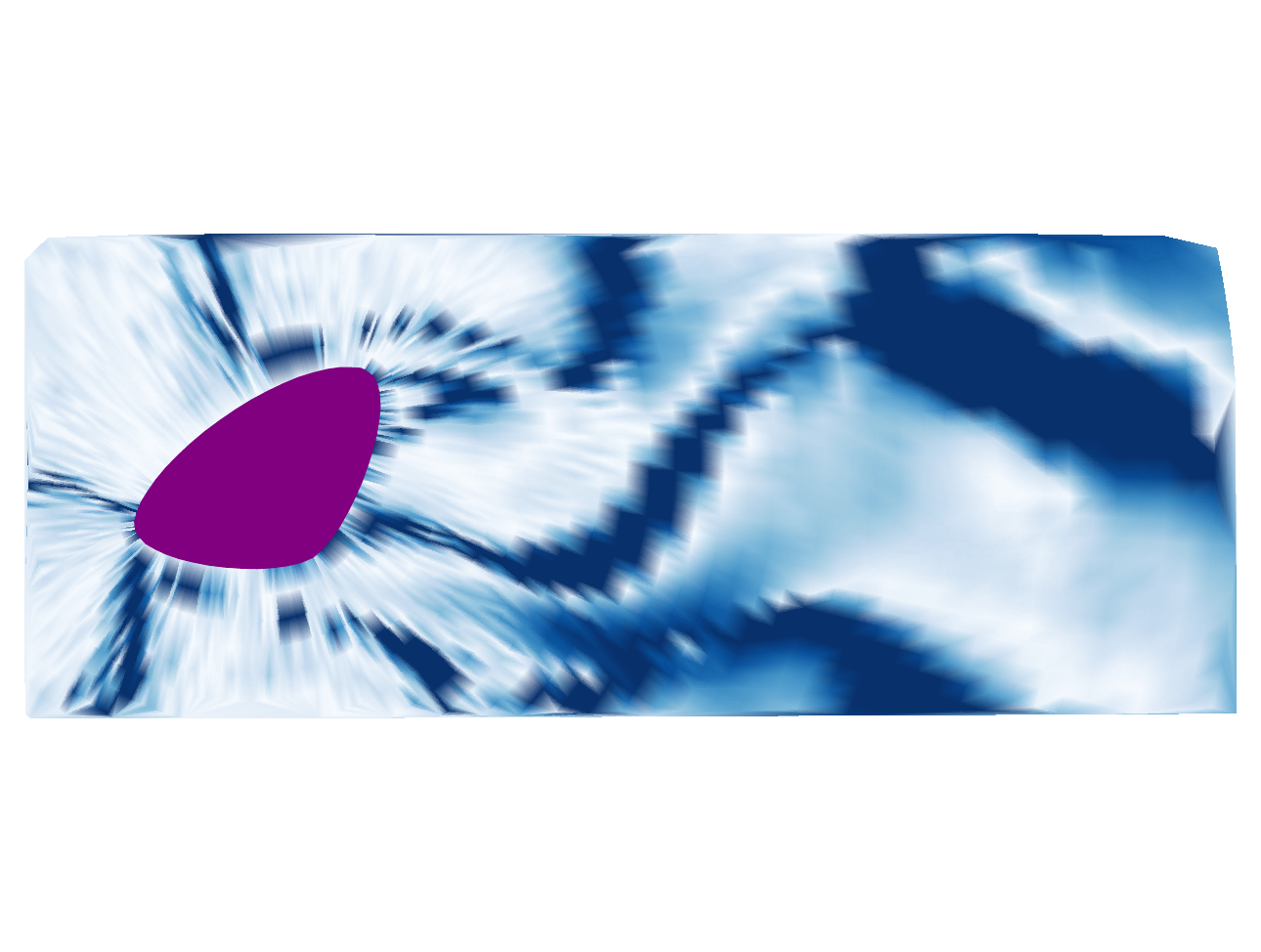}
    \end{minipage}                             &  \begin{minipage}{\spatialfigwidth\textwidth}
      \includegraphics[width=\linewidth]{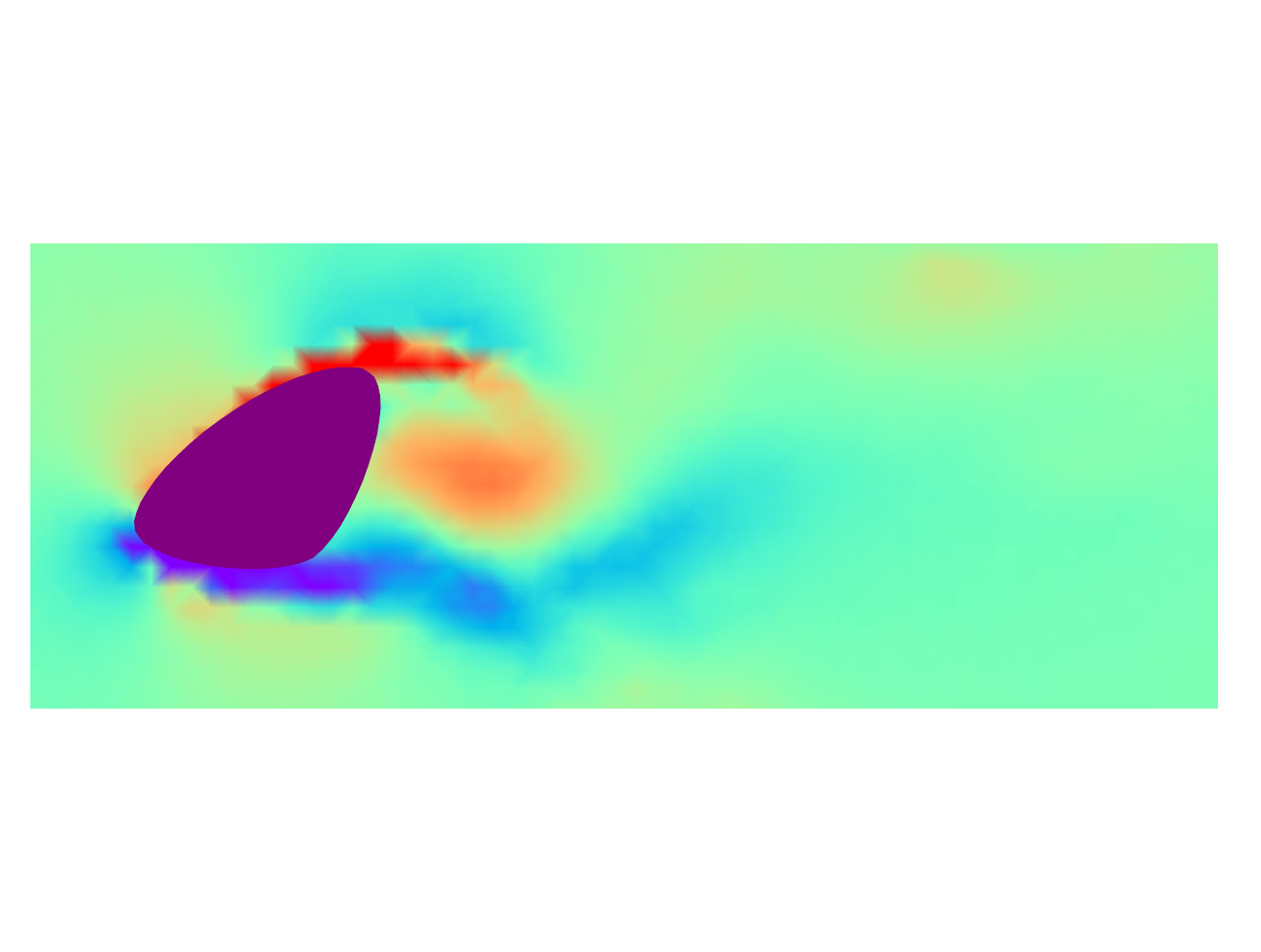}
    \end{minipage}                             & \begin{minipage}{\spatialfigwidth\textwidth}
      \includegraphics[width=\linewidth]{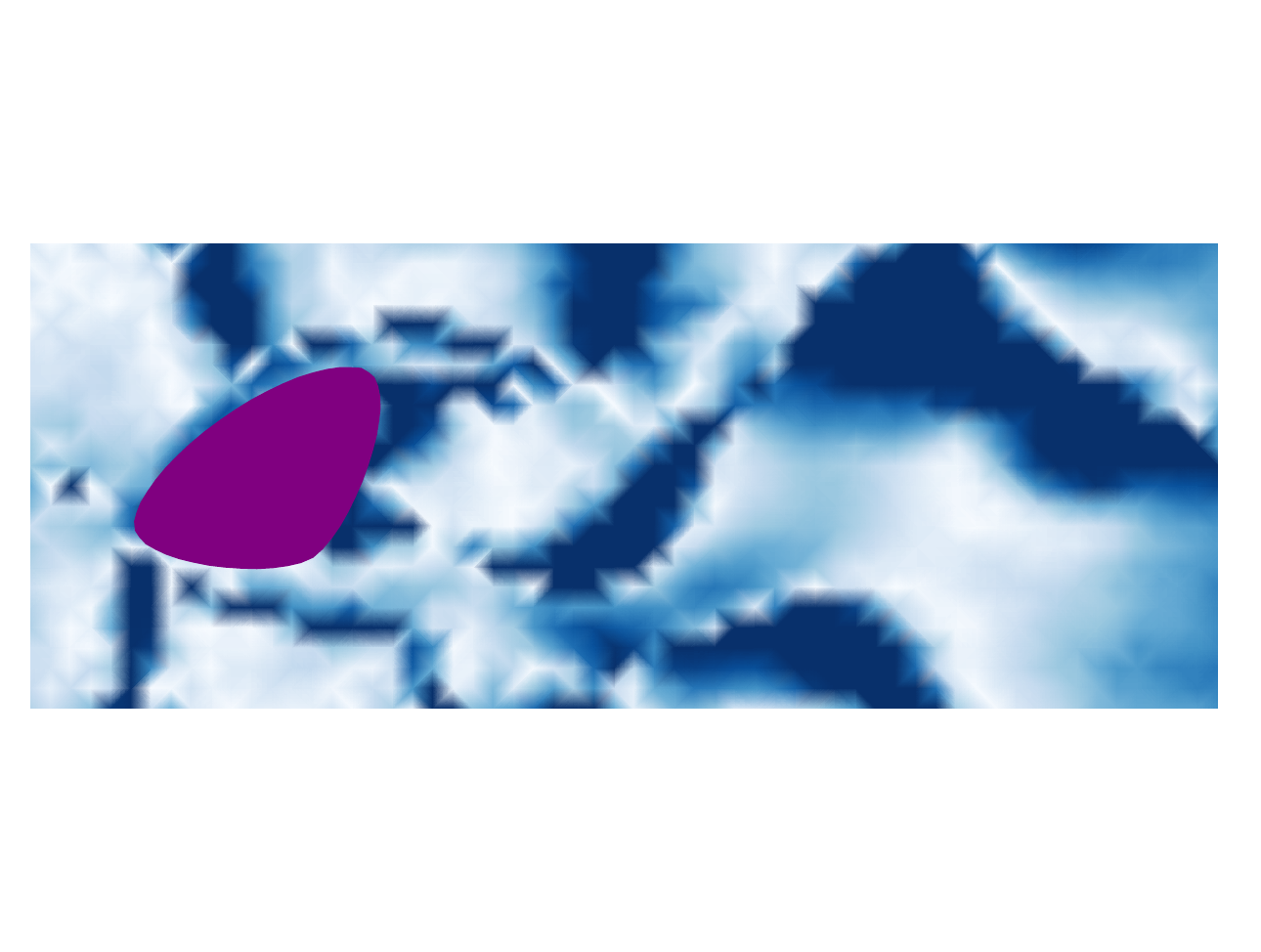}
    \end{minipage}          \\
\% Err    & \multicolumn{2}{c}{MAPE \textbf{\textit{33.78\%}}, HV-MAPE {\ul \textbf{24.97\%}}} & \multicolumn{2}{c}{MAPE 49.81\%, HV-MAPE 42,89\%}\\ \bottomrule
\end{tabular}
    
    \caption{Ground truth (top) and predicted (bottom) vorticity fields and percentage error maps for a snapshot from Shape A, using the large sensor setup. Vorticity colormap constrained to 7.5\% of $\max(|\tilde{\omega}_{t,i}|)$.}
    \label{fig:spatialex_s1660_large}
\end{figure}

\begin{figure}
    \centering
    \includegraphics[width=0.40\textwidth]{images/spatial/annulus_large_s1660_gt_vorticity.pdf}
    \includegraphics[width=0.045\textwidth]{images/spatial/pcterrorbarr0.pdf}
\begin{tabular}{@{}lcccc@{}}
\toprule
\multirow{2}{*}{Model} & \multicolumn{2}{c|}{Annular Sampling}                        & \multicolumn{2}{c}{Cartesian Sampling}      \\ \cmidrule(l){2-5} 
                       & \multicolumn{1}{c|}{Vorticity} & \multicolumn{1}{c|}{\% Error} & \multicolumn{1}{c|}{Vorticity} & \% Error \\ \midrule
SD                     &    \begin{minipage}{\spatialfigwidth\textwidth}
      \includegraphics[width=\linewidth]{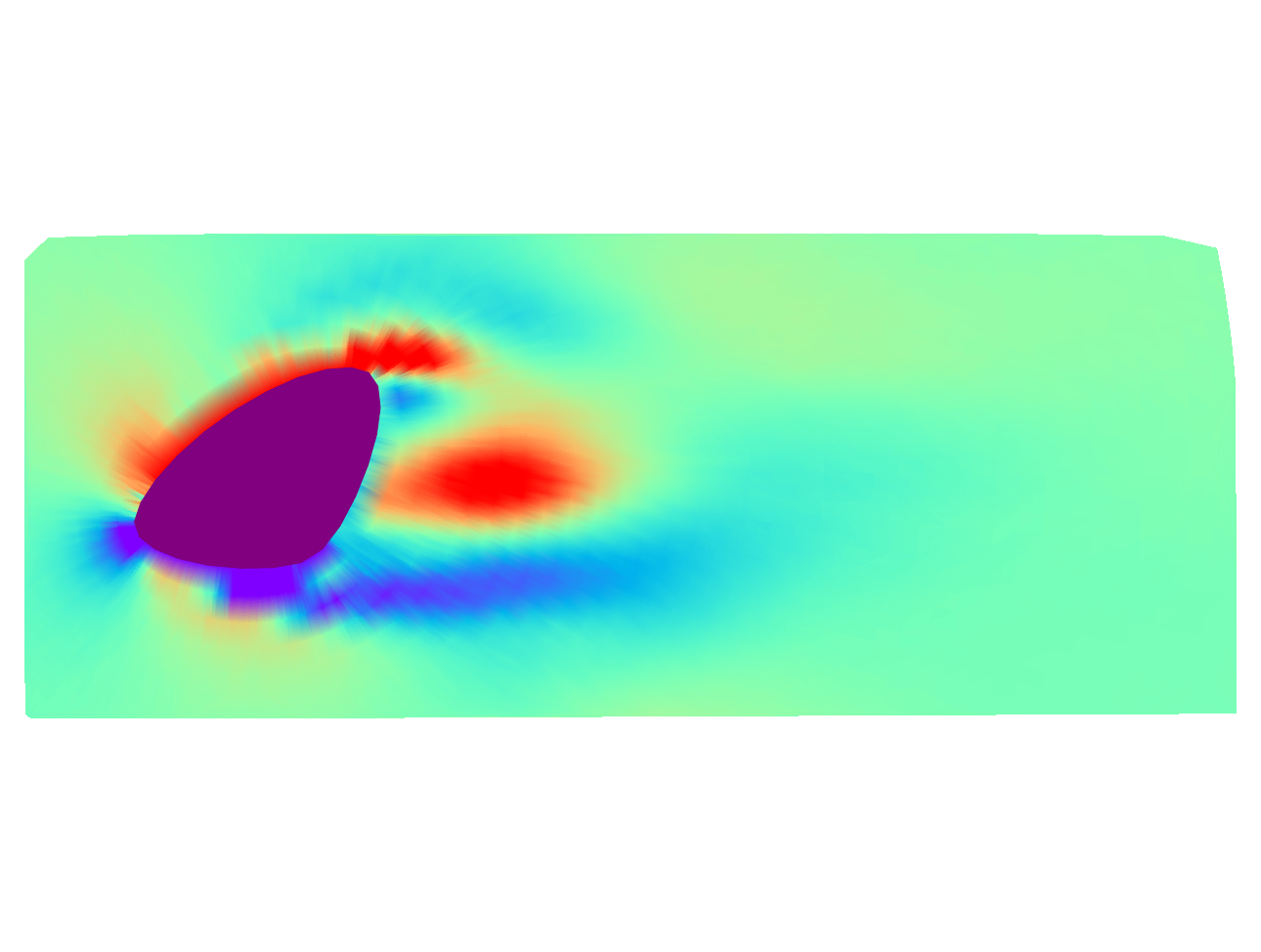}
    \end{minipage}                            &  \begin{minipage}{\spatialfigwidth\textwidth}
      \includegraphics[width=\linewidth]{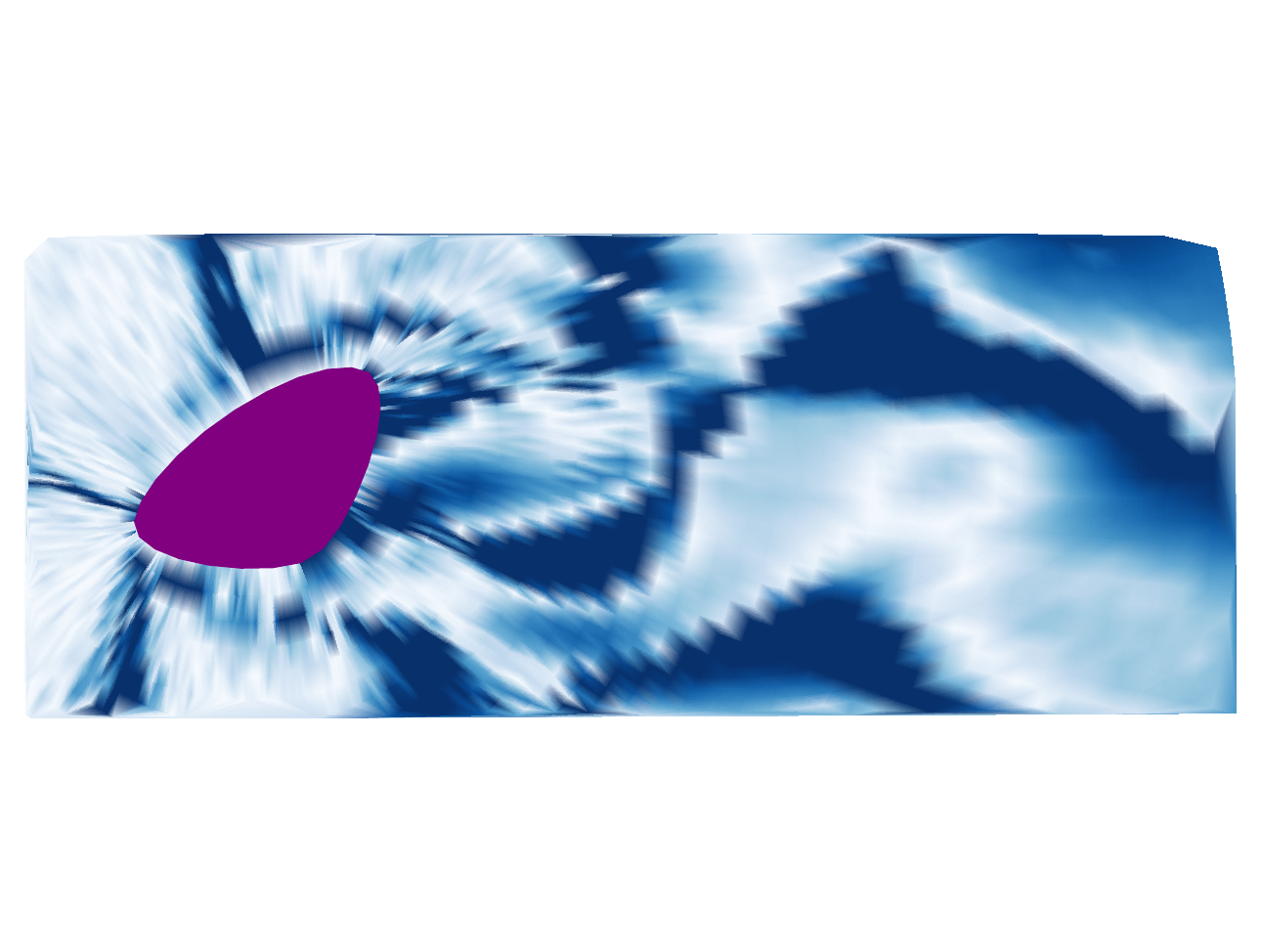}
    \end{minipage}                             &  \begin{minipage}{\spatialfigwidth\textwidth}
      \includegraphics[width=\linewidth]{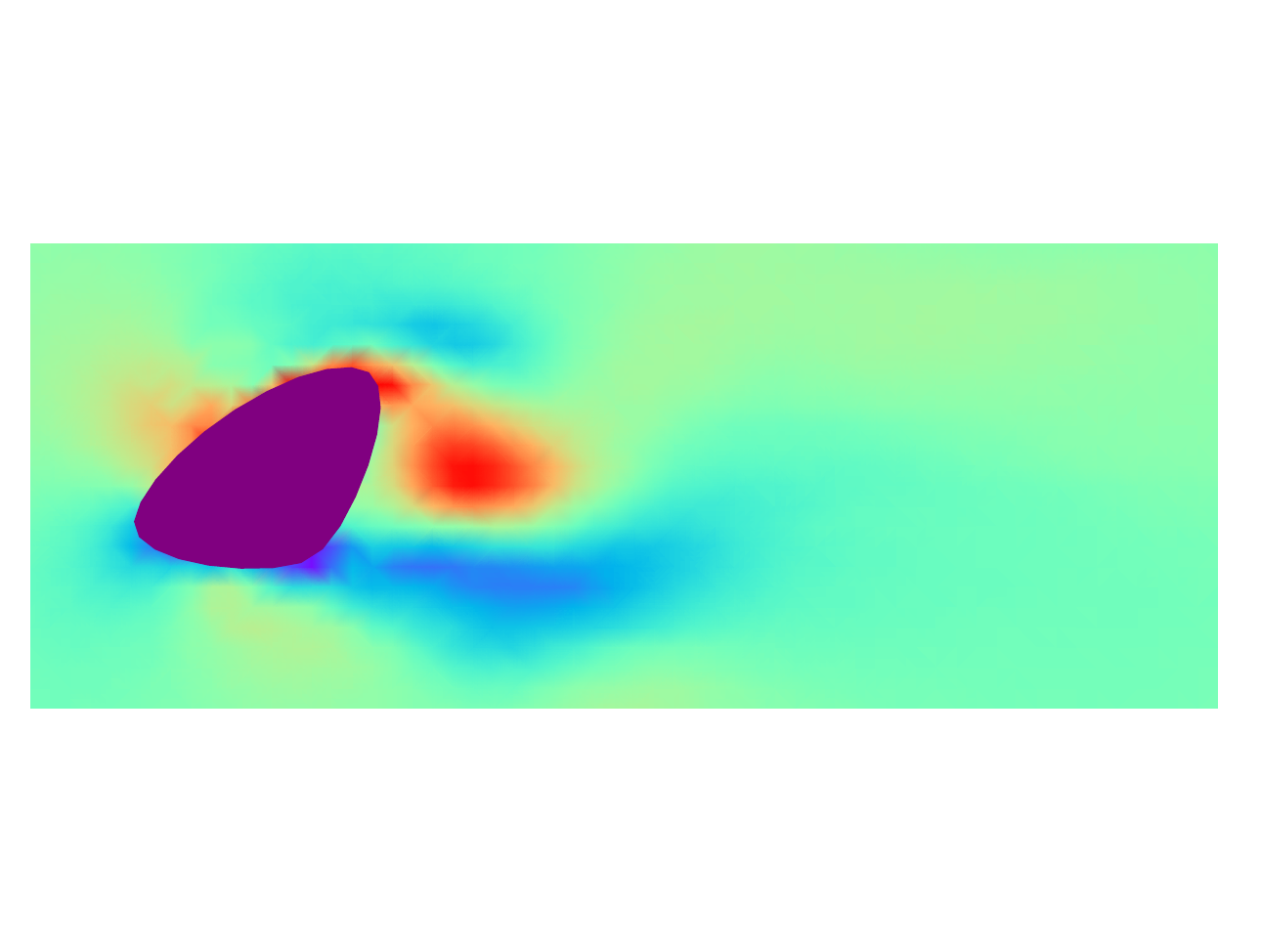}
    \end{minipage}                             & \begin{minipage}{\spatialfigwidth\textwidth}
      \includegraphics[width=\linewidth]{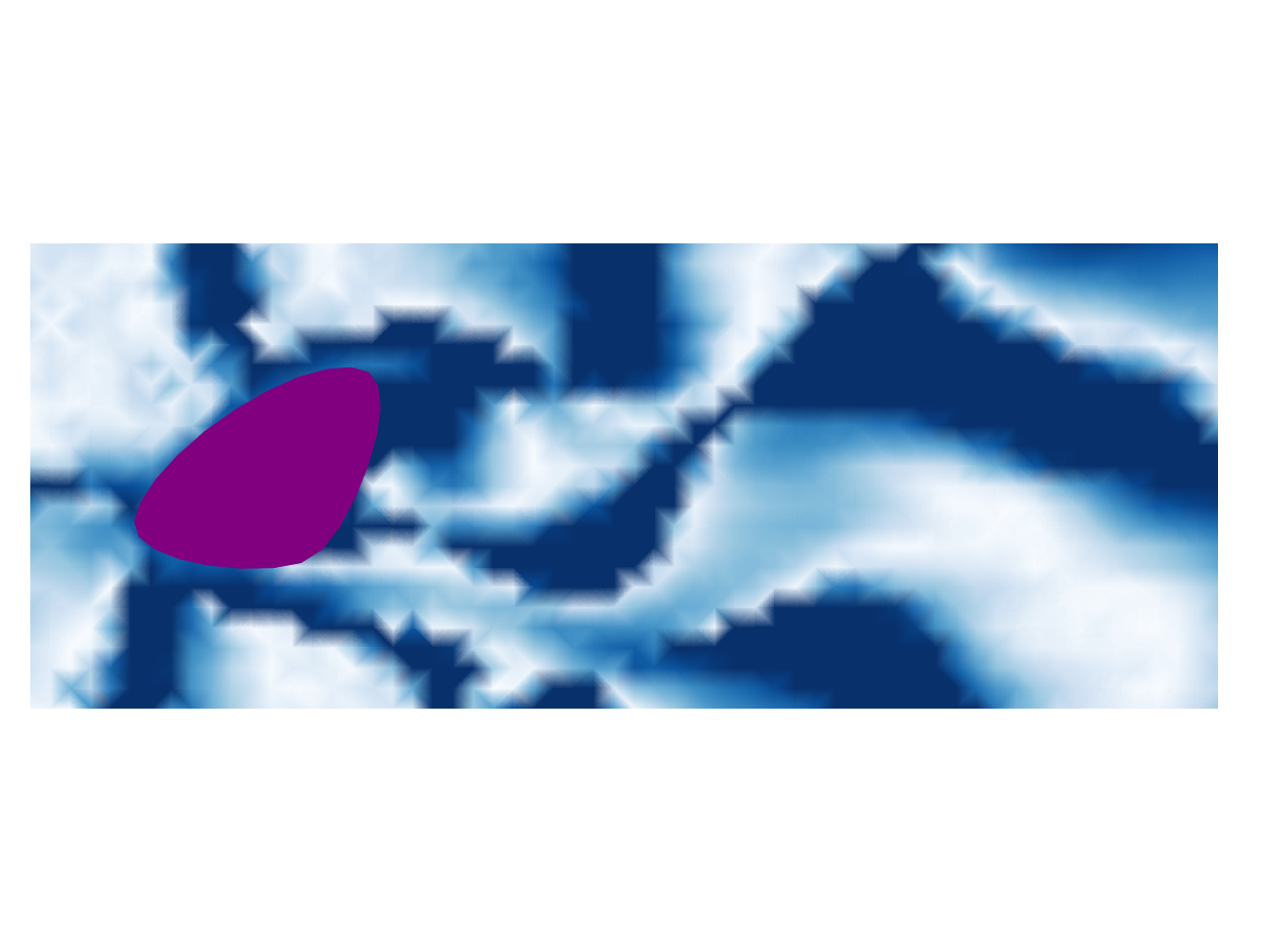}
    \end{minipage}         \\ 
\% Err    & \multicolumn{2}{c}{MAPE \textbf{41.59\%}, HV-MAPE \textbf{35.92\%}} & \multicolumn{2}{c}{MAPE 54.26\%, HV-MAPE 52.30\%}\\ \midrule
SD-Large               &    \begin{minipage}{\spatialfigwidth\textwidth}
      \includegraphics[width=\linewidth]{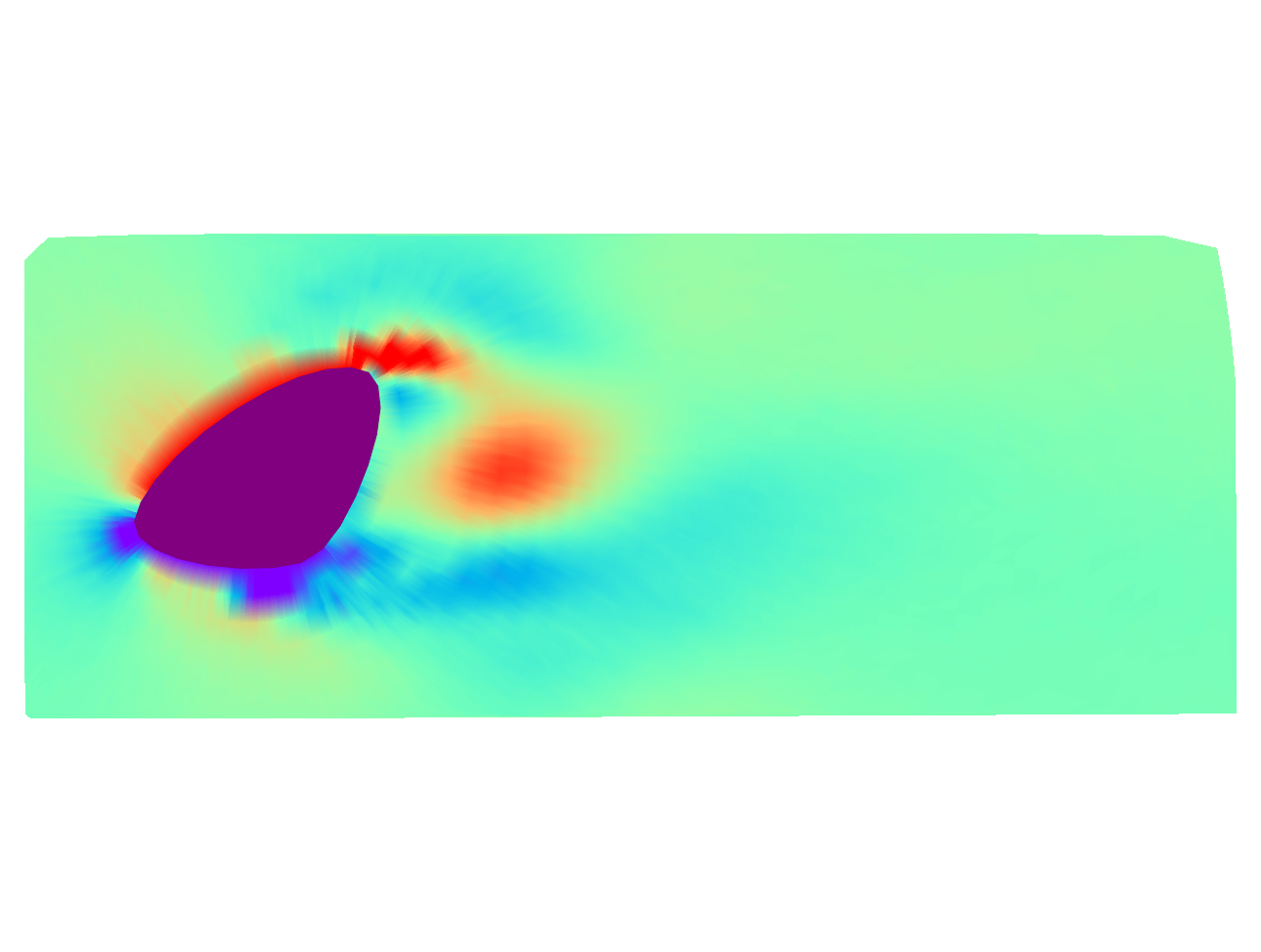}
    \end{minipage}                            &  \begin{minipage}{\spatialfigwidth\textwidth}
      \includegraphics[width=\linewidth]{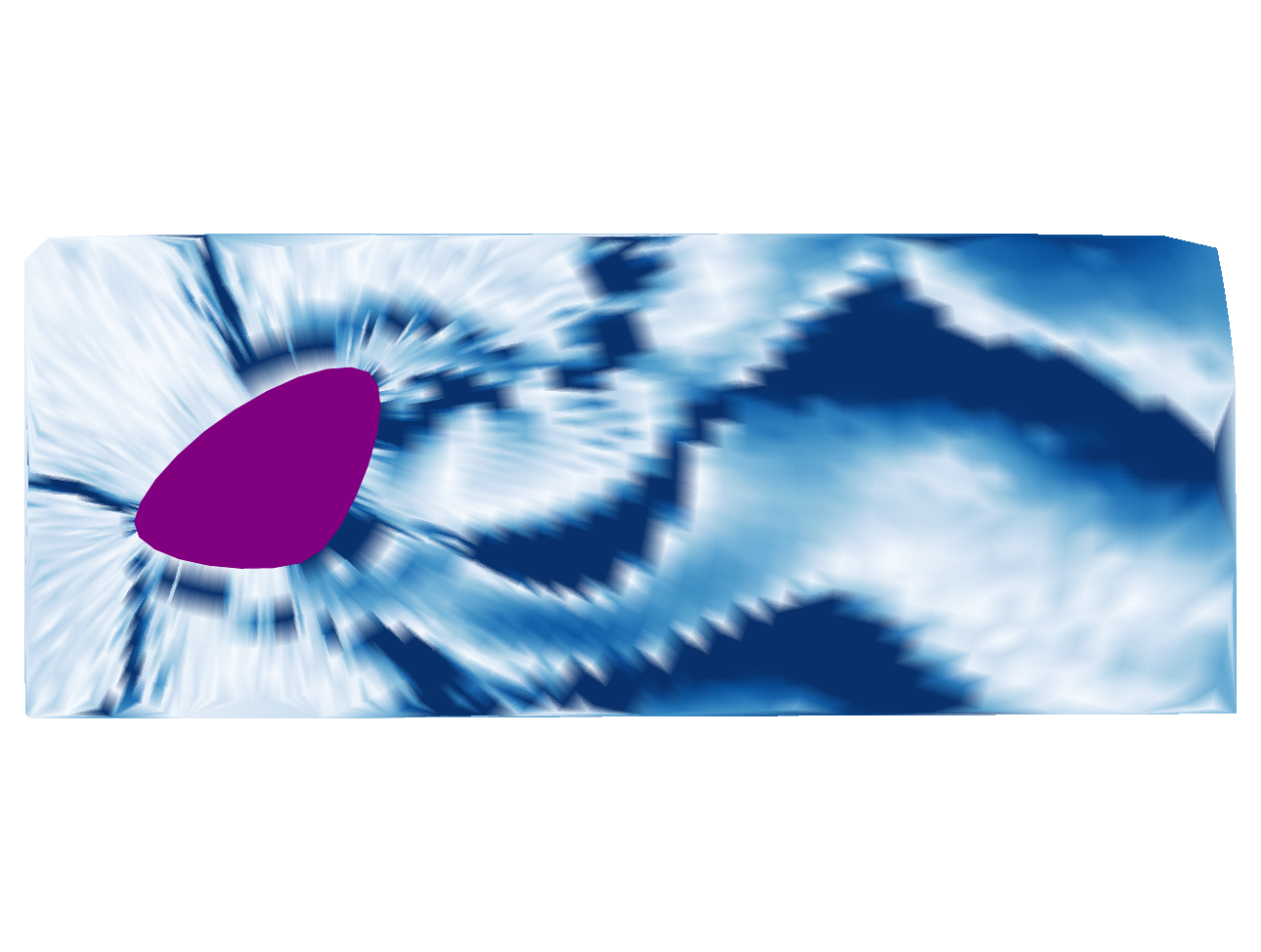}
    \end{minipage}                             &  \begin{minipage}{\spatialfigwidth\textwidth}
      \includegraphics[width=\linewidth]{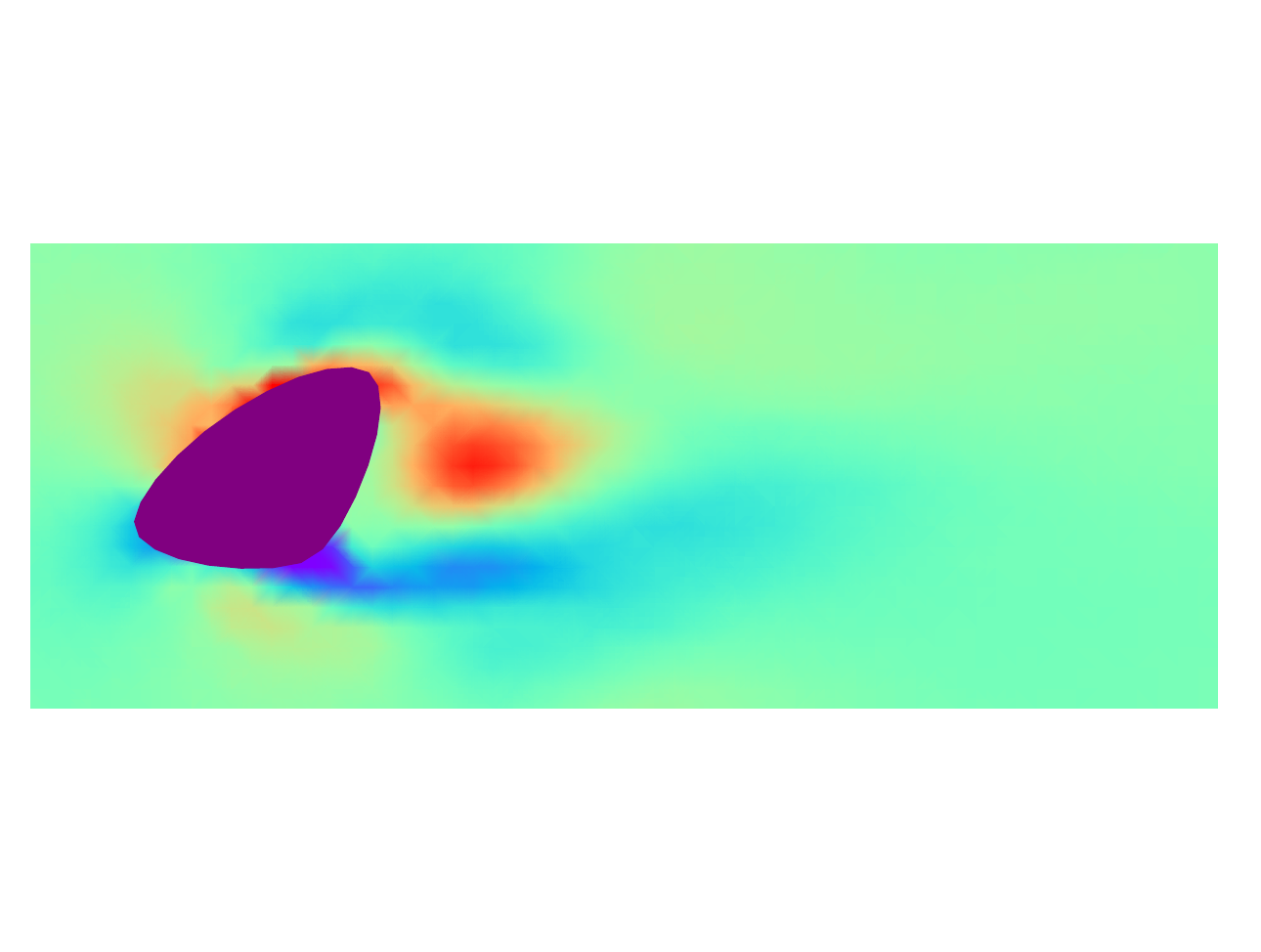}
    \end{minipage}                             & \begin{minipage}{\spatialfigwidth\textwidth}
      \includegraphics[width=\linewidth]{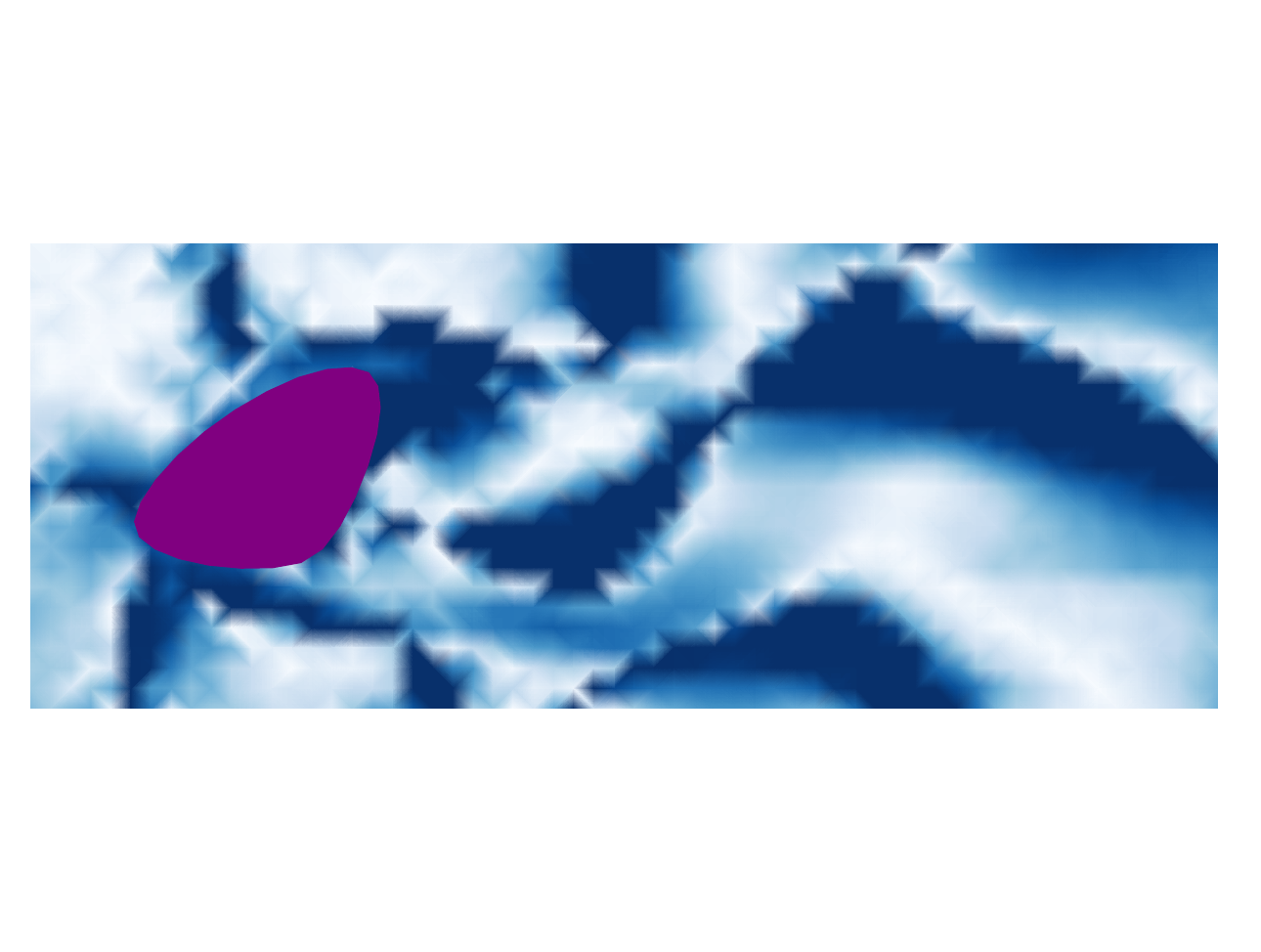}
    \end{minipage}          \\
\% Err    & \multicolumn{2}{c}{MAPE \textbf{38.90\%}, HV-MAPE \textbf{34.72\%}} & \multicolumn{2}{c}{MAPE 52.60\%, HV-MAPE 52.94\%}\\ \midrule
SD-UNet                &    \begin{minipage}{\spatialfigwidth\textwidth}
      \includegraphics[width=\linewidth]{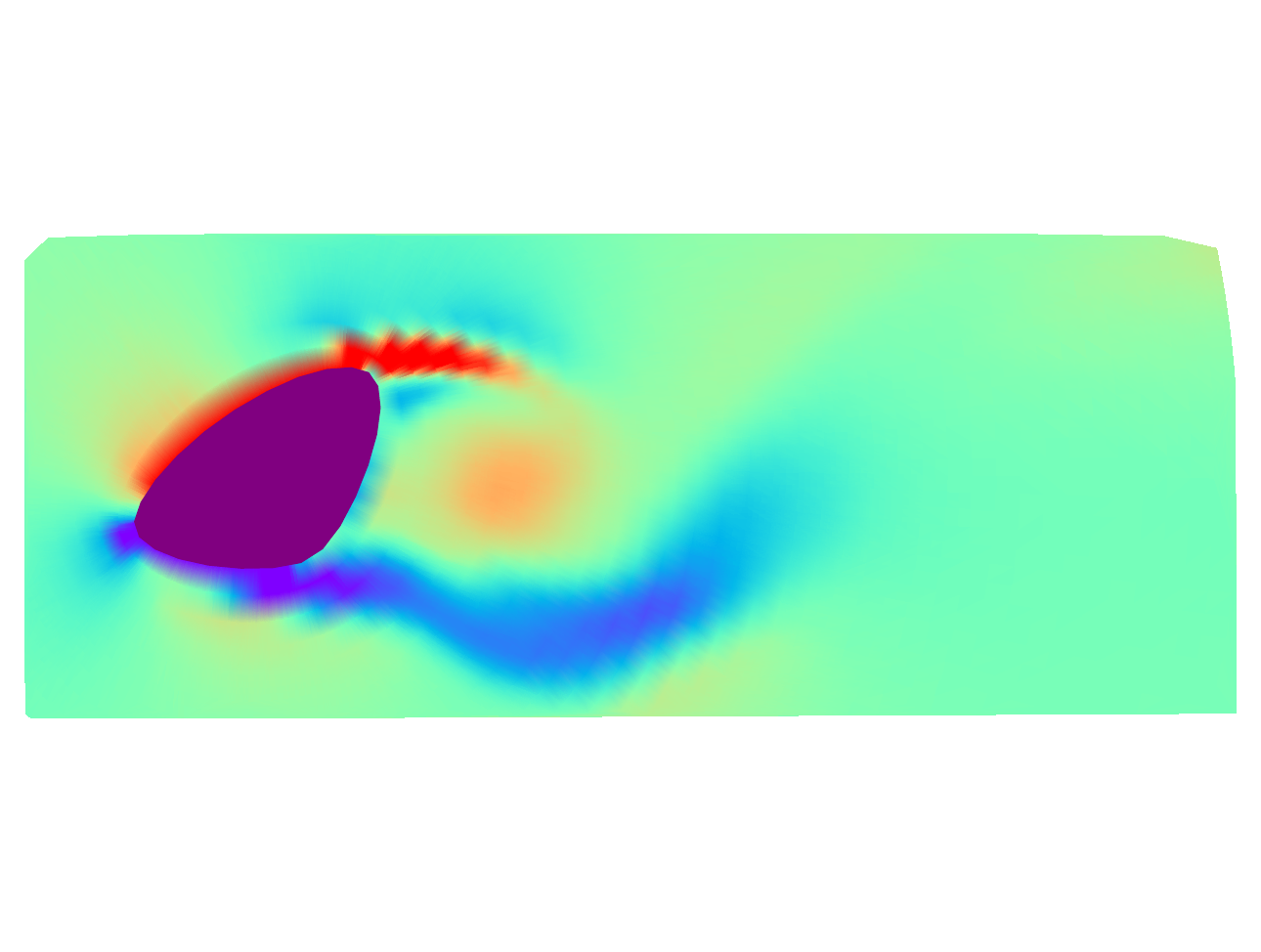}
    \end{minipage}                            &  \begin{minipage}{\spatialfigwidth\textwidth}
      \includegraphics[width=\linewidth]{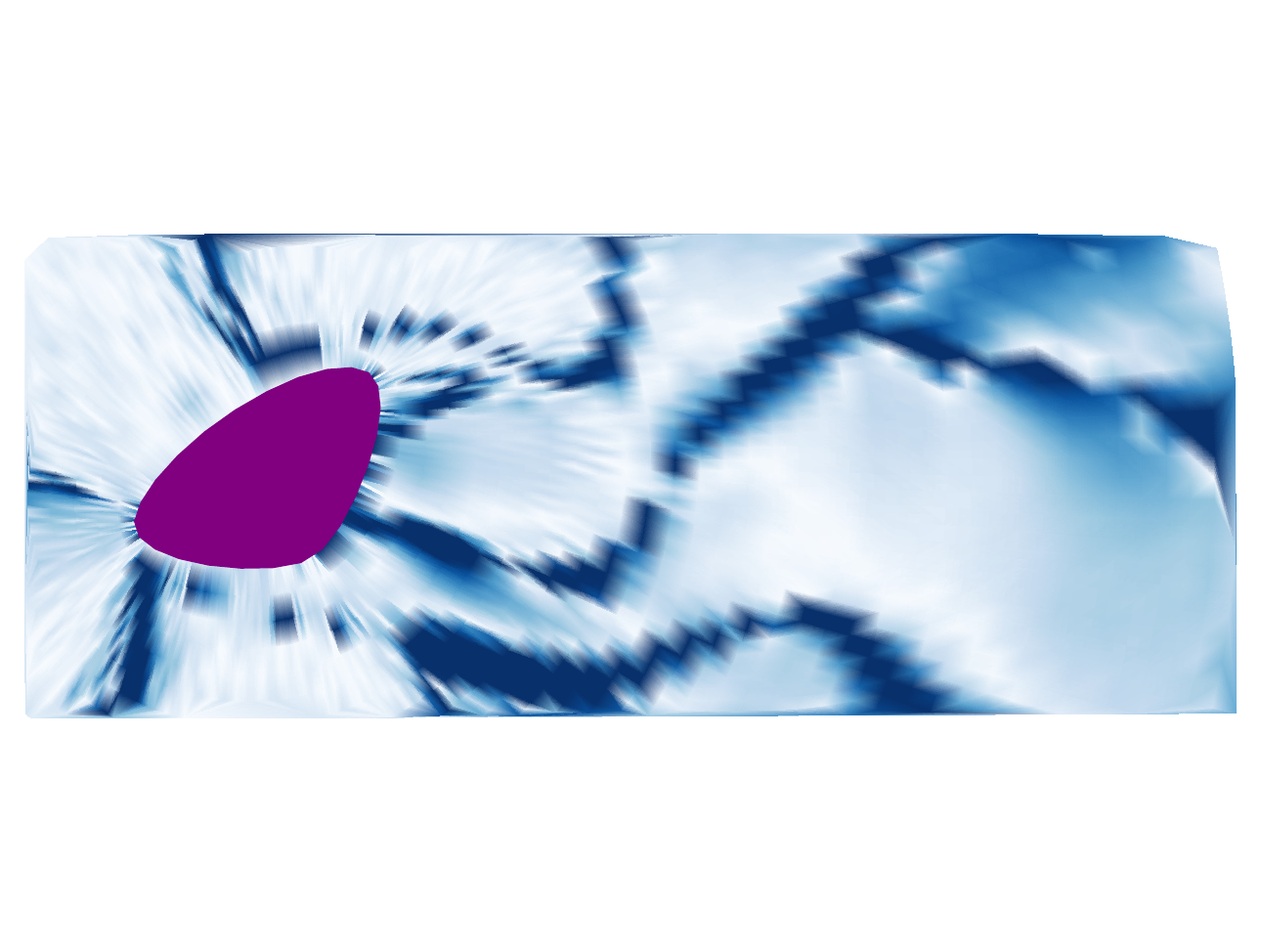}
    \end{minipage}                             &  \begin{minipage}{\spatialfigwidth\textwidth}
      \includegraphics[width=\linewidth]{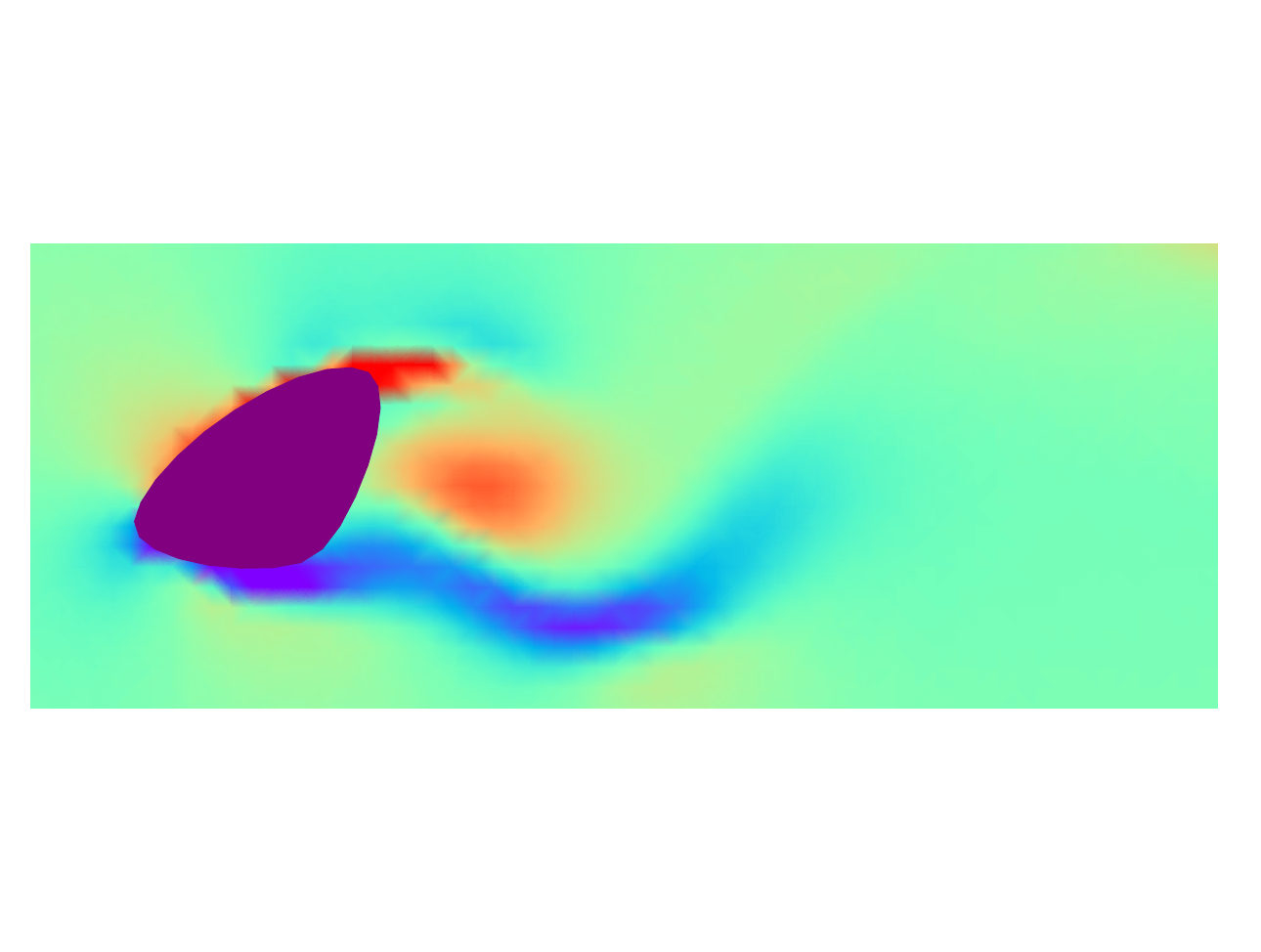}
    \end{minipage}                             & \begin{minipage}{\spatialfigwidth\textwidth}
      \includegraphics[width=\linewidth]{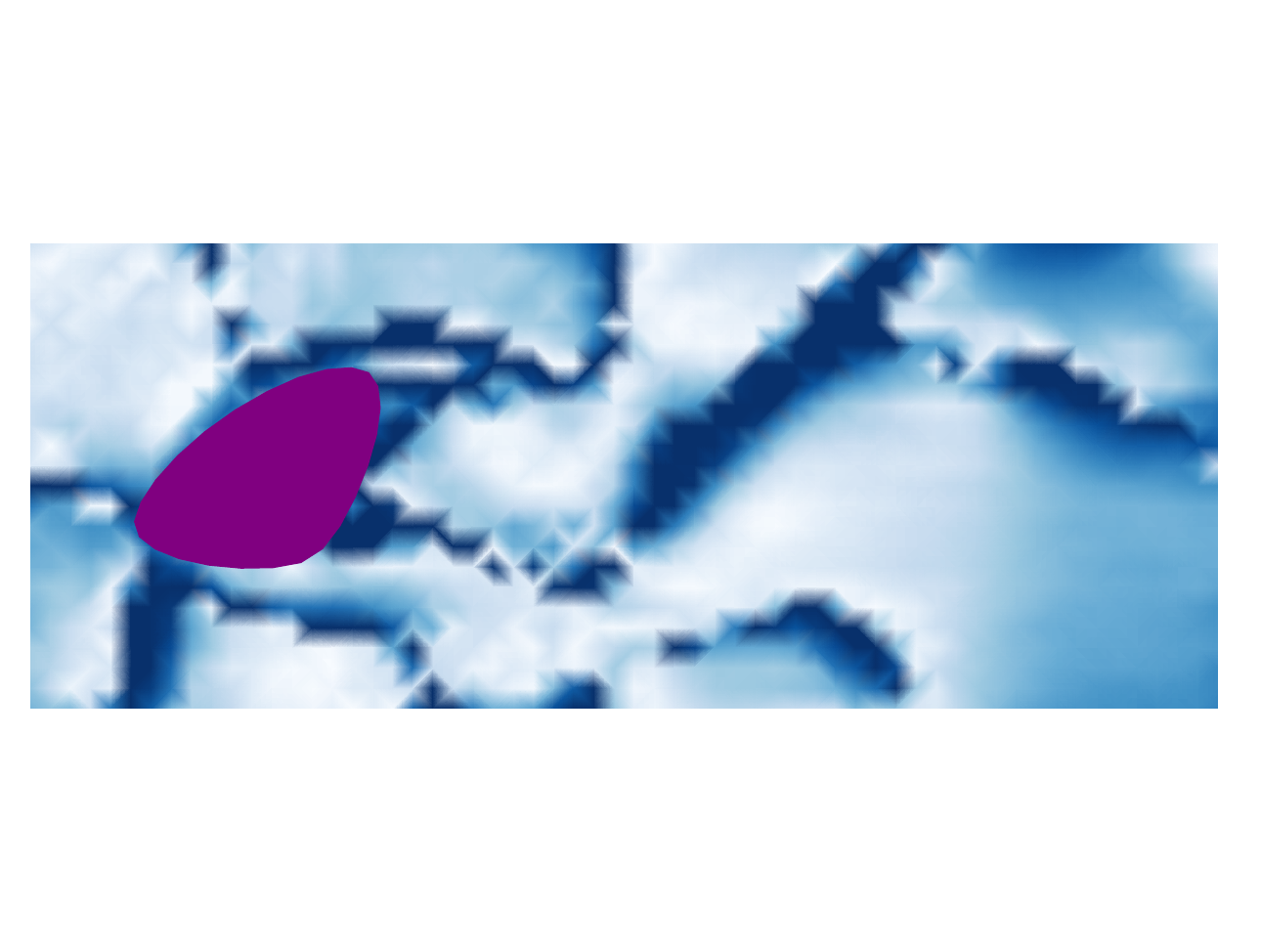}
    \end{minipage}          \\
\% Err    & \multicolumn{2}{c}{MAPE \textit{\textbf{30.06\%}}, HV-MAPE {\ul \textbf{23.23\%}}} & \multicolumn{2}{c}{MAPE 43.25\%, HV-MAPE 39.12\%}\\ \midrule
SD-FNO                 &    \begin{minipage}{\spatialfigwidth\textwidth}
      \includegraphics[width=\linewidth]{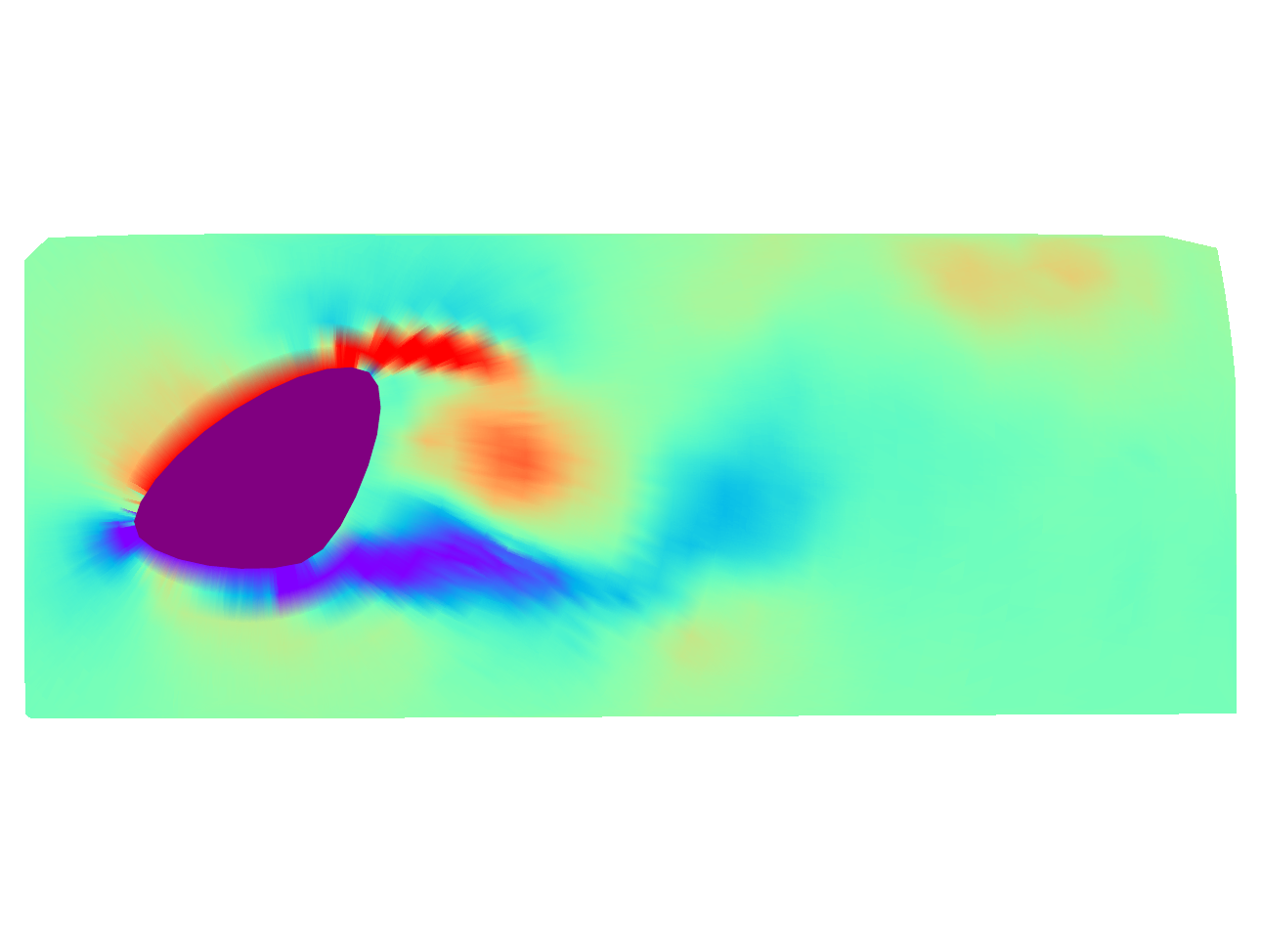}
    \end{minipage}                            &  \begin{minipage}{\spatialfigwidth\textwidth}
      \includegraphics[width=\linewidth]{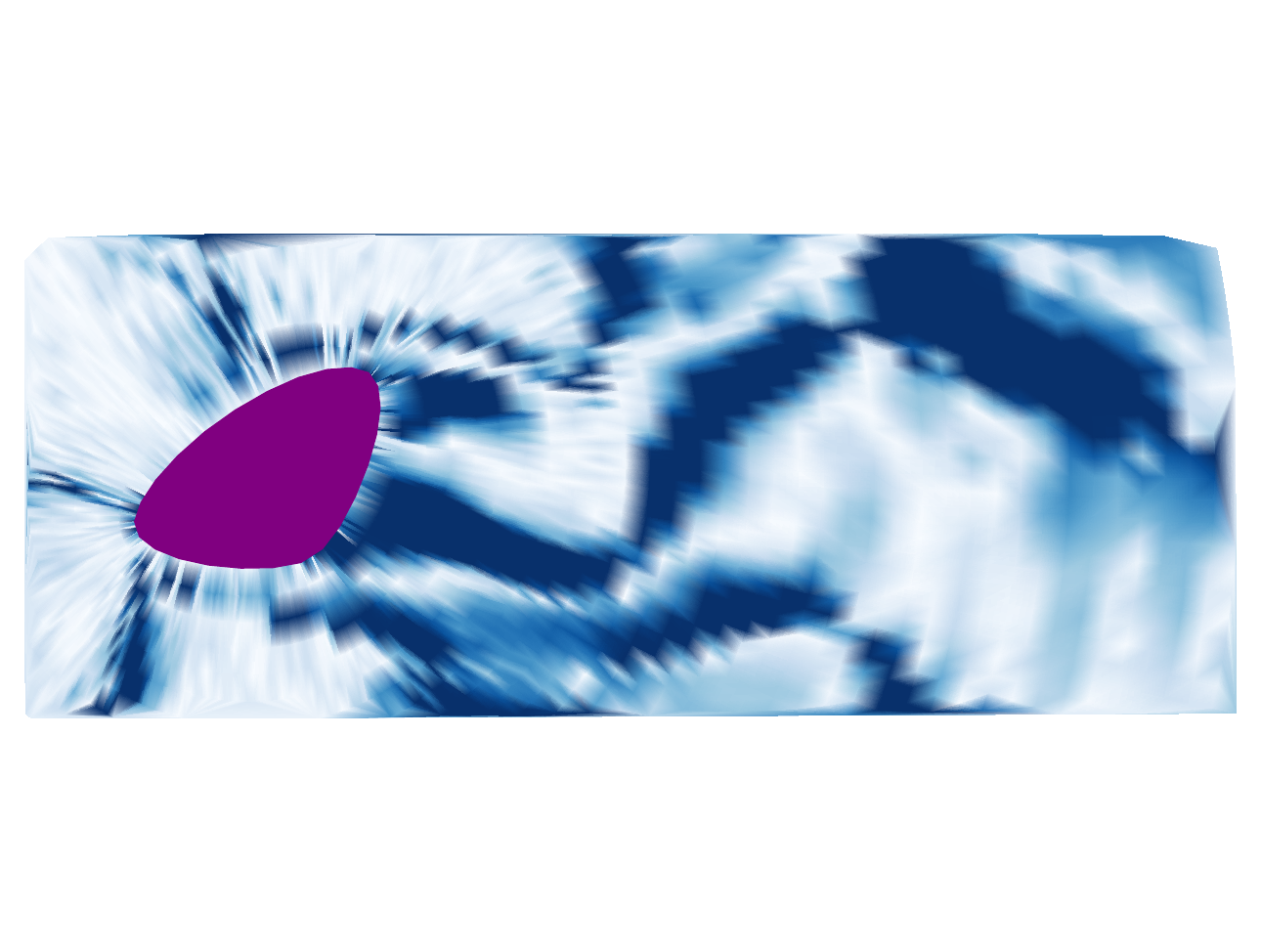}
    \end{minipage}                             &  \begin{minipage}{\spatialfigwidth\textwidth}
      \includegraphics[width=\linewidth]{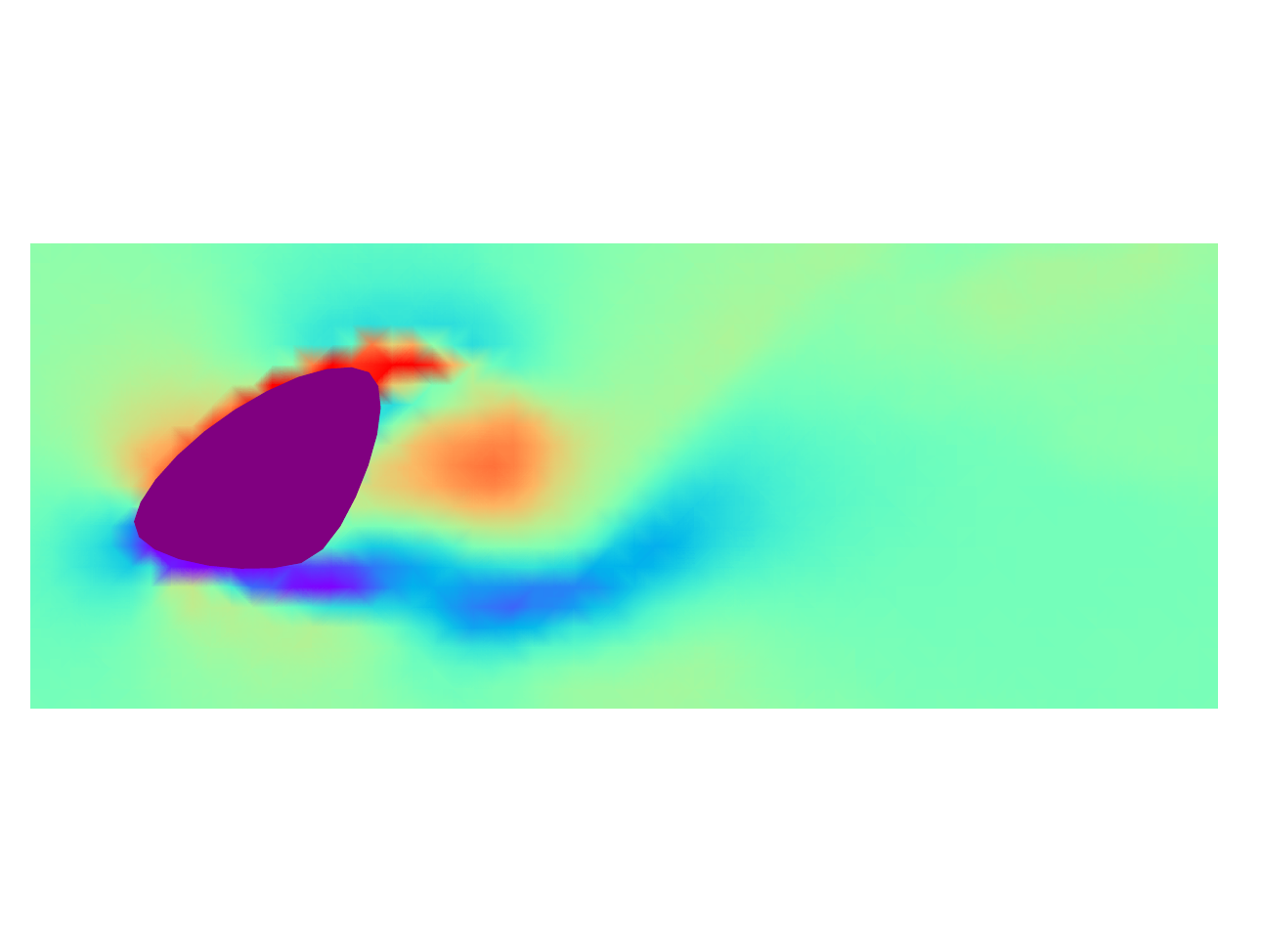}
    \end{minipage}                             & \begin{minipage}{\spatialfigwidth\textwidth}
      \includegraphics[width=\linewidth]{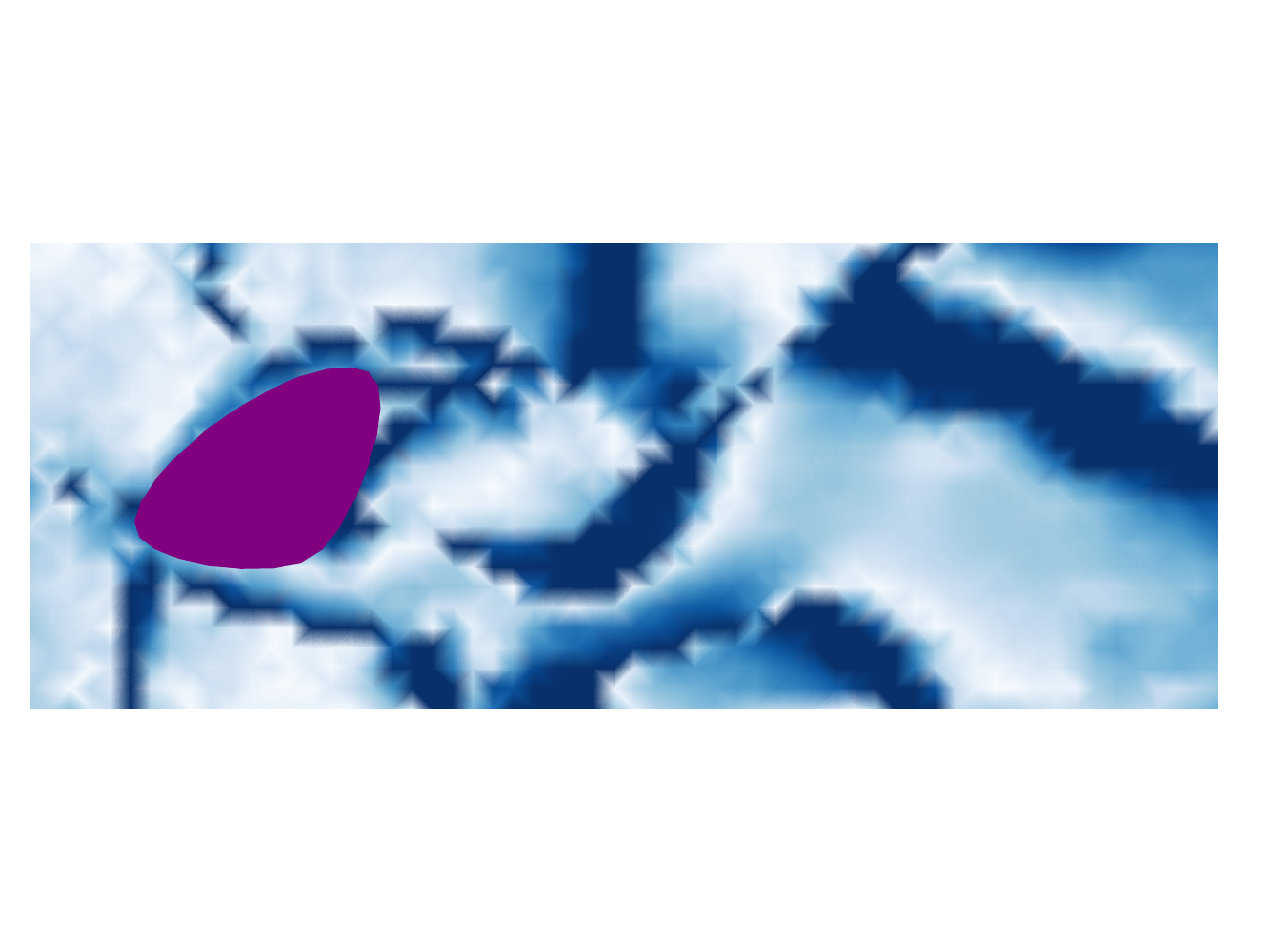}
    \end{minipage}          \\
\% Err    & \multicolumn{2}{c}{MAPE \textbf{37.76\%}, HV-MAPE \textbf{34.61\%}} & \multicolumn{2}{c}{MAPE 48.19\%, HV-MAPE 43.76\%}\\ \bottomrule
\end{tabular}
    
    \caption{Ground truth (top) and predicted (bottom) vorticity fields and percentage error maps for a snapshot from Shape A, using the medium sensor setup. Vorticity colormap constrained to 7.5\% of $\max(|\tilde{\omega}_{t,i}|)$.}
    \label{fig:spatialex_s1660_medium}
\end{figure}

\begin{figure}
    \centering
    \includegraphics[width=0.40\textwidth]{images/spatial/annulus_large_s1660_gt_vorticity.pdf}
    \includegraphics[width=0.045\textwidth]{images/spatial/pcterrorbarr0.pdf}
\begin{tabular}{@{}lcccc@{}}
\toprule
\multirow{2}{*}{Model} & \multicolumn{2}{c|}{Annular Sampling}                        & \multicolumn{2}{c}{Cartesian Sampling}      \\ \cmidrule(l){2-5} 
                       & \multicolumn{1}{c|}{Vorticity} & \multicolumn{1}{c|}{\% Error} & \multicolumn{1}{c|}{Vorticity} & \% Error \\ \midrule
SD                     &    \begin{minipage}{\spatialfigwidth\textwidth}
      \includegraphics[width=\linewidth]{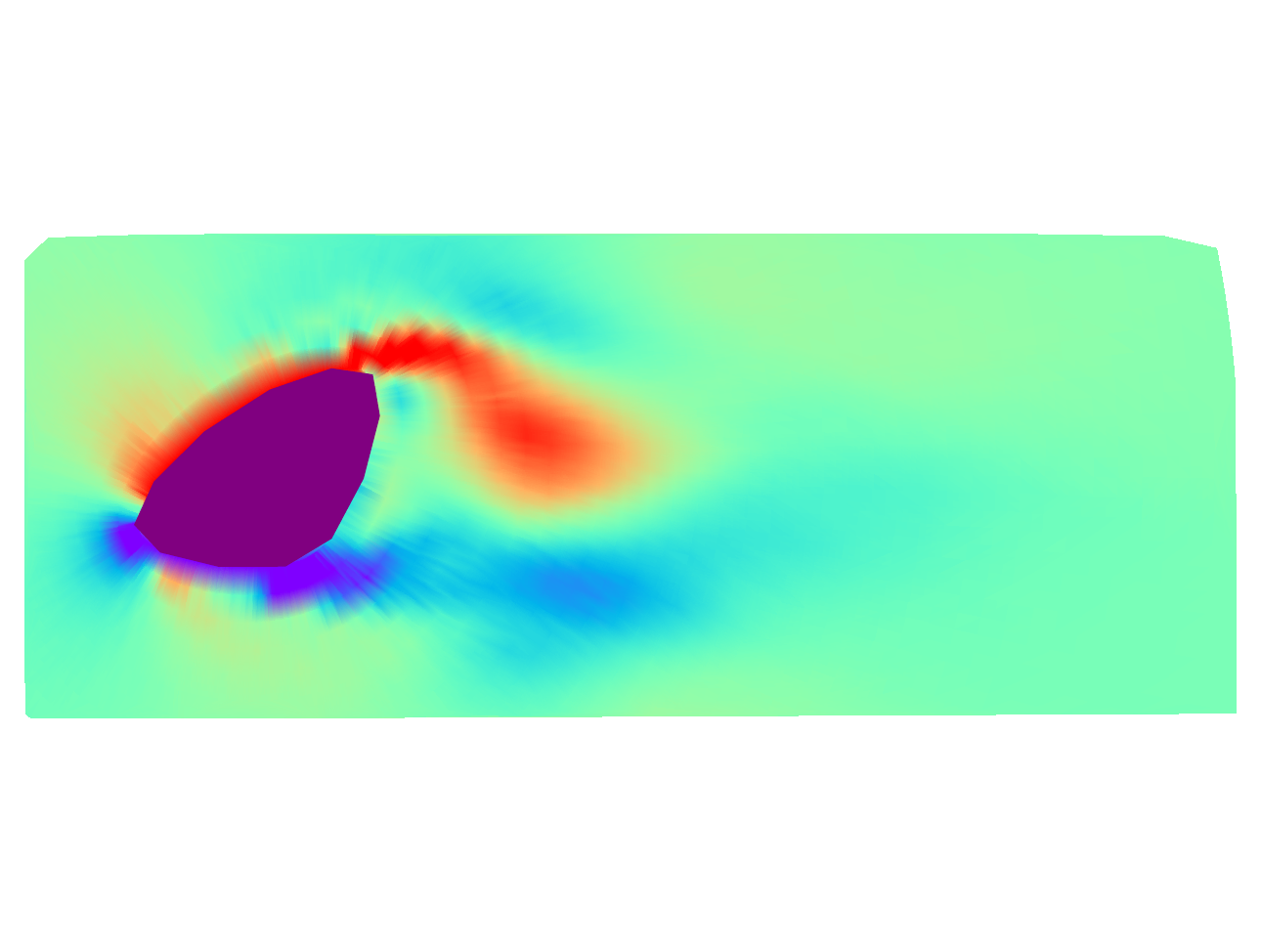}
    \end{minipage}                            &  \begin{minipage}{\spatialfigwidth\textwidth}
      \includegraphics[width=\linewidth]{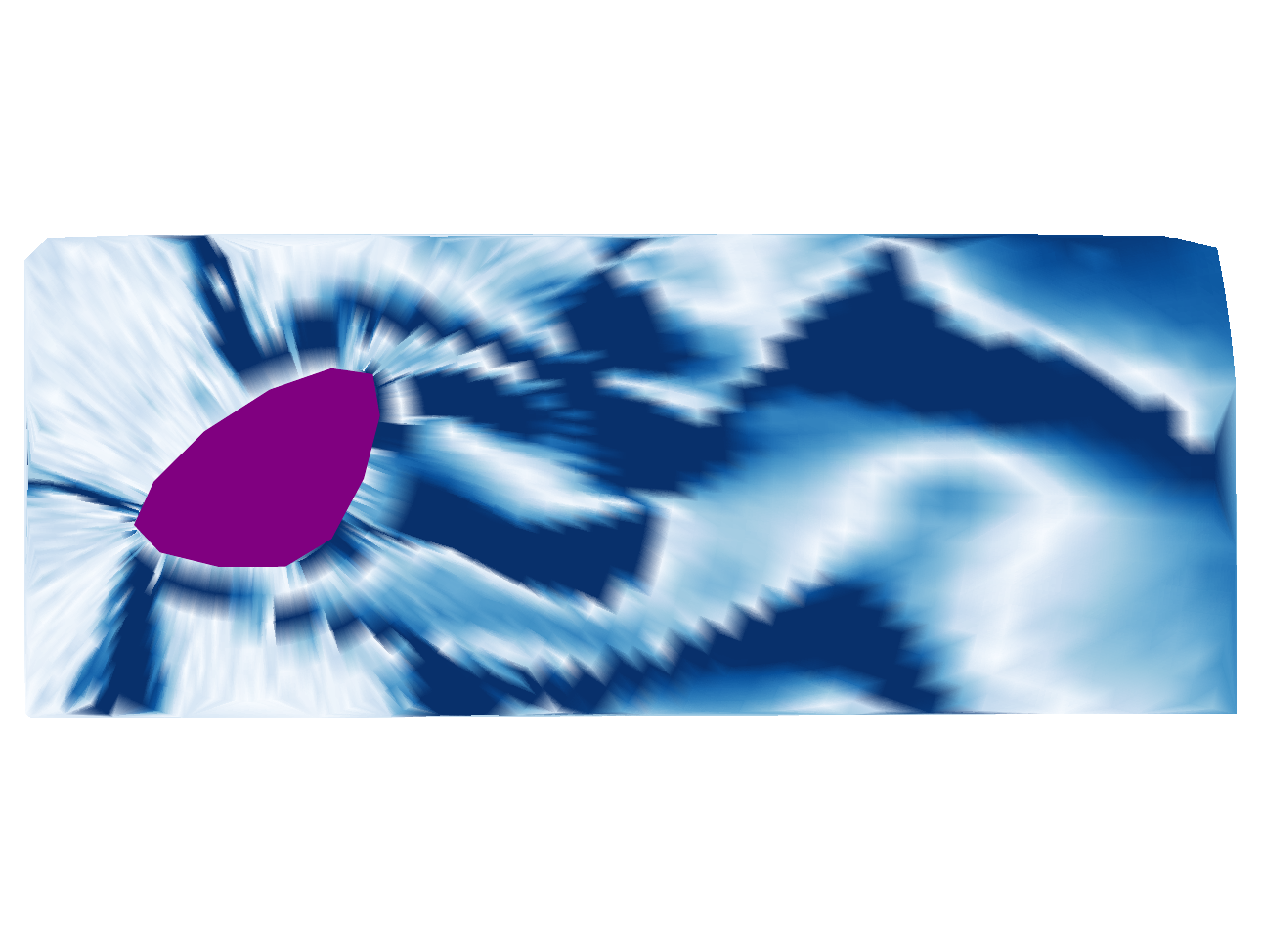}
    \end{minipage}                             &  \begin{minipage}{\spatialfigwidth\textwidth}
      \includegraphics[width=\linewidth]{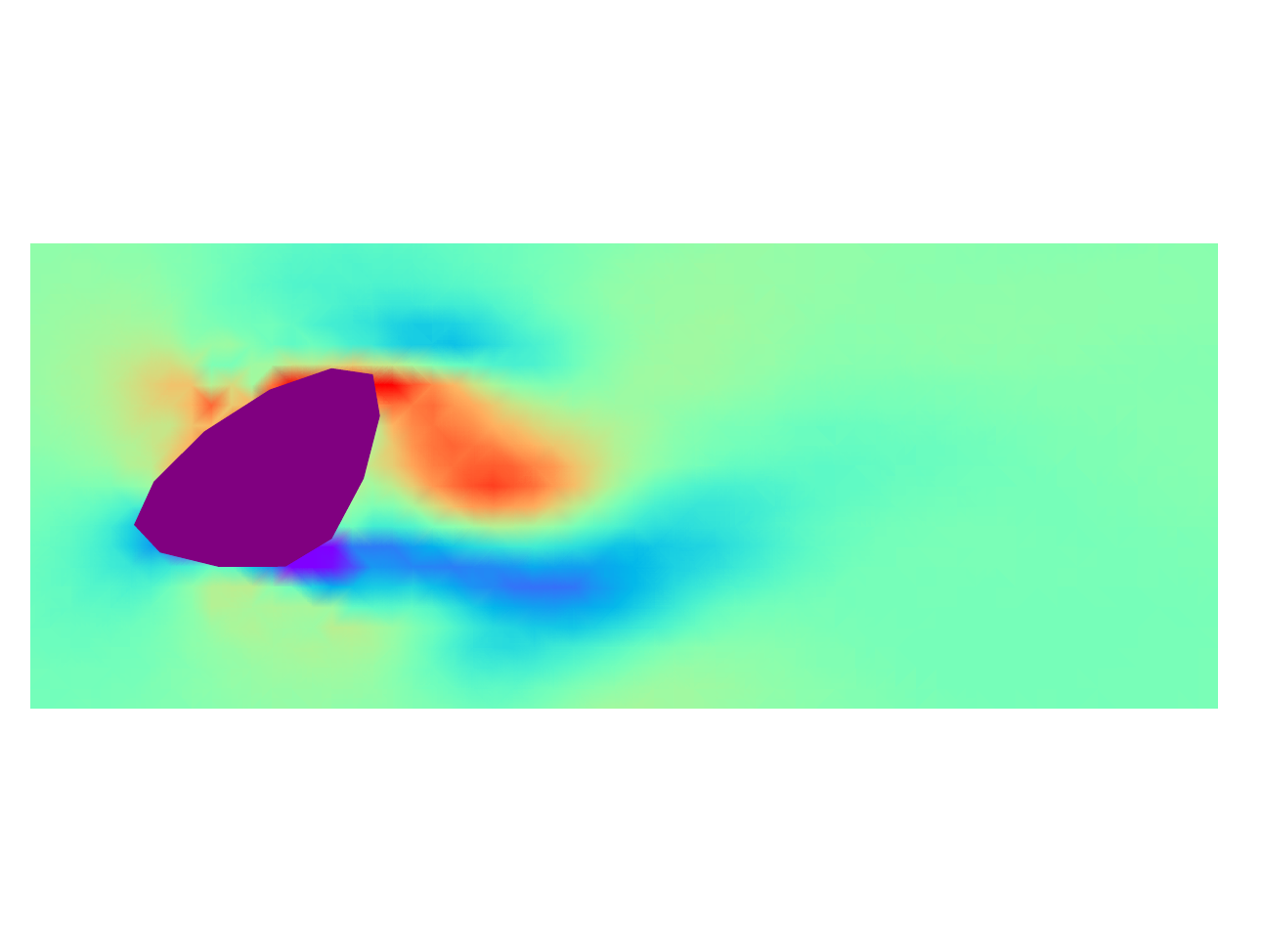}
    \end{minipage}                             & \begin{minipage}{\spatialfigwidth\textwidth}
      \includegraphics[width=\linewidth]{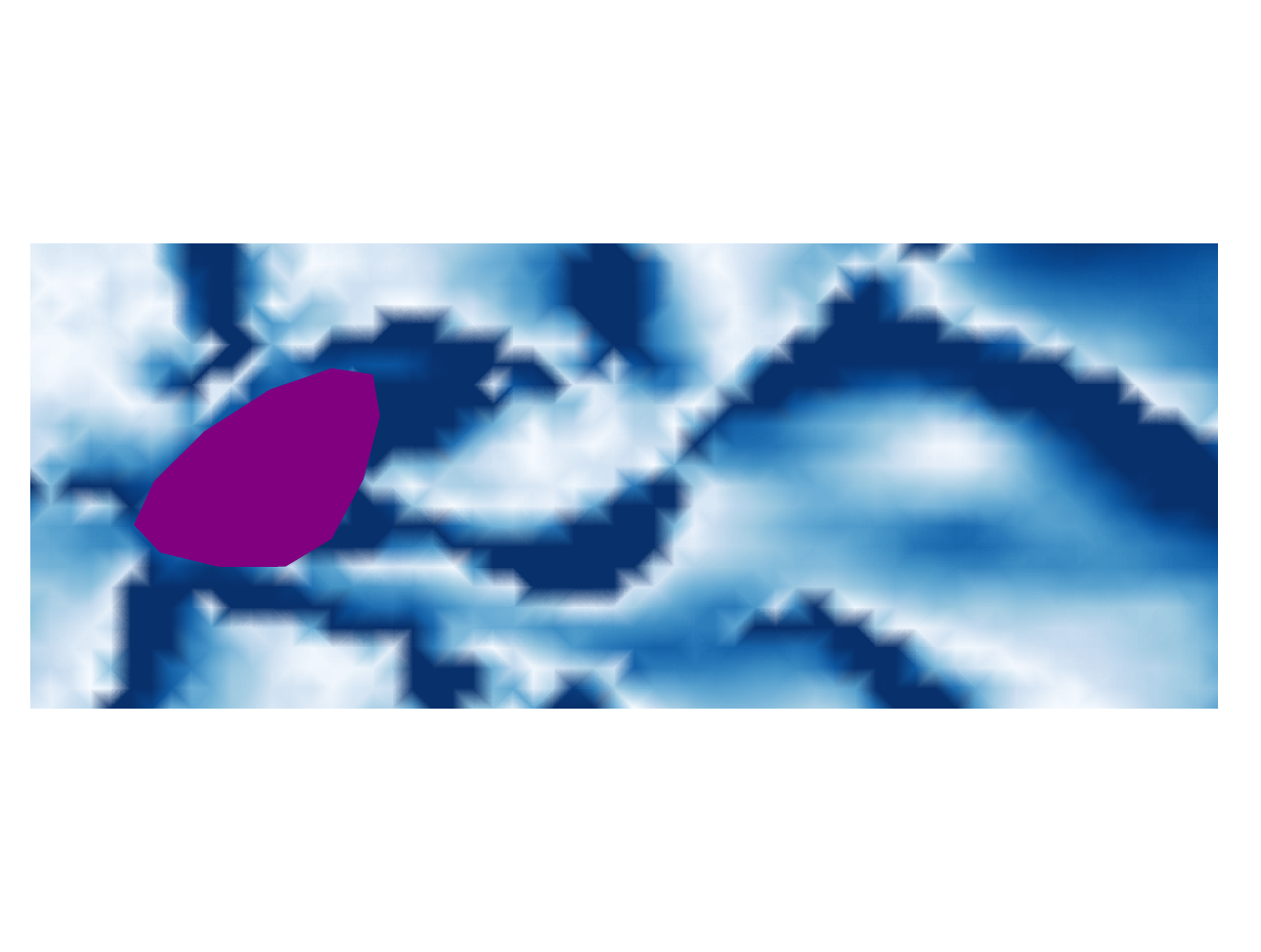}
    \end{minipage}         \\ 
\% Err    & \multicolumn{2}{c}{MAPE \textbf{47.19\%}, HV-MAPE \textbf{43.91\%}} & \multicolumn{2}{c}{MAPE 54.63\%, HV-MAPE 54.52\%}\\ \midrule
SD-Large               &    \begin{minipage}{\spatialfigwidth\textwidth}
      \includegraphics[width=\linewidth]{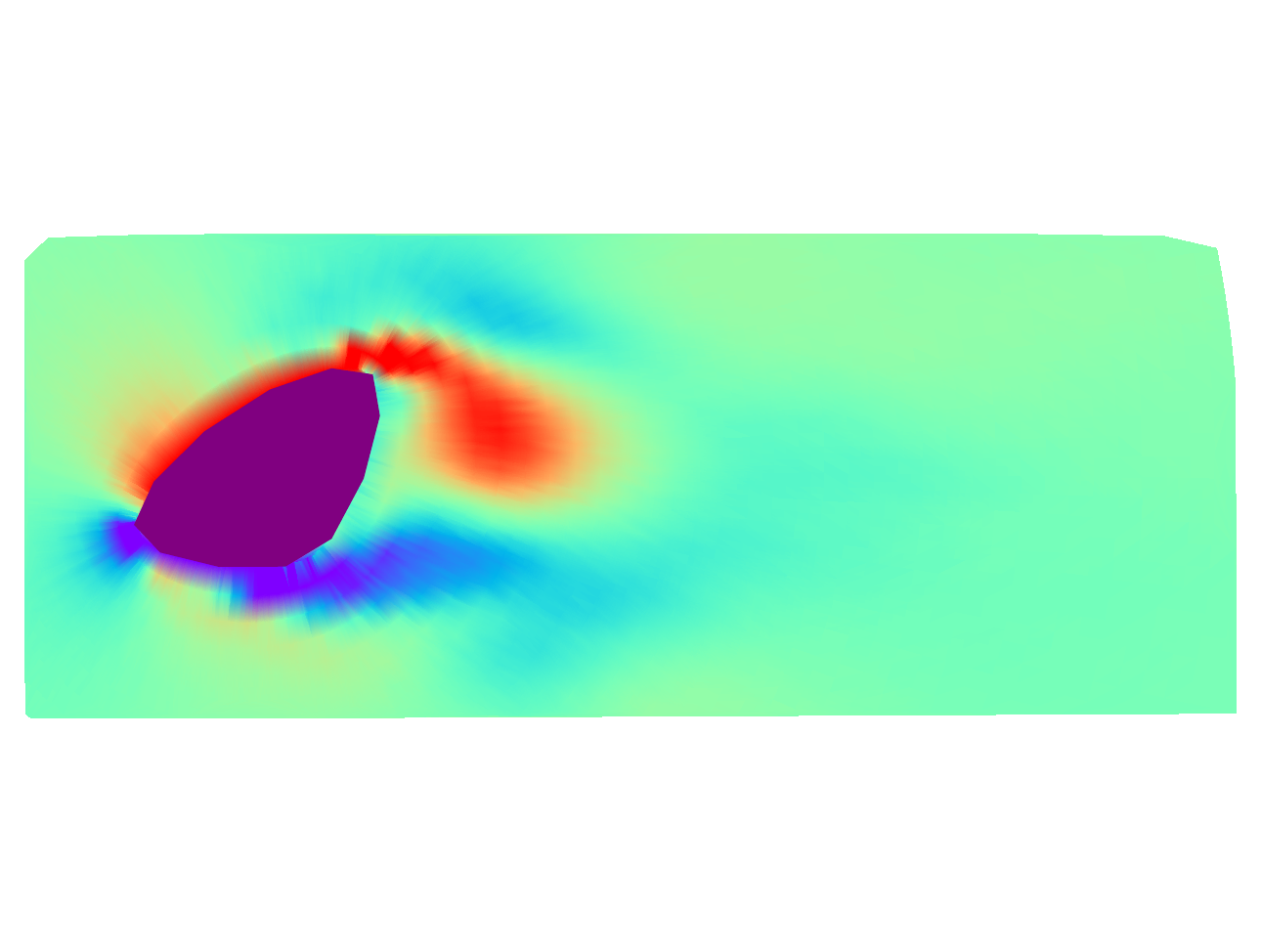}
    \end{minipage}                            &  \begin{minipage}{\spatialfigwidth\textwidth}
      \includegraphics[width=\linewidth]{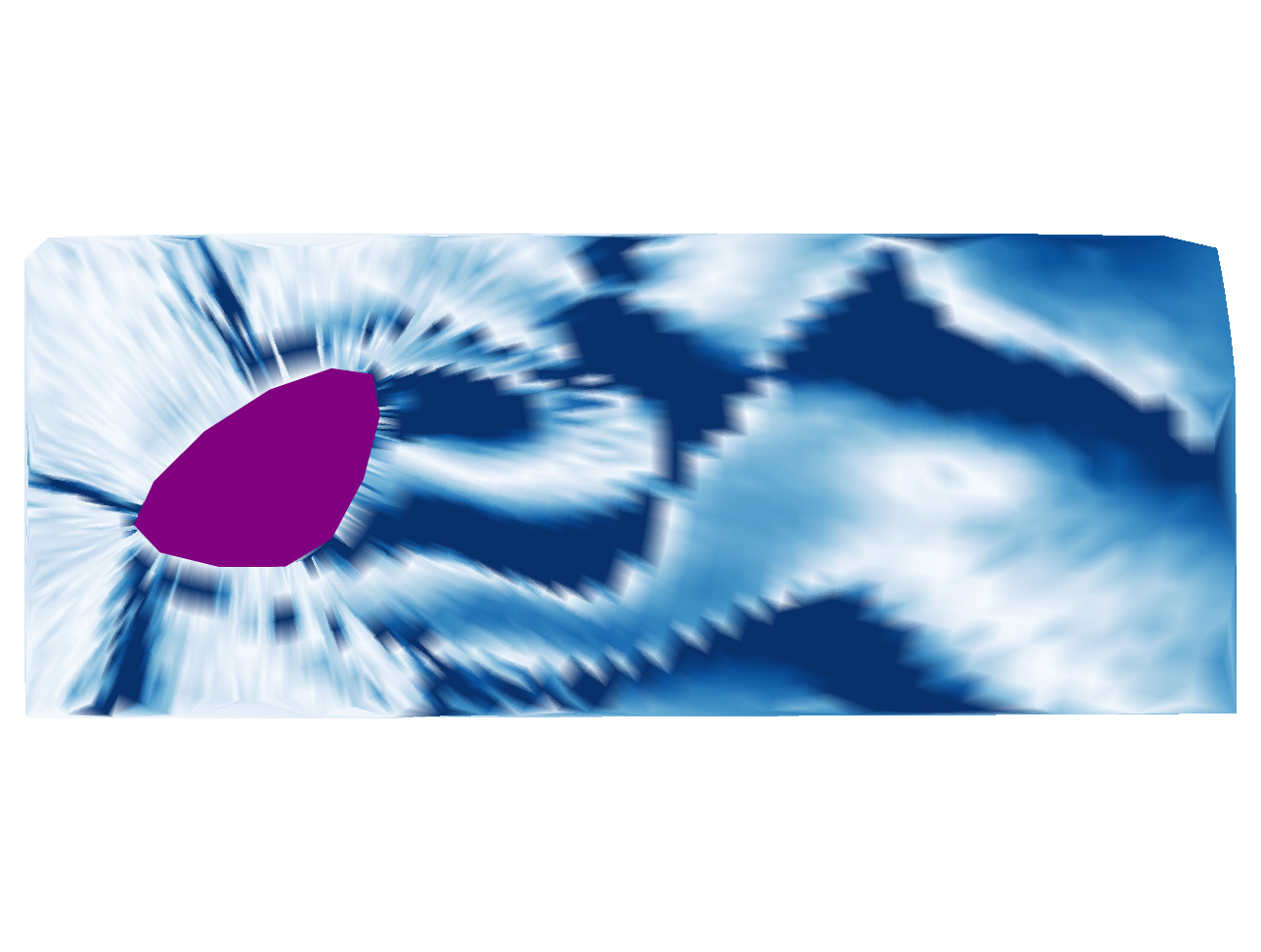}
    \end{minipage}                             &  \begin{minipage}{\spatialfigwidth\textwidth}
      \includegraphics[width=\linewidth]{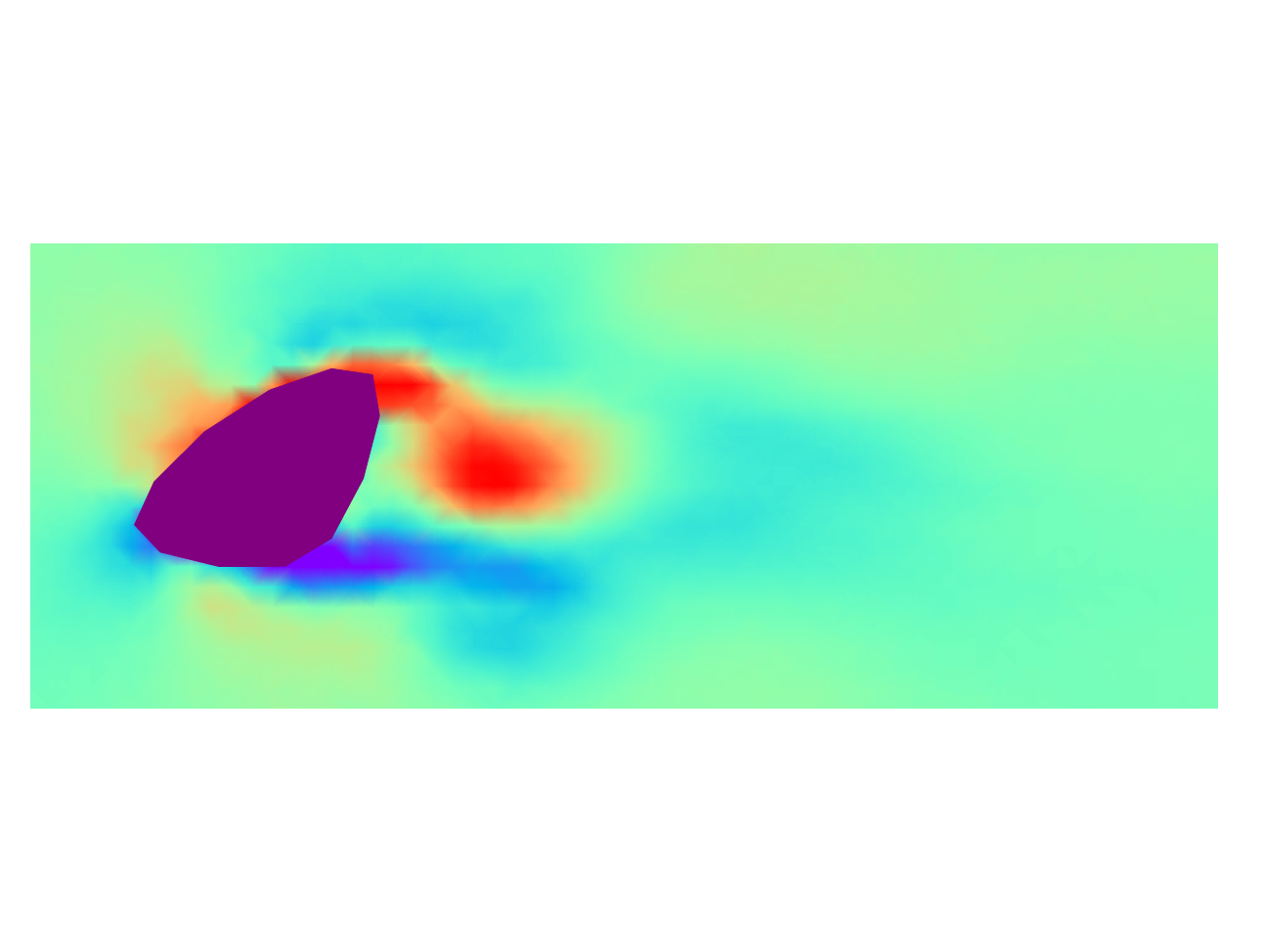}
    \end{minipage}                             & \begin{minipage}{\spatialfigwidth\textwidth}
      \includegraphics[width=\linewidth]{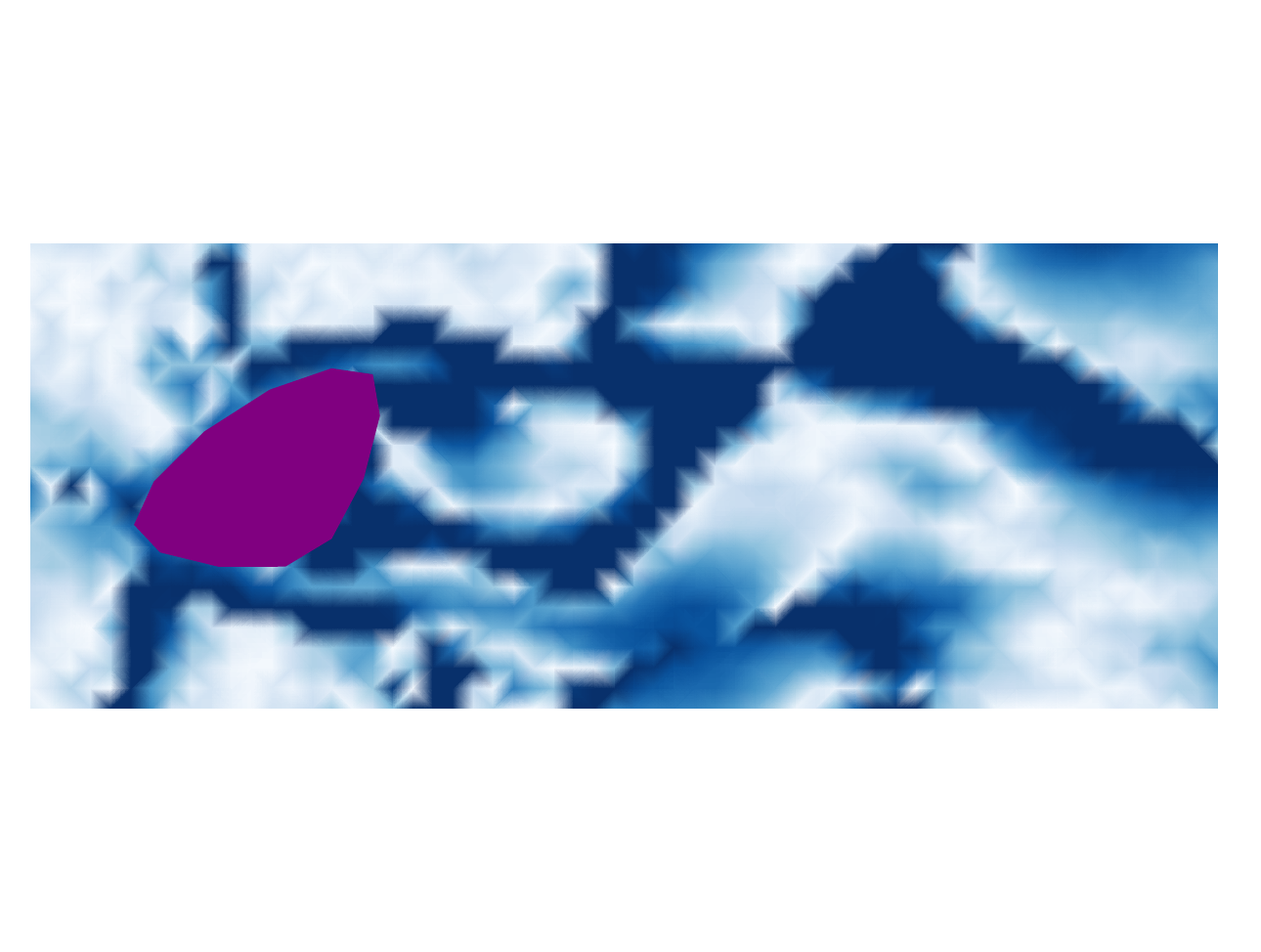}
    \end{minipage}          \\
\% Err    & \multicolumn{2}{c}{MAPE \textbf{42.69\%}, HV-MAPE \textbf{37.49\%}} & \multicolumn{2}{c}{MAPE 53.92\%, HV-MAPE 49.89\%}\\ \midrule
SD-UNet                &    \begin{minipage}{\spatialfigwidth\textwidth}
      \includegraphics[width=\linewidth]{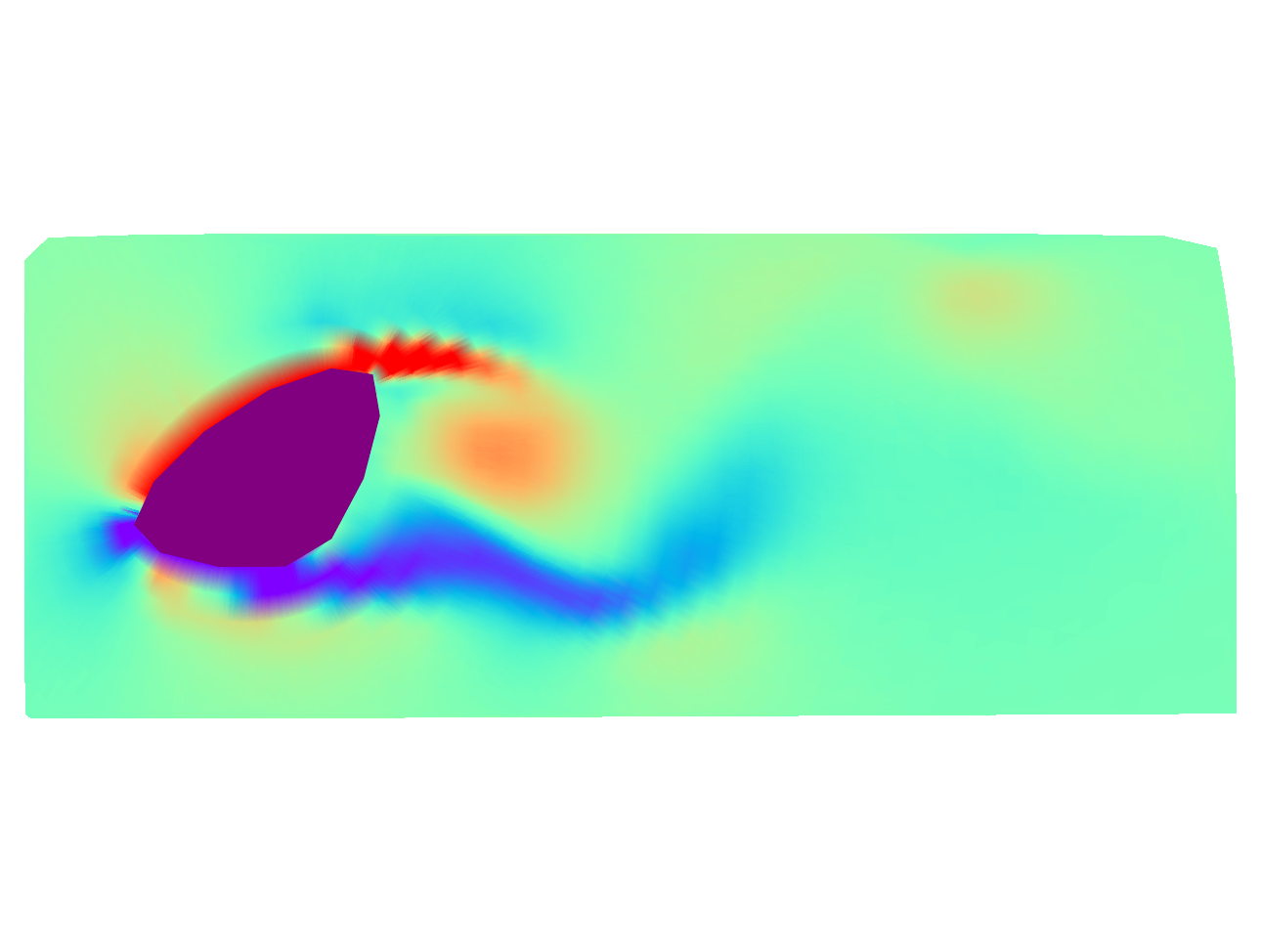}
    \end{minipage}                            &  \begin{minipage}{\spatialfigwidth\textwidth}
      \includegraphics[width=\linewidth]{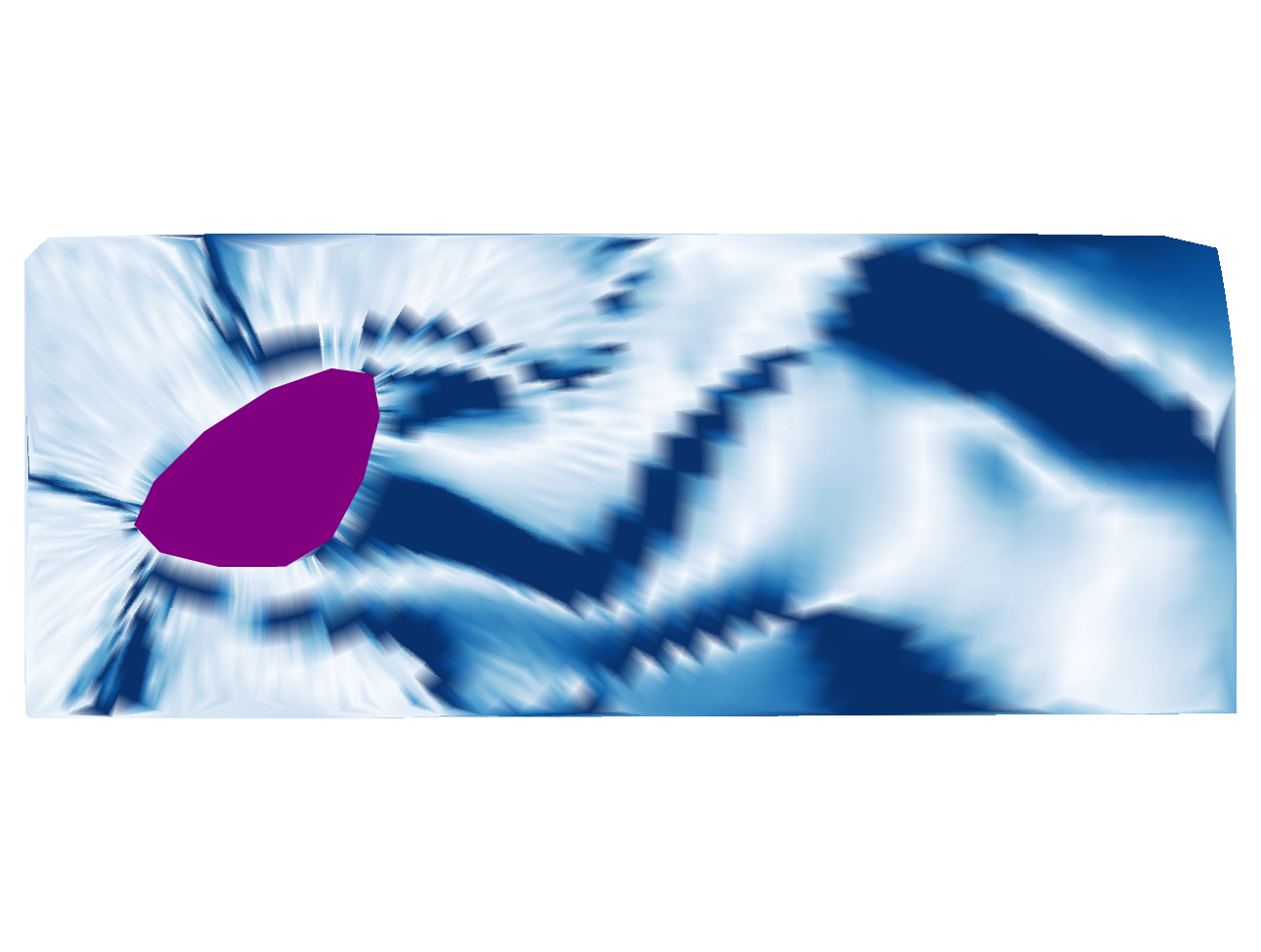}
    \end{minipage}                             &  \begin{minipage}{\spatialfigwidth\textwidth}
      \includegraphics[width=\linewidth]{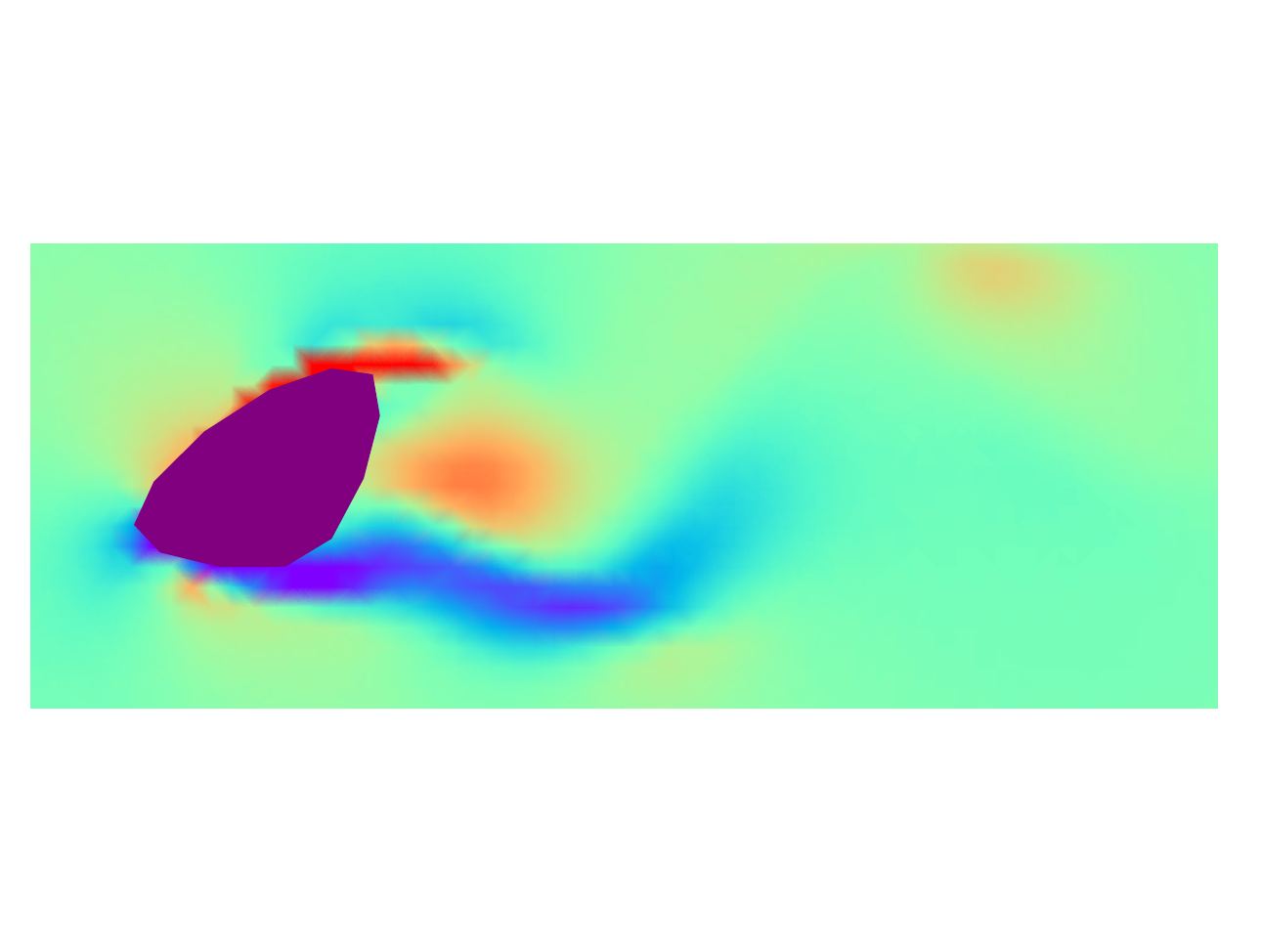}
    \end{minipage}                             & \begin{minipage}{\spatialfigwidth\textwidth}
      \includegraphics[width=\linewidth]{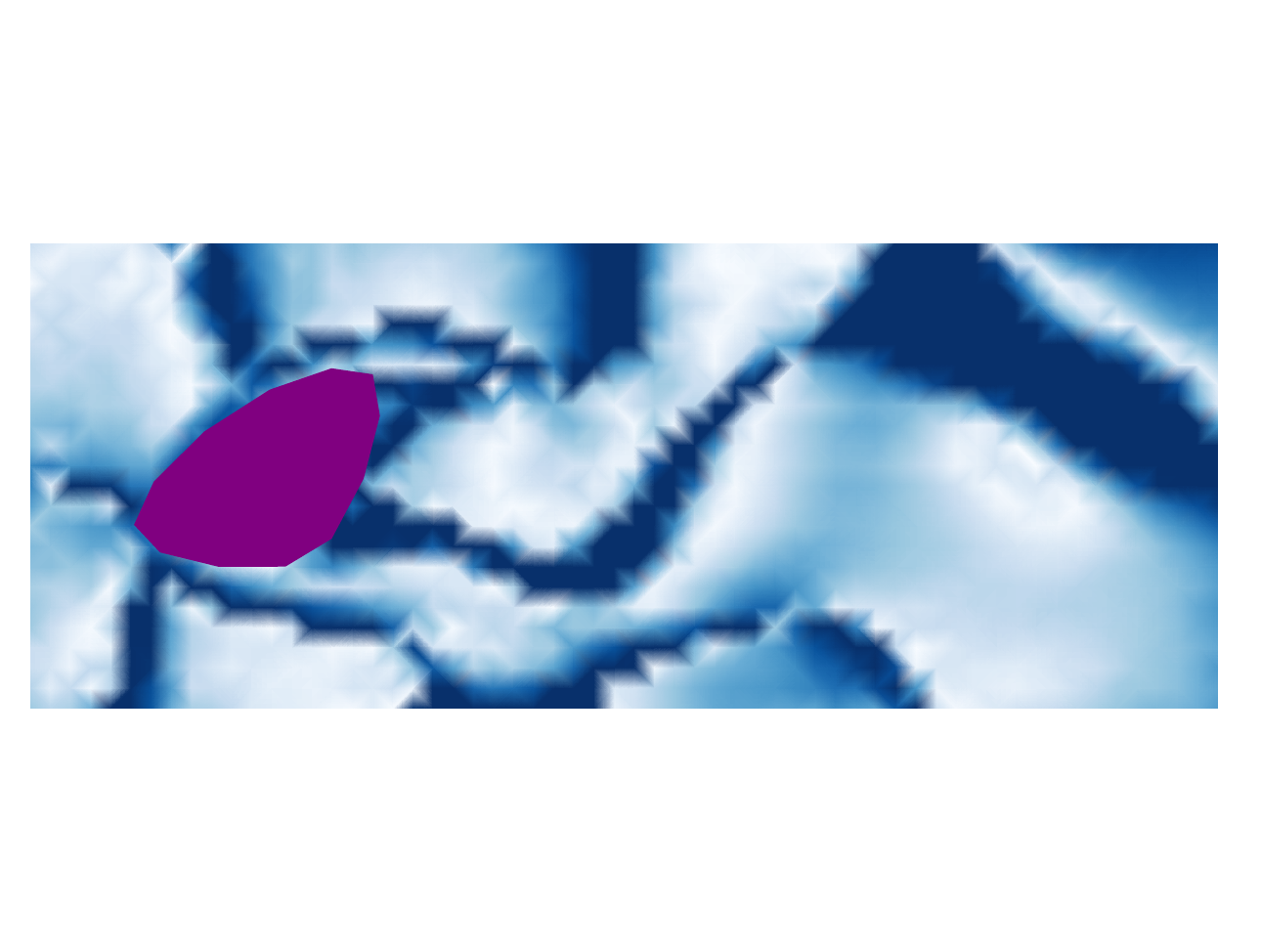}
    \end{minipage}          \\
\% Err    & \multicolumn{2}{c}{MAPE \textbf{37.16\%}, HV-MAPE {\ul \textbf{ 34.38\%}}} & \multicolumn{2}{c}{MAPE 46.98\%, HV-MAPE 44.51\%}\\ \midrule
SD-FNO                 &    \begin{minipage}{\spatialfigwidth\textwidth}
      \includegraphics[width=\linewidth]{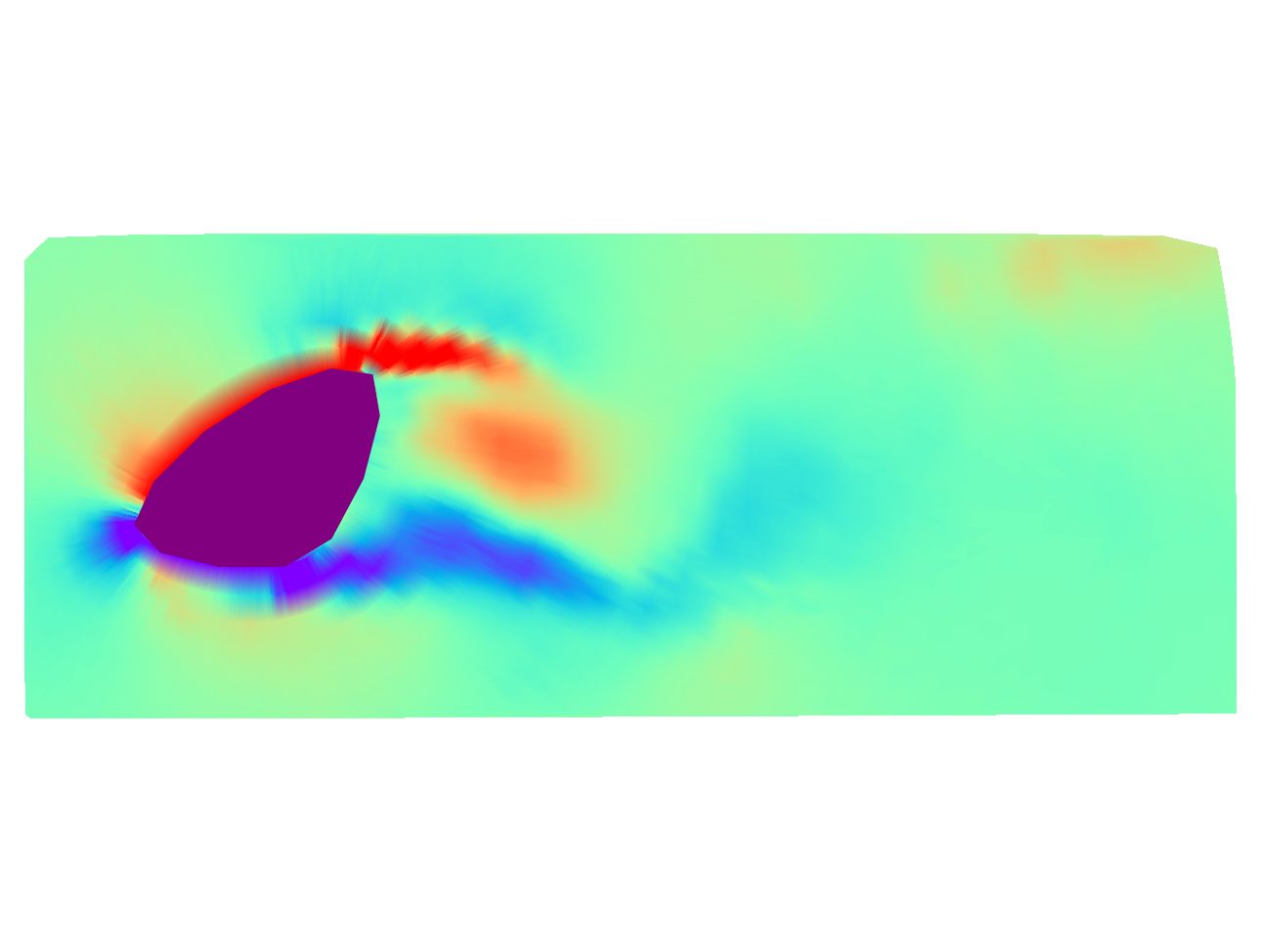}
    \end{minipage}                            &  \begin{minipage}{\spatialfigwidth\textwidth}
      \includegraphics[width=\linewidth]{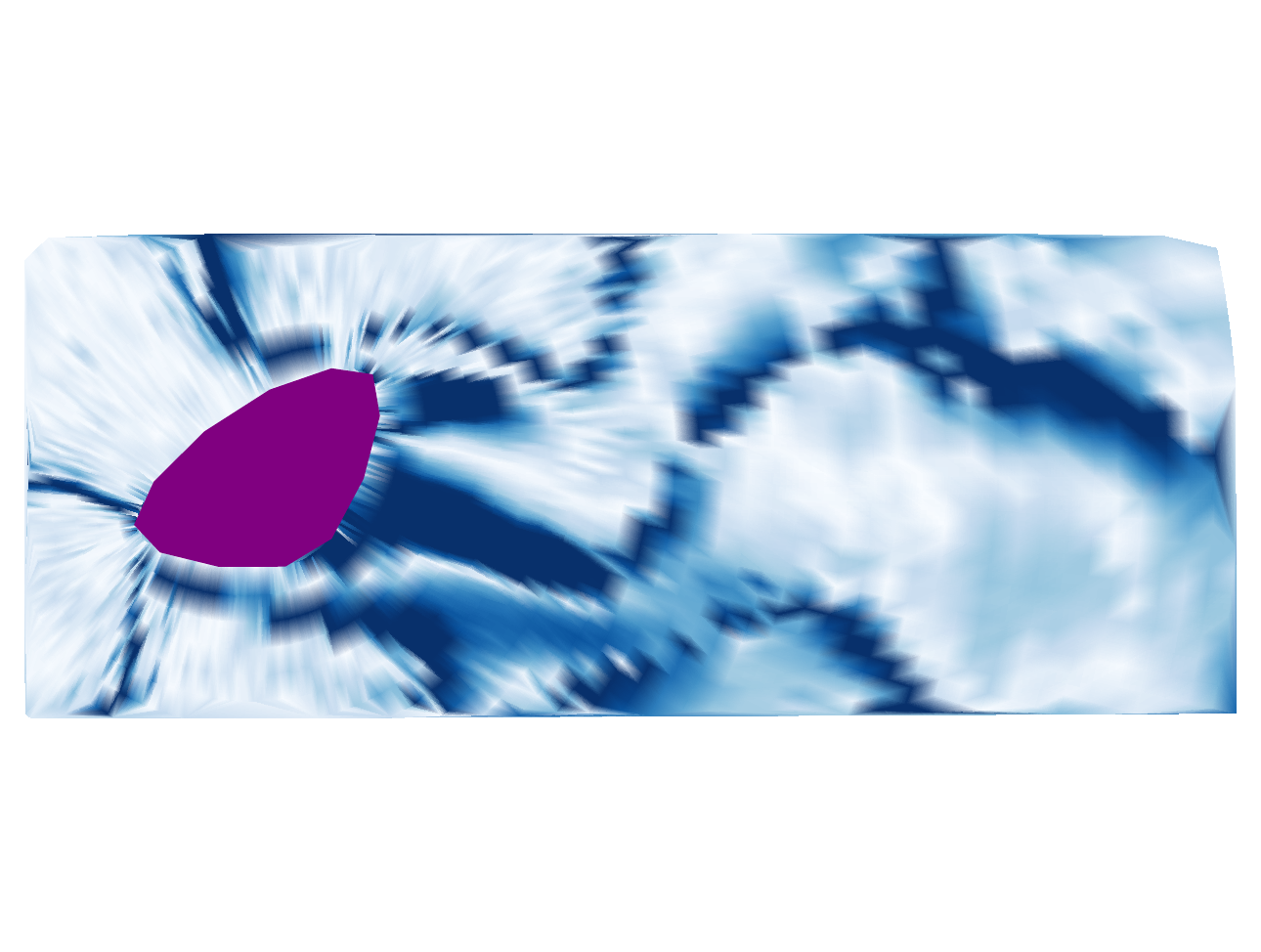}
    \end{minipage}                             &  \begin{minipage}{\spatialfigwidth\textwidth}
      \includegraphics[width=\linewidth]{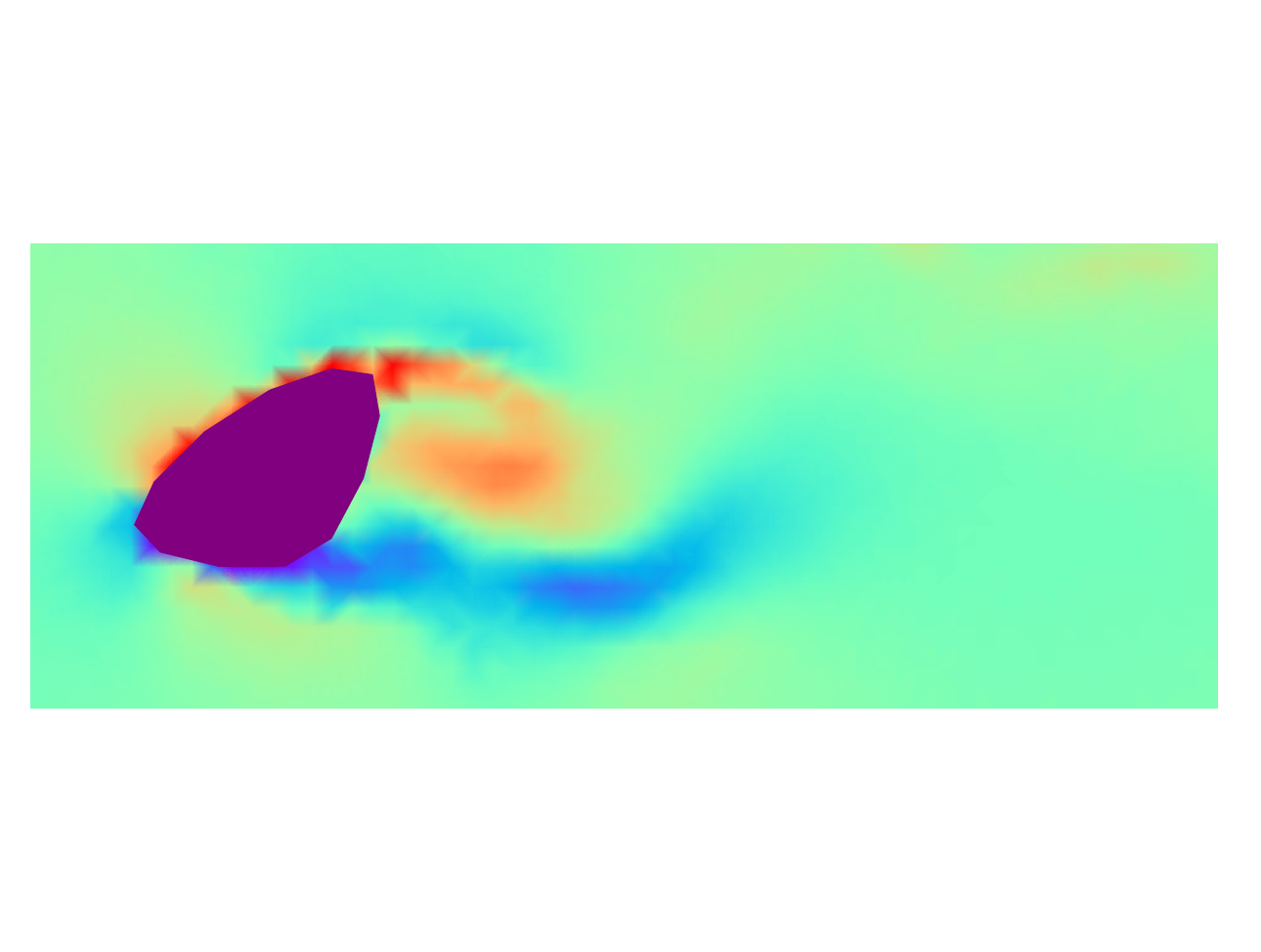}
    \end{minipage}                             & \begin{minipage}{\spatialfigwidth\textwidth}
      \includegraphics[width=\linewidth]{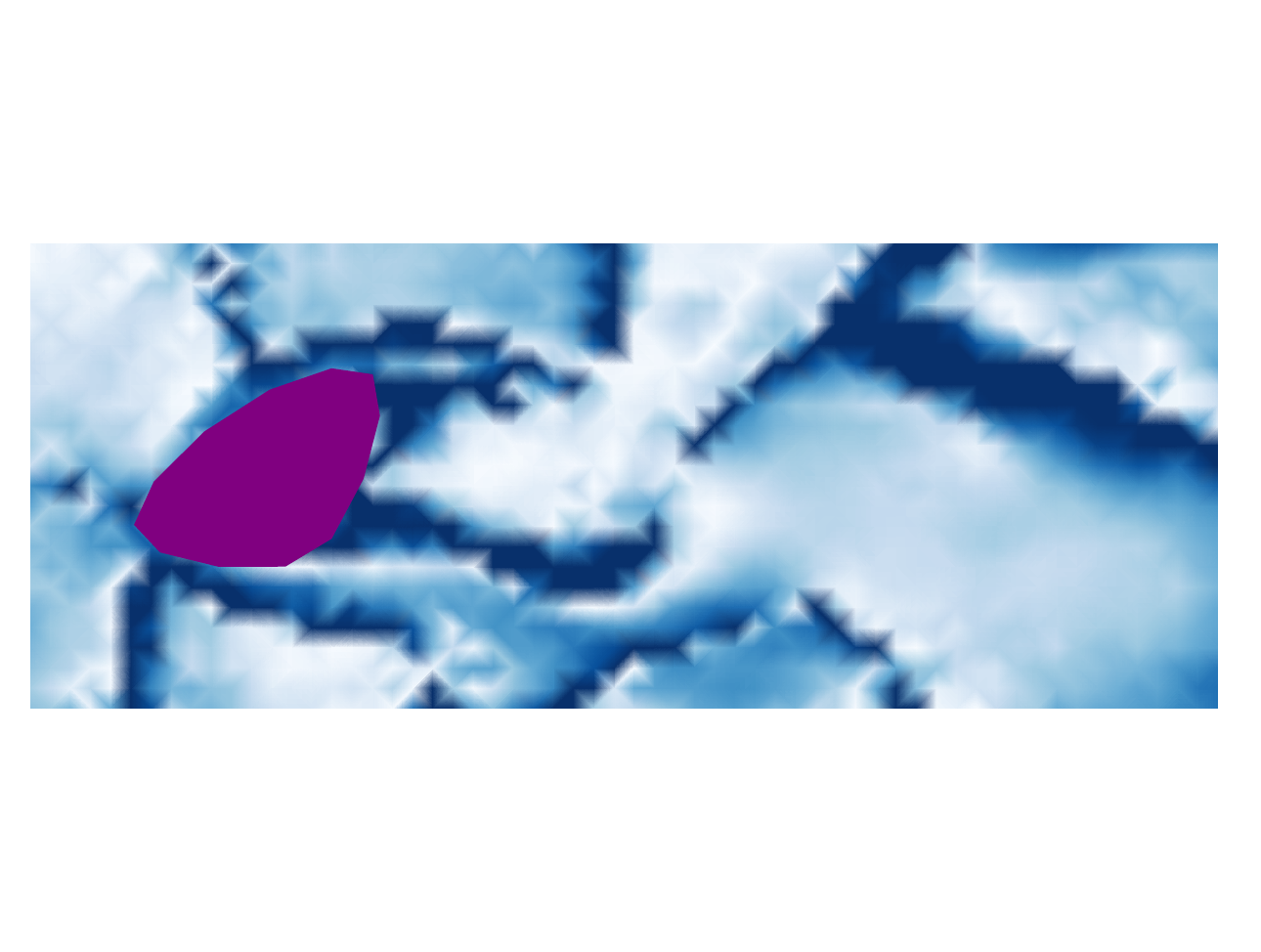}
    \end{minipage}          \\
\% Err    & \multicolumn{2}{c}{MAPE \textit{\textbf{36.20\%}}, HV-MAPE \textbf{ 35.23\%}} & \multicolumn{2}{c}{MAPE 43.92\%, HV-MAPE 46.26\%}\\ \bottomrule
\end{tabular}
    
    \caption{Ground truth (top) and predicted (bottom) vorticity fields and percentage error maps for a snapshot from Shape A, using the small sensor setup. Vorticity colormap constrained to 7.5\% of $\max(|\tilde{\omega}_{t,i}|)$.}
    \label{fig:spatialex_s1660_small}
\end{figure}

\begin{figure}
    \centering
    \includegraphics[width=0.40\textwidth]{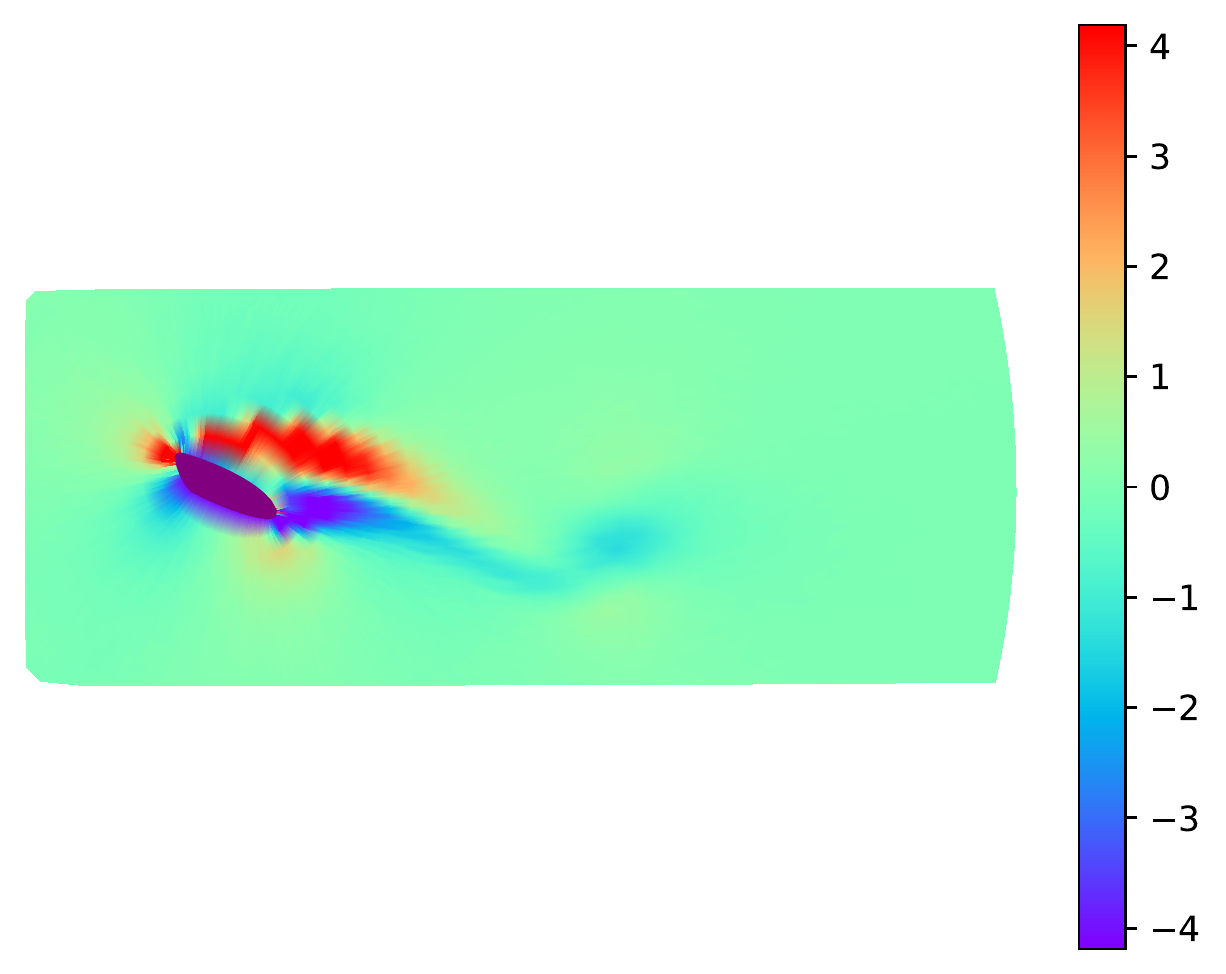}
    \includegraphics[width=0.045\textwidth]{images/spatial/pcterrorbarr0.pdf}
\begin{tabular}{@{}lcccc@{}}
\toprule
\multirow{2}{*}{Model} & \multicolumn{2}{c|}{Annular Sampling}                        & \multicolumn{2}{c}{Cartesian Sampling}      \\ \cmidrule(l){2-5} 
                       & \multicolumn{1}{c|}{Vorticity} & \multicolumn{1}{c|}{\% Error} & \multicolumn{1}{c|}{Vorticity} & \% Error \\ \midrule
SD                     &    \begin{minipage}{\spatialfigwidth\textwidth}
      \includegraphics[width=\linewidth]{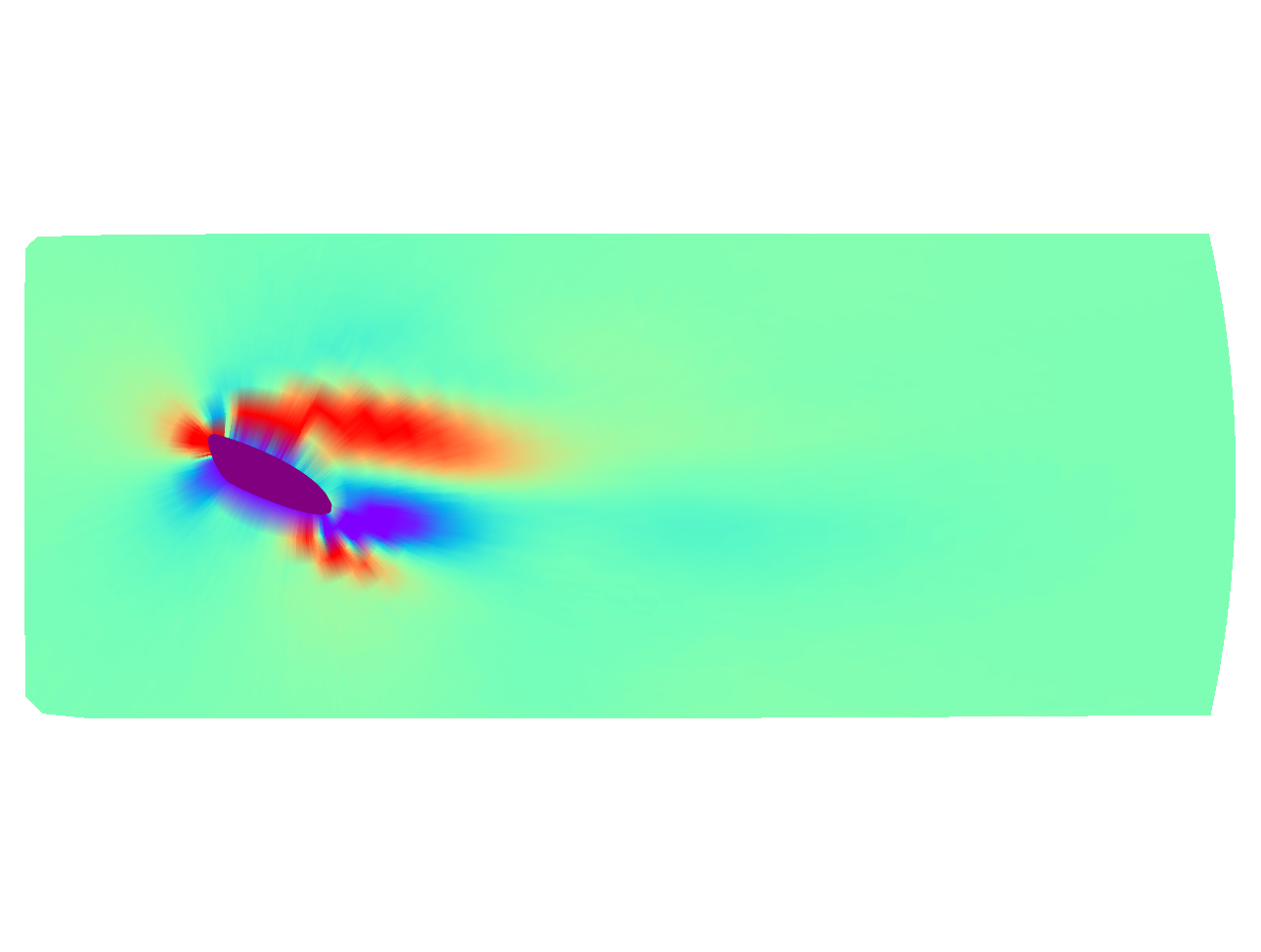}
    \end{minipage}                            &  \begin{minipage}{\spatialfigwidth\textwidth}
      \includegraphics[width=\linewidth]{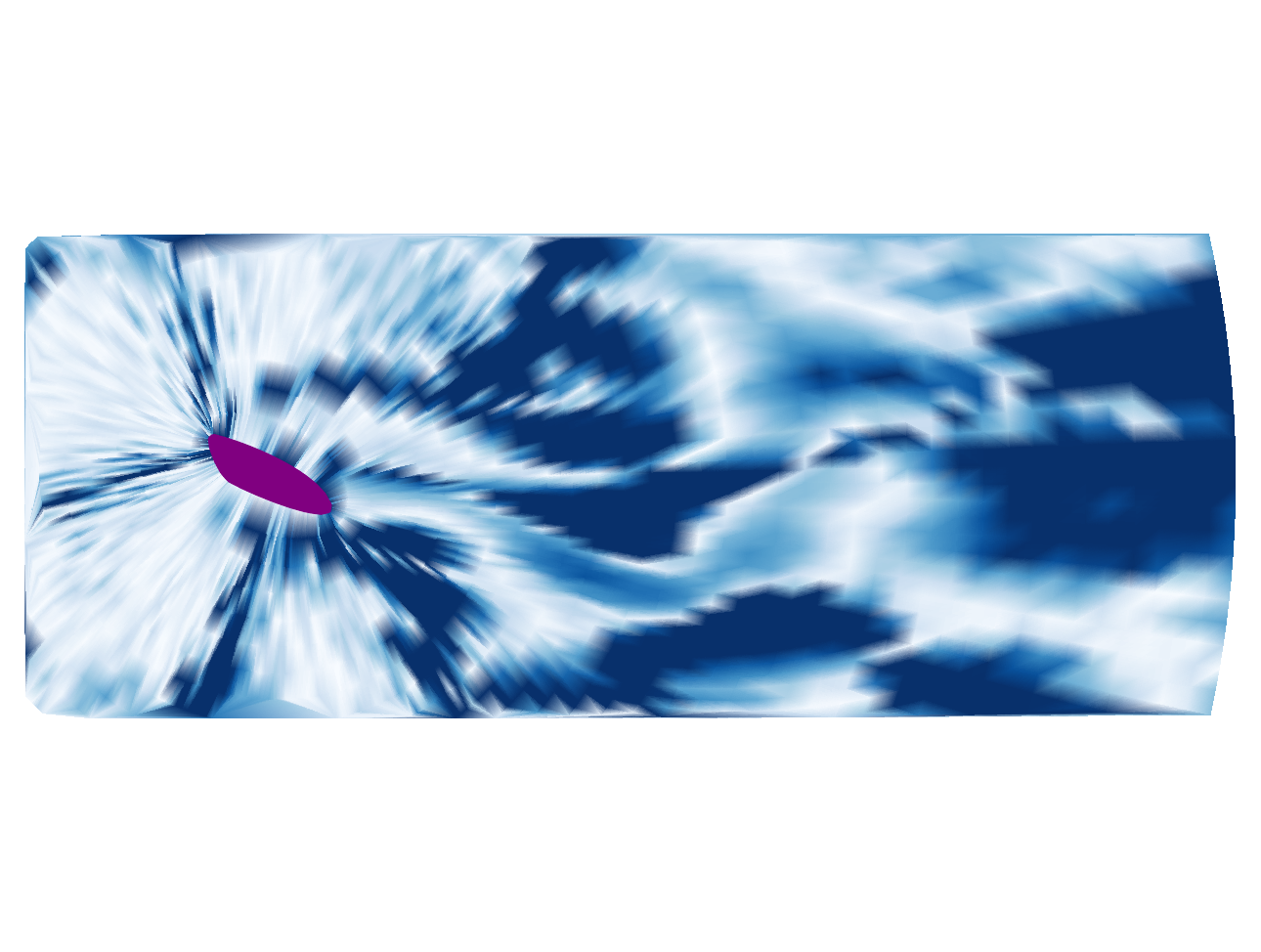}
    \end{minipage}                             &  \begin{minipage}{\spatialfigwidth\textwidth}
      \includegraphics[width=\linewidth]{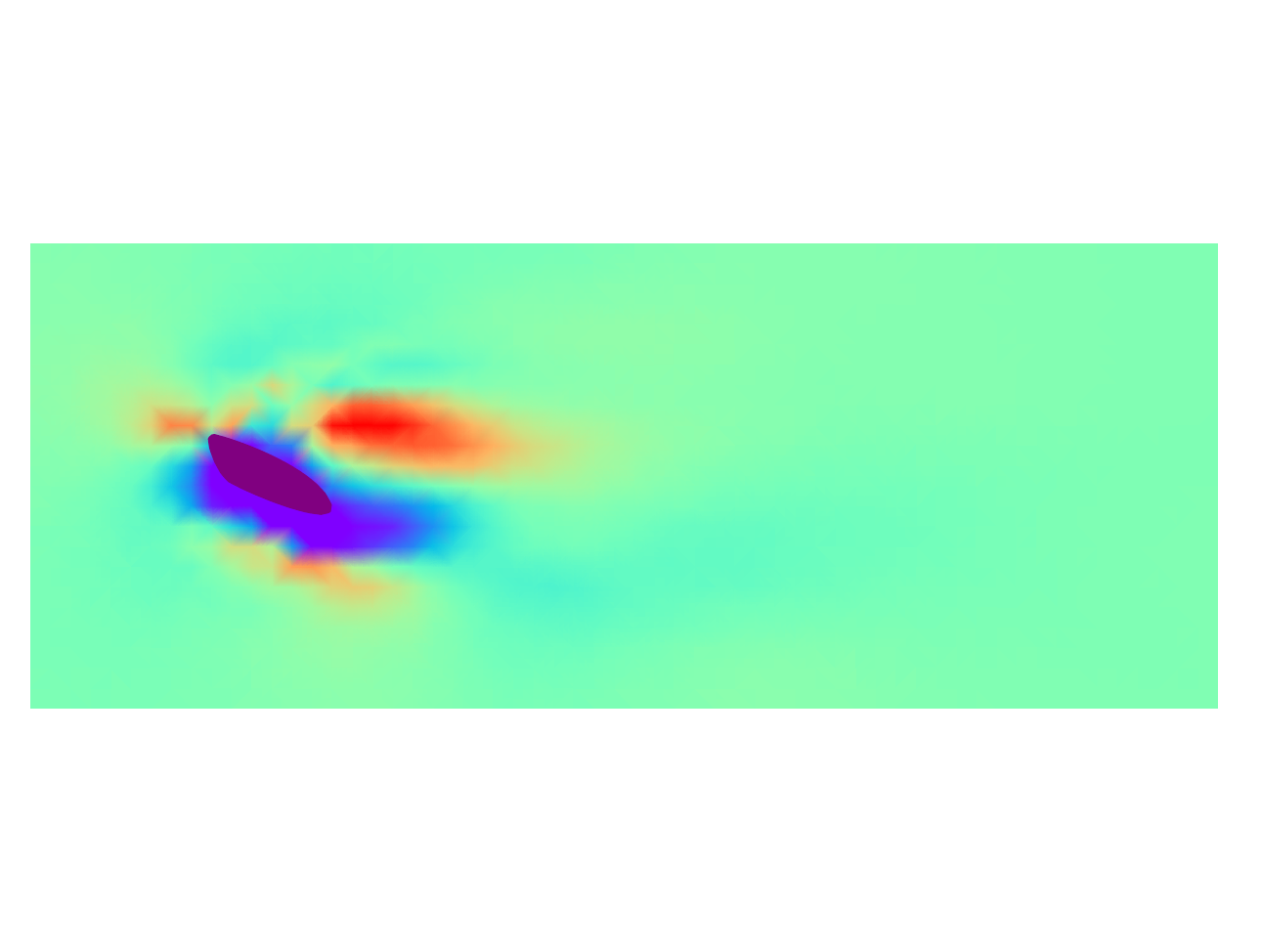}
    \end{minipage}                             & \begin{minipage}{\spatialfigwidth\textwidth}
      \includegraphics[width=\linewidth]{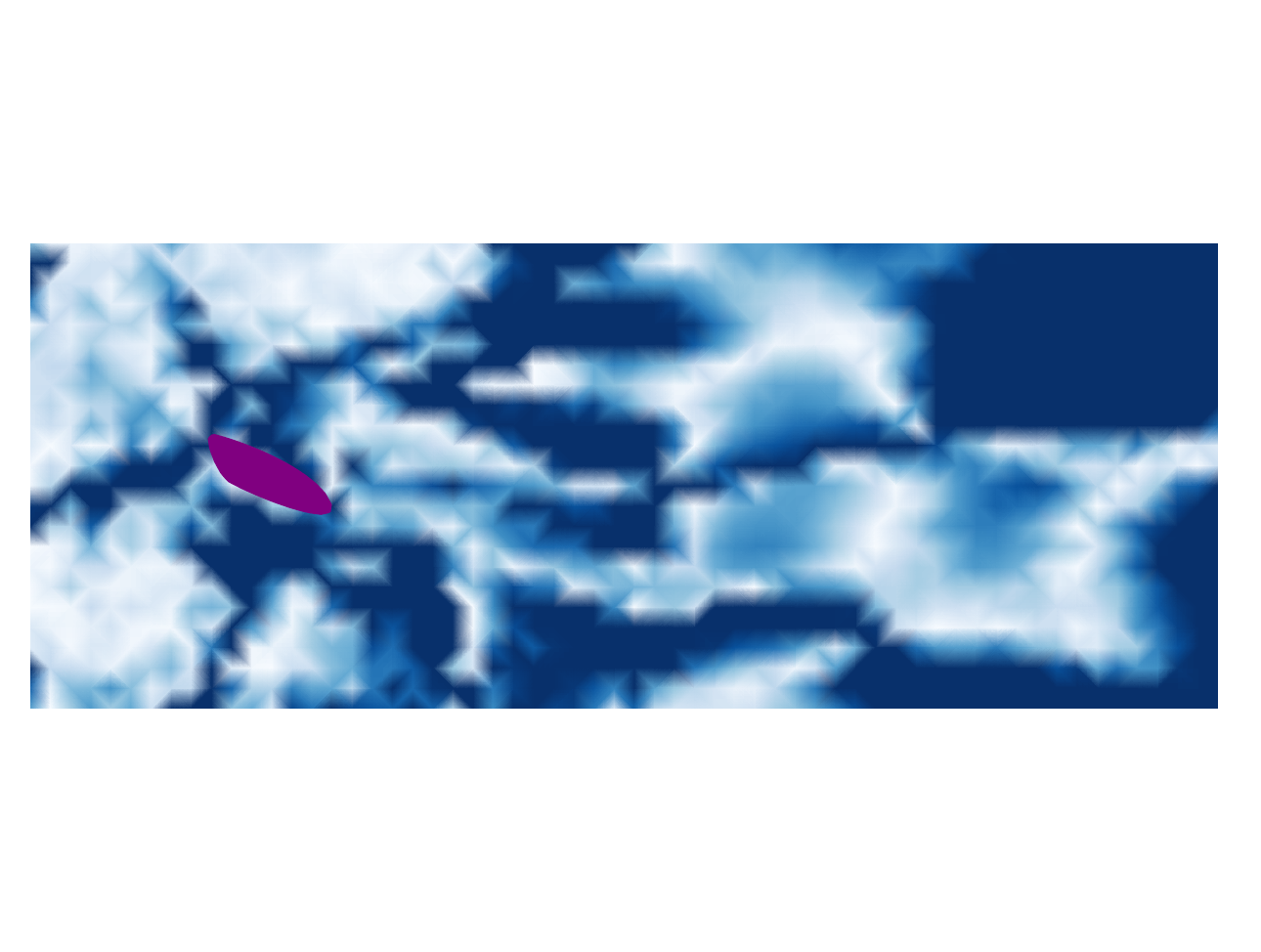}
    \end{minipage}         \\ 
\% Err    & \multicolumn{2}{c}{MAPE \textbf{44.79\%}, HV-MAPE \textbf{34.39\%}} & \multicolumn{2}{c}{MAPE 58.82\%, HV-MAPE 54.76\%}\\ \midrule
SD-Large               &    \begin{minipage}{\spatialfigwidth\textwidth}
      \includegraphics[width=\linewidth]{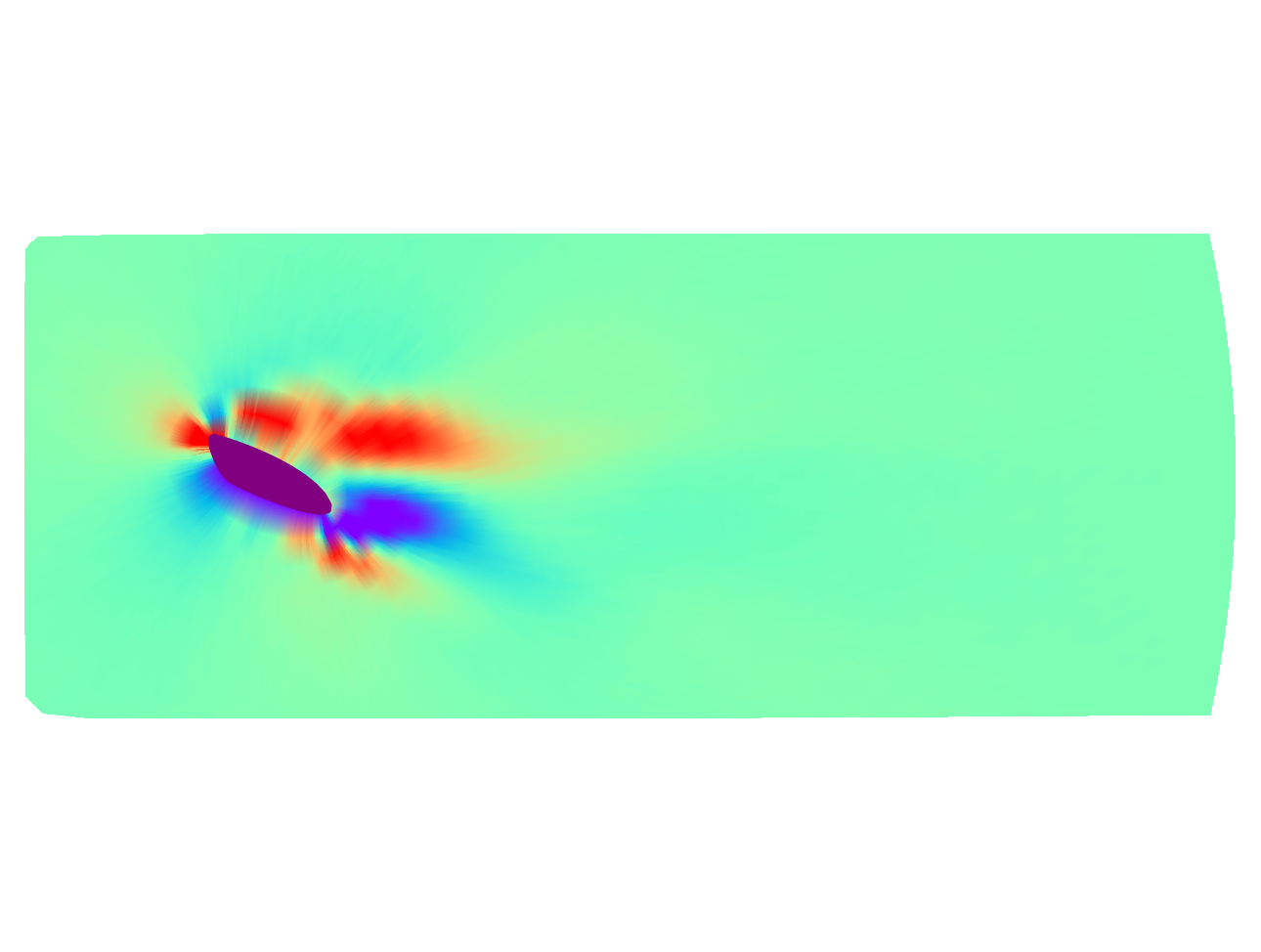}
    \end{minipage}                            &  \begin{minipage}{\spatialfigwidth\textwidth}
      \includegraphics[width=\linewidth]{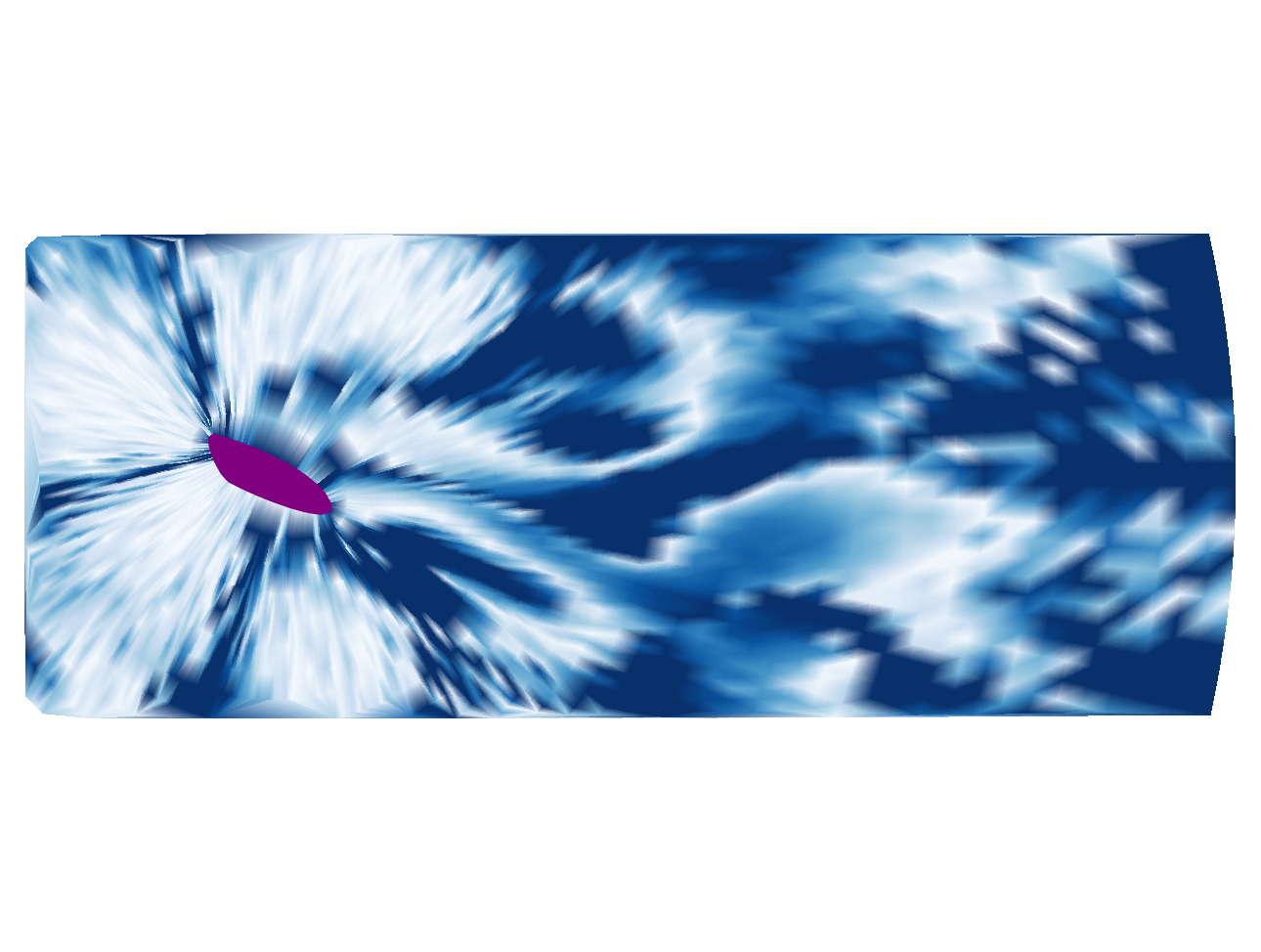}
    \end{minipage}                             &  \begin{minipage}{\spatialfigwidth\textwidth}
      \includegraphics[width=\linewidth]{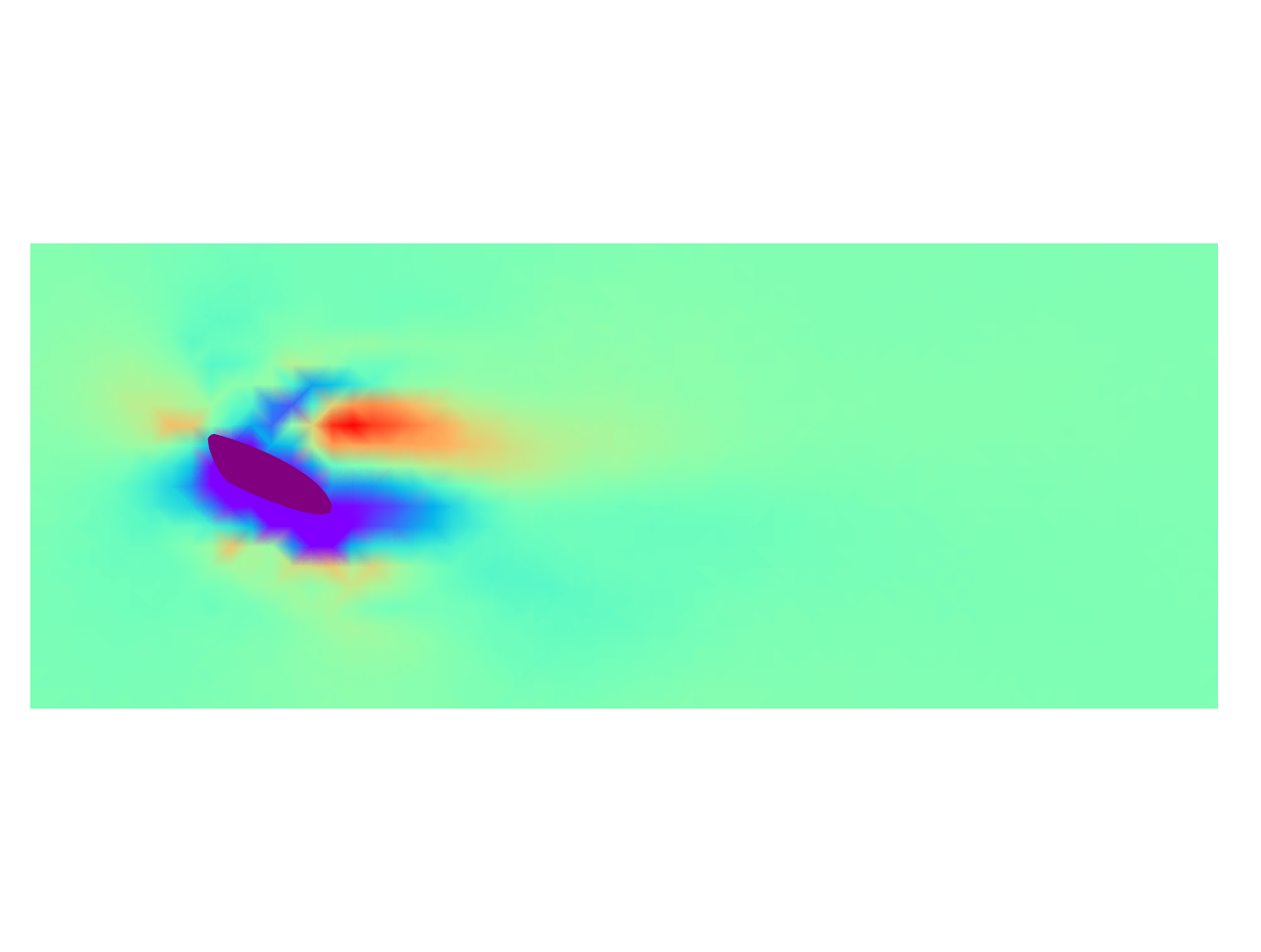}
    \end{minipage}                             & \begin{minipage}{\spatialfigwidth\textwidth}
      \includegraphics[width=\linewidth]{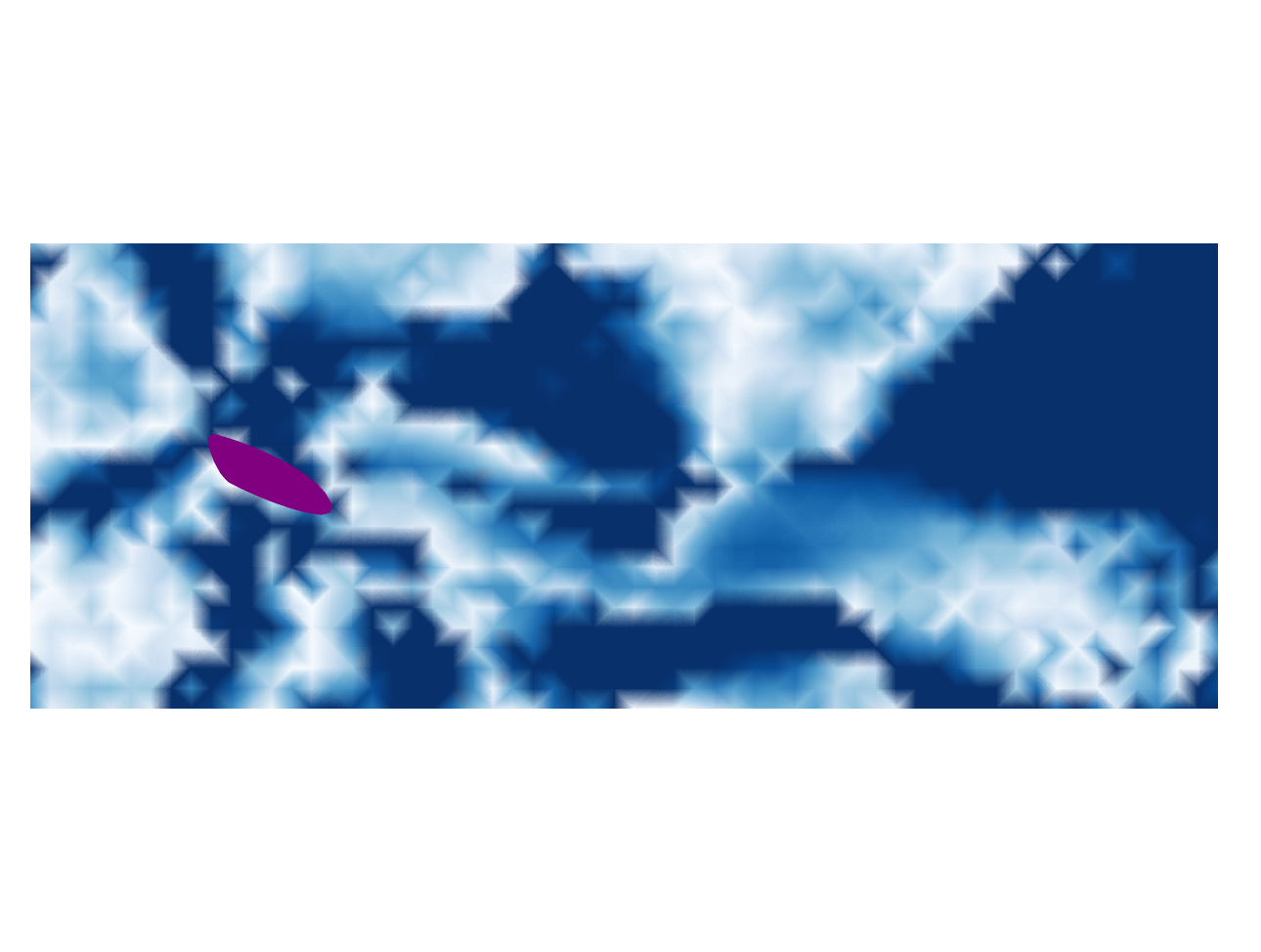}
    \end{minipage}          \\
\% Err    & \multicolumn{2}{c}{MAPE \textbf{48.44\%}, HV-MAPE \textbf{40.20\%}} & \multicolumn{2}{c}{MAPE 61.51\%, HV-MAPE 64.99\%}\\ \midrule
SD-UNet                &    \begin{minipage}{\spatialfigwidth\textwidth}
      \includegraphics[width=\linewidth]{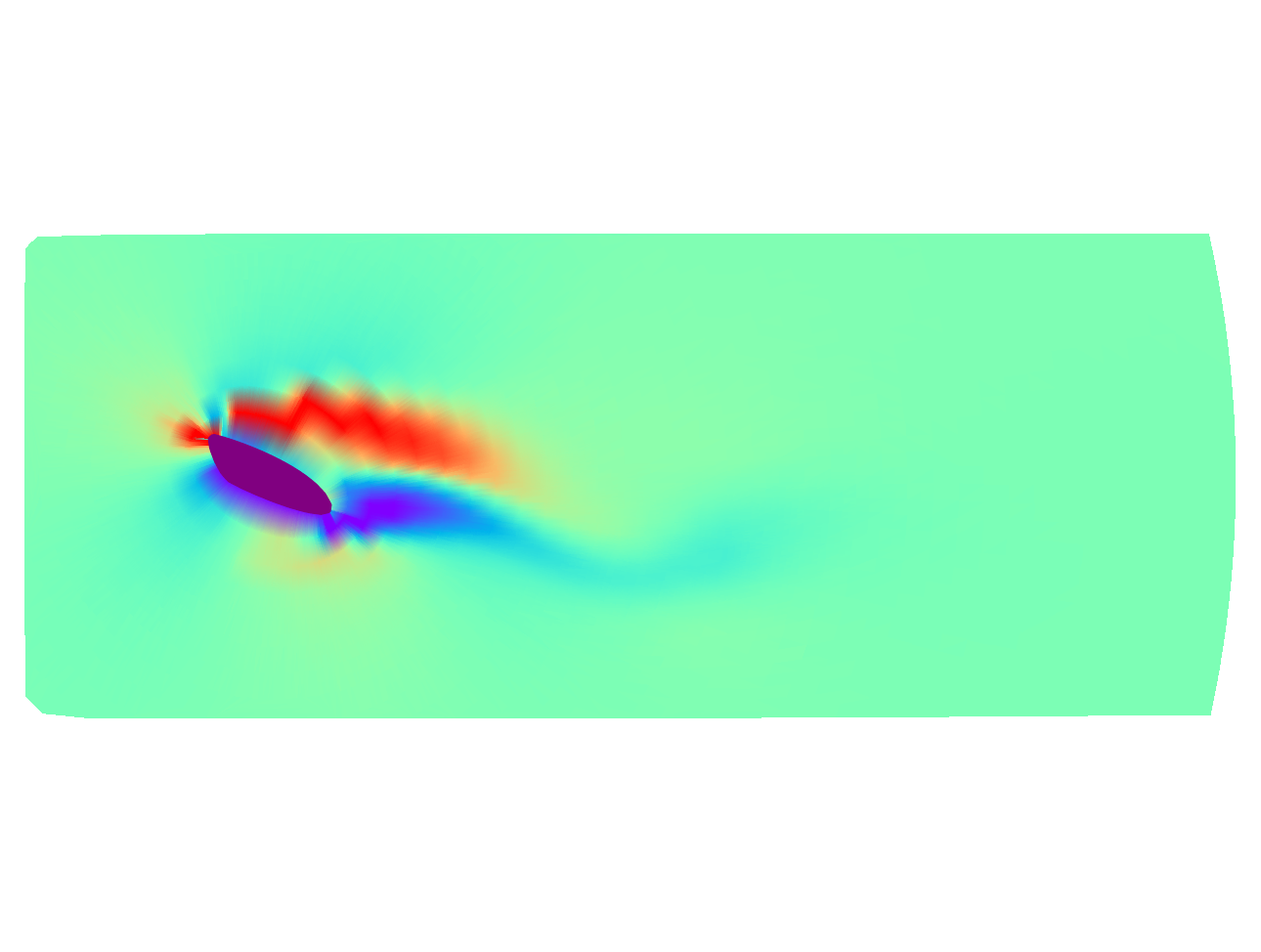}
    \end{minipage}                            &  \begin{minipage}{\spatialfigwidth\textwidth}
      \includegraphics[width=\linewidth]{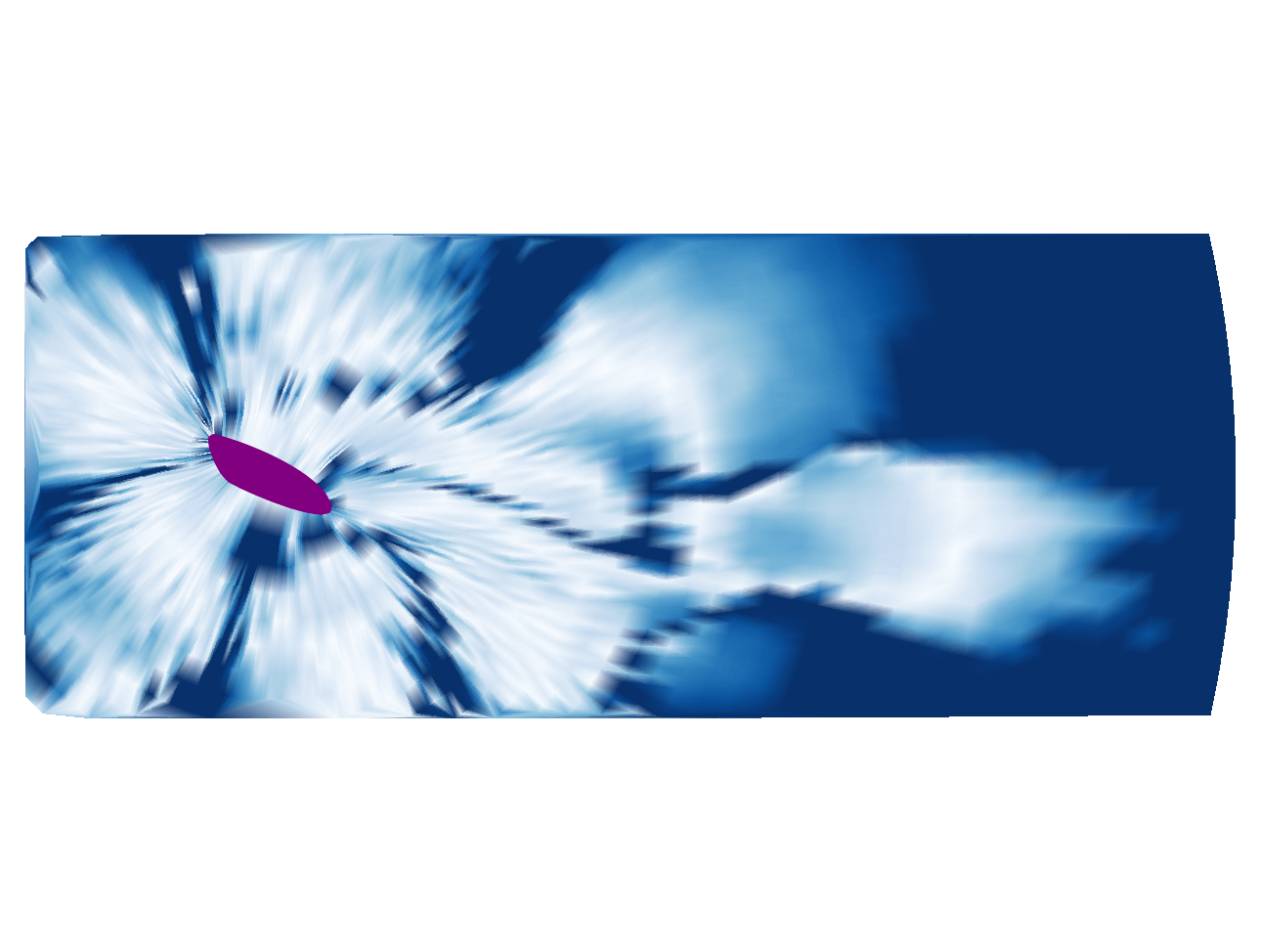}
    \end{minipage}                             &  \begin{minipage}{\spatialfigwidth\textwidth}
      \includegraphics[width=\linewidth]{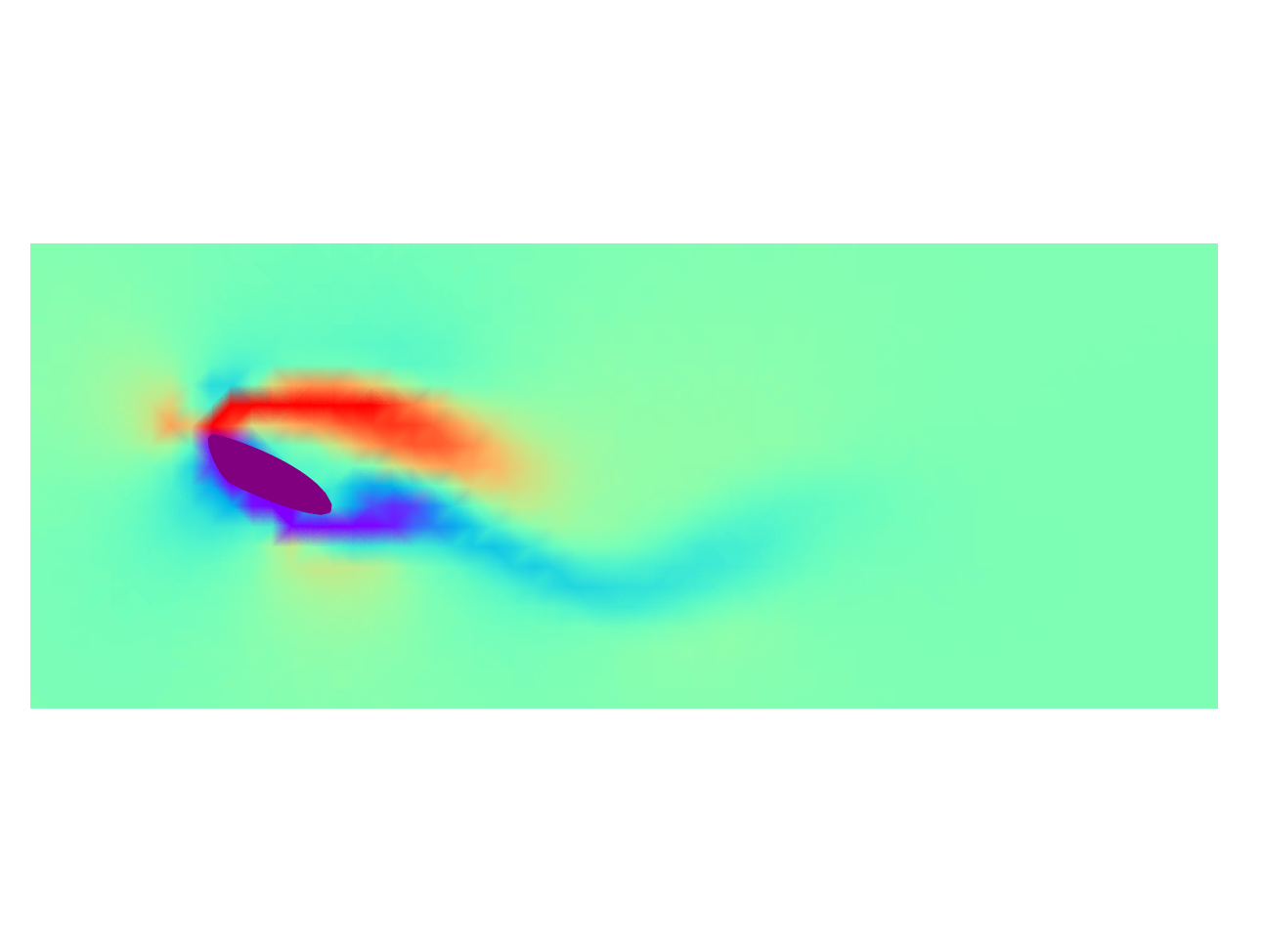}
    \end{minipage}                             & \begin{minipage}{\spatialfigwidth\textwidth}
      \includegraphics[width=\linewidth]{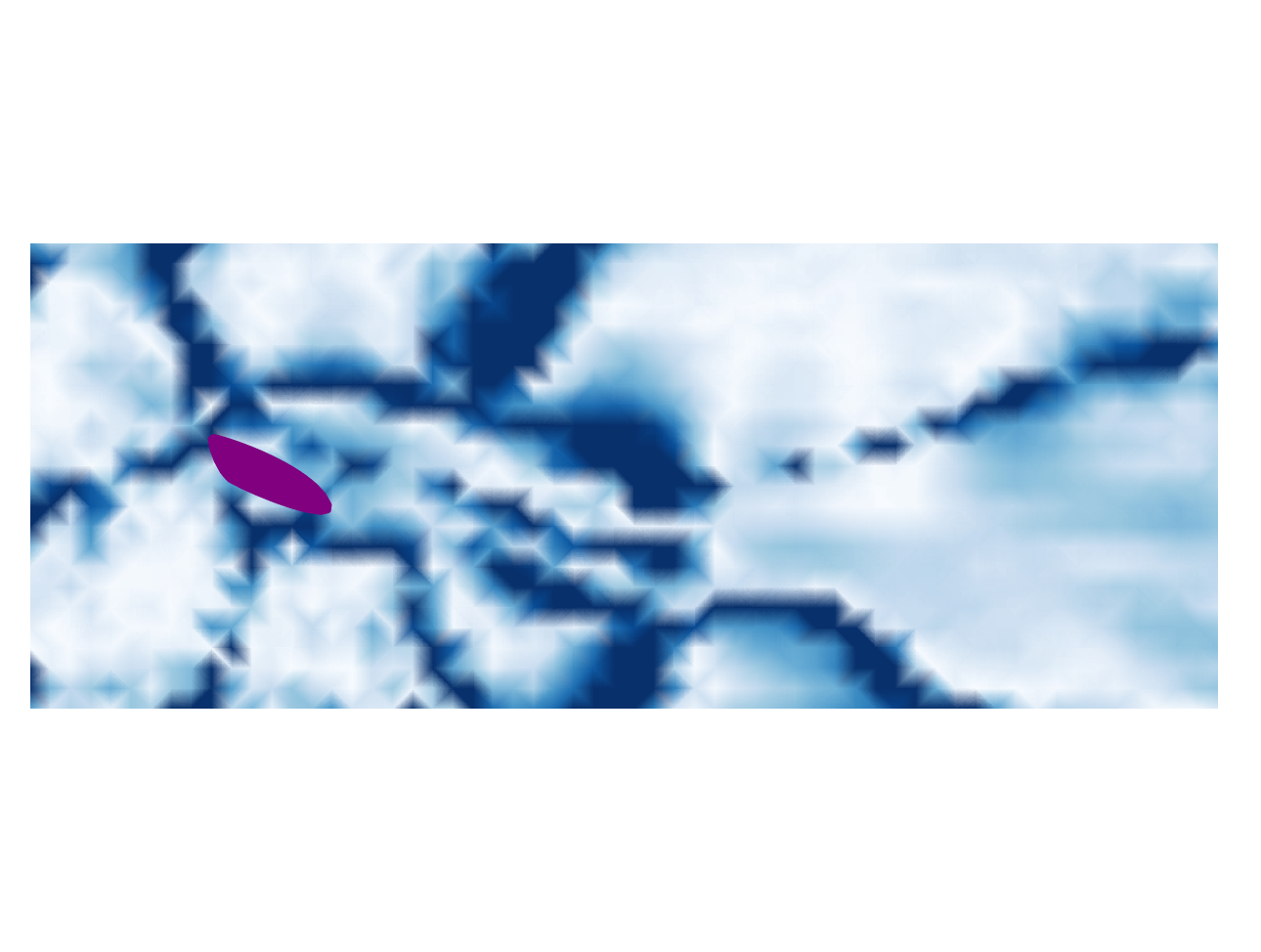}
    \end{minipage}          \\
\% Err    & \multicolumn{2}{c}{MAPE 40.94\%, HV-MAPE {\ul \textbf{24.37\%}}} & \multicolumn{2}{c}{MAPE \textit{\textbf{35.87\%}}, HV-MAPE 39.18\%}\\ \midrule
SD-FNO                 &    \begin{minipage}{\spatialfigwidth\textwidth}
      \includegraphics[width=\linewidth]{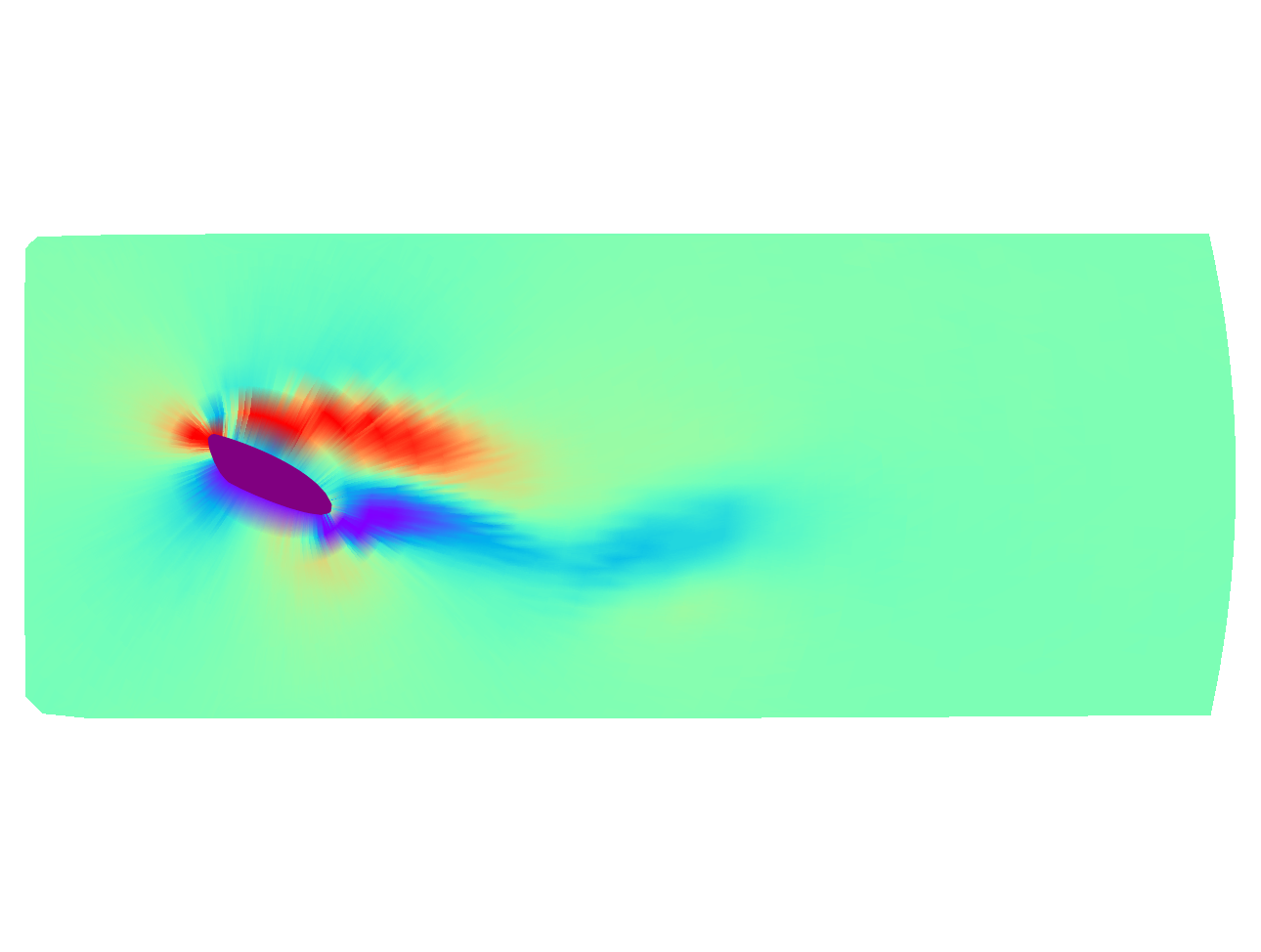}
    \end{minipage}                            &  \begin{minipage}{\spatialfigwidth\textwidth}
      \includegraphics[width=\linewidth]{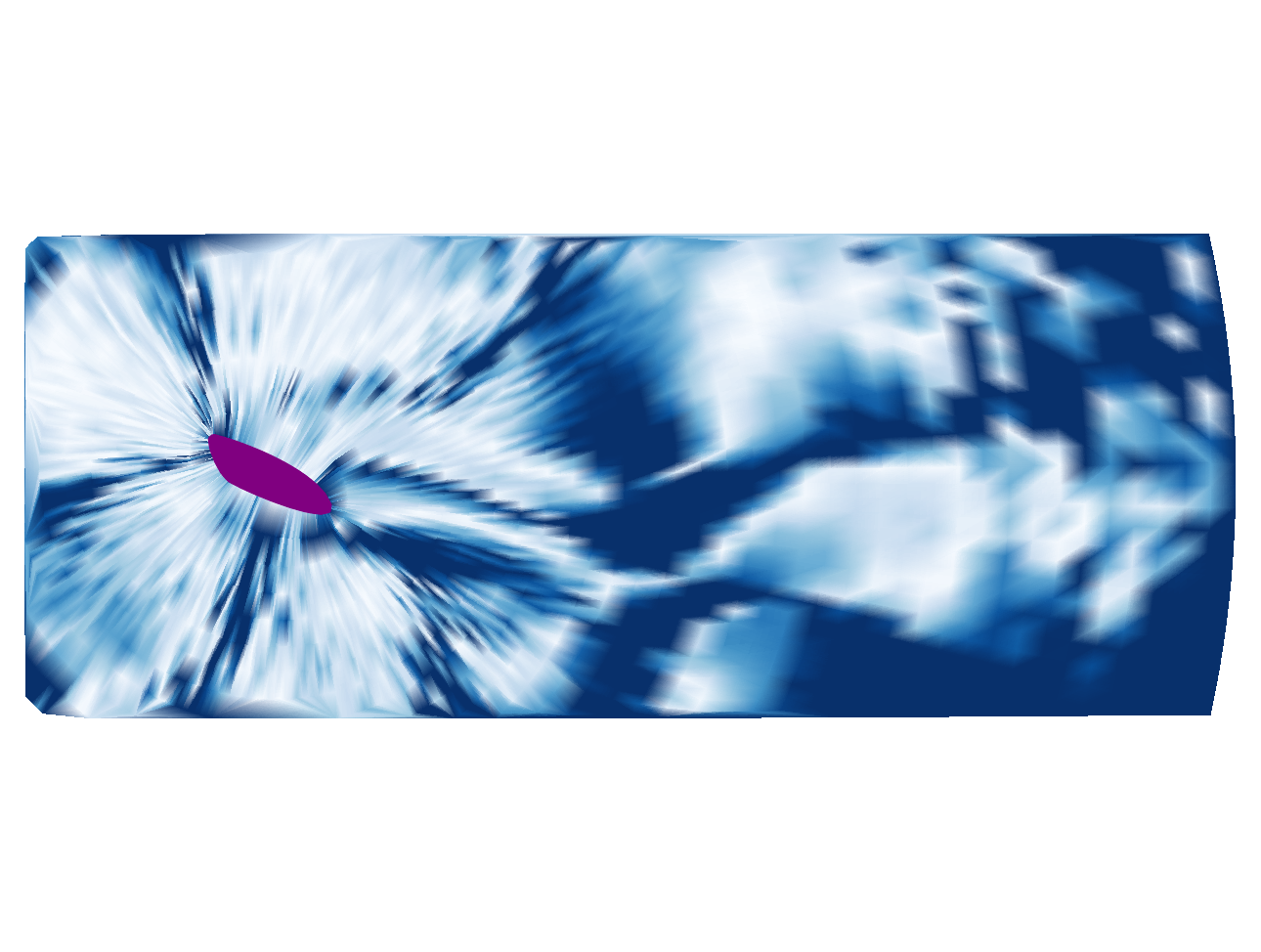}
    \end{minipage}                             &  \begin{minipage}{\spatialfigwidth\textwidth}
      \includegraphics[width=\linewidth]{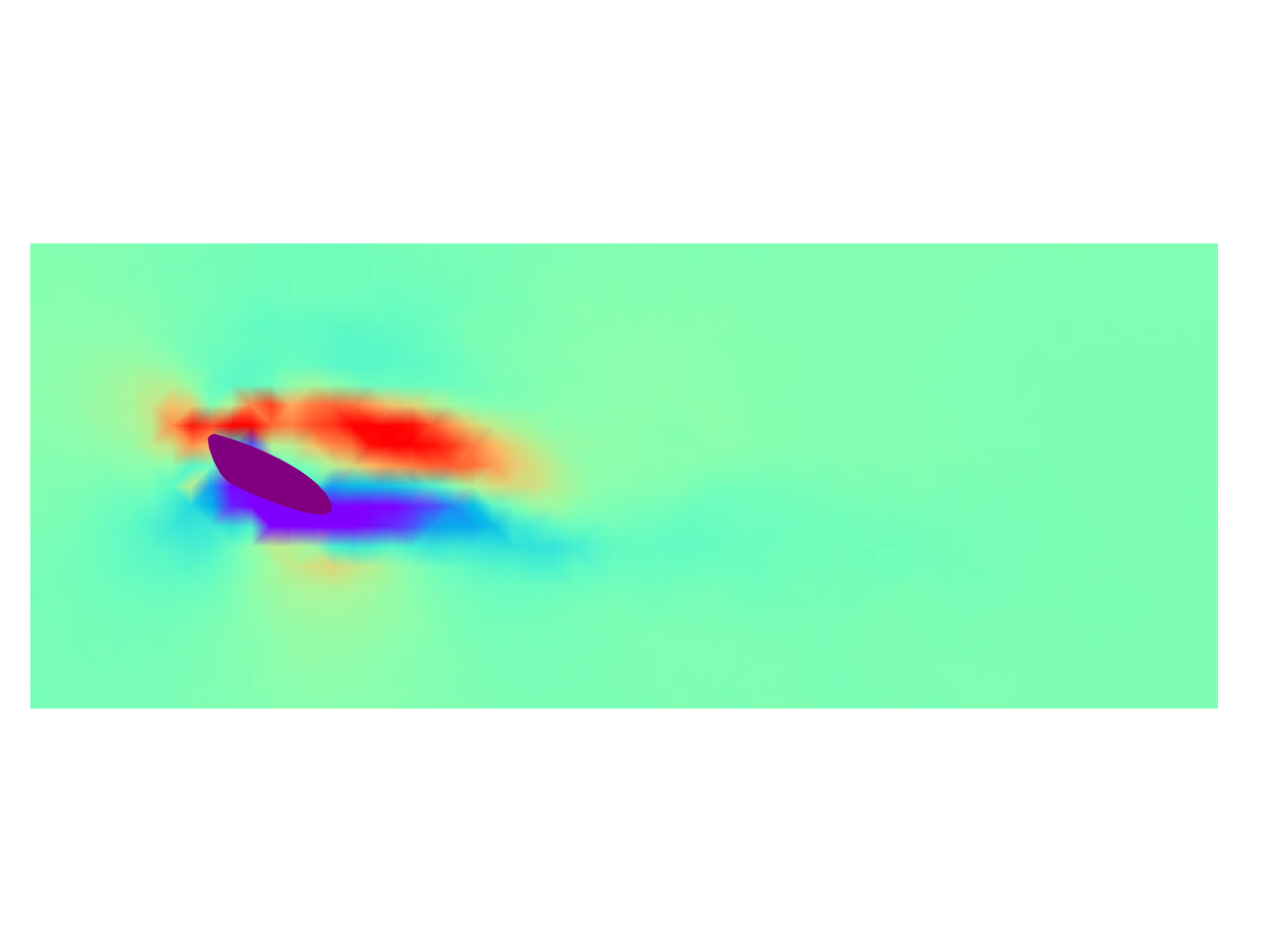}
    \end{minipage}                             & \begin{minipage}{\spatialfigwidth\textwidth}
      \includegraphics[width=\linewidth]{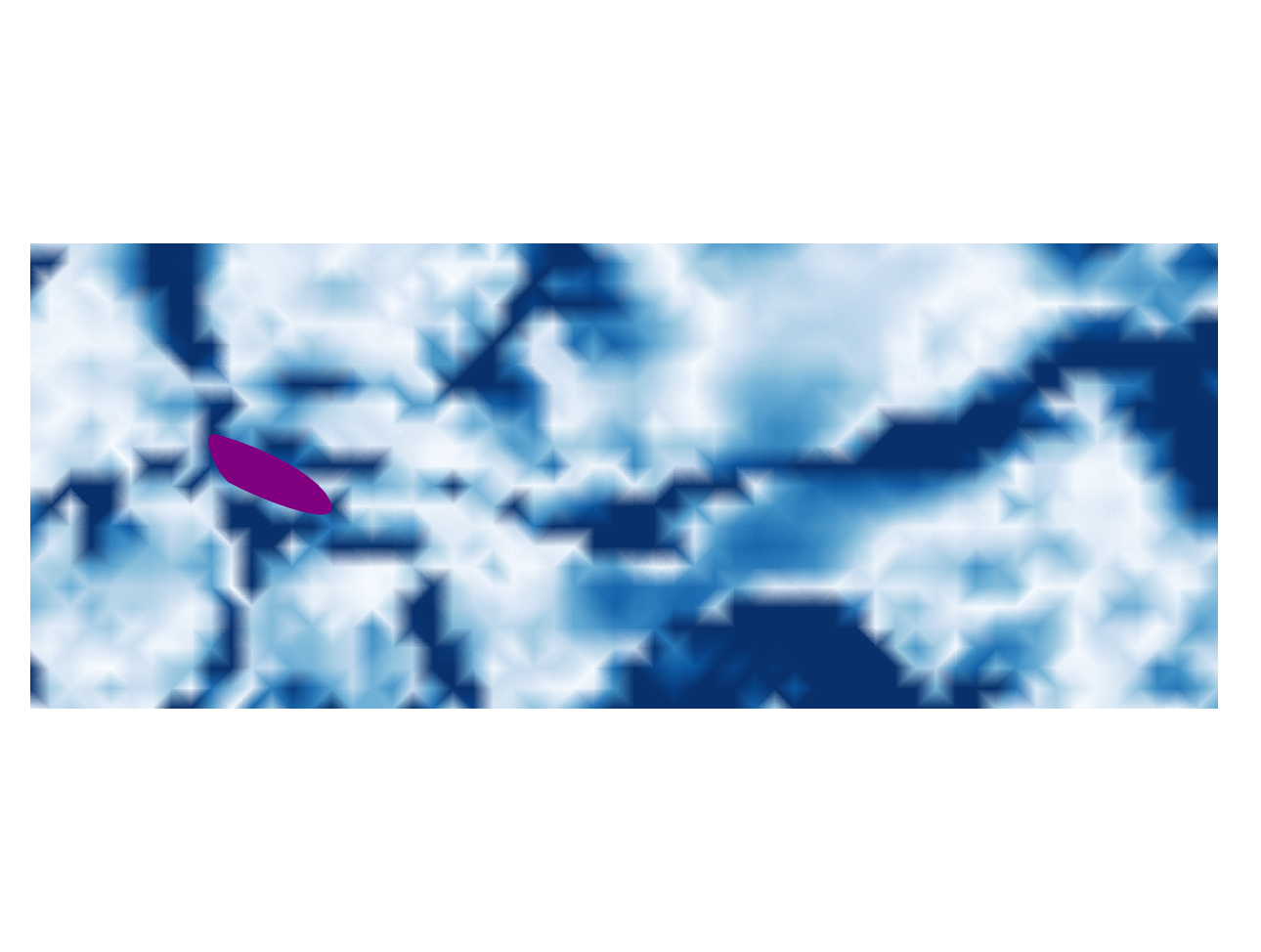}
    \end{minipage}          \\
\% Err    & \multicolumn{2}{c}{MAPE \textbf{43.70\%}, \textbf{HV-MAPE 34.35\%}} & \multicolumn{2}{c}{MAPE 44.99\%, HV-MAPE 42.72\%}\\ \bottomrule
\end{tabular}
    
    \caption{Ground truth (top) and predicted (bottom) vorticity fields and percentage error maps for a snapshot from Shape B, using the large sensor setup. Vorticity colormap constrained to 3\% of $\max(|\tilde{\omega}_{t,i}|)$.}
    \label{fig:spatialex_s1400_large}
\end{figure}

\begin{figure}
    \centering
    \includegraphics[width=0.40\textwidth]{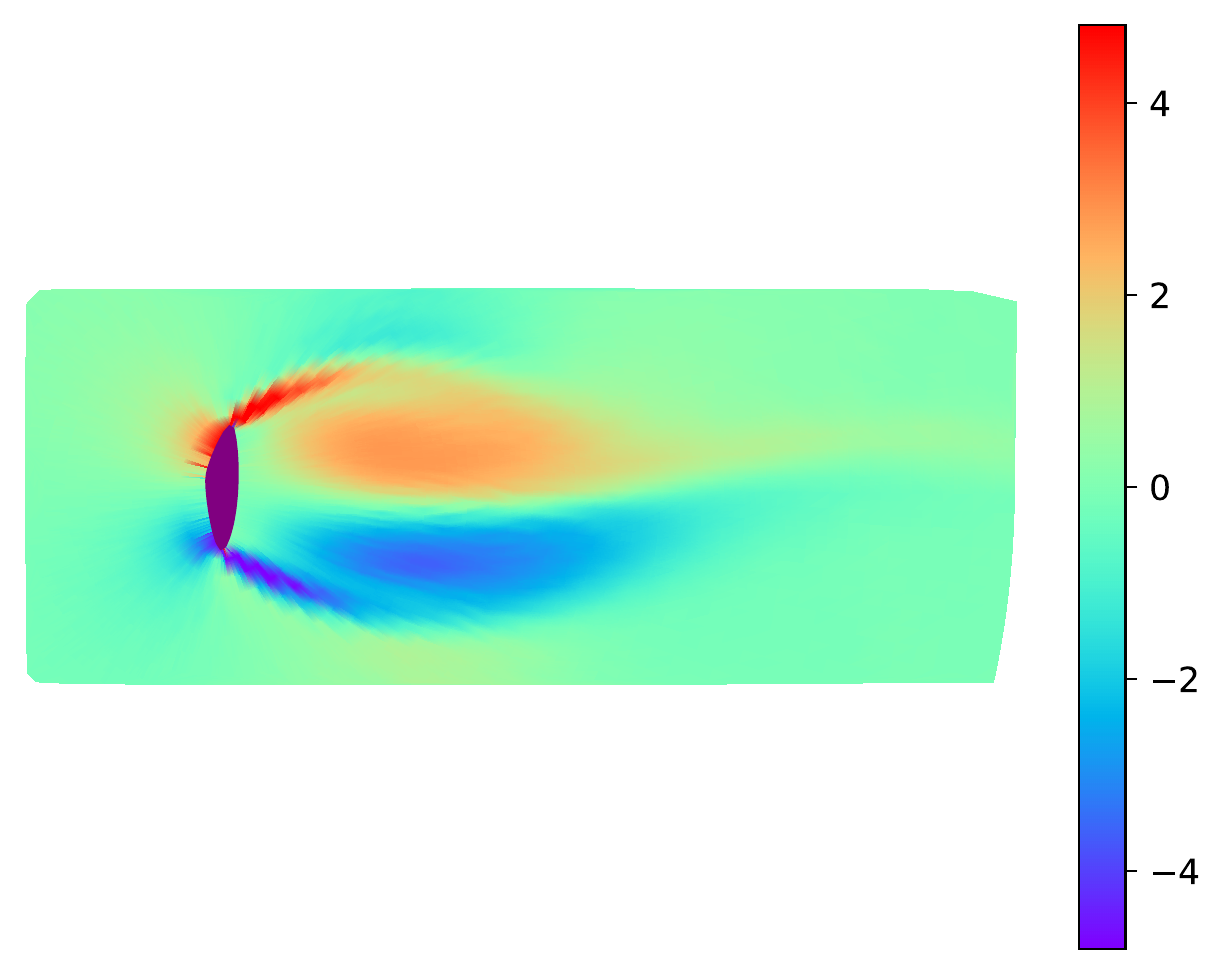}
    \includegraphics[width=0.045\textwidth]{images/spatial/pcterrorbarr0.pdf}
\begin{tabular}{@{}lcccc@{}}
\toprule
\multirow{2}{*}{Model} & \multicolumn{2}{c|}{Annular Sampling}                        & \multicolumn{2}{c}{Cartesian Sampling}      \\ \cmidrule(l){2-5} 
                       & \multicolumn{1}{c|}{Vorticity} & \multicolumn{1}{c|}{\% Error} & \multicolumn{1}{c|}{Vorticity} & \% Error \\ \midrule
SD                     &    \begin{minipage}{\spatialfigwidth\textwidth}
      \includegraphics[width=\linewidth]{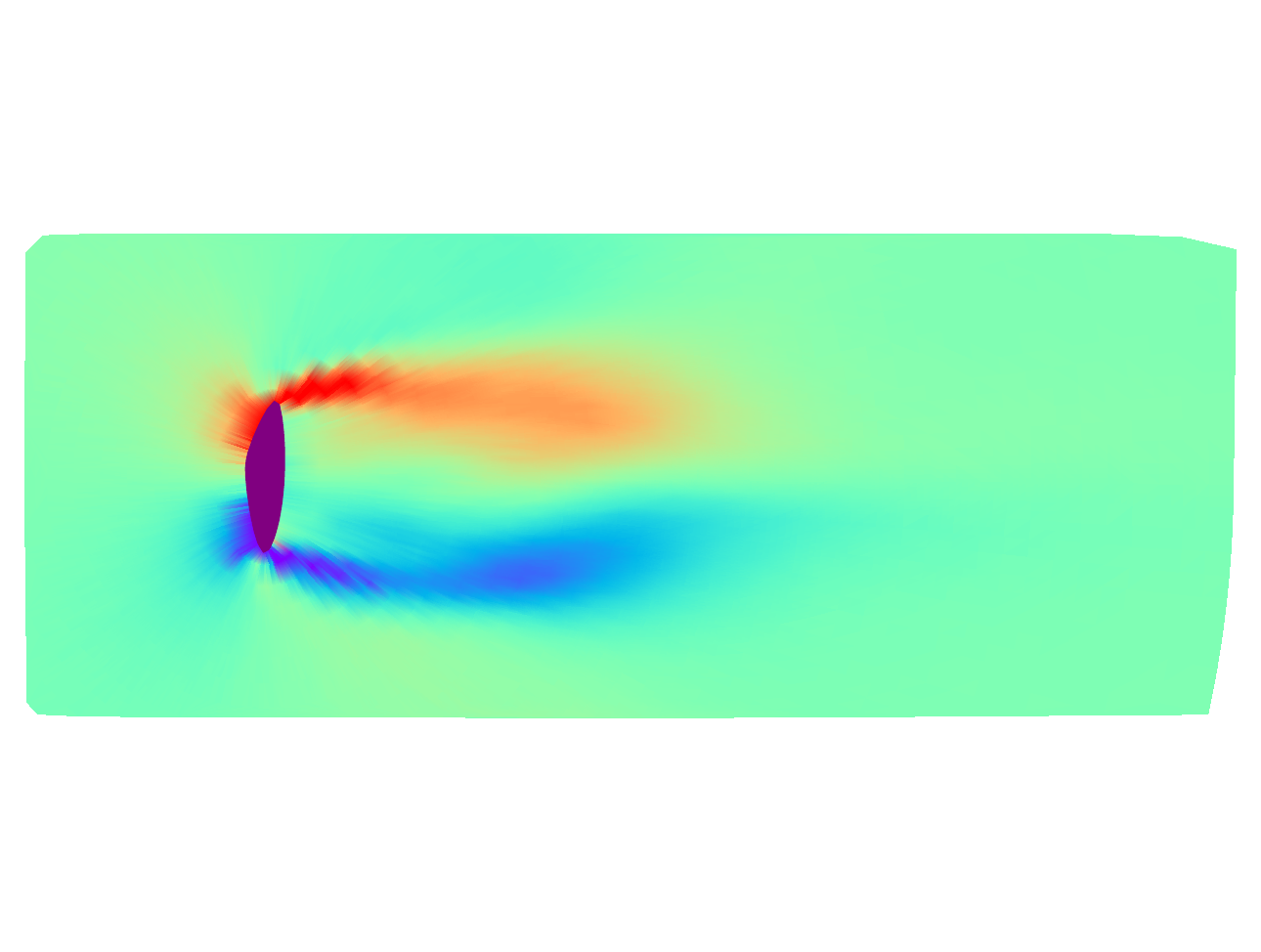}
    \end{minipage}                            &  \begin{minipage}{\spatialfigwidth\textwidth}
      \includegraphics[width=\linewidth]{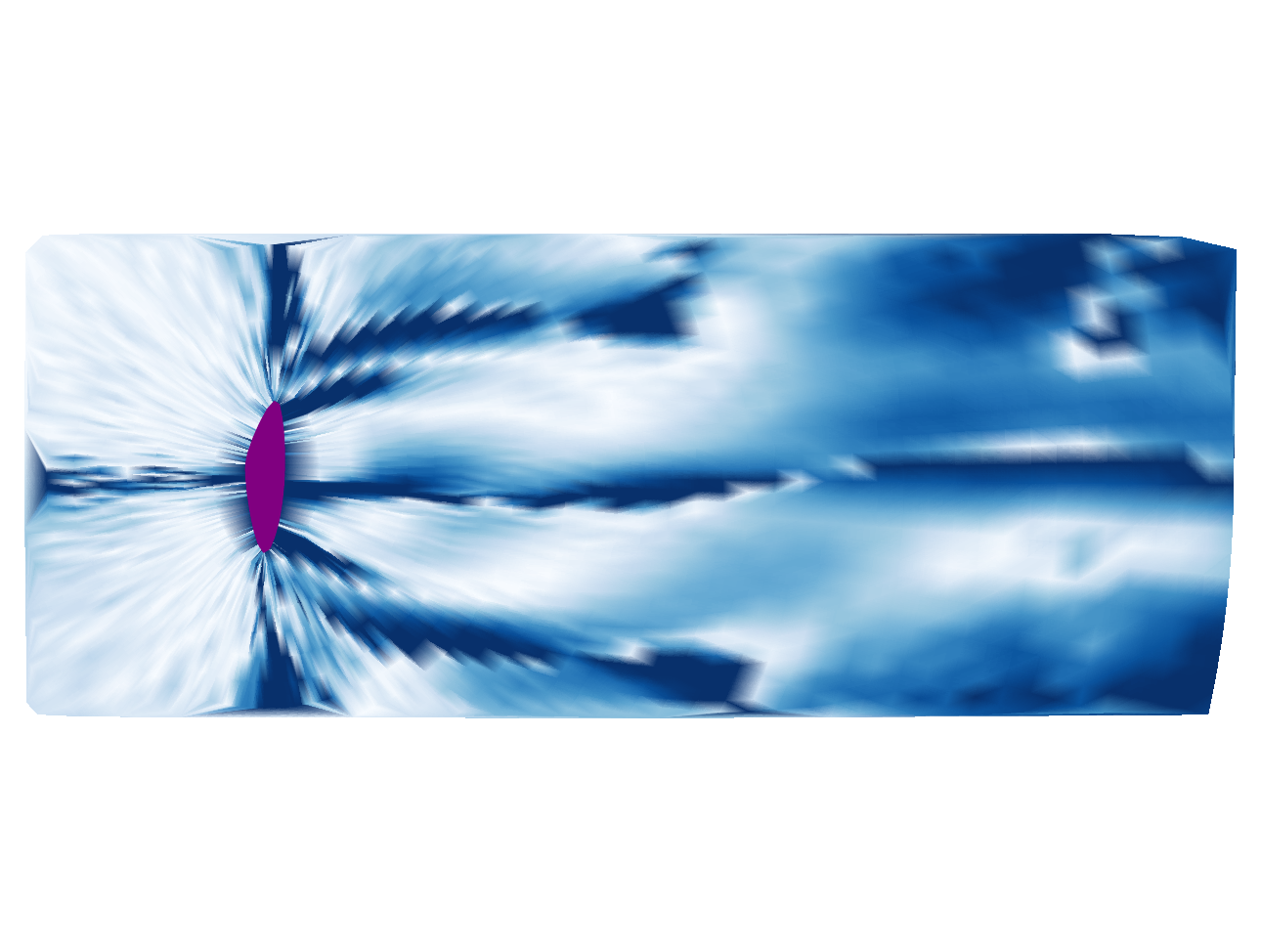}
    \end{minipage}                             &  \begin{minipage}{\spatialfigwidth\textwidth}
      \includegraphics[width=\linewidth]{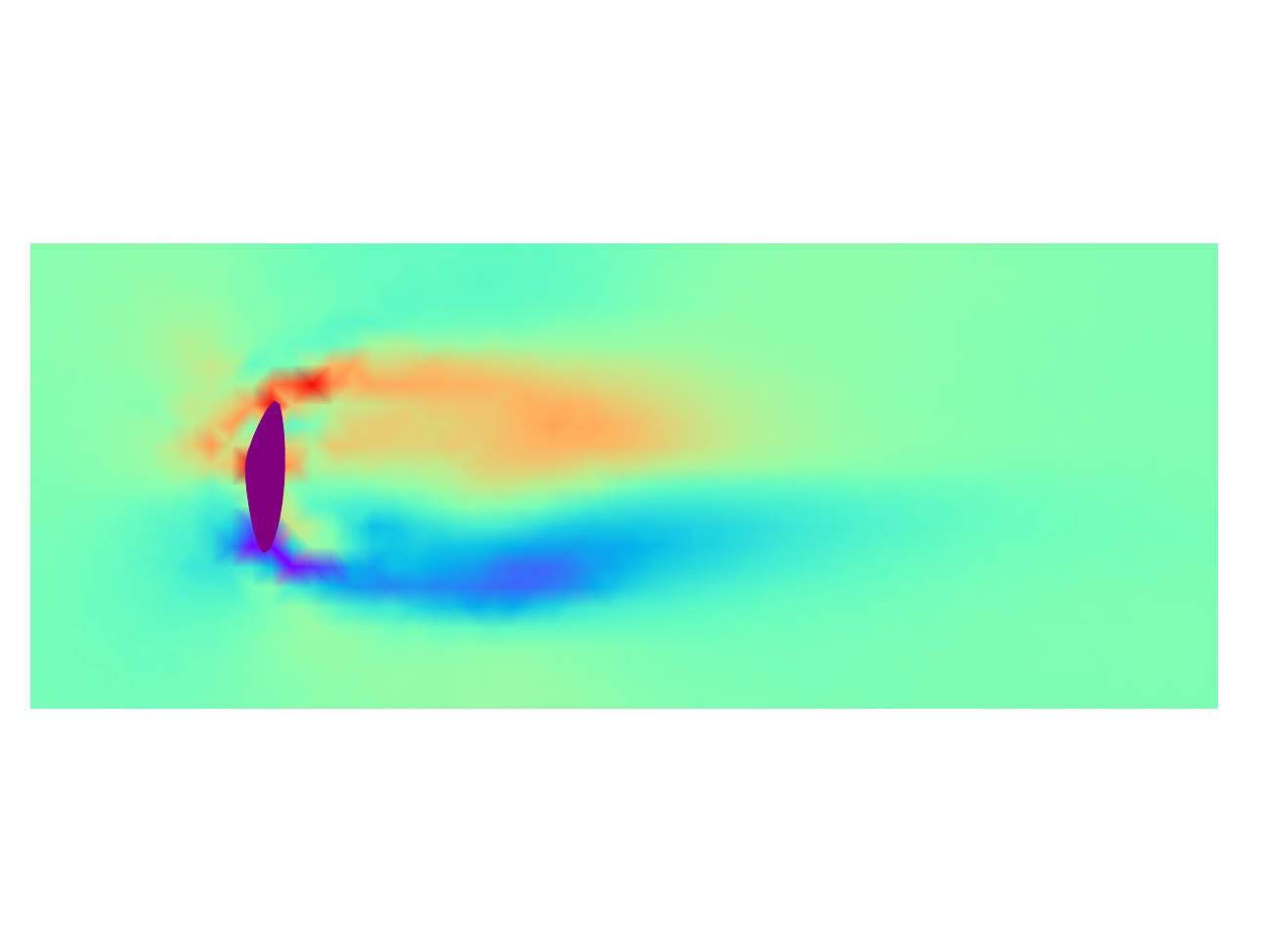}
    \end{minipage}                             & \begin{minipage}{\spatialfigwidth\textwidth}
      \includegraphics[width=\linewidth]{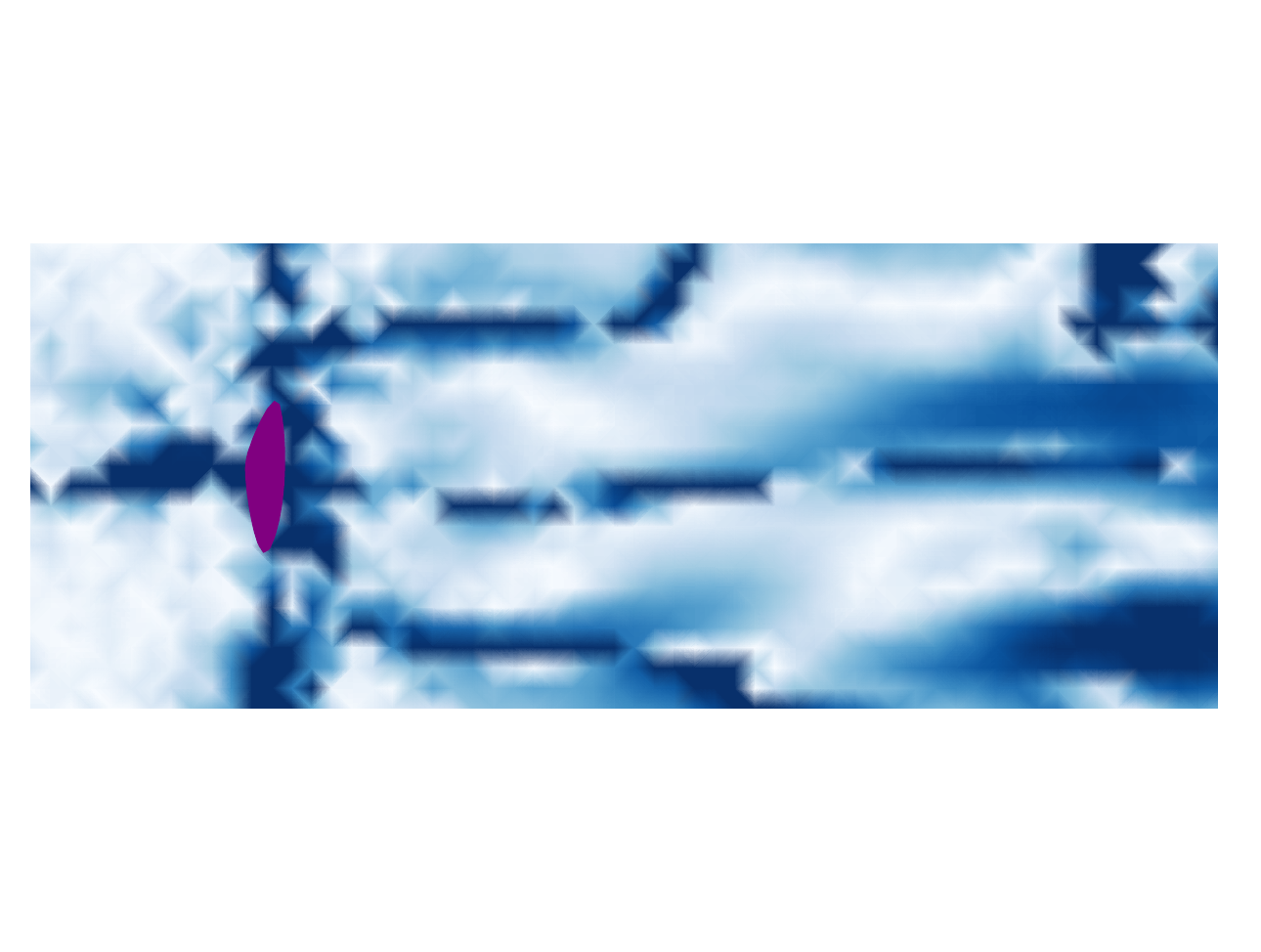}
    \end{minipage}         \\ 
\% Err    & \multicolumn{2}{c}{MAPE \textbf{47.19\%}, HV-MAPE \textbf{37.69\%}} & \multicolumn{2}{c}{MAPE 54.63\%, HV-MAPE 40.30\%}\\ \midrule
SD-Large               &    \begin{minipage}{\spatialfigwidth\textwidth}
      \includegraphics[width=\linewidth]{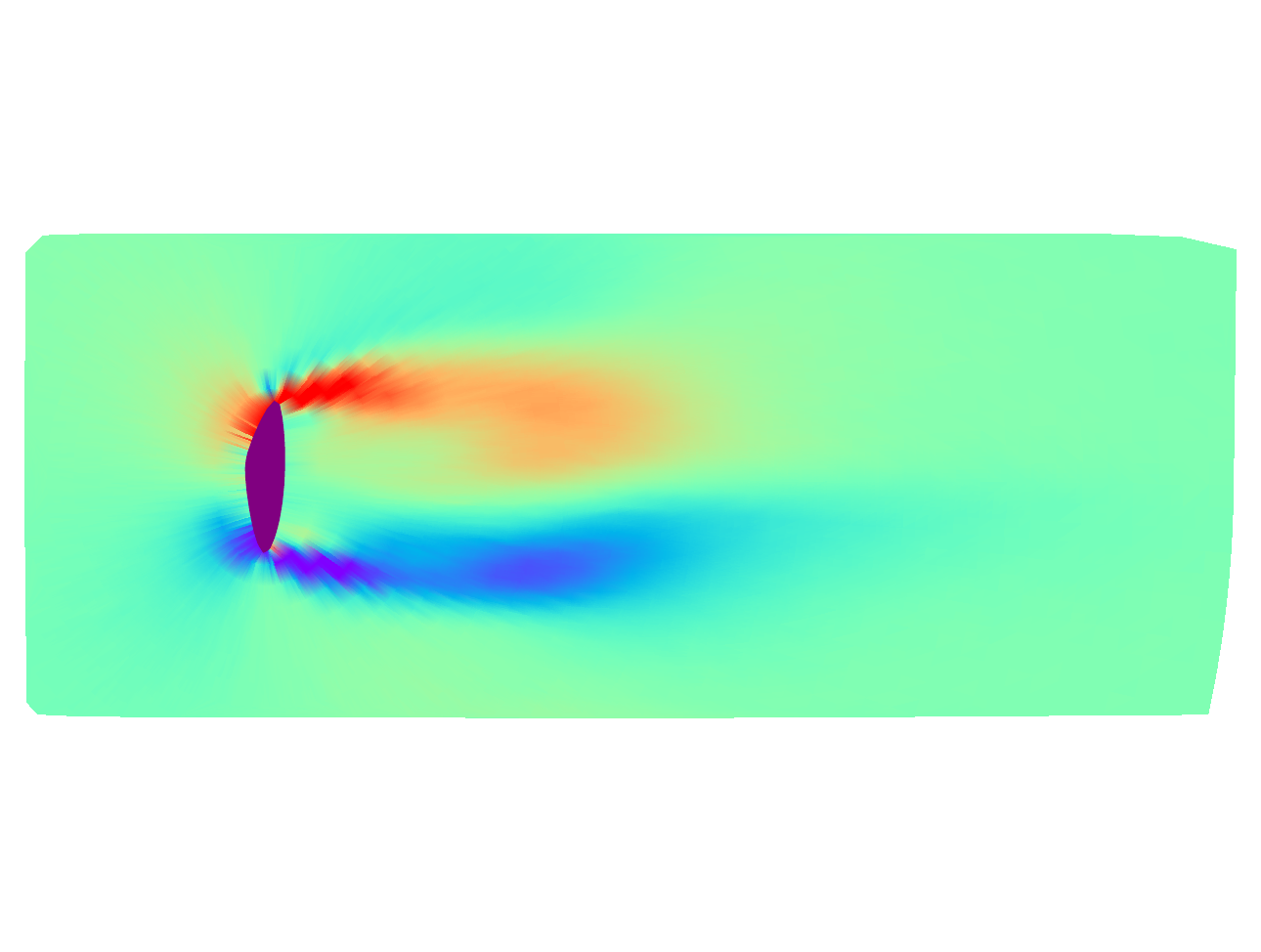}
    \end{minipage}                            &  \begin{minipage}{\spatialfigwidth\textwidth}
      \includegraphics[width=\linewidth]{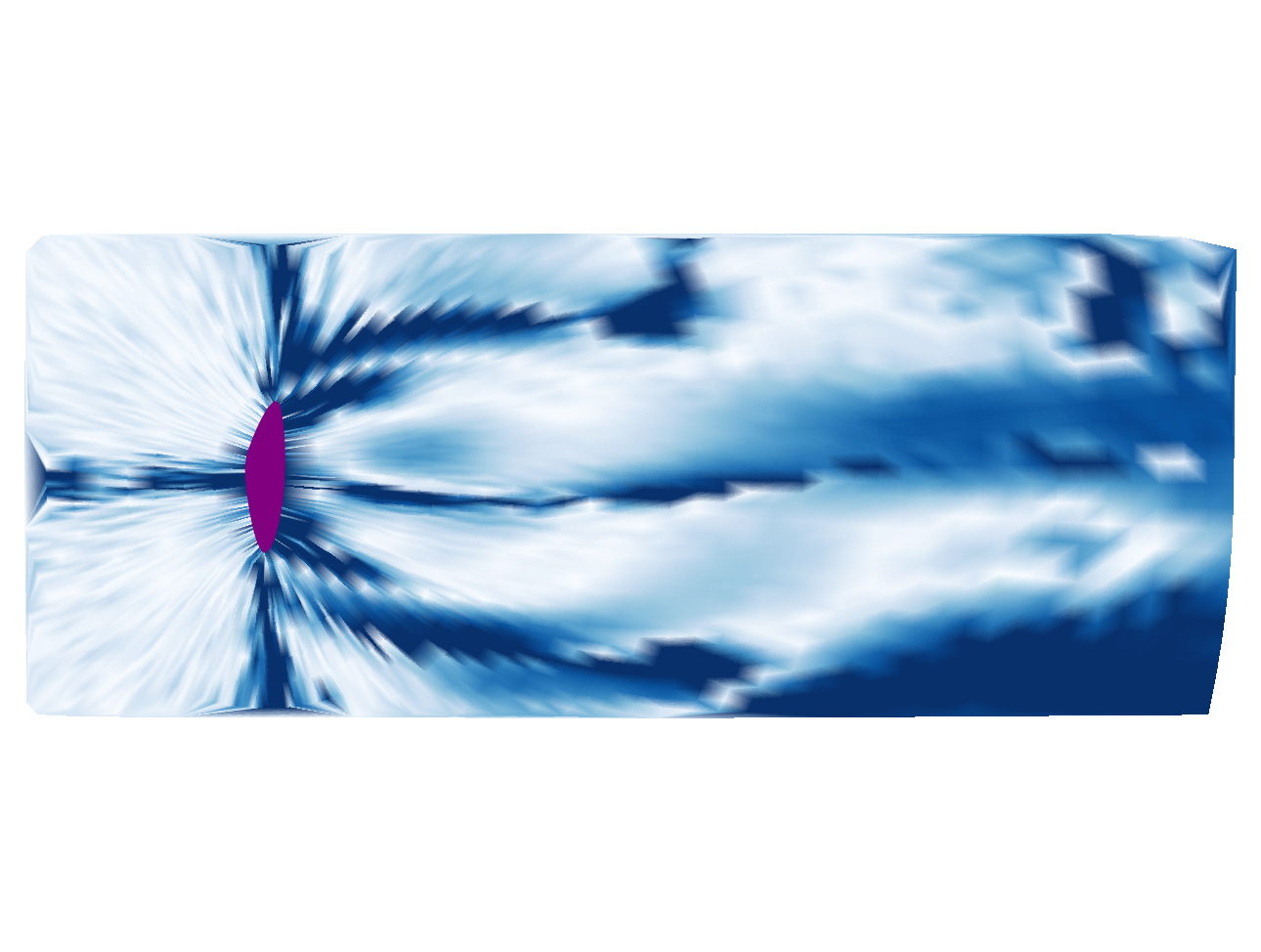}
    \end{minipage}                             &  \begin{minipage}{\spatialfigwidth\textwidth}
      \includegraphics[width=\linewidth]{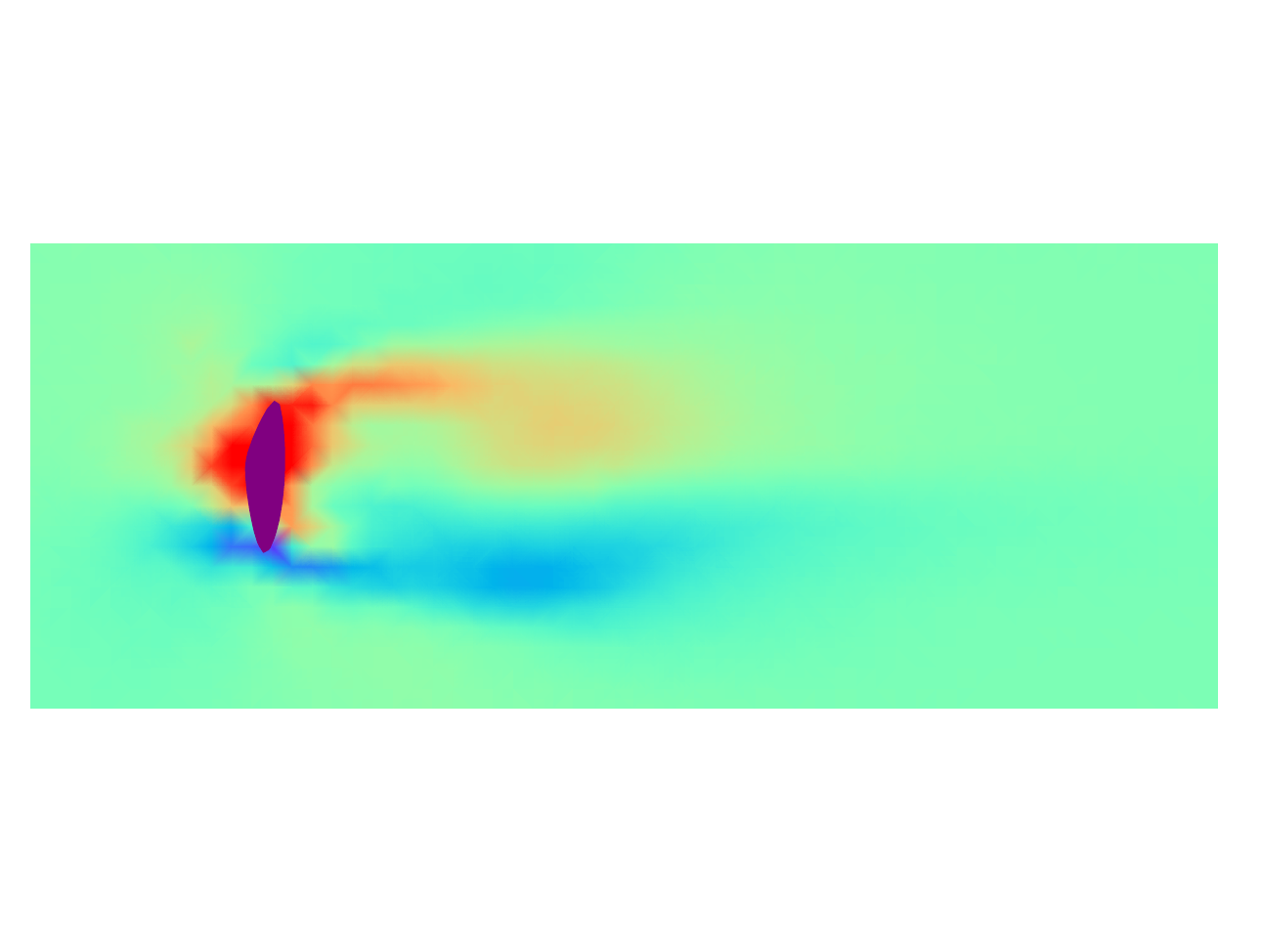}
    \end{minipage}                             & \begin{minipage}{\spatialfigwidth\textwidth}
      \includegraphics[width=\linewidth]{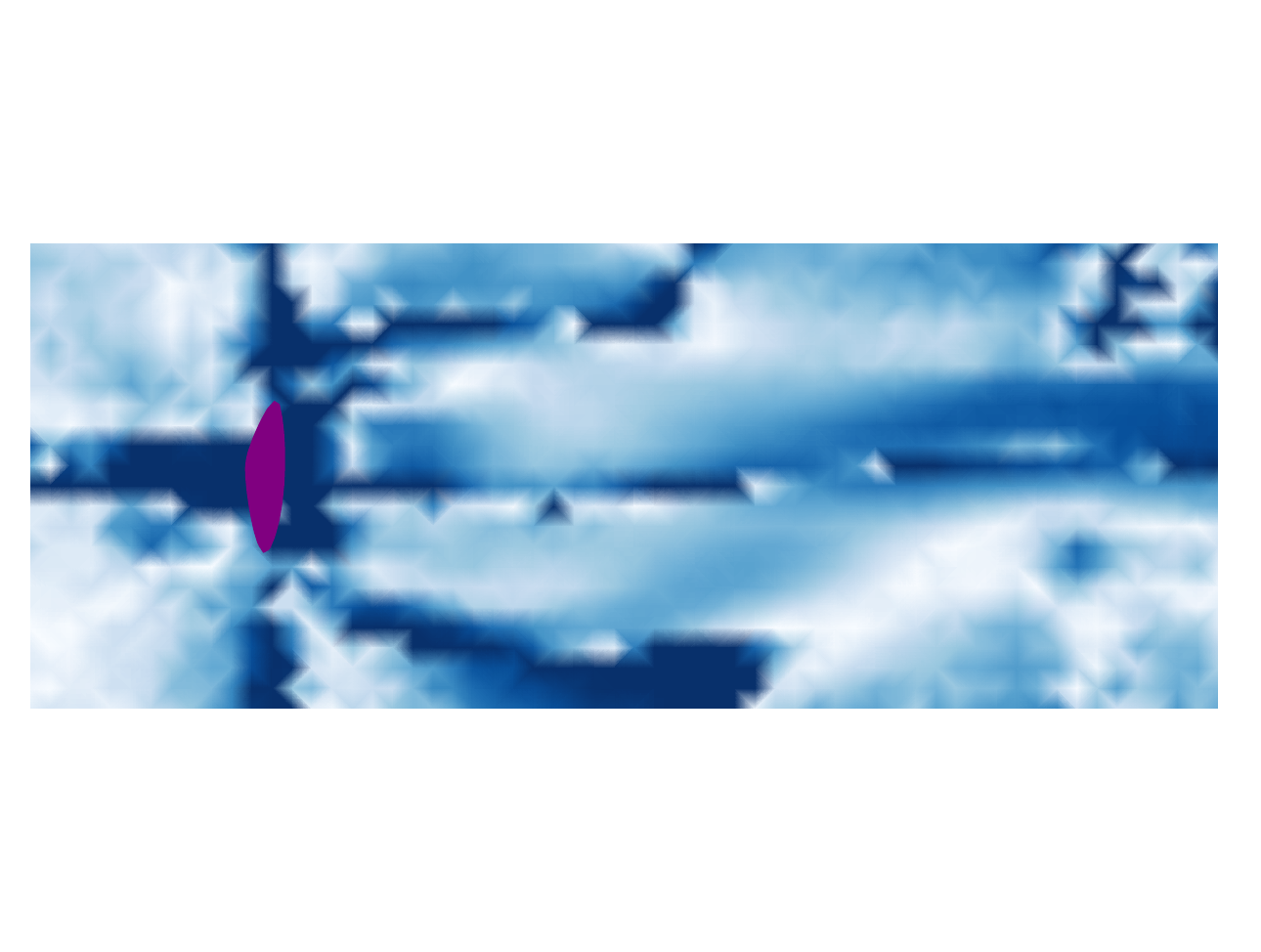}
    \end{minipage}          \\
\% Err    & \multicolumn{2}{c}{MAPE \textbf{39.81\%}, HV-MAPE \textbf{40.20\%}} & \multicolumn{2}{c}{MAPE 47.16\%, HV-MAPE 46.77}\\ \midrule
SD-UNet                &    \begin{minipage}{\spatialfigwidth\textwidth}
      \includegraphics[width=\linewidth]{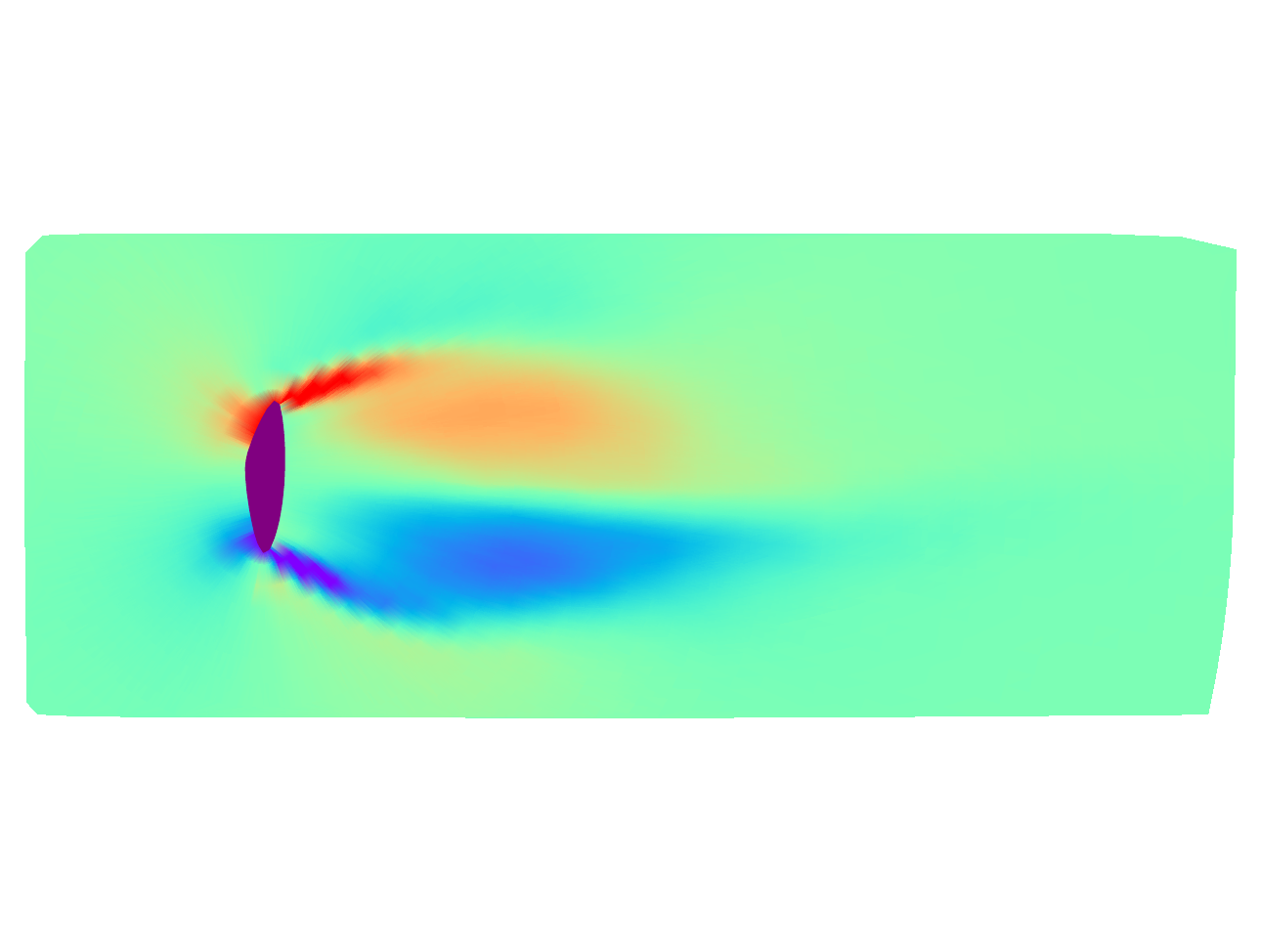}
    \end{minipage}                            &  \begin{minipage}{\spatialfigwidth\textwidth}
      \includegraphics[width=\linewidth]{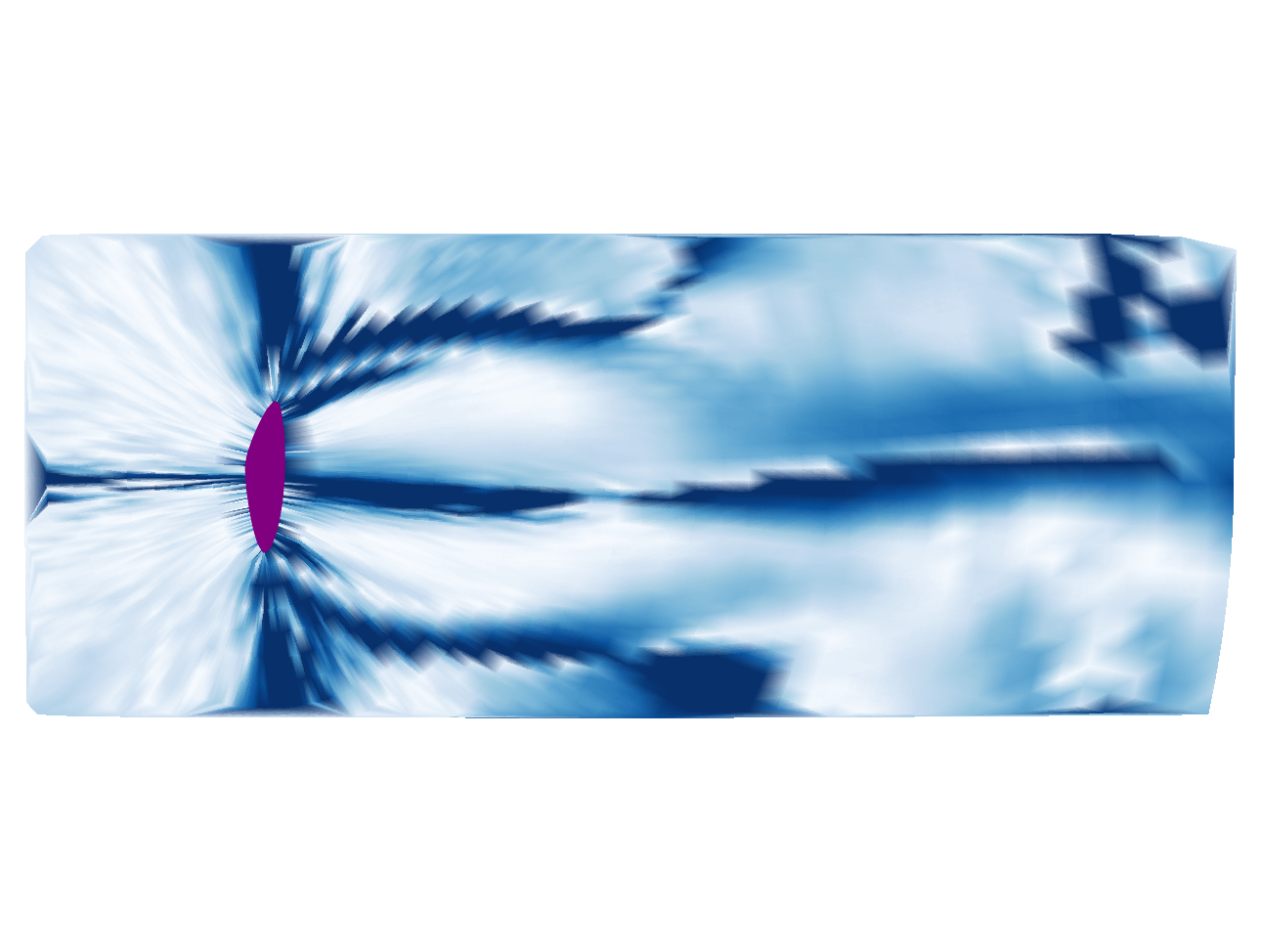}
    \end{minipage}                             &  \begin{minipage}{\spatialfigwidth\textwidth}
      \includegraphics[width=\linewidth]{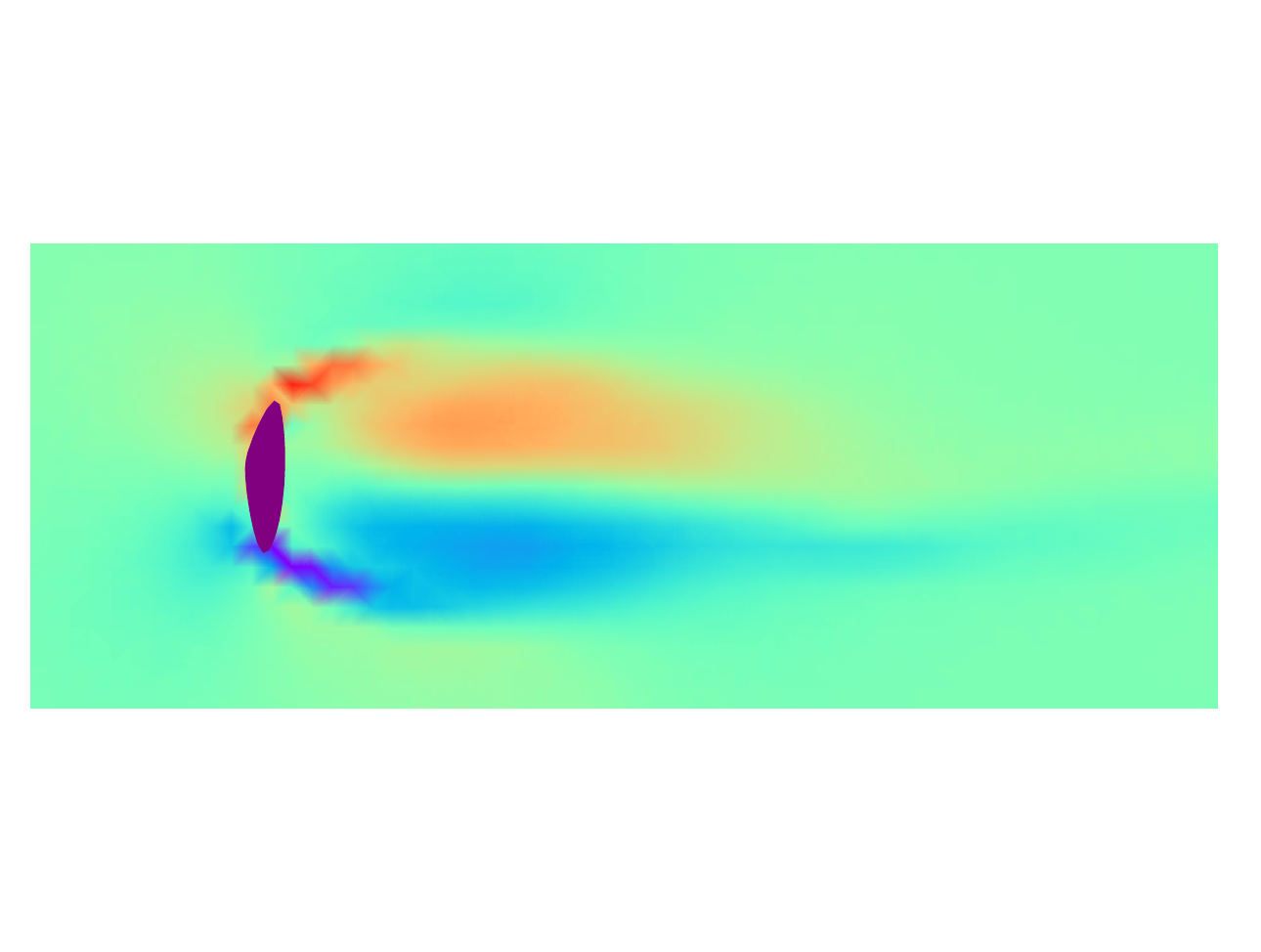}
    \end{minipage}                             & \begin{minipage}{\spatialfigwidth\textwidth}
      \includegraphics[width=\linewidth]{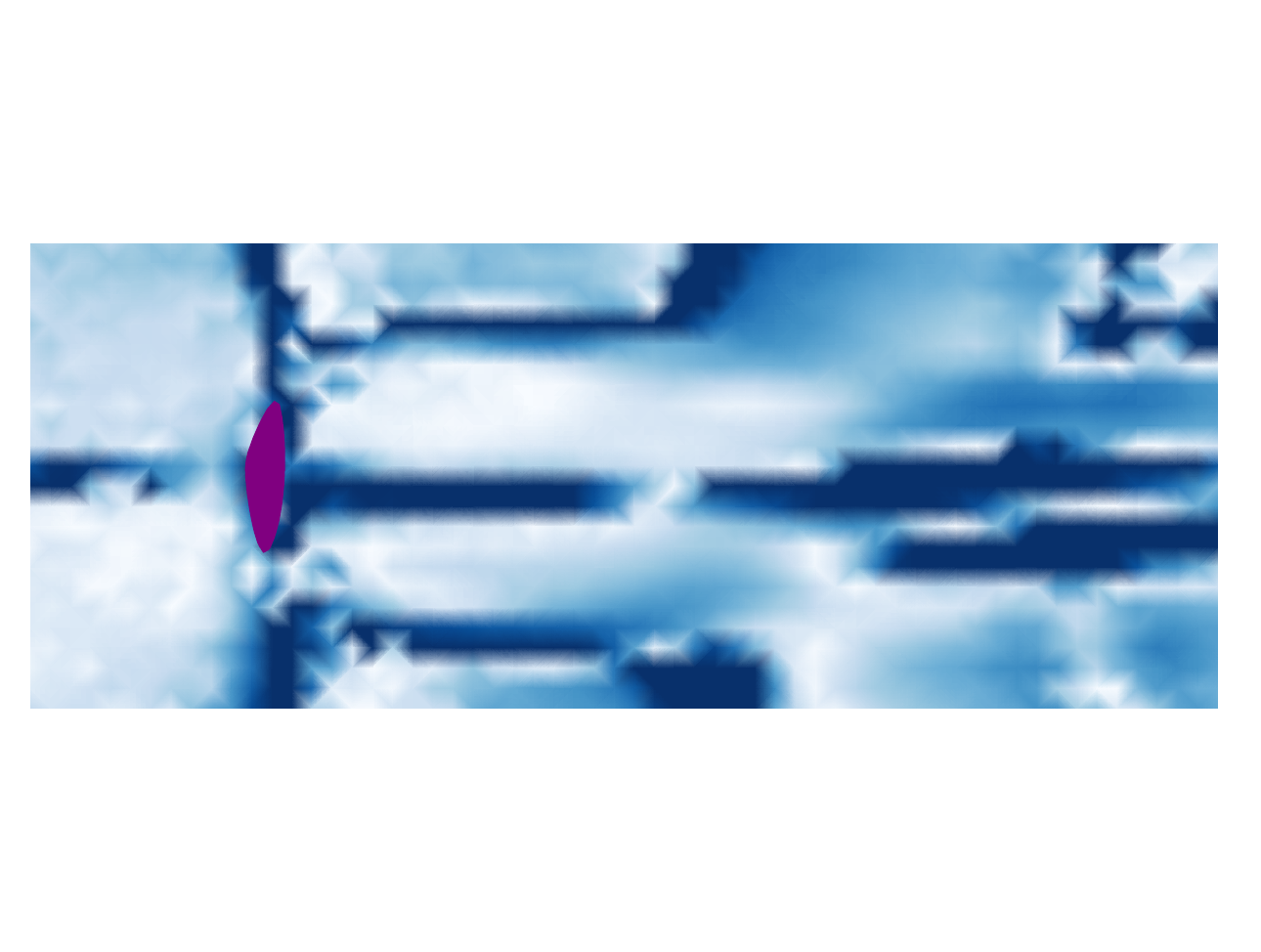}
    \end{minipage}          \\
\% Err    & \multicolumn{2}{c}{MAPE \textit{\textbf{36.86\%}}, HV-MAPE {\ul \textbf{35.95\%}}} & \multicolumn{2}{c}{MAPE 46.19\%, 43.25\%}\\ \midrule
SD-FNO                 &    \begin{minipage}{\spatialfigwidth\textwidth}
      \includegraphics[width=\linewidth]{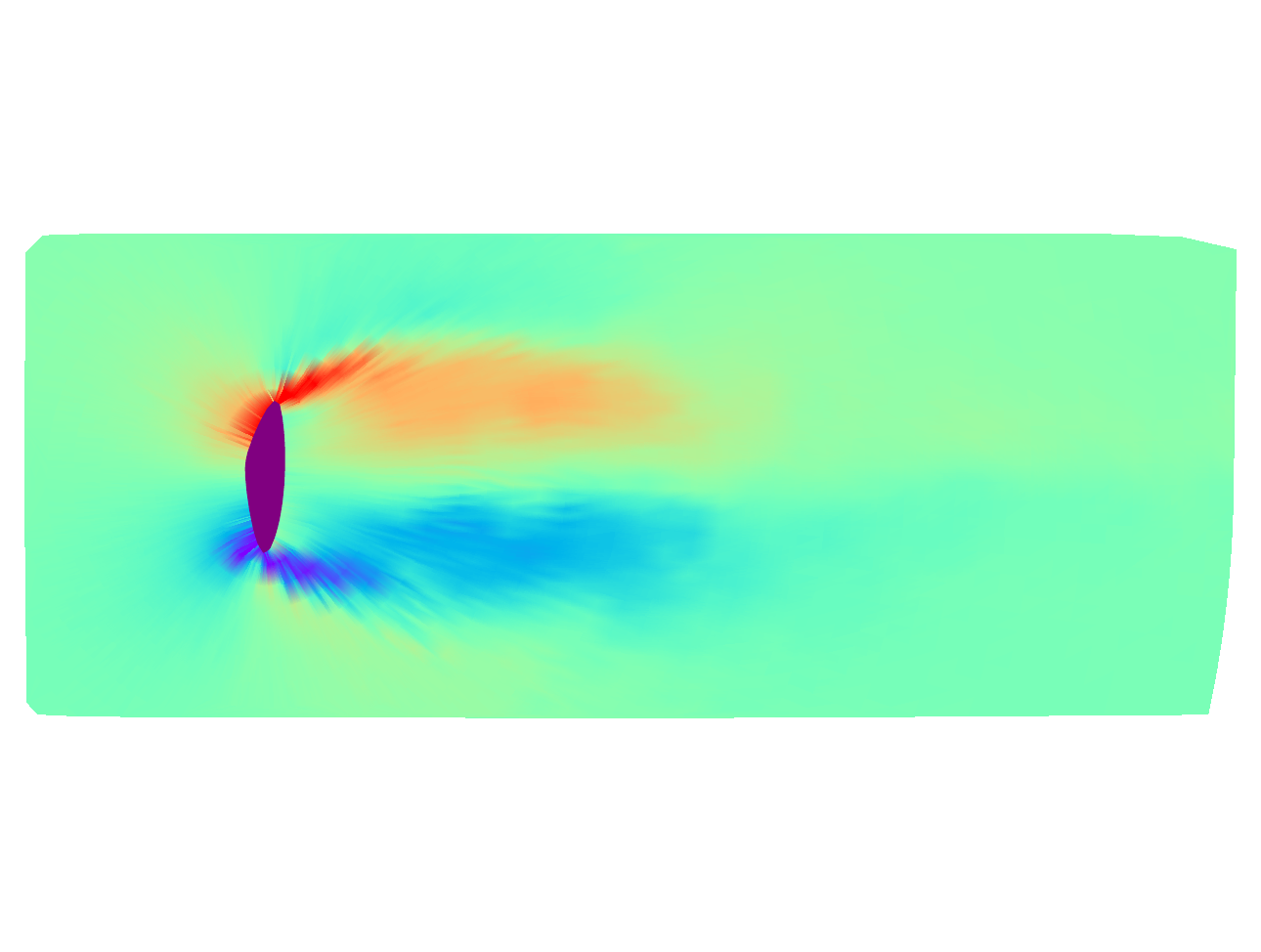}
    \end{minipage}                            &  \begin{minipage}{\spatialfigwidth\textwidth}
      \includegraphics[width=\linewidth]{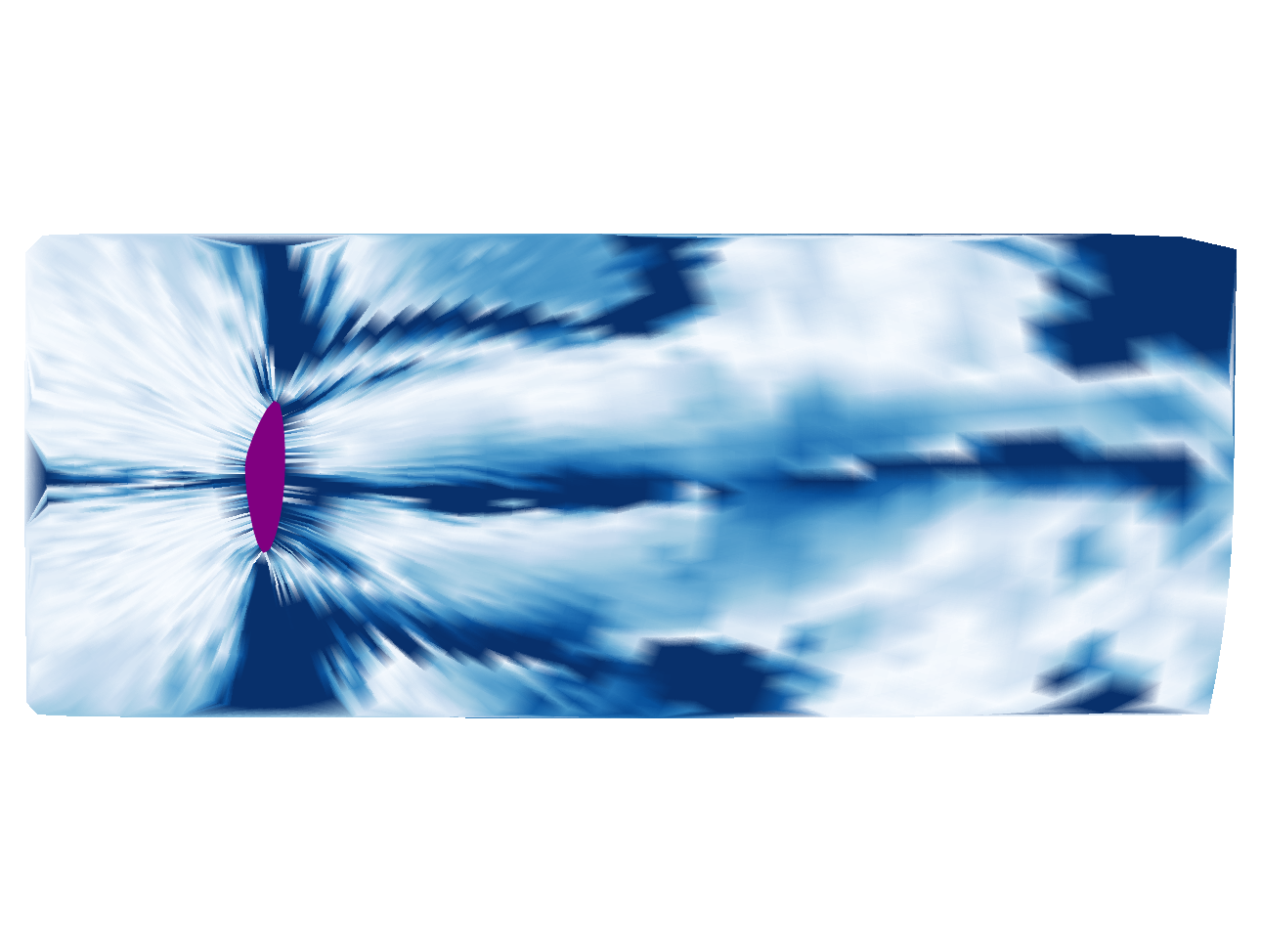}
    \end{minipage}                             &  \begin{minipage}{\spatialfigwidth\textwidth}
      \includegraphics[width=\linewidth]{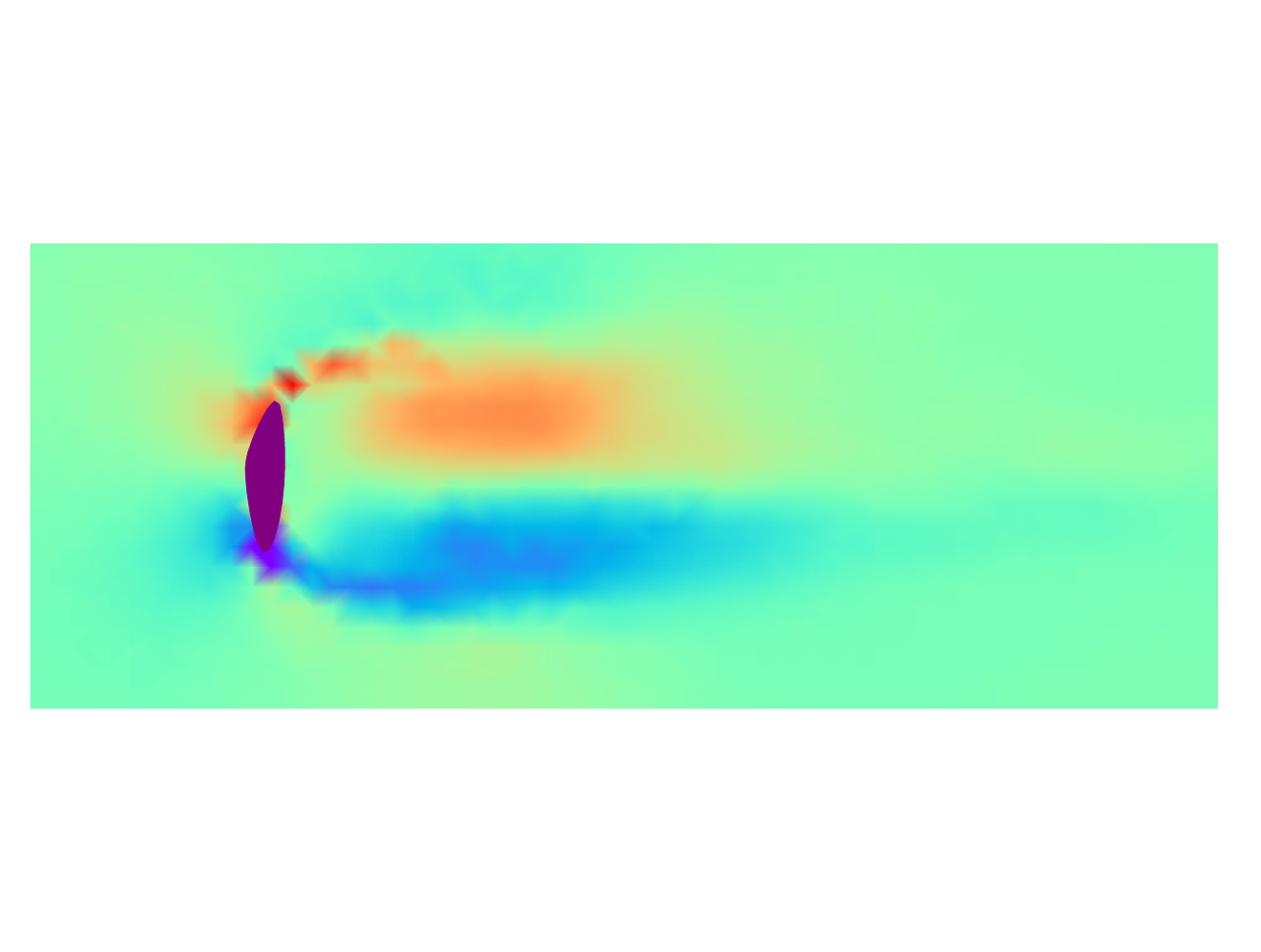}
    \end{minipage}                             & \begin{minipage}{\spatialfigwidth\textwidth}
      \includegraphics[width=\linewidth]{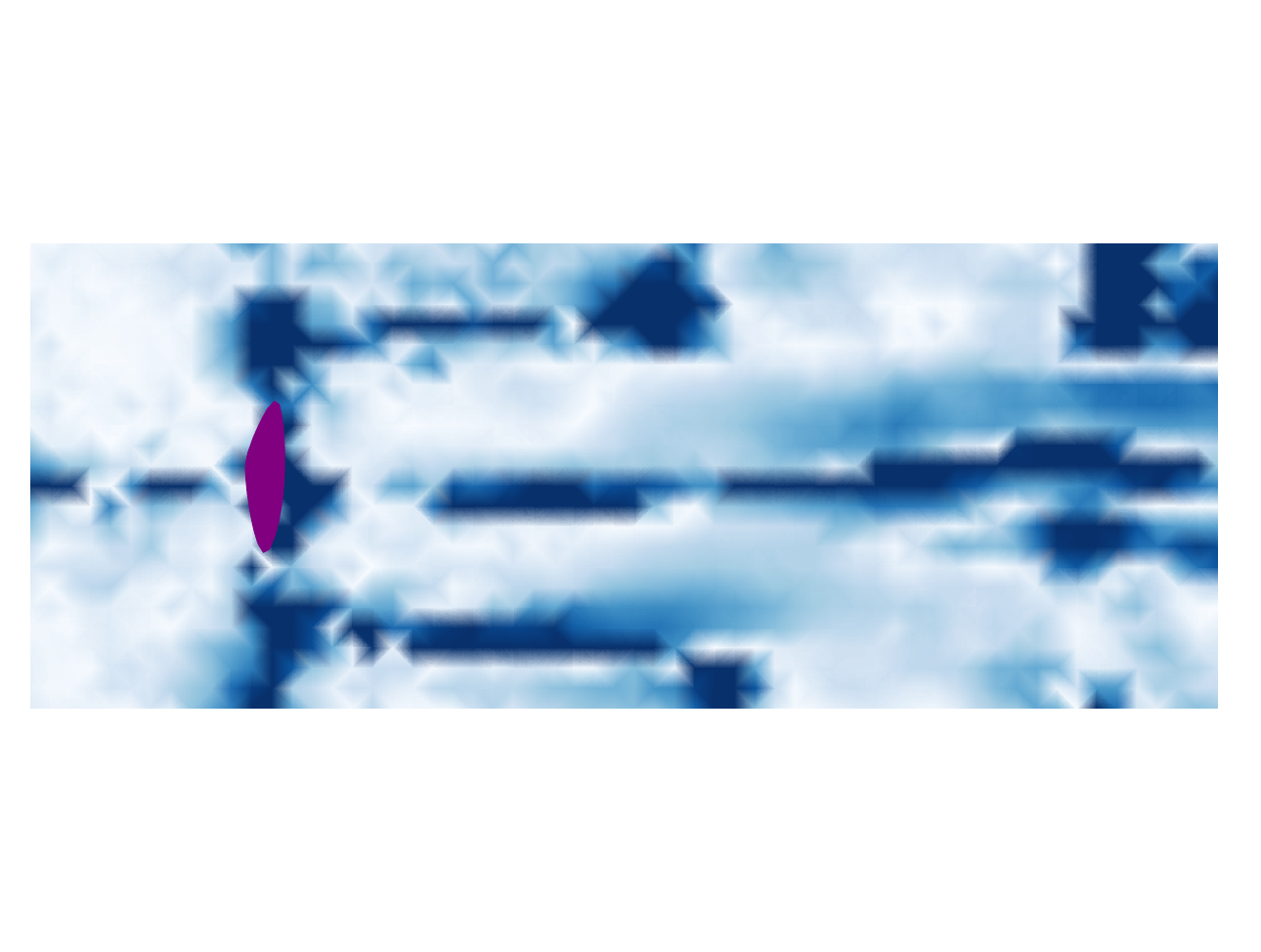}
    \end{minipage}          \\
\% Err    & \multicolumn{2}{c}{MAPE 38.69\%, HV-MAPE 38.83\%} & \multicolumn{2}{c}{MAPE \textbf{37.43\%}, HV-MAPE \textbf{36.15\%}}\\ \bottomrule
\end{tabular}
    
    \caption{Ground truth (top) and predicted (bottom) vorticity fields and percentage error maps for a snapshot from Shape C, using the large sensor setup. Vorticity colormap constrained to 5\% of $\max(|\tilde{\omega}_{t,i}|)$.}
    \label{fig:spatialex_s500_large}
\end{figure}

To provide deeper insight into the performance of the models with the different sensor and sampling setups beyond the overall error values, a gallery of predictions is provided. Figures \ref{fig:spatialex_s1660_large}, \ref{fig:spatialex_s1660_medium} and \ref{fig:spatialex_s1660_small} depict the predictions for a bluff body-like shape, dubbed `Shape A', using the large, medium and small sensor setups, respectively. Furthermore, \figref{fig:spatialex_s1400_large} and \figref{fig:spatialex_s500_large} showcase the predictions for `Shape B', an oval-like object set at a high incidence angle relative to the incident flow displaying separation over its upper surface, and `Shape C', a thick flat plate-like object with characteristic counter-rotating vortices immediately downstream of the object, for the large sensor setup. Juxtaposing the models' performance over a range of flows showing a diverse range of dynamics, the three snapshots are chosen to illustrate different levels of relative performance between the sampling strategies and models; Shape A's snapshot is chosen to display the relative strengths of the top performing combination (SD-UNet with annular sampling), Shape B displays a case where Cartesian sampling wins in terms of MAPE (albeit losing by a far greater margin in terms of HV-MAPE), and in Shape C's case the top performers are closely matched.


Starting with Shape A in Figures \ref{fig:spatialex_s1660_large}-\ref{fig:spatialex_s1660_small}, the focus is on the differences between the Cartesian and annular sampling methods. The boundary layers stand out as the areas of greatest difference between the sampling methods. With Cartesian sampling, none of the models can accurately predict the existence of the high vorticity concentrations near the stagnation point, the highly positive concentration of vorticity along the upper surface of the object, or the location of the separation on the lower surface. In contrast, all three of these features are correctly predicted with the annular sampling, even by the relatively simple feedforward SD and SD-Large models, regardless of the sensor setup. The two sampling strategies are closer in terms of performance in the wake, as evidenced by the greater similarity in the error maps, but the Cartesian models have higher error downstream of the object.

Additionally, comparing the predictions from different sensor setups for Shape A, the above conclusion regarding the limited impact of the setup on model performance is reinforced. Error levels do not show a strong tendency to decline with larger sensor setups, exemplified by the lowest errors for the SD-UNet being displayed for the medium sensor setup in \figref{fig:spatialex_s1660_medium}. Randomness in model training, caused by model initialization and stochastic model architecture features like dropout, has a substantially greater impact on performance in particular snapshots. This effect disappears when averaged over the whole dataset and many epochs of training, resulting in the paradoxical looking situations like predictions from models using fewer sensors being more accurate for specific samples despite higher overall error.

Moving forward to Shape B in \figref{fig:spatialex_s1400_large}, the difference is most striking for the SD and SD-Large, where the Cartesian versions of the models predict a high concentration of negative vorticity entirely engulfing the object, which is highly non physical. Additionally, the location of the high positive vorticity blob near the `leading edge' of the object is predicted as \textit{detached} from the object surface. In contrast, the same models with annular sampling correctly predict these key flow features, appropriately placing the concentrations of vorticity near the stagnation point, the separation near the leading edge, and the starting vortex-like formation near the rear of the object, though also performing poorly with respect to the vortex shed further downstream of the object. These issues are not as prevalent for the more complicated SD-UNet model, with the Cartesian sampling ekeing out a rare quantitative win in terms of the MAPE (by 5\%) against annular for the SD-UNet owing to the high error with annular sampling far downstream of the object despite more accurate predictions near the object. Lower performance further away from the object in the case of annular sampling is caused by the same radial and angular spacing, which translates to a lower grid point density further away from the object (see \secref{sec:gtvalues}). As a consequence, as with the other snapshots, annular sampling displays greater performance nearer the object and achieves a decisive win versus Cartesian sampling once the low-vorticity regions far downstream of the object are filtered out, achieving a 15\% lower HV-MAPE. 

The trends for the other shapes continue in the case of Shape C in \figref{fig:spatialex_s500_large}. Similar to both feedforward models for Shape B, the SD-Large model with Cartesian sampling spuriously predicts a large concentration of very high vorticity surrounding the object while also substantially under-estimating the intensity of the downstream vortices. The SD-UNet with annular sampling performs the best but with Cartesian sampling the error is much higher, also due to an underestimation of the intensity of the two counter-rotating vortices behind the object. Finally, the SD-FNO is the trend-breaker, with Cartesian sampling managing a narrow quantitative win thanks to a better estimation of the vortex intensity. However, from a qualitative perspective, both SD-FNO predictions are noisy and do not accurately reproduce the smoothness of the vorticity field unlike the SD-UNet.


As a final remark, we draw attention to the presence of high error along the same contours for different sampling strategies and models among the images for each snapshot. This is due to the presence of very high percentage error (despite low absolute error) near the zero contours of the target field due to very small denominators. Visible in low vorticity areas across all geometries, it is ultimately caused by the objective function as explained above. It is also the main reason why the MAPE and HV-MAPE may appear high with values consistently between 20\% and 60\%.

\clearpage

\clearpage

\clearpage

\clearpage

\clearpage

\clearpage

\subsection{Spatio-temporal multi-geometry vorticity reconstruction}
\label{sec:results-spatio-temporal}
%
Since the SD-UNet model, used alongside annular sampling and the large sensor setup, was identified as the best performing combination in \secref{sec:results-spatial}, this combination was chosen for the spatial model in this work's approach to the STMGFR task. The performance of the temporal model is summarized in \tabref{tab:spatio-temporal_mape} for different values of the temporal gap $k$ (refer to \tabref{tab:spatial_mape} for the spatial model's performance). A set of results with $k=0$, at which point the temporal model (FNO) effectively acts like an autoencoder since ground truth snapshots are provided as inputs, are included in addition to the $k=20$ and $k=80$ values mentioned in \secref{sec:model_spatio-temporal_reconstruction} to provide greater insight into the temporal model's behavior. 

\begin{table}[h!]
\caption{\% errors of the temporal model predictions, averaged over the validation set, given ground truth snapshots and reconstructed snapshots (i.e. spatial model outputs) as inputs. Outliers filtered identically to \tabref{tab:spatial_mape}. Only the temporal model was re-trained for different values of $k$.}
\label{tab:spatio-temporal_mape}
\begin{tabular}{@{}lcc|cc|cc@{}}
\toprule
\multirow{2}{*}{$k$ ($\Delta \tau^*$)} & \multicolumn{2}{c|}{0 (0.0)} & \multicolumn{2}{c|}{20 (0.667)} & \multicolumn{2}{c}{80 (2.667)} \\
                                     & MAPE          & HV-MAPE      & MAPE           & HV-MAPE        & MAPE           & HV-MAPE       \\ \midrule
From ground truth                    & 19.76\%       & 10.75\%      & 23.40\%        & 11.75\%        & 29.58\%        & 19.53\%       \\
From reconstruction                  & 28.89\%       & 17.86\%      & 31.02\%        & 17.86\%        & 31.88\%        & 21.97\%       \\ \bottomrule
\end{tabular}
\end{table}

As expected, the error declines as the temporal gap $k$ is reduced and the model performs better when ground truth snapshots are provided as the inputs as opposed to reconstructed inputs. Surprisingly, however, MAPE levels for this task are substantially lower than results for the purely spatial task in \tabref{tab:spatial_mape} despite its greater difficulty. The reason for this is clearly illustrated by the $k=0$ results: the additional FNO model placed `at the end' of the SD-UNet model, with a parameter count exceeding the entirety of the SD-UNet, serves as a denoising autoencoder\cite{denoising_autoencoder} which substantially reduces the error of the SD-UNet output. Effectively acting like a model employed in SMGFR as opposed to STMGFR in this setting, examples of reconstructed snapshots from this $k=0$ `SD-UNet + FNO' configuration are presented in Appendix \ref{sec:app:spatio-temporalpred-k0}. 

\begin{figure}
    \centering
    \begin{tabular}{@{}cccc@{}}
\toprule
Ground truth at $t$                      & \multicolumn{2}{c}{Reconstruction at $t$}                              & \% Error \\ \midrule
\multicolumn{1}{c}{
    \begin{minipage}{\spatiotemporalfigwidth\textwidth}
      \includegraphics[width=\linewidth]{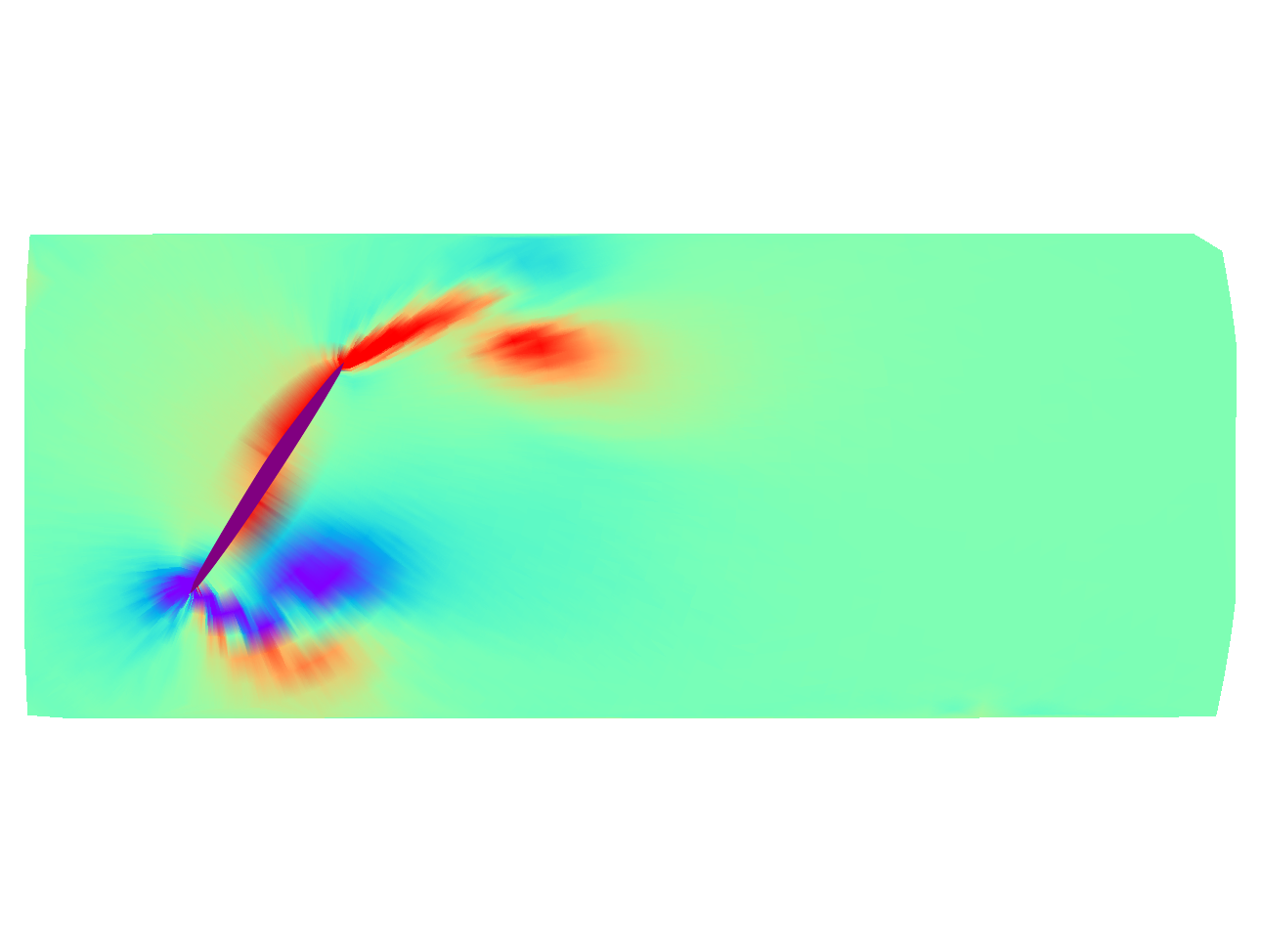}
    \end{minipage}}                  & \multicolumn{2}{c}{
    \begin{minipage}{\spatiotemporalfigwidth\textwidth}
      \includegraphics[width=\linewidth]{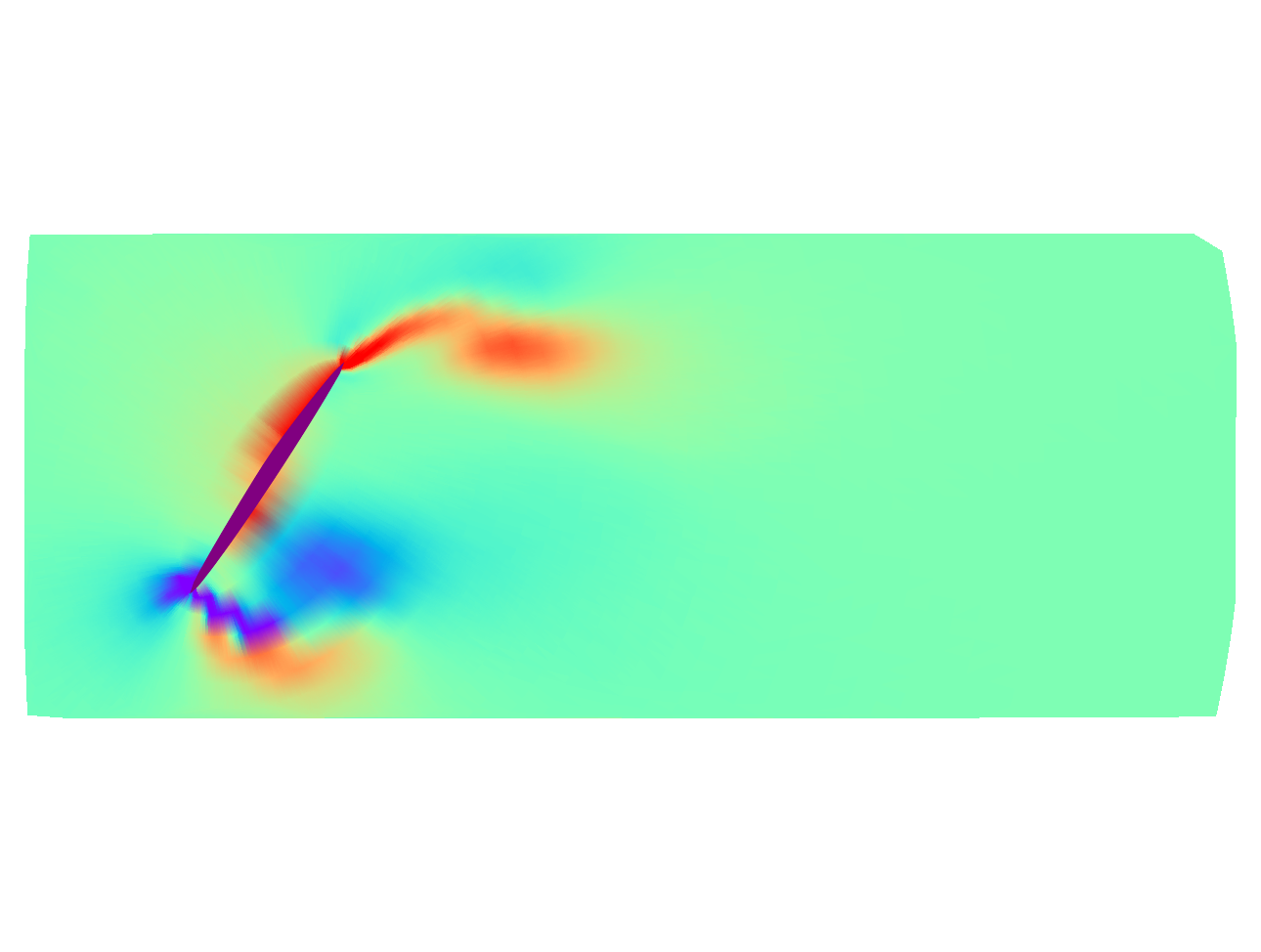}
    \end{minipage}}                                                &  
    \begin{minipage}{\spatiotemporalfigwidth\textwidth}
      \includegraphics[width=\linewidth]{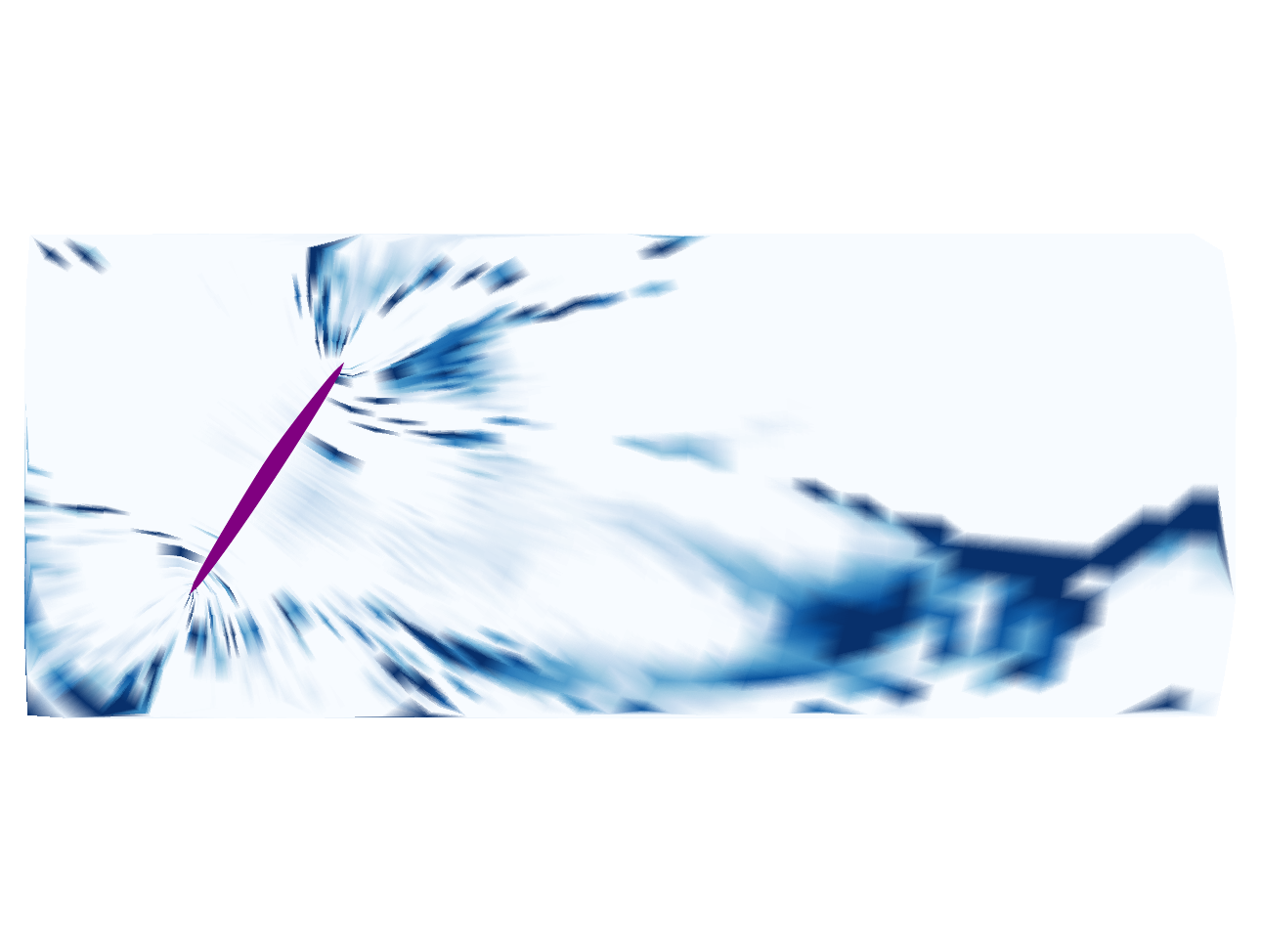}
    \end{minipage} \\
    \begin{minipage}{\spatiotemporalfigwidth\textwidth}
      \includegraphics[width=\linewidth]{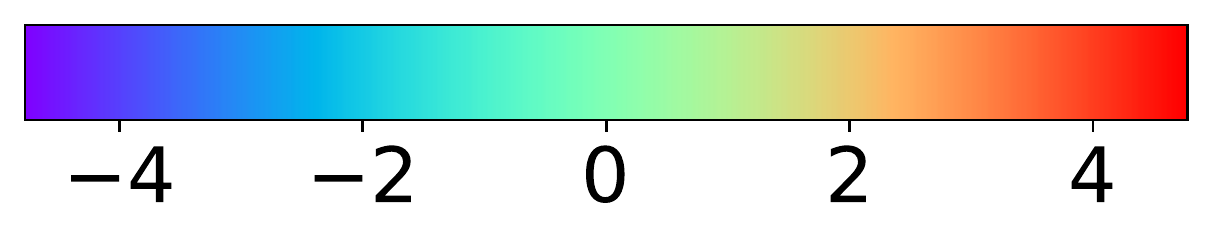}
    \end{minipage} &&&
    \begin{minipage}{\spatiotemporalfigwidth\textwidth}
      \includegraphics[width=\linewidth]{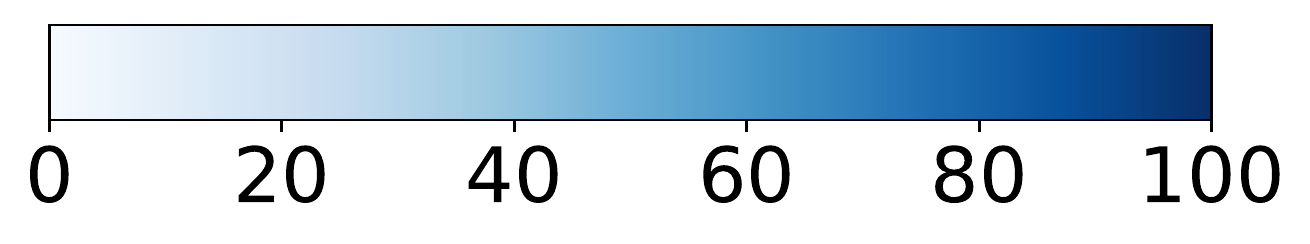}
    \end{minipage} \\
    &&&MAPE: 28.54\% \\ \midrule
Ground truth at $t+k$                    & \multicolumn{2}{c}{Reconstruction at $t+k$}                            & \% Error \\ \midrule
\multicolumn{1}{c}{\multirow{4}{*}{\begin{minipage}{\spatiotemporalfigwidth\textwidth}
      \vspace{1.25cm}
      \includegraphics[width=\linewidth]{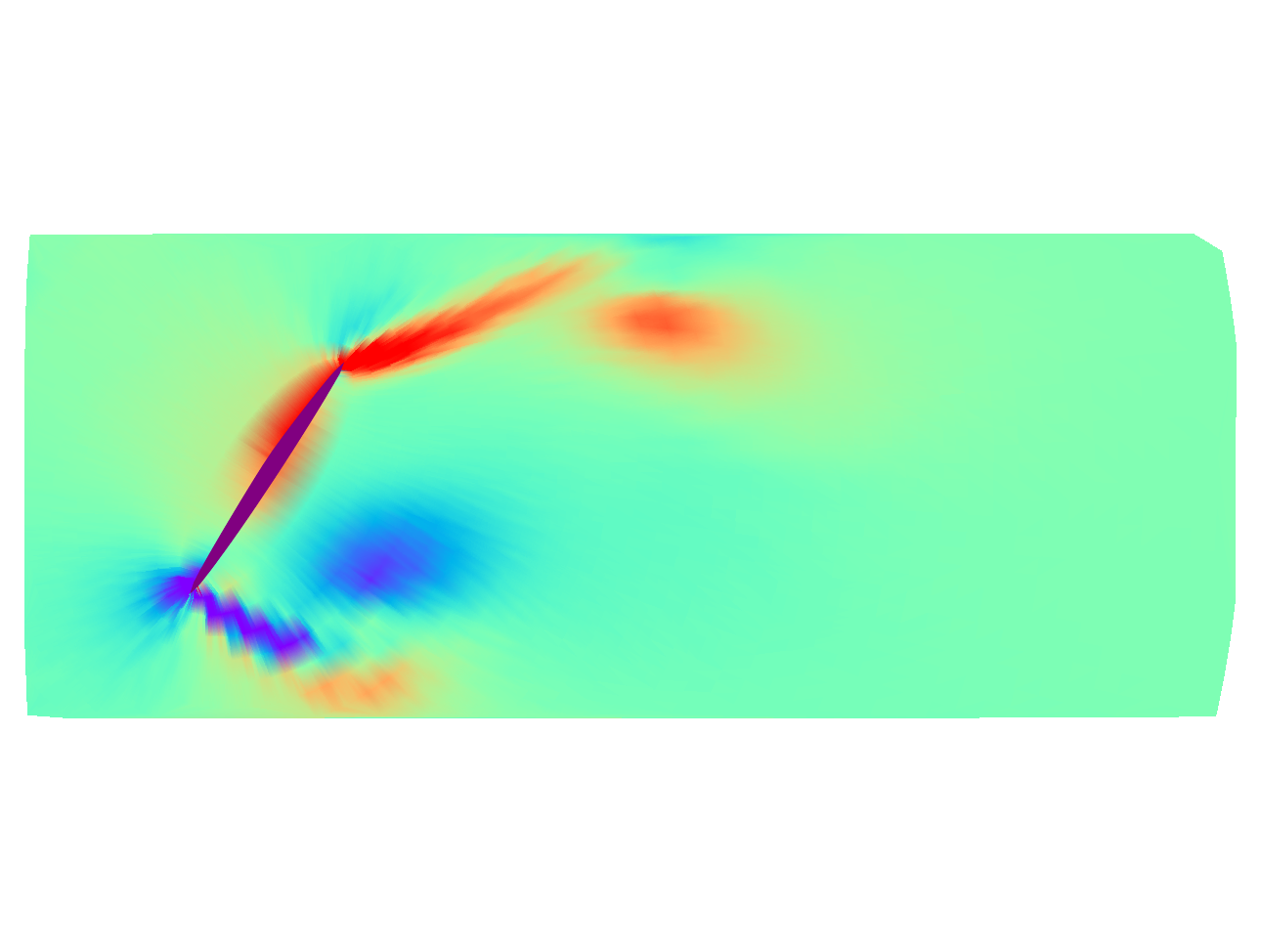}
    \end{minipage}}} & \multicolumn{1}{l}{From ground truth at $t$}   & \multicolumn{1}{c}{
    \begin{minipage}{\spatiotemporalfigwidth\textwidth}
      \includegraphics[width=\linewidth]{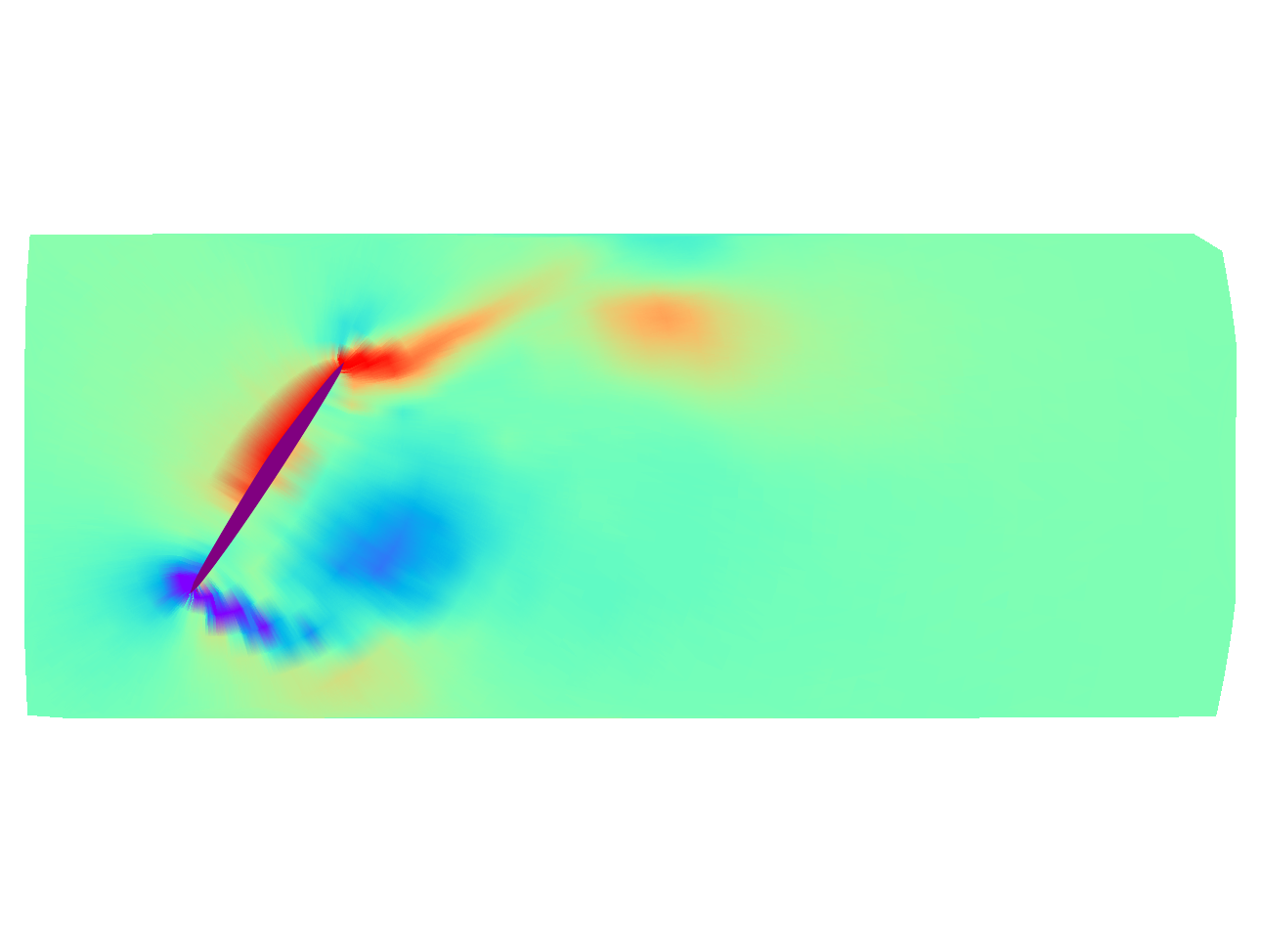}
    \end{minipage}} &  
    \begin{minipage}{\spatiotemporalfigwidth\textwidth}
      \includegraphics[width=\linewidth]{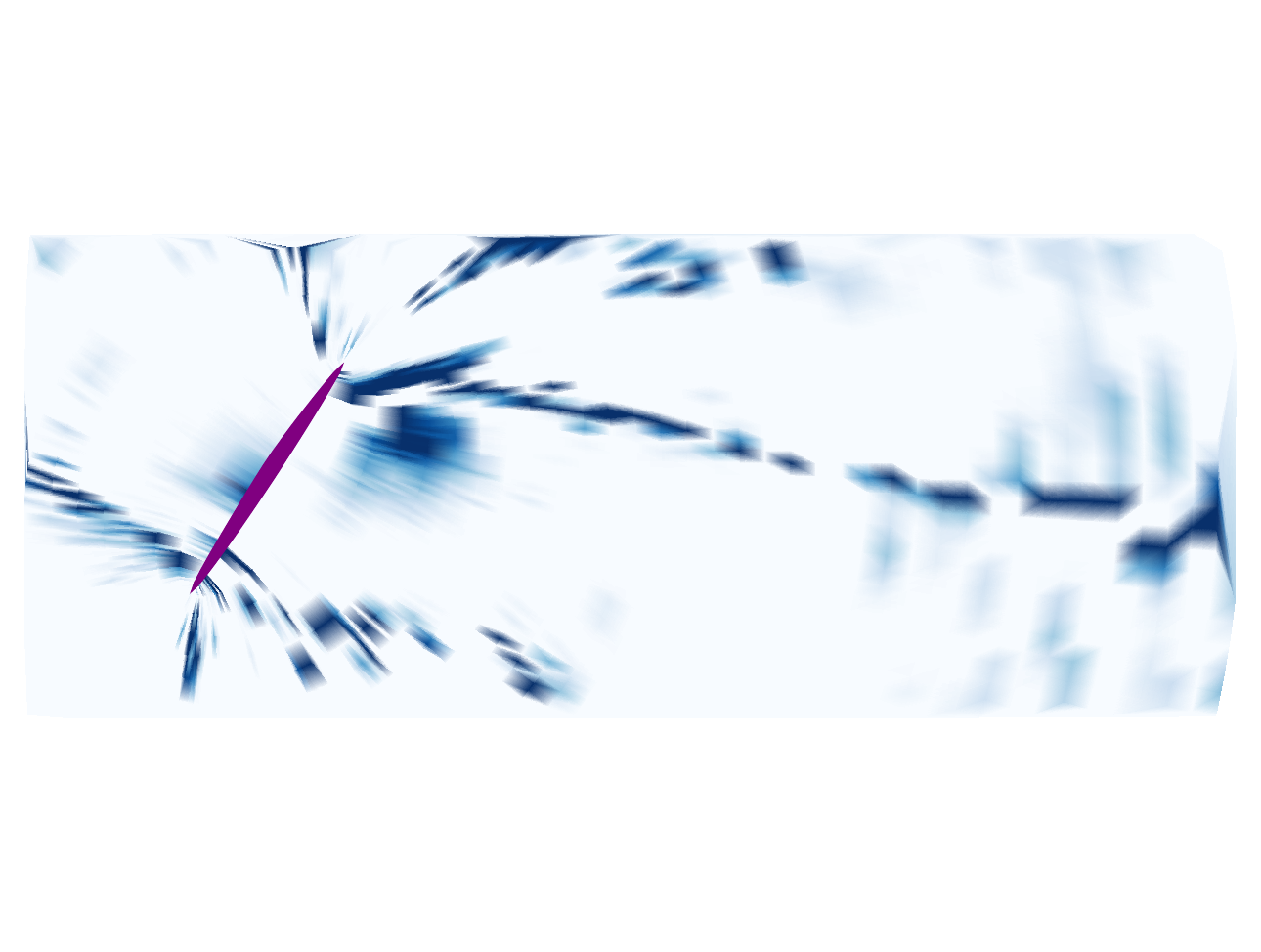}
    \end{minipage}  \\
    &&& MAPE: 27.28\% \\ \cmidrule(l){2-4} 
    \multicolumn{1}{c}{} & \multicolumn{1}{l}{From Reconstruction at $t$} & \multicolumn{1}{c}{
    \begin{minipage}{\spatiotemporalfigwidth\textwidth}
      \includegraphics[width=\linewidth]{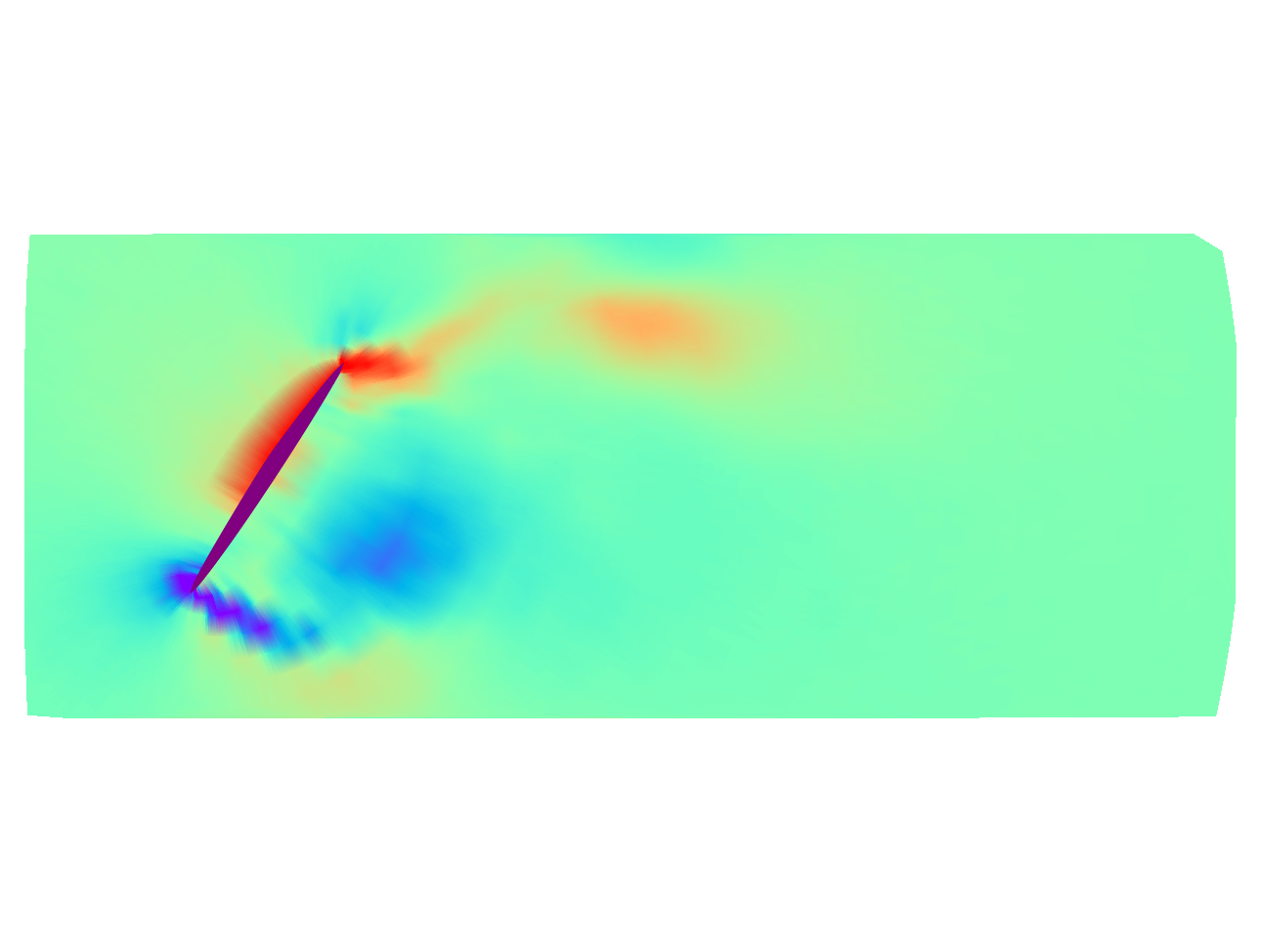}
    \end{minipage}} &  \begin{minipage}{\spatiotemporalfigwidth\textwidth}
      \includegraphics[width=\linewidth]{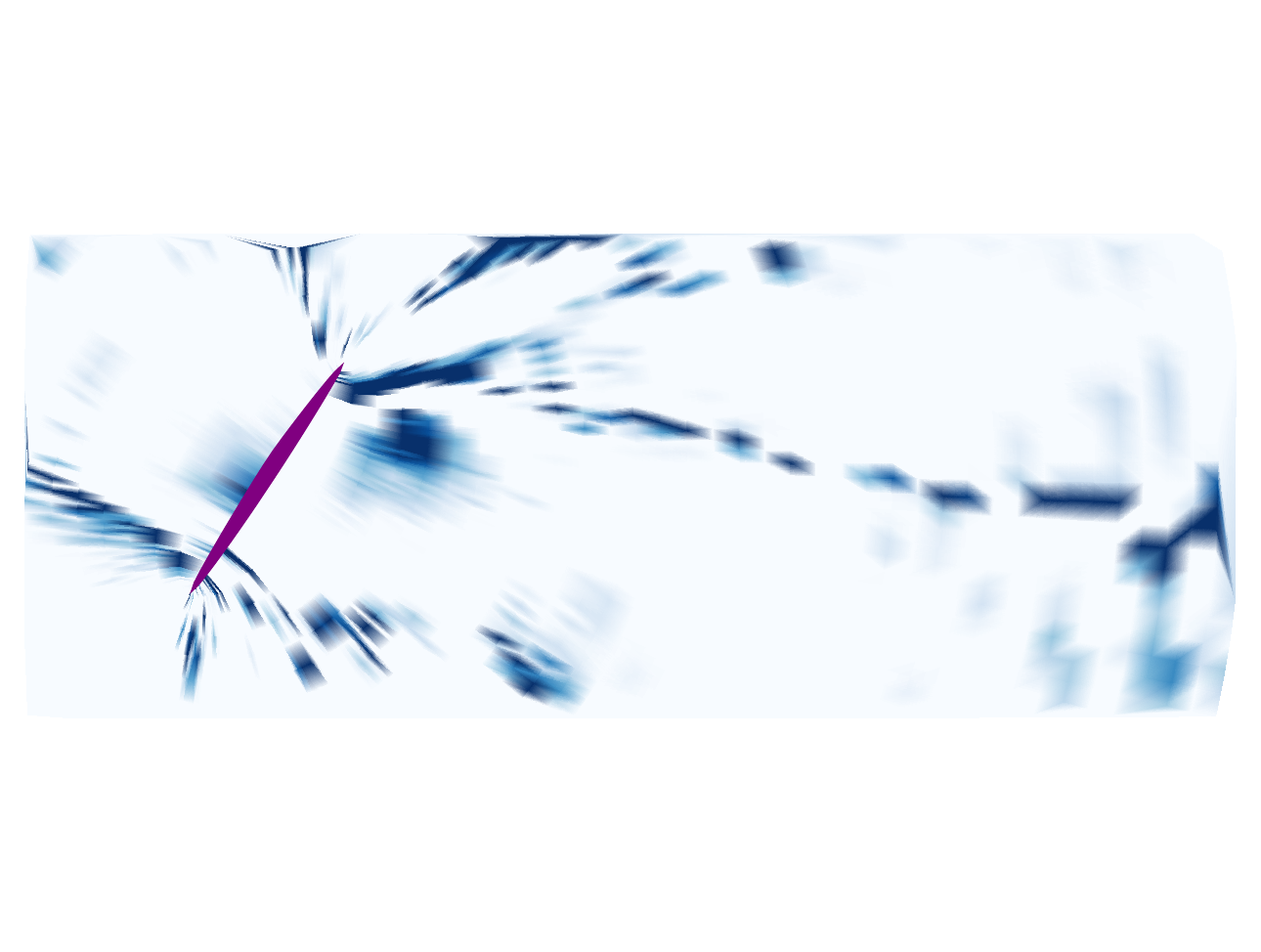}
    \end{minipage}   \\
    &&& MAPE: 27.41\% \\ \bottomrule
\end{tabular}
    \caption{STMGFR as applied to a snapshot from Shape D. The left column shows the ground truth vorticity fields at time $t$ (top) and $t+k$ (bottom); the middle column depicts the reconstruction of the snapshot at $t$ from sensors by the spatial model (top), the reconstruction of the snapshot at $t+k$ given the ground truth snapshot by the temporal model (middle) and the reconstruction of the snapshot at $t+k$ given the spatial model reconstruction (bottom); the right column displays the corresponding error maps. Vorticity colormap constrained to 5\% of $\max(|\tilde{\omega}_{t+k,i}|)$.}
    \label{fig:spatio-temporalex_s3700_large}
\end{figure}

\begin{figure}
    \centering
    \begin{tabular}{@{}cccc@{}}
\toprule
Ground truth at $t$                      & \multicolumn{2}{c}{Reconstruction at $t$}                              & \% Error \\ \midrule
\multicolumn{1}{c}{
    \begin{minipage}{\spatiotemporalfigwidth\textwidth}
      \includegraphics[width=\linewidth]{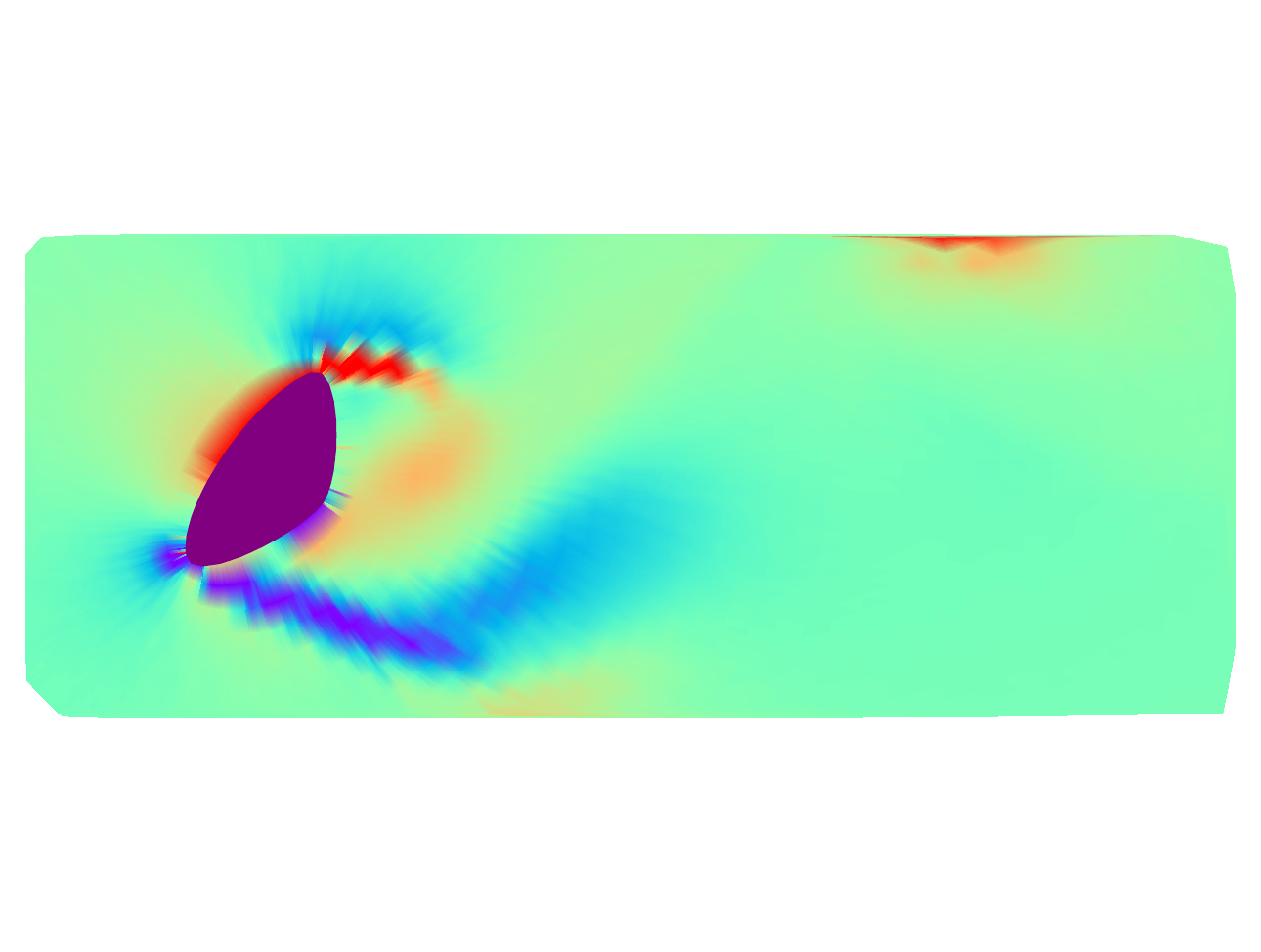}
    \end{minipage}}                  & \multicolumn{2}{c}{
    \begin{minipage}{\spatiotemporalfigwidth\textwidth}
      \includegraphics[width=\linewidth]{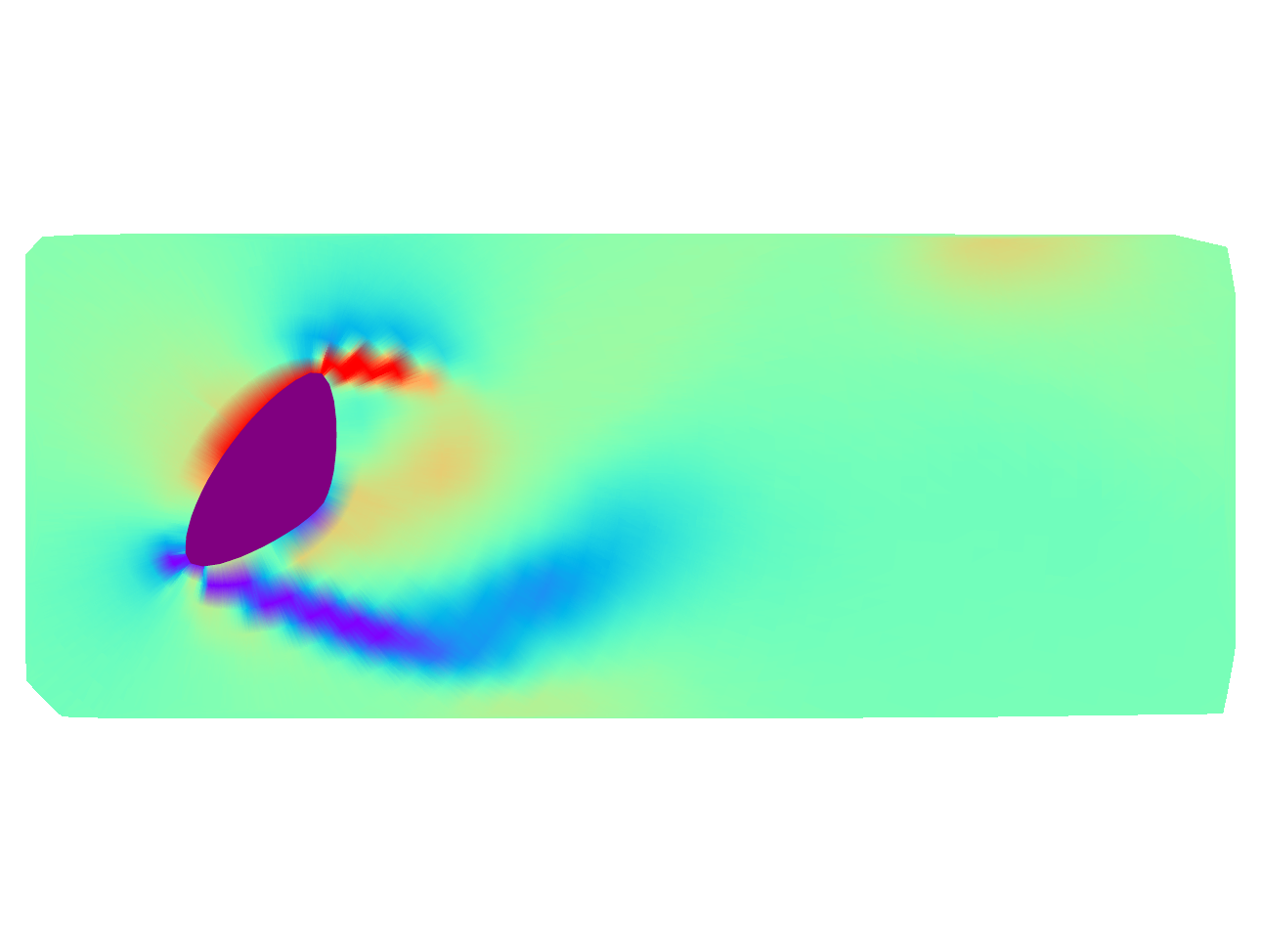}
    \end{minipage}}                                                &  
    \begin{minipage}{\spatiotemporalfigwidth\textwidth}
      \includegraphics[width=\linewidth]{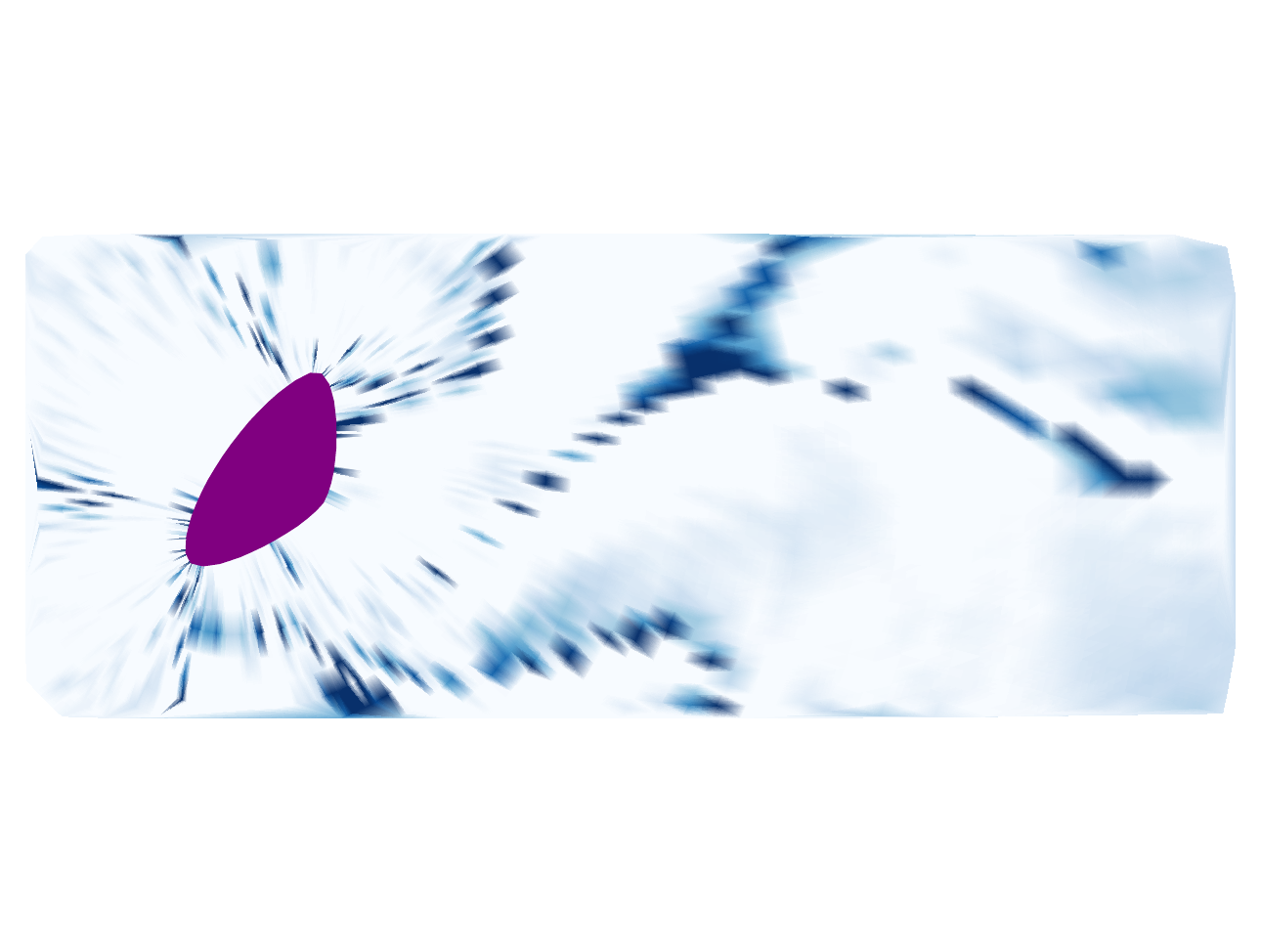}
    \end{minipage} \\
    \begin{minipage}{\spatiotemporalfigwidth\textwidth}
      \includegraphics[width=\linewidth]{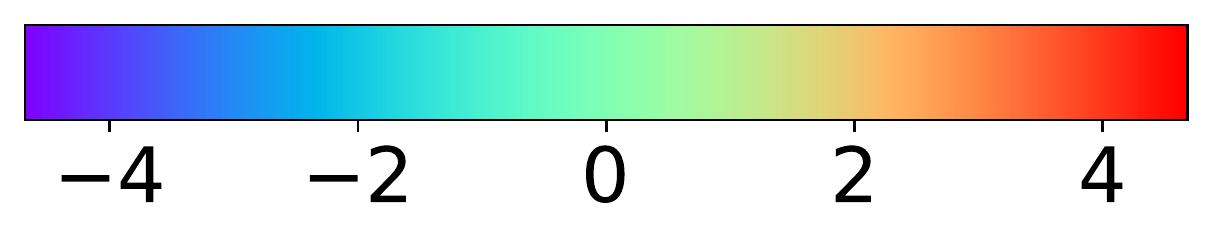}
    \end{minipage} &&&
    \begin{minipage}{\spatiotemporalfigwidth\textwidth}
      \includegraphics[width=\linewidth]{images/spatiotemporal/pcterrorcbarr0.pdf}
    \end{minipage} \\
    &&&MAPE: 24.25\% \\ \midrule
Ground truth at $t+k$                    & \multicolumn{2}{c}{Reconstruction at $t+k$}                            & \% Error \\ \midrule
\multicolumn{1}{c}{\multirow{4}{*}{\begin{minipage}{\spatiotemporalfigwidth\textwidth}
      \vspace{1.25cm}
      \includegraphics[width=\linewidth]{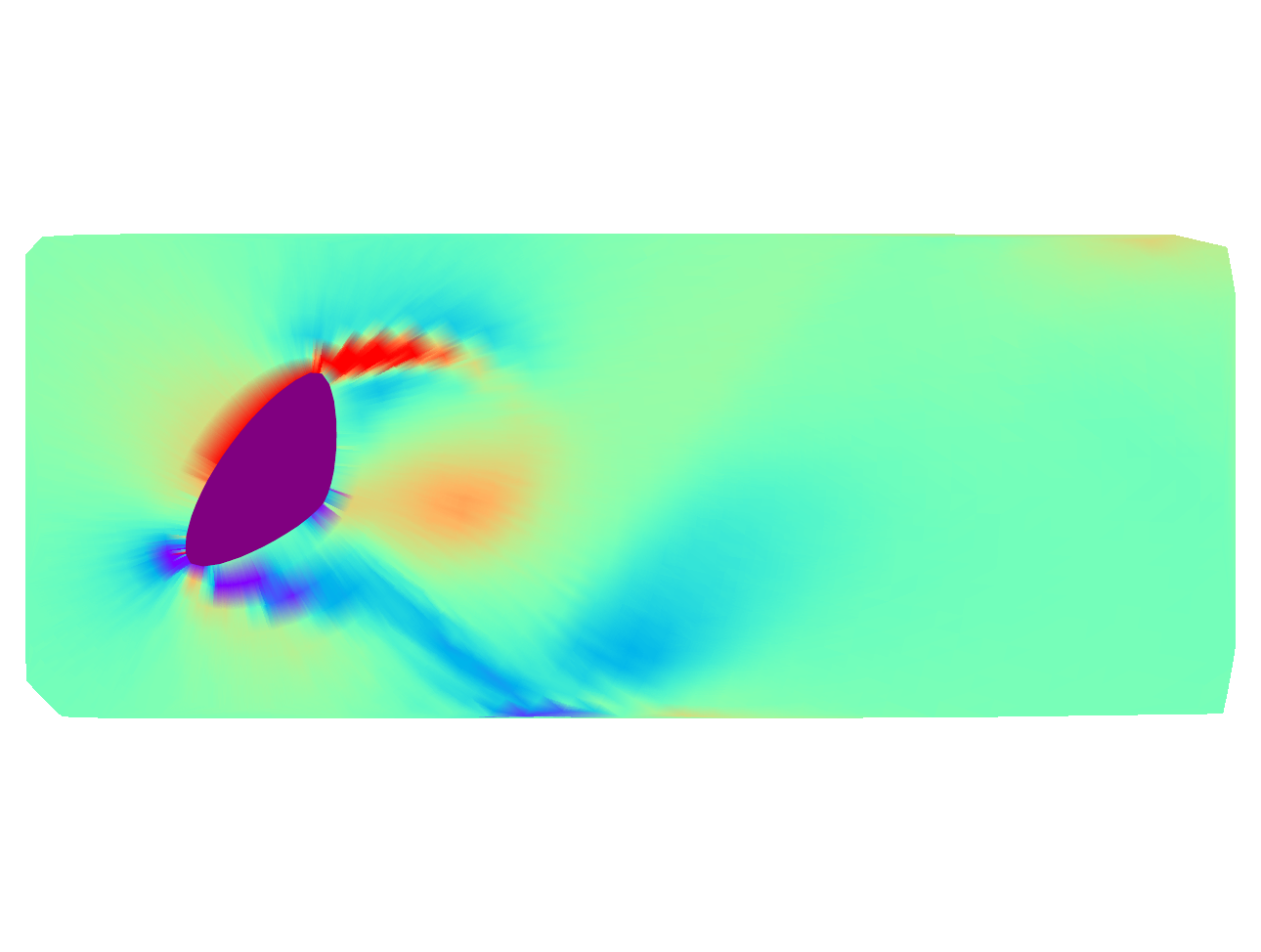}
    \end{minipage}}} & \multicolumn{1}{l}{From ground truth at $t$}   & \multicolumn{1}{c}{
    \begin{minipage}{\spatiotemporalfigwidth\textwidth}
      \includegraphics[width=\linewidth]{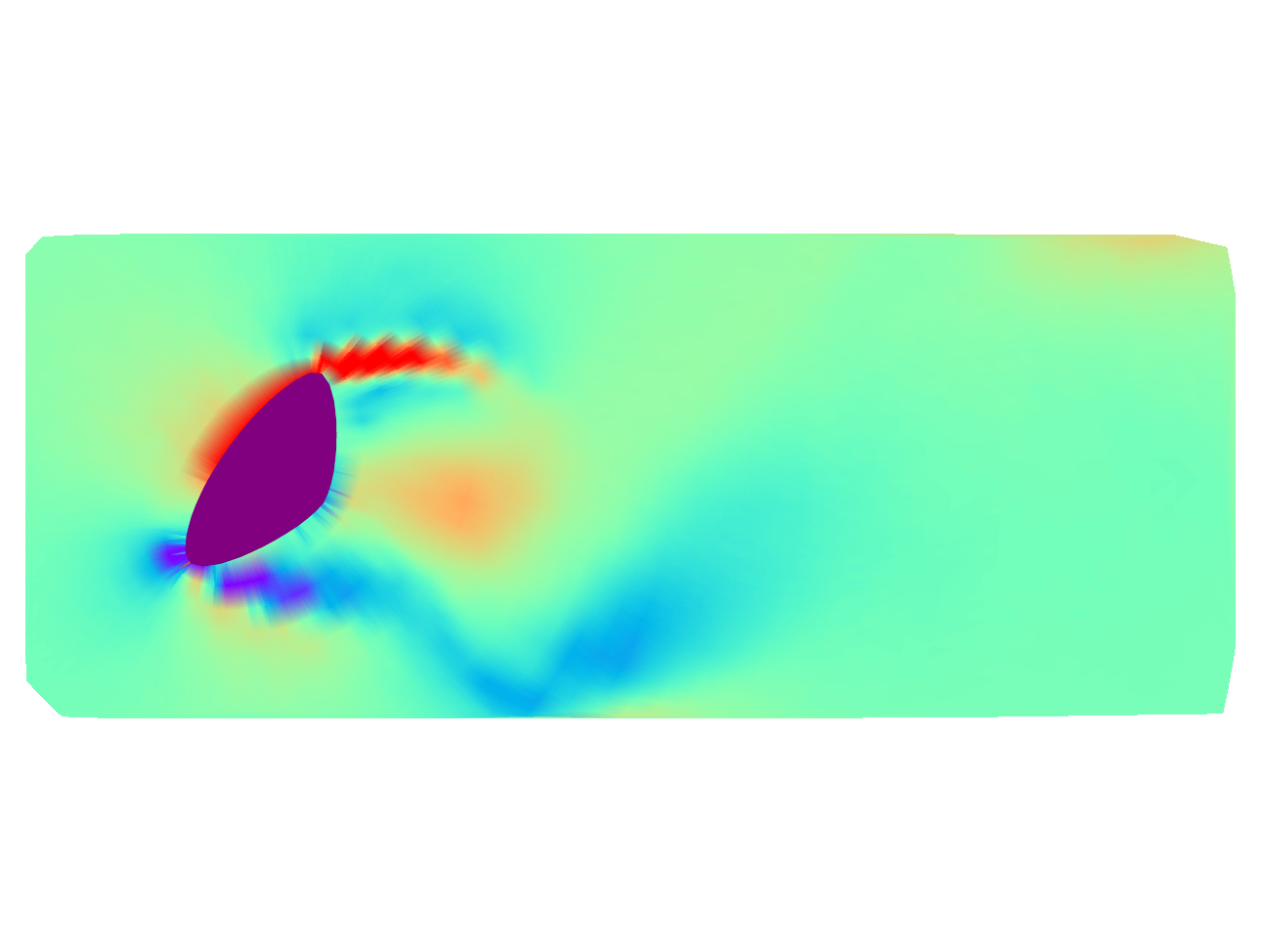}
    \end{minipage}} &  
    \begin{minipage}{\spatiotemporalfigwidth\textwidth}
      \includegraphics[width=\linewidth]{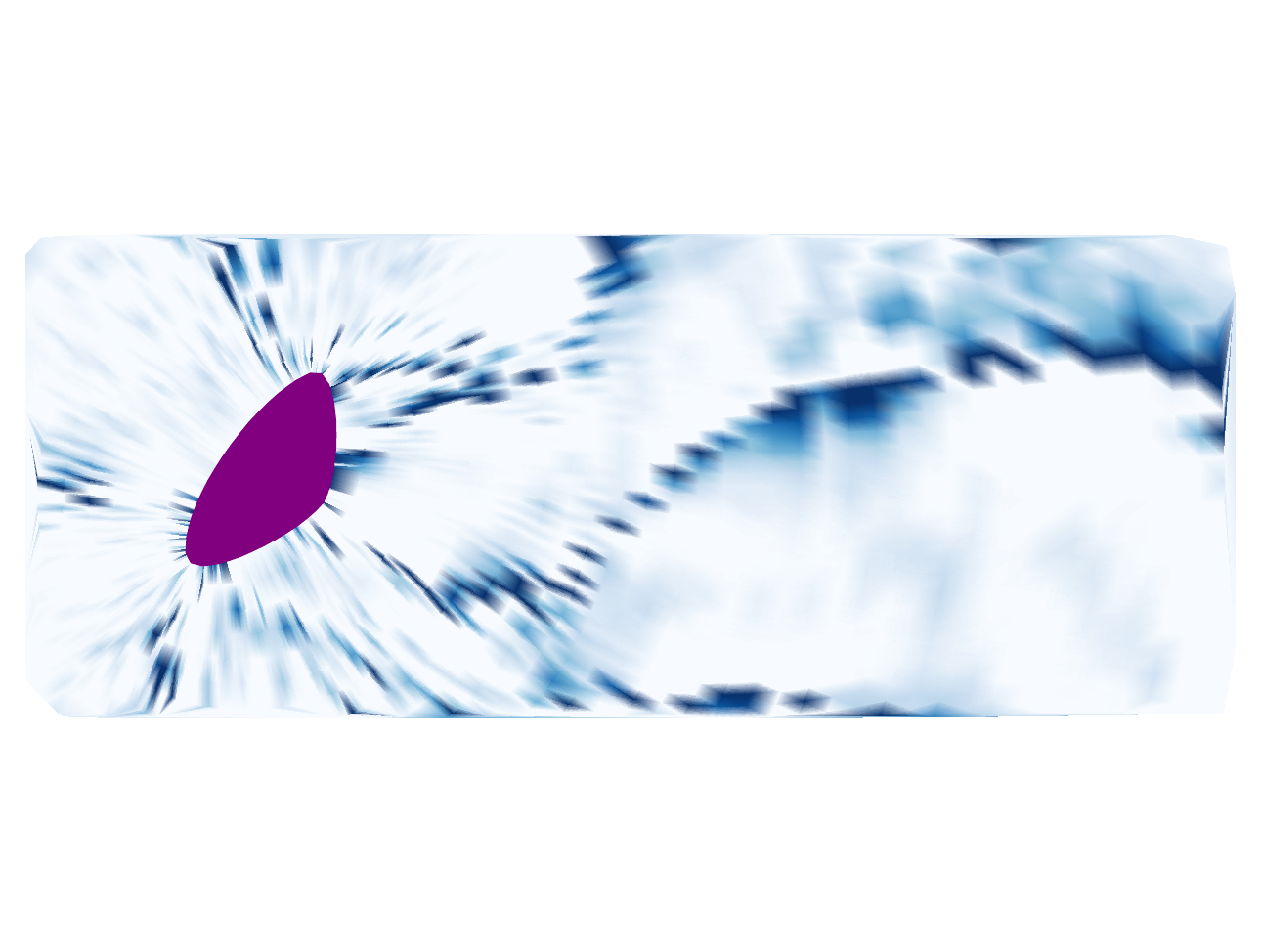}
    \end{minipage}  \\
    &&& MAPE: 22.83\% \\ \cmidrule(l){2-4} 
    \multicolumn{1}{c}{} & \multicolumn{1}{l}{From Reconstruction at $t$} & \multicolumn{1}{c}{
    \begin{minipage}{\spatiotemporalfigwidth\textwidth}
      \includegraphics[width=\linewidth]{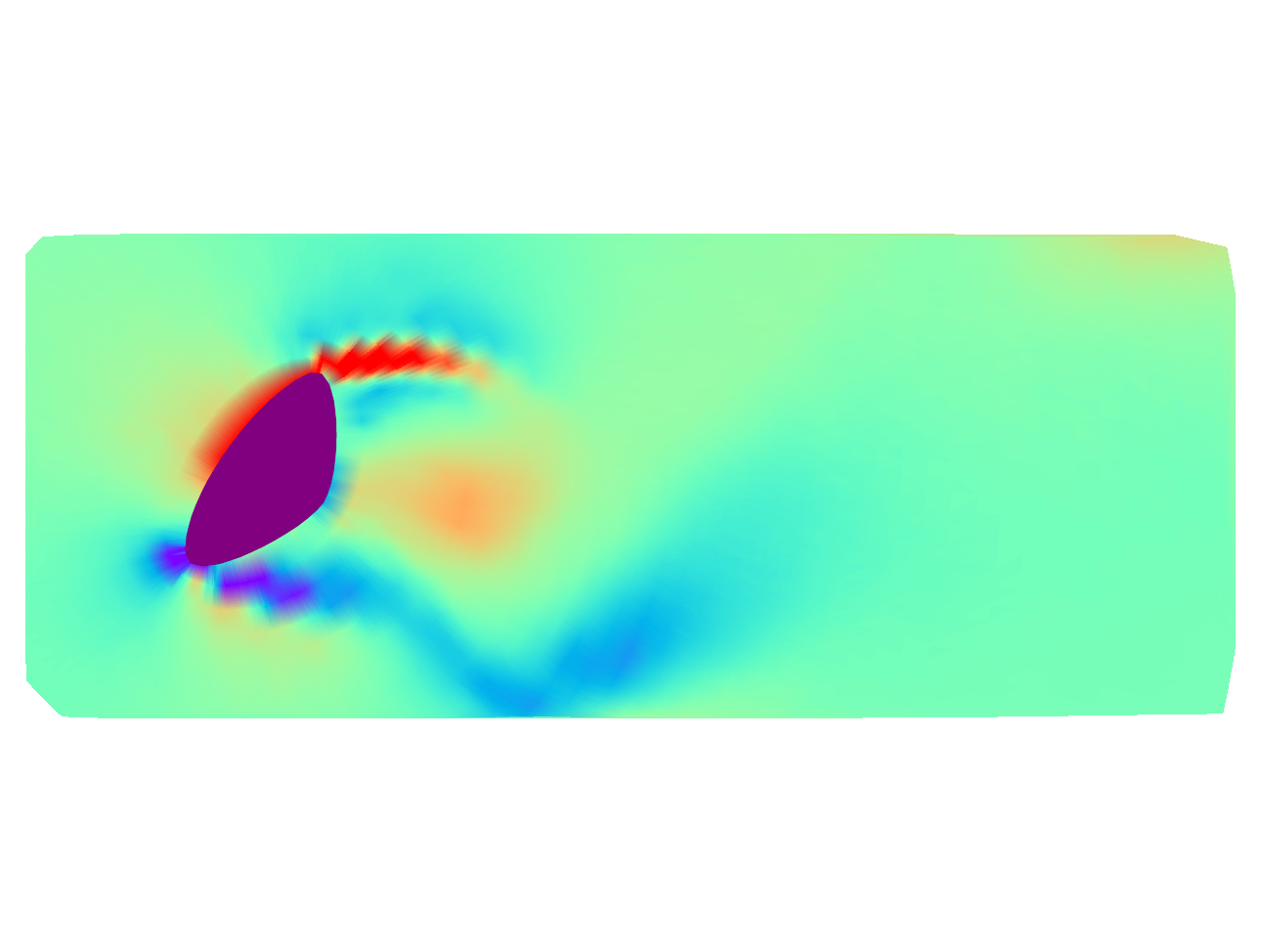}
    \end{minipage}} &  \begin{minipage}{\spatiotemporalfigwidth\textwidth}
      \includegraphics[width=\linewidth]{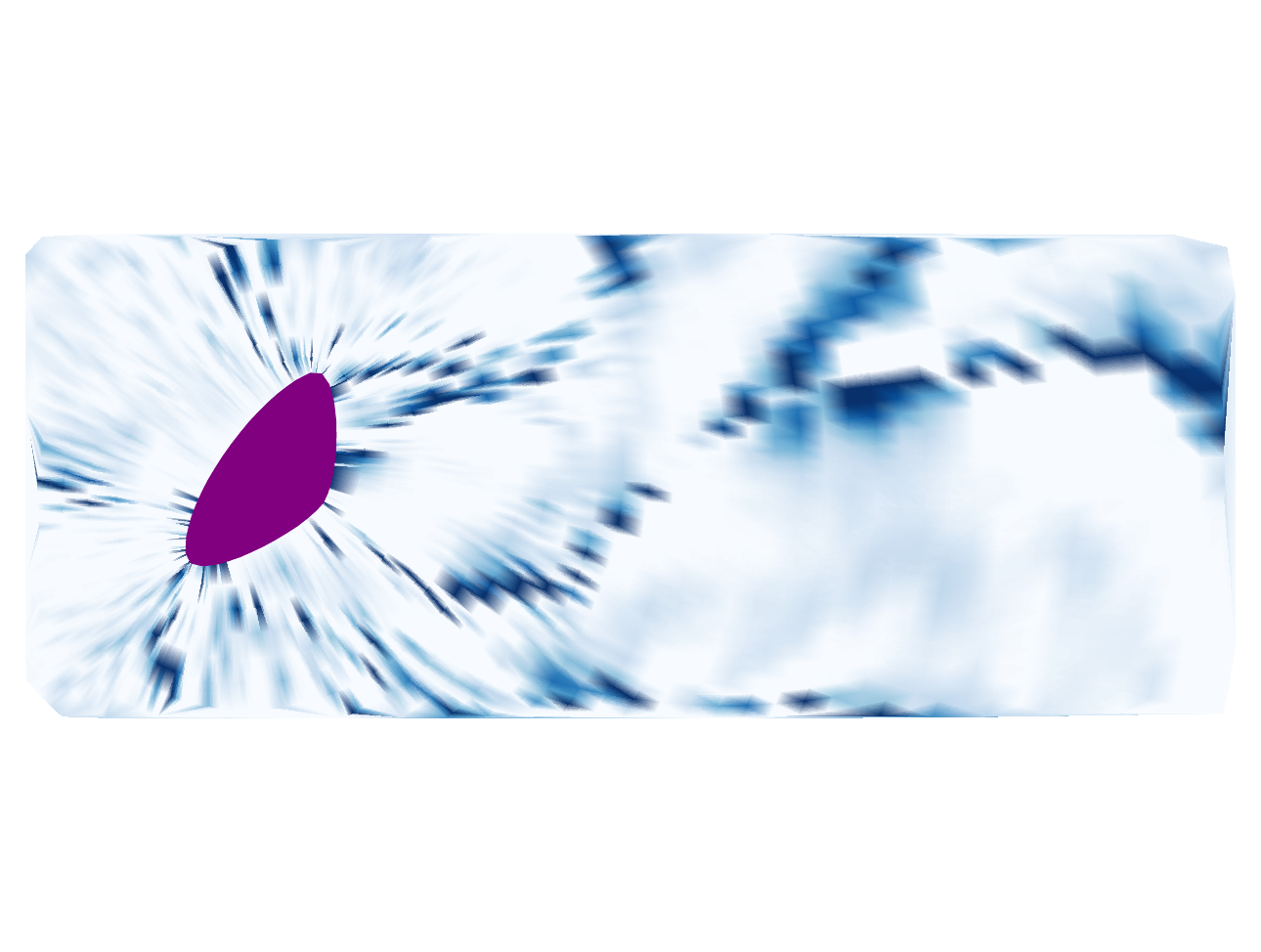}
    \end{minipage}   \\
    &&& MAPE: 23.88\% \\ \bottomrule
\end{tabular}
    \caption{STMGFR as applied to a snapshot from Shape E. The left column shows the ground truth vorticity fields at time $t$ (top) and $t+k$ (bottom); the middle column depicts the reconstruction of the snapshot at $t$ from sensors by the spatial model (top), the reconstruction of the snapshot at $t+k$ given the ground truth snapshot by the temporal model (middle) and the reconstruction of the snapshot at $t+k$ given the spatial model reconstruction (bottom); the right column displays the corresponding error maps. Vorticity colormap constrained to 5\% of $\max(|\tilde{\omega}_{t+k,i}|)$.}
    \label{fig:spatio-temporalex_s5530_large}
\end{figure}

\begin{figure}
    \centering
    \begin{tabular}{@{}cccc@{}}
\toprule
Ground truth at $t$                      & \multicolumn{2}{c}{Reconstruction at $t$}                              & \% Error \\ \midrule
\multicolumn{1}{c}{
    \begin{minipage}{\spatiotemporalfigwidth\textwidth}
      \includegraphics[width=\linewidth]{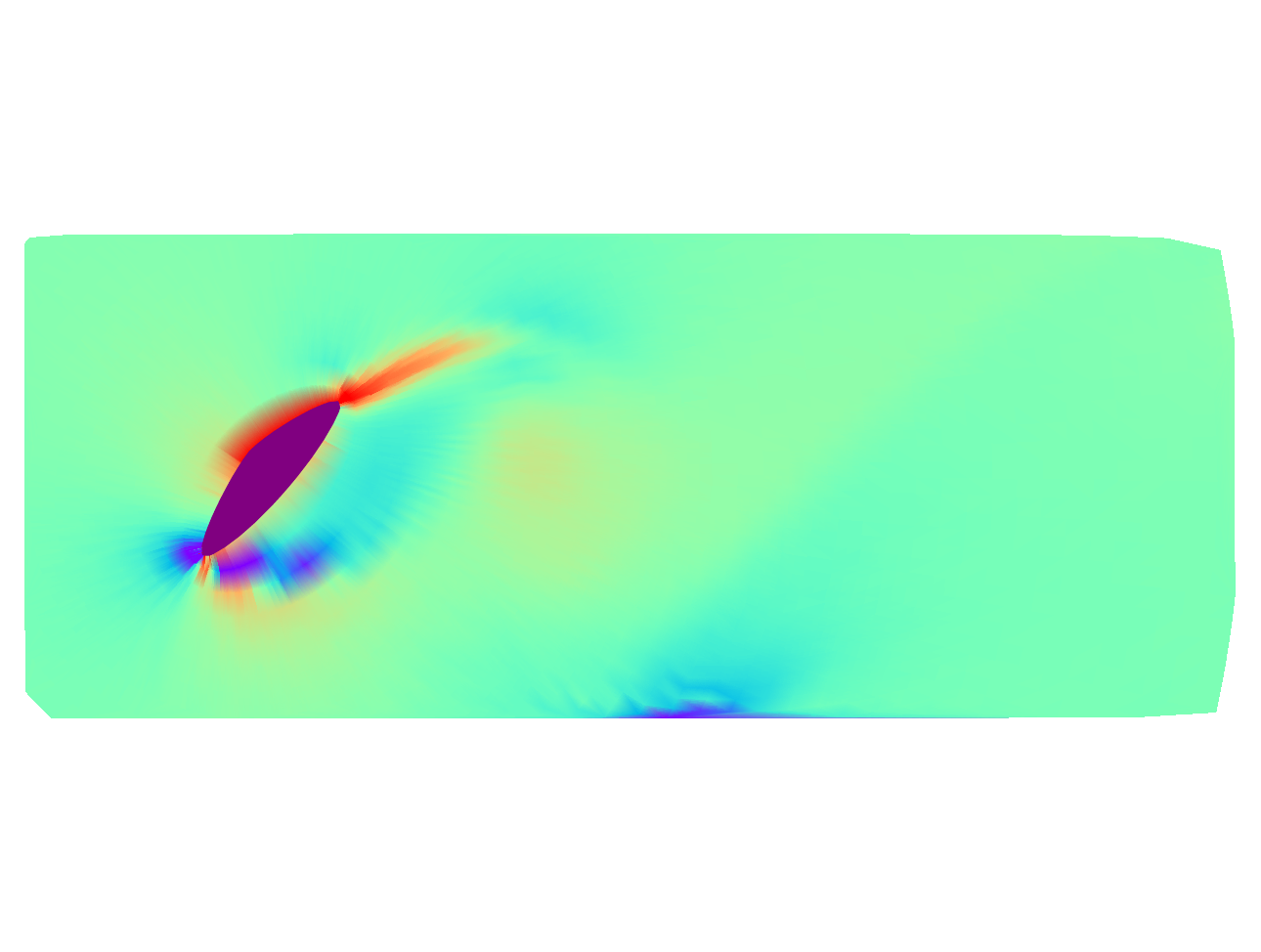}
    \end{minipage}}                  & \multicolumn{2}{c}{
    \begin{minipage}{\spatiotemporalfigwidth\textwidth}
      \includegraphics[width=\linewidth]{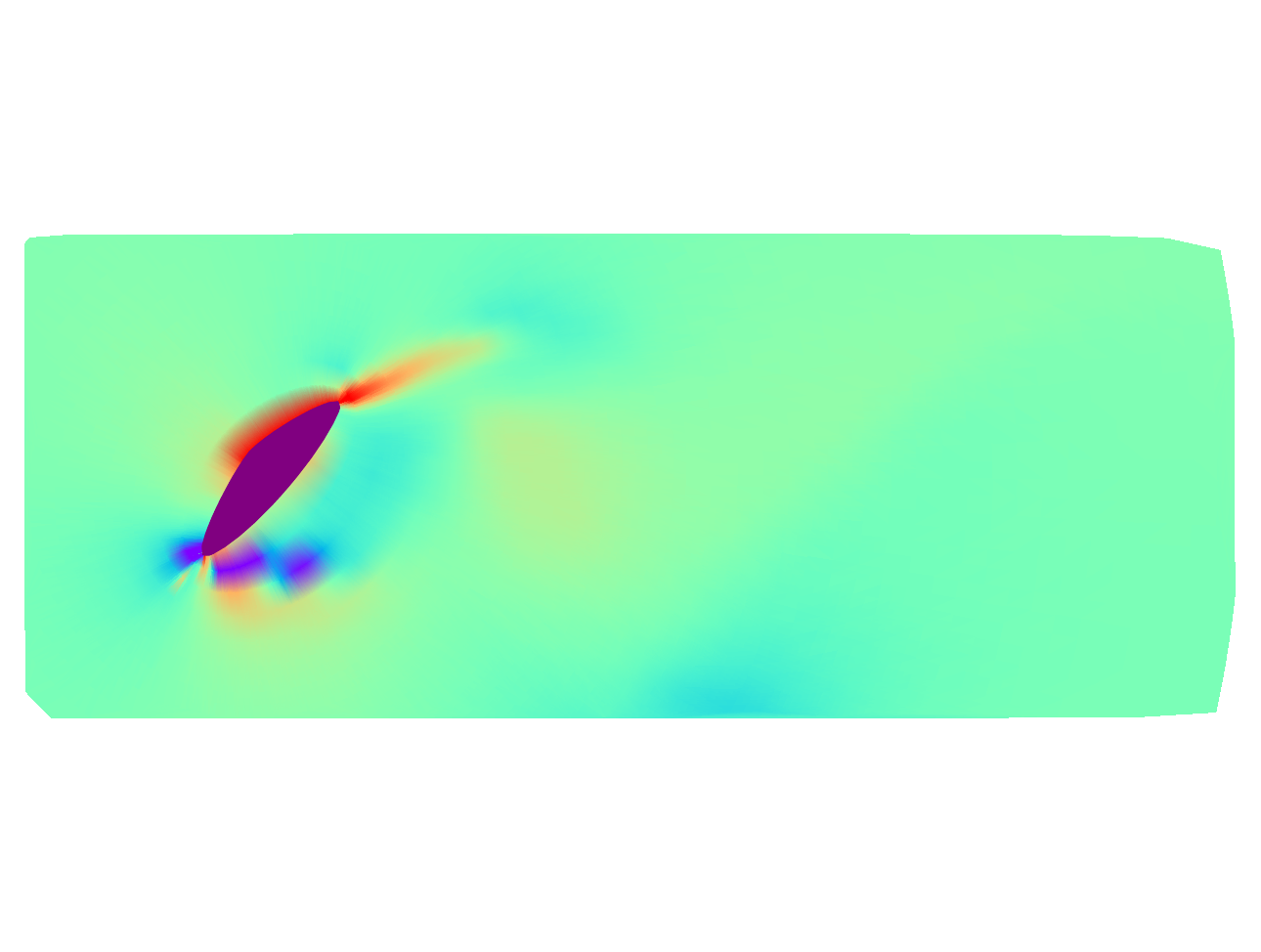}
    \end{minipage}}                                                &  
    \begin{minipage}{\spatiotemporalfigwidth\textwidth}
      \includegraphics[width=\linewidth]{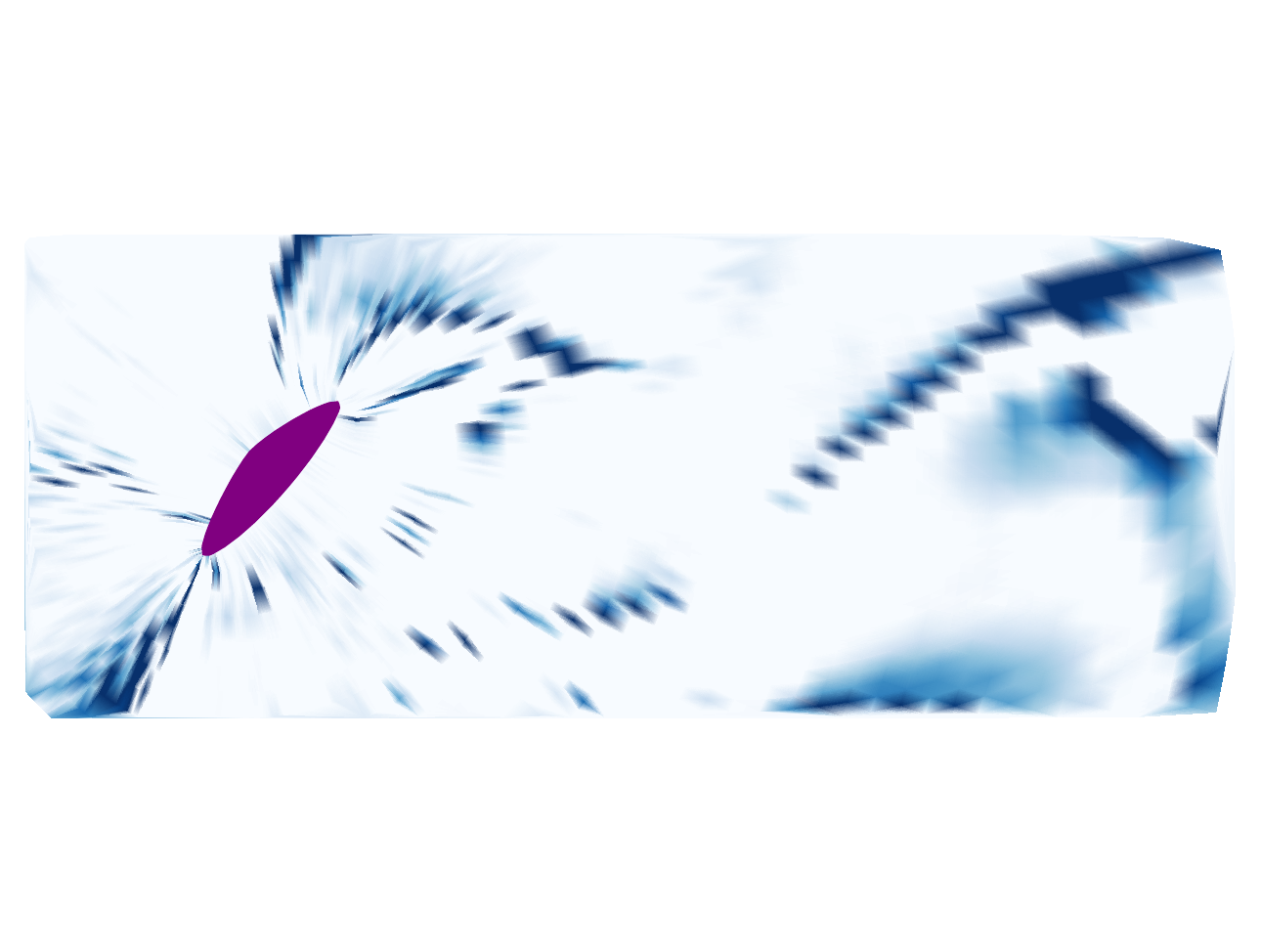}
    \end{minipage} \\
    \begin{minipage}{\spatiotemporalfigwidth\textwidth}
      \includegraphics[width=\linewidth]{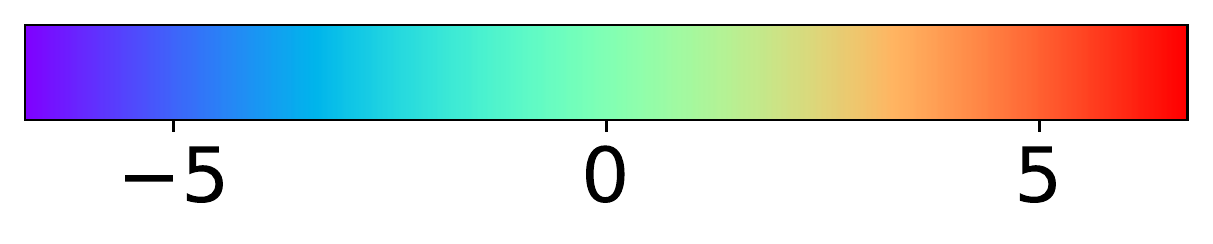}
    \end{minipage} &&&
    \begin{minipage}{\spatiotemporalfigwidth\textwidth}
      \includegraphics[width=\linewidth]{images/spatiotemporal/pcterrorcbarr0.pdf}
    \end{minipage} \\
    &&&MAPE: 34.75\% \\ \midrule
Ground truth at $t+k$                    & \multicolumn{2}{c}{Reconstruction at $t+k$}                            & \% Error \\ \midrule
\multicolumn{1}{c}{\multirow{4}{*}{\begin{minipage}{\spatiotemporalfigwidth\textwidth}
      \vspace{1.25cm}
      \includegraphics[width=\linewidth]{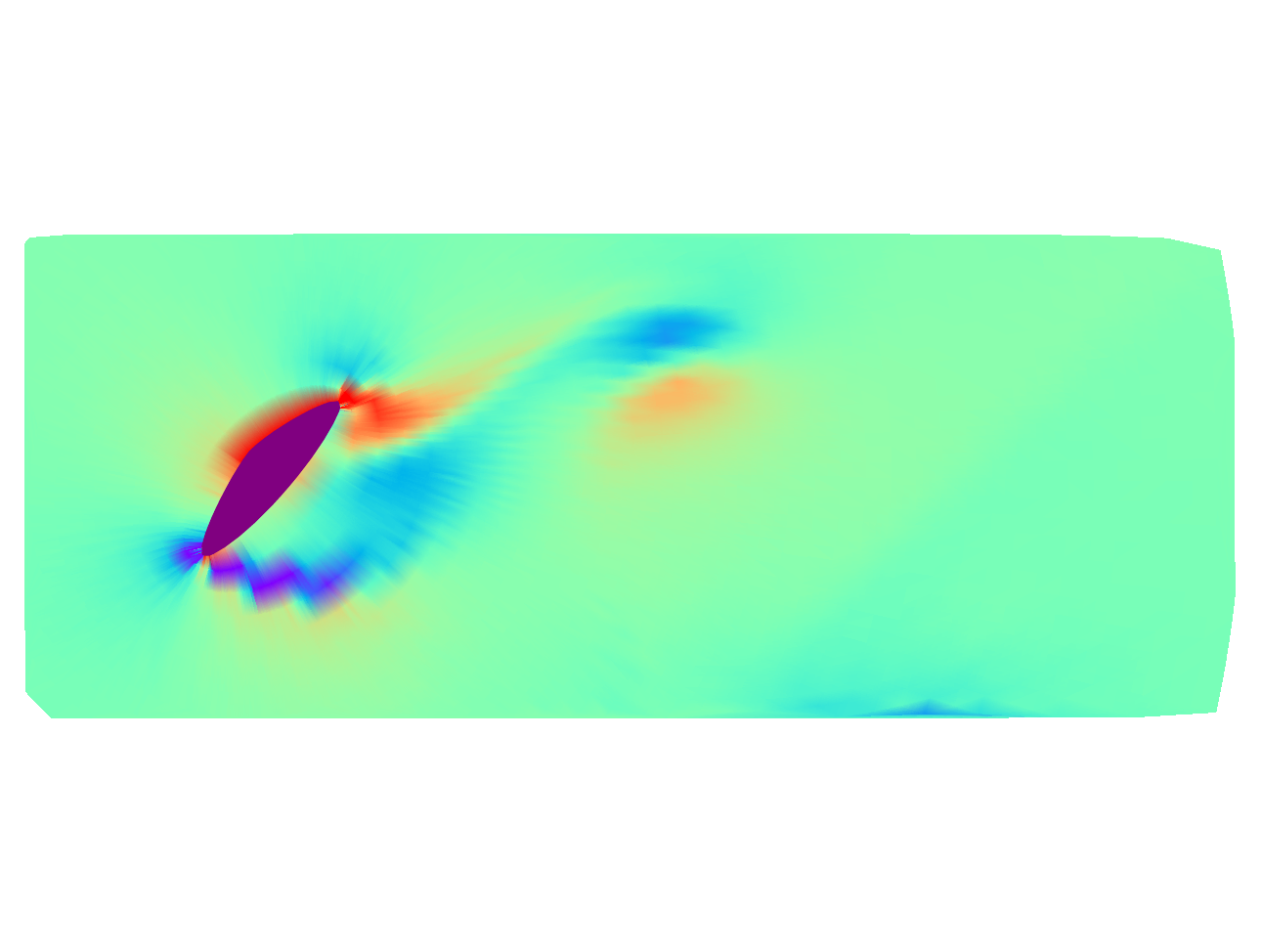}
    \end{minipage}}} & \multicolumn{1}{l}{From ground truth at $t$}   & \multicolumn{1}{c}{
    \begin{minipage}{\spatiotemporalfigwidth\textwidth}
      \includegraphics[width=\linewidth]{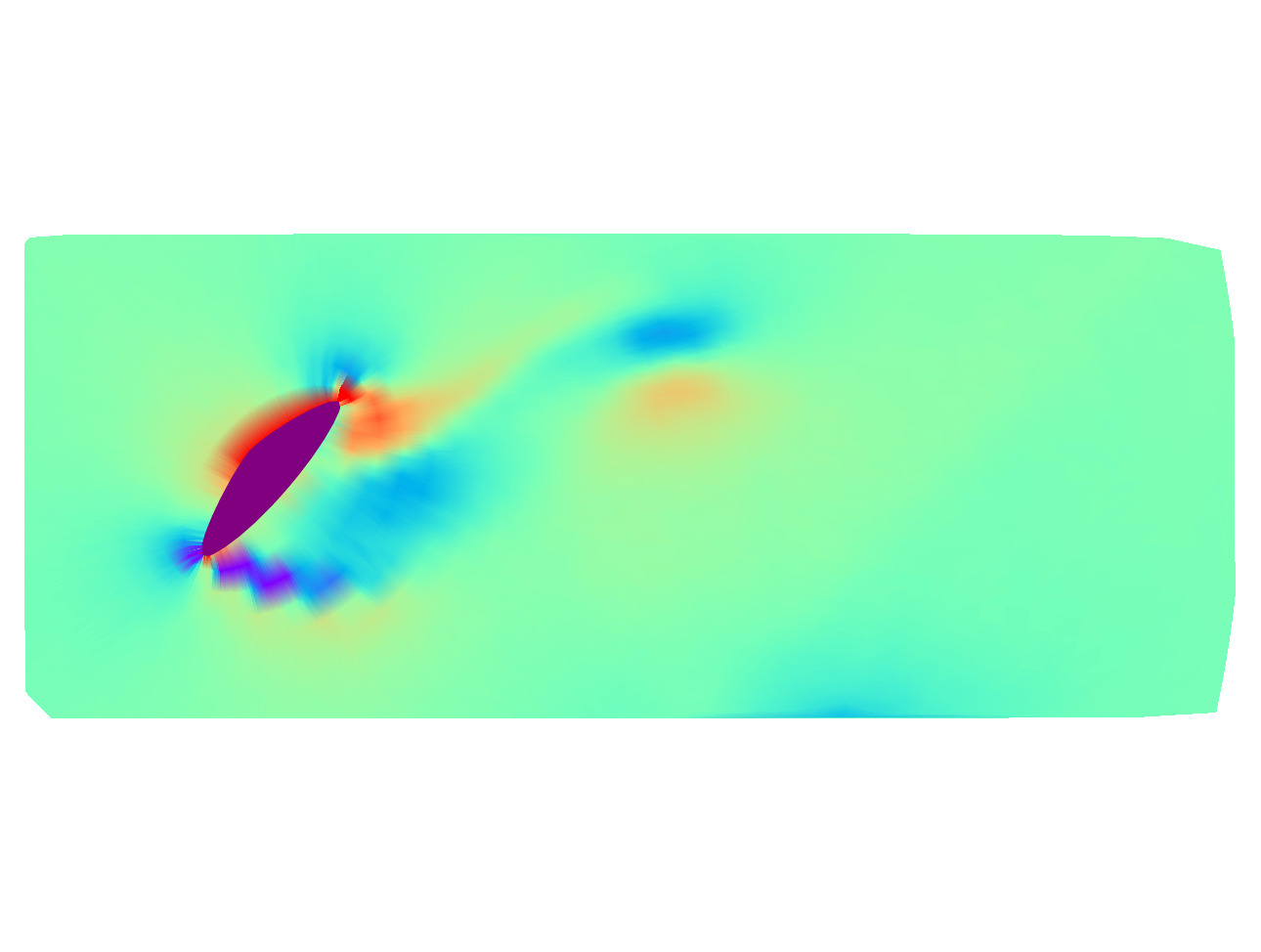}
    \end{minipage}} &  
    \begin{minipage}{\spatiotemporalfigwidth\textwidth}
      \includegraphics[width=\linewidth]{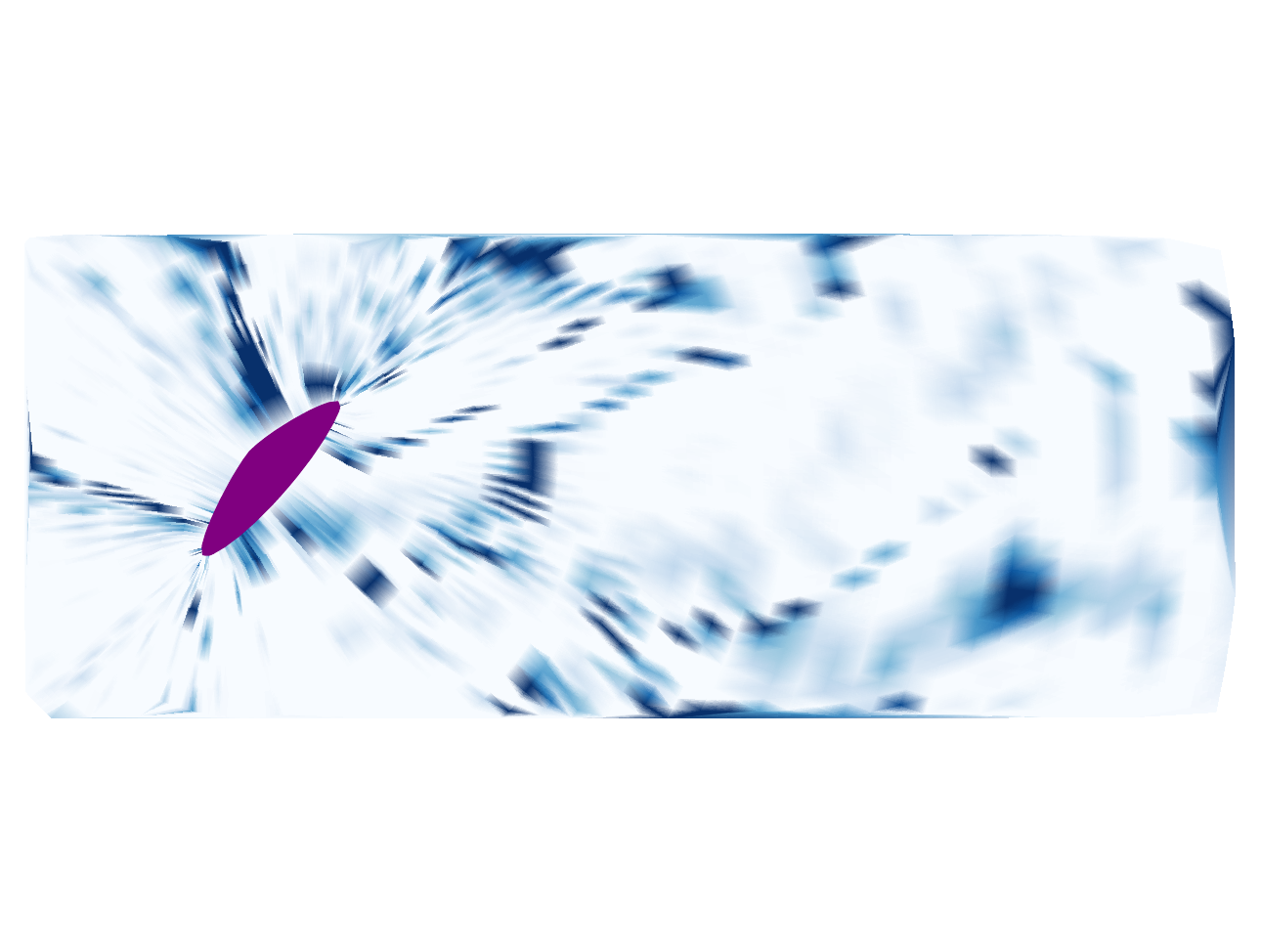}
    \end{minipage}  \\
    &&& MAPE: 33.89\% \\ \cmidrule(l){2-4} 
    \multicolumn{1}{c}{} & \multicolumn{1}{l}{From Reconstruction at $t$} & \multicolumn{1}{c}{
    \begin{minipage}{\spatiotemporalfigwidth\textwidth}
      \includegraphics[width=\linewidth]{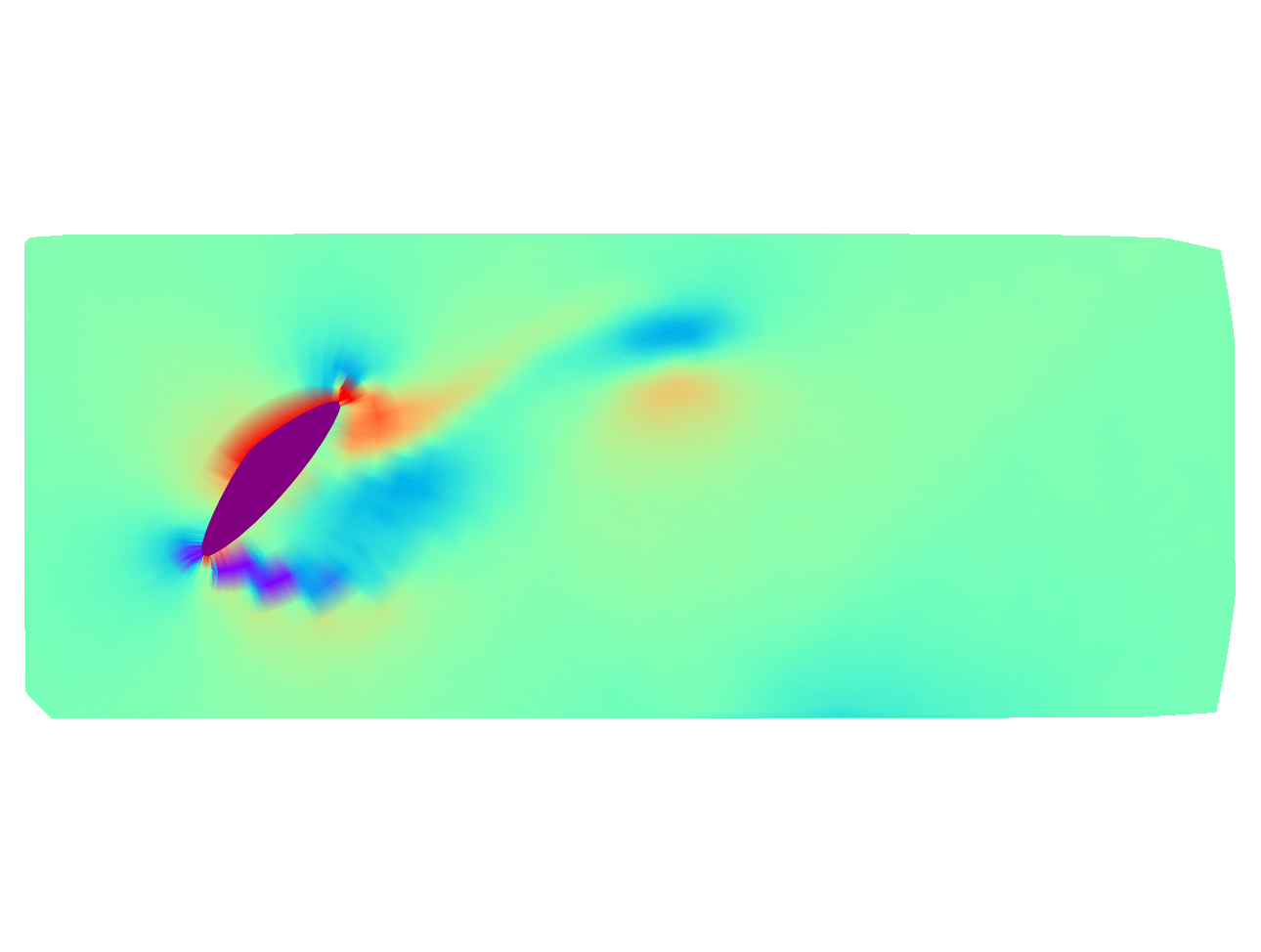}
    \end{minipage}} &  \begin{minipage}{\spatiotemporalfigwidth\textwidth}
      \includegraphics[width=\linewidth]{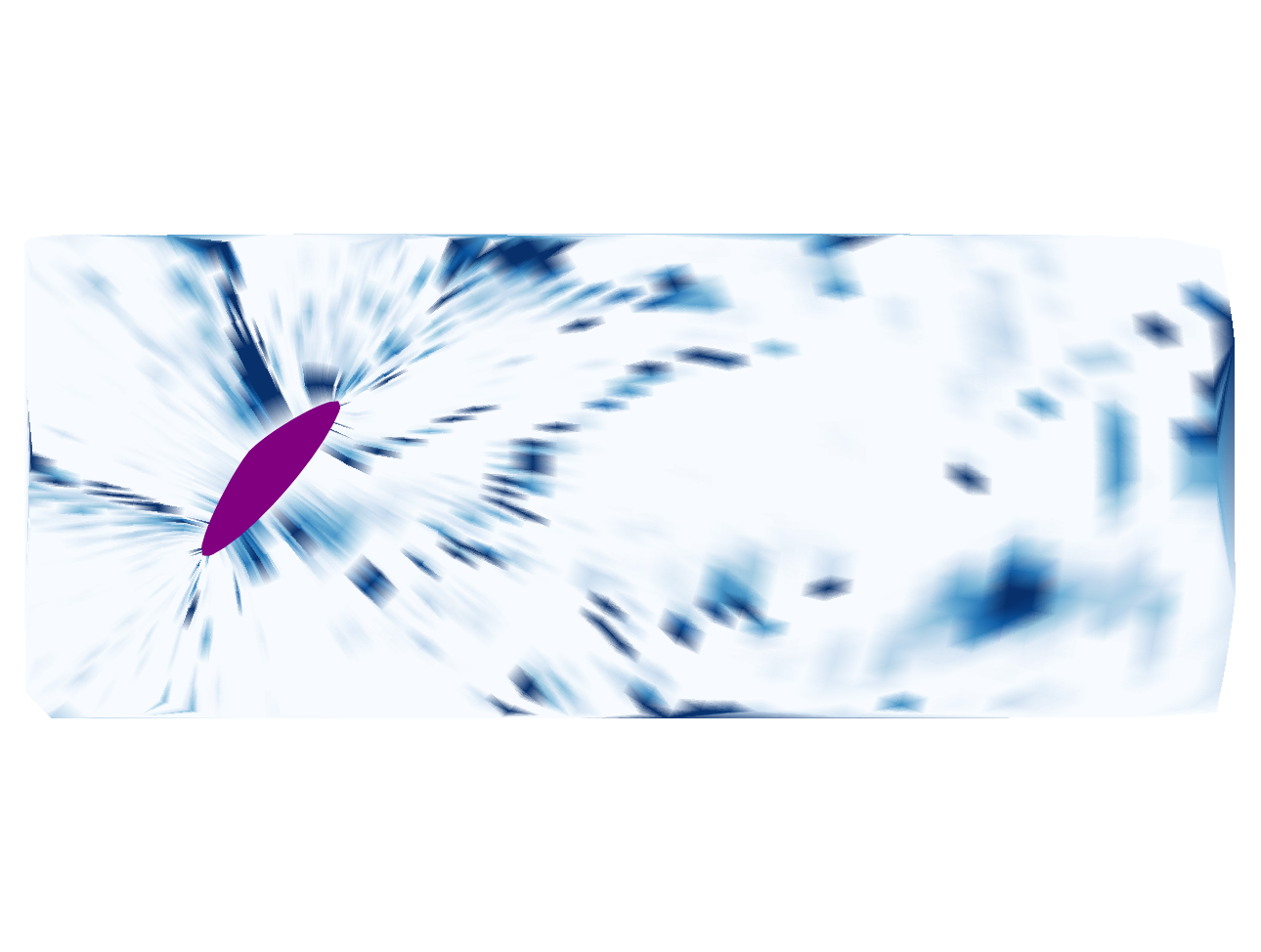}
    \end{minipage}   \\
    &&& MAPE: 35.51\% \\ \bottomrule
\end{tabular}
    \caption{STMGFR as applied to a snapshot from Shape F. The left column shows the ground truth vorticity fields at time $t$ (top) and $t+k$ (bottom); the middle column depicts the reconstruction of the snapshot at $t$ from sensors by the spatial model (top), the reconstruction of the snapshot at $t+k$ given the ground truth snapshot by the temporal model (middle) and the reconstruction of the snapshot at $t+k$ given the spatial model reconstruction (bottom); the right column displays the corresponding error maps. Vorticity colormap constrained to 3\% of $\max(|\tilde{\omega}_{t+k,i}|)$.}
    \label{fig:spatio-temporalex_s7200_large}
\end{figure}

As $k$ increases, the denoising effect declines monotonically as the temporal model must simultaneously correct for the errors in its input and predict the time evolution of the input snapshot. Hence, to dissect the sources of error, we focus on results with $k=80$ over a number of example snapshots; similar to \secref{sec:model_spatial_reconstruction}, Figures \ref{fig:spatio-temporalex_s3700_large} to \ref{fig:spatio-temporalex_s7200_large} provide example ground truth and predicted snapshots from three further shapes -- dubbed Shapes D, E and F -- chosen from the validation dataset for displaying a diverse range of flow dynamics.

Shape D, in \figref{fig:spatio-temporalex_s3700_large}, is a thin flat plate like object set at a high angle of attack relative to the incoming flow. The temporal model predictions replicate key flow features successfully, correctly predicting the location of the two shed vortices behind the object, demonstrating that the model is effective at capturing the mechanisms of advection in this flow. Some notable sources of error are the under-estimation of the intensity of these vortices and the missing concentration of high positive vorticity on the lower surface of the object, which is uncharacteristic given the good performance of SD-FNO predictions in \secref{sec:model_spatial_reconstruction} in this regard.  

Shape E (\figref{fig:spatio-temporalex_s5530_large}) is a bluff-body like object similar to Shape A  (\figref{fig:spatialex_s1660_large}). The temporal reconstruction in the figure has the lowest MAPE among the three snapshots presented in the figures. The phenomenon of note in this example is that, unlike the previous example where the most challenging aspect of the problem was predicting the advection of vortices, the model has to predict the \textit{formation} of a vortex at time $t+k$ from the snapshot at time $t$, in this instance immediately behind the object. This is executed very well by the model, as evidenced by the error map, which shows very low error (below 10\% throughout) for the region corresponding to the vortex.

Finally, Shape F in \figref{fig:spatio-temporalex_s7200_large} is a thick airfoil like shape, also set at a high incidence angle relative to the incoming flow. The challenge present in this example is essentially the combination of the those in Figures \ref{fig:spatio-temporalex_s3700_large} and \ref{fig:spatio-temporalex_s5530_large}; there is no vortical structure present at time $t$, but by time $t+k$ vortical structures have already formed and advected downstream from the object. This more complicated challenge translates to relatively higher quantitative error metrics for the snapshot depicted in the figure compared to the previous examples, but critical flow features are very well replicated at time $t+k$. The two shed counter-rotating vortices are predicted at the correct location with correct intensity, the shape of the interface between high and low vorticity regions immediately downstream of the object is correct, and the highly rotational intense separation regions past the leading and trailing edges of the object are present. Overall, despite the substantially more difficult challenge in this example, the model runs well and offers a good understanding of key fluid dynamics phenomena in this flow.

The good performance of the temporal model when given ground truth snapshots as inputs is largely consistent with the previous literature on the FNO\cite{fourier_neural_operator}, where the capability of the FNO to time-march the vorticity field for turbulence-in-a-box settings were demonstrated with low error levels. The additional insight, however, is that these results demonstrate that the FNO model coupled with our training methodology (whereby 50\% of the inputs are randomly replaced with spatial model predictions) is robust to noisy inputs, with a loss of accuracy of the order of 2\% (as shown in \tabref{tab:spatio-temporal_mape}) despite average errors in the input approaching 40\%. In fact, in two of the three cases displayed in the figures, the MAPE of the temporal model prediction from the spatial model reconstruction relative to the ground truth at time $t+k$ is lower than the MAPE of the spatial model reconstruction relative to the ground truth at time $t$. In a physical experimental setting, where measurement noise is a real concern, this is a key capability as the impact of the measurement errors on the spatial reconstruction will not catastrophically degrade the accuracy of the temporal reconstruction.

\clearpage

\clearpage

\clearpage

\subsection{Training and inference time}
\label{sec:runtime}
The final topic of discussion for comparing the relative merits of the different model architectures in the previous sections is the computational cost of training and using each model. \tabref{tab:runtime} outlines the wall-clock runtimes for running training (conducted on an IBM AC922 system with two 20-core POWER9 CPUs and two Nvidia V100 GPUs) and inference (conducted with an AMD EPYC 7443 CPU and a single Nvidia A100 GPU) using each model, using the single-precision floating point format.

\begin{table}[h!]
\caption{Wall-clock runtimes for training and running the models in Sections \ref{sec:results-spatial} and \ref{sec:results-spatio-temporal} for the large sensor setup. Training runtime statistics are compiled for the training dataset (consisting of $601 \times 64 = 38464$ snapshots); inference statistics are for the validation dataset (consisting of $601 \times 16 = 9616$ snapshots). The training time column shows the \textit{total} training time (equal to [Time/epoch] $\times$ [No. of epochs]).}
\label{tab:runtime}
\begin{tabular}{@{}l|lcccccc@{}}
\toprule
\multicolumn{2}{c}{Model}           & Batch size & Batches/epoch & Time/epoch (s) & No. of epochs & Training time (s) & Inference time (s) \\ \midrule
\multirow{4}{*}{\rotatebox[origin=c]{90}{Spatial}} & SD       & 500        & 77                & 5                  & 813           & 4065           & 0.76               \\
                         & SD-Large & 500        & 77                & 7                  & 560           & 3920           & 1.52               \\
                         & SD-UNet  & 48         & 802               & 157                & 183           & 28731          & 8.98               \\
                         & SD-FNO   & 50         & 770               & 147                & 196           & 28812          & 6.91               \\ \midrule
\rotatebox[origin=c]{90}{Temporal}  & \begin{tabular}{l}FNO\\ ($k=0$)\end{tabular}     & 100        & 372               & 187                & 248           & 46376          & 44                 \\ \bottomrule
\end{tabular}
\end{table}

While the SD-UNet and SD-FNO consistently displayed better performance in terms of numerical error metrics and qualitative factors such as correct predictions of locations and intensities of vortical structures in \secref{sec:results-spatial}, this improvement comes at a substantial cost in terms of training and model runtime. This can substantially hinder the relative usefulness of the SD-UNet and SD-FNO in a laboratory setting where real-time reconstruction of flow on lower-power devices is desired. 

The runtime costs for the (temporal) FNO model used as the temporal component of the approach to STMGFR are even higher than the spatial models, though it is likely that an implementation-specific issue was present as GPU power draw was observed to be substantially lower for this model during inference than the four spatial architectures despite high utilization statistics. 

\section{Conclusion and future work}
\label{sec:conclusion}

This work introduced the Spatial and Spatio-temporal Multi-Geometry Flow Reconstruction (SMGFR, STMGFR) tasks for reconstructing dense contemporaneous or future vorticity fields of flows past arbitrary objects from current sparse sensor measurements, respectively, without geometry-specific training. To achieve optimal performance in these tasks, the use of Schwarz-Christoffel mappings to choose the sampling points of the dense fields was explored. 

The performance of four different models was investigated on the SMGFR task, using datasets generated via both the novel mapping aided sampling strategy and a more traditional Cartesian sampling strategy, with three different sensor setups. The results showed that the mapping aided approach provides a substantial boost in accuracy for all model and sensor setup configurations, enabling percentage errors under 3\%, 10\% and 30\% for reconstructions of pressure, velocity and vorticity fields, respectively. Improvements in terms of mean absolute percentage error exceed 15 percentage points in select cases for the challenging vorticity reconstruction tasks, while the impact of the size of the sensor setup is modest. The best performing model architecture was a convolutional architecture based on the U-Net\cite{unet}. Comparisons of snapshots from the different configurations revealed that the usage of the mapping approach substantially boosted the accuracy of the predictions in the immediate vicinity of the objects.

For the novel STMGFR task involving the prediction of future snapshots given current sensor measurements, an innovative approach separating the task to spatial and temporal components was developed, whereby a spatial model first reconstructs the current snapshot given the current measurements (equivalent to the SMGFR task) and subsequently a temporal model predicts the future snapshot given the predicted current snapshot. A stack of Fourier Neural Operator \cite{fourier_neural_operator} layers acting as the temporal model was coupled with the best performing configuration from the SMGFR experiments used as the spatial model. The temporal model was trained to be robust to input noise caused by inaccuracies in spatial model predictions, by randomly providing it spatial model predictions or ground truth snapshots for each sample during training. Experimentation indicated that this approach is capable of accurately reconstructing future vorticity snapshots with mean absolute percentage error levels on the order of 30\%. Furthermore, using a temporal gap of zero, the same two model setup can be used to improve accuracy of models used in SMGFR, also bringing their MAPE levels below 30\%.

We hope to expand the investigations in this work in the future by:
\begin{itemize}
    \item Developing models that perform well over a range of Reynolds numbers. The present work focused on predictions for flows at $\Rey \approx 300$; the models presented are not expected to perform well for other $\Rey$ and likely require re-training to adjust to different Reynolds numbers. Two potential ways of overcoming this are using a dataset containing snapshots from a range of Reynolds numbers, and using `physics-informed' loss functions.
    \item Experimentation with more advanced neural network architectures. While this work focused on the relatively simple case of supervised training of feedforward and convolutional architectures to focus on investigating the mapping approach presented, techniques such as generative adversarial networks\cite{gan_flow_superresolution} and variational autoencoders\cite{flow_reconstruction_8} are gaining traction in FR literature. Adoption of such techniques may lead to higher accuracy in SMGFR and STMGFR tasks.
    \item Extending the methodology to 3D fluid flows. The mapping approach in this work relies on the Schwarz-Christoffel mapping, which is defined for the complex plane $\mathbb{C}$ only. A version of this work for $\mathbb{R}^3$ will require an alternative mapping approach.
    \item Prediction of lift and drag coefficients. Although this work focused on the reconstruction of vorticity fields for the ease of comparison with previous works, changing the target fields to velocity and pressure can permit the prediction of the lift and drag coefficients. The mapping approach is especially conducive to this as, unlike Cartesian sampling, it eliminates the need to \textcolor{black}{further process} the vorticity and pressure fields to \textcolor{black}{obtain values at the} object boundary.
\end{itemize}

\section{Acknowledgements}

This work was supported by a PhD studentship funded by the Department of Aeronautics, Imperial College London and an Academic Hardware Grant provided by Nvidia. The authors would like to thank Sean Chai, Neil Ashton and Thomas Delillo for fruitful discussions at the start of this study.

\section{Code repositories}
The \texttt{\href{https://github.com/aligirayhanozbay/pydscpack}{pydscpack}} library for computing Schwarz-Christoffel mappings and \href{https://github.com/aligirayhanozbay/flow_prediction}{the code} for replicating the results in this work can be found on GitHub.

\bibliography{aipsamp}

\begin{thebibliography}{54}%
\makeatletter
\providecommand \@ifxundefined [1]{%
 \@ifx{#1\undefined}
}%
\providecommand \@ifnum [1]{%
 \ifnum #1\expandafter \@firstoftwo
 \else \expandafter \@secondoftwo
 \fi
}%
\providecommand \@ifx [1]{%
 \ifx #1\expandafter \@firstoftwo
 \else \expandafter \@secondoftwo
 \fi
}%
\providecommand \natexlab [1]{#1}%
\providecommand \enquote  [1]{``#1''}%
\providecommand \bibnamefont  [1]{#1}%
\providecommand \bibfnamefont [1]{#1}%
\providecommand \citenamefont [1]{#1}%
\providecommand \href@noop [0]{\@secondoftwo}%
\providecommand \href [0]{\begingroup \@sanitize@url \@href}%
\providecommand \@href[1]{\@@startlink{#1}\@@href}%
\providecommand \@@href[1]{\endgroup#1\@@endlink}%
\providecommand \@sanitize@url [0]{\catcode `\\12\catcode `\$12\catcode
  `\&12\catcode `\#12\catcode `\^12\catcode `\_12\catcode `\%12\relax}%
\providecommand \@@startlink[1]{}%
\providecommand \@@endlink[0]{}%
\providecommand \url  [0]{\begingroup\@sanitize@url \@url }%
\providecommand \@url [1]{\endgroup\@href {#1}{\urlprefix }}%
\providecommand \urlprefix  [0]{URL }%
\providecommand \Eprint [0]{\href }%
\providecommand \doibase [0]{http://dx.doi.org/}%
\providecommand \selectlanguage [0]{\@gobble}%
\providecommand \bibinfo  [0]{\@secondoftwo}%
\providecommand \bibfield  [0]{\@secondoftwo}%
\providecommand \translation [1]{[#1]}%
\providecommand \BibitemOpen [0]{}%
\providecommand \bibitemStop [0]{}%
\providecommand \bibitemNoStop [0]{.\EOS\space}%
\providecommand \EOS [0]{\spacefactor3000\relax}%
\providecommand \BibitemShut  [1]{\csname bibitem#1\endcsname}%
\let\auto@bib@innerbib\@empty
\bibitem [{\citenamefont {Grant}(1997)}]{piv_review}%
  \BibitemOpen
  \bibfield  {author} {\bibinfo {author} {\bibfnamefont {I.}~\bibnamefont
  {Grant}},\ }\bibfield  {title} {\enquote {\bibinfo {title} {Particle image
  velocimetry: A review},}\ }\href {\doibase 10.1243/0954406971521665}
  {\bibfield  {journal} {\bibinfo  {journal} {Proceedings of the Institution of
  Mechanical Engineers, Part C: Journal of Mechanical Engineering Science}\
  }\textbf {\bibinfo {volume} {211}},\ \bibinfo {pages} {55--76} (\bibinfo
  {year} {1997})},\ \Eprint
  {http://arxiv.org/abs/https://doi.org/10.1243/0954406971521665}
  {https://doi.org/10.1243/0954406971521665} \BibitemShut {NoStop}%
\bibitem [{\citenamefont {Elkins}\ and\ \citenamefont
  {Alley}(2007)}]{mrv_review}%
  \BibitemOpen
  \bibfield  {author} {\bibinfo {author} {\bibfnamefont {C.~J.}\ \bibnamefont
  {Elkins}}\ and\ \bibinfo {author} {\bibfnamefont {M.~T.}\ \bibnamefont
  {Alley}},\ }\bibfield  {title} {\enquote {\bibinfo {title} {Magnetic
  resonance velocimetry: applications of magnetic resonance imaging in the
  measurement of fluid motion},}\ }\href@noop {} {\bibfield  {journal}
  {\bibinfo  {journal} {Experiments in Fluids}\ }\textbf {\bibinfo {volume}
  {43}},\ \bibinfo {pages} {823--858} (\bibinfo {year} {2007})}\BibitemShut
  {NoStop}%
\bibitem [{\citenamefont {Rajan}\ \emph {et~al.}(2009)\citenamefont {Rajan},
  \citenamefont {Varghese}, \citenamefont {van Leeuwen},\ and\ \citenamefont
  {Steenbergen}}]{ldf_review}%
  \BibitemOpen
  \bibfield  {author} {\bibinfo {author} {\bibfnamefont {V.}~\bibnamefont
  {Rajan}}, \bibinfo {author} {\bibfnamefont {B.}~\bibnamefont {Varghese}},
  \bibinfo {author} {\bibfnamefont {T.~G.}\ \bibnamefont {van Leeuwen}}, \ and\
  \bibinfo {author} {\bibfnamefont {W.}~\bibnamefont {Steenbergen}},\
  }\bibfield  {title} {\enquote {\bibinfo {title} {Review of methodological
  developments in laser doppler flowmetry},}\ }\href@noop {} {\bibfield
  {journal} {\bibinfo  {journal} {Lasers in medical science}\ }\textbf
  {\bibinfo {volume} {24}},\ \bibinfo {pages} {269--283} (\bibinfo {year}
  {2009})}\BibitemShut {NoStop}%
\bibitem [{\citenamefont {Erichson}\ \emph {et~al.}(2020)\citenamefont
  {Erichson}, \citenamefont {Mathelin}, \citenamefont {Yao}, \citenamefont
  {Brunton}, \citenamefont {Mahoney},\ and\ \citenamefont
  {Kutz}}]{flow_reconstruction_2}%
  \BibitemOpen
  \bibfield  {author} {\bibinfo {author} {\bibfnamefont {N.~B.}\ \bibnamefont
  {Erichson}}, \bibinfo {author} {\bibfnamefont {L.}~\bibnamefont {Mathelin}},
  \bibinfo {author} {\bibfnamefont {Z.}~\bibnamefont {Yao}}, \bibinfo {author}
  {\bibfnamefont {S.~L.}\ \bibnamefont {Brunton}}, \bibinfo {author}
  {\bibfnamefont {M.~W.}\ \bibnamefont {Mahoney}}, \ and\ \bibinfo {author}
  {\bibfnamefont {J.~N.}\ \bibnamefont {Kutz}},\ }\bibfield  {title} {\enquote
  {\bibinfo {title} {Shallow neural networks for fluid flow reconstruction with
  limited sensors},}\ }\href {\doibase 10.1098/rspa.2020.0097} {\bibfield
  {journal} {\bibinfo  {journal} {Proceedings of the Royal Society A:
  Mathematical, Physical and Engineering Sciences}\ }\textbf {\bibinfo {volume}
  {476}},\ \bibinfo {pages} {20200097} (\bibinfo {year} {2020})},\ \Eprint
  {http://arxiv.org/abs/https://royalsocietypublishing.org/doi/pdf/10.1098/rspa.2020.0097}
  {https://royalsocietypublishing.org/doi/pdf/10.1098/rspa.2020.0097}
  \BibitemShut {NoStop}%
\bibitem [{\citenamefont {Everson}\ and\ \citenamefont
  {Sirovich}(1995)}]{gappy_pod}%
  \BibitemOpen
  \bibfield  {author} {\bibinfo {author} {\bibfnamefont {R.}~\bibnamefont
  {Everson}}\ and\ \bibinfo {author} {\bibfnamefont {L.}~\bibnamefont
  {Sirovich}},\ }\bibfield  {title} {\enquote {\bibinfo {title}
  {Karhunen--lo\`{e}ve procedure for gappy data},}\ }\href {\doibase
  10.1364/JOSAA.12.001657} {\bibfield  {journal} {\bibinfo  {journal} {J. Opt.
  Soc. Am. A}\ }\textbf {\bibinfo {volume} {12}},\ \bibinfo {pages}
  {1657--1664} (\bibinfo {year} {1995})}\BibitemShut {NoStop}%
\bibitem [{\citenamefont {Venturi}\ and\ \citenamefont
  {Karniadakis}(2004)}]{gappy_pod_ex1}%
  \BibitemOpen
  \bibfield  {author} {\bibinfo {author} {\bibfnamefont {D.}~\bibnamefont
  {Venturi}}\ and\ \bibinfo {author} {\bibfnamefont {G.~E.}\ \bibnamefont
  {Karniadakis}},\ }\bibfield  {title} {\enquote {\bibinfo {title} {Gappy data
  and reconstruction procedures for flow past a cylinder},}\ }\href {\doibase
  10.1017/S0022112004001338} {\bibfield  {journal} {\bibinfo  {journal}
  {Journal of Fluid Mechanics}\ }\textbf {\bibinfo {volume} {519}},\ \bibinfo
  {pages} {315–336} (\bibinfo {year} {2004})}\BibitemShut {NoStop}%
\bibitem [{\citenamefont {Adrian}\ and\ \citenamefont
  {Moin}(1988)}]{fr_history_lse}%
  \BibitemOpen
  \bibfield  {author} {\bibinfo {author} {\bibfnamefont {R.~J.}\ \bibnamefont
  {Adrian}}\ and\ \bibinfo {author} {\bibfnamefont {P.}~\bibnamefont {Moin}},\
  }\bibfield  {title} {\enquote {\bibinfo {title} {Stochastic estimation of
  organized turbulent structure: homogeneous shear flow},}\ }\href {\doibase
  10.1017/S0022112088001442} {\bibfield  {journal} {\bibinfo  {journal}
  {Journal of Fluid Mechanics}\ }\textbf {\bibinfo {volume} {190}},\ \bibinfo
  {pages} {531–559} (\bibinfo {year} {1988})}\BibitemShut {NoStop}%
\bibitem [{\citenamefont {Dubois}\ \emph {et~al.}(2022)\citenamefont {Dubois},
  \citenamefont {Gomez}, \citenamefont {Planckaert},\ and\ \citenamefont
  {Perret}}]{flow_reconstruction_3}%
  \BibitemOpen
  \bibfield  {author} {\bibinfo {author} {\bibfnamefont {P.}~\bibnamefont
  {Dubois}}, \bibinfo {author} {\bibfnamefont {T.}~\bibnamefont {Gomez}},
  \bibinfo {author} {\bibfnamefont {L.}~\bibnamefont {Planckaert}}, \ and\
  \bibinfo {author} {\bibfnamefont {L.}~\bibnamefont {Perret}},\ }\bibfield
  {title} {\enquote {\bibinfo {title} {Machine learning for fluid flow
  reconstruction from limited measurements},}\ }\href {\doibase
  https://doi.org/10.1016/j.jcp.2021.110733} {\bibfield  {journal} {\bibinfo
  {journal} {Journal of Computational Physics}\ }\textbf {\bibinfo {volume}
  {448}},\ \bibinfo {pages} {110733} (\bibinfo {year} {2022})}\BibitemShut
  {NoStop}%
\bibitem [{\citenamefont {Fukami}, \citenamefont {Fukagata},\ and\
  \citenamefont {Taira}(2020)}]{flow_reconstruction_4}%
  \BibitemOpen
  \bibfield  {author} {\bibinfo {author} {\bibfnamefont {K.}~\bibnamefont
  {Fukami}}, \bibinfo {author} {\bibfnamefont {K.}~\bibnamefont {Fukagata}}, \
  and\ \bibinfo {author} {\bibfnamefont {K.}~\bibnamefont {Taira}},\ }\bibfield
   {title} {\enquote {\bibinfo {title} {Assessment of supervised machine
  learning methods for fluid flows},}\ }\href@noop {} {\bibfield  {journal}
  {\bibinfo  {journal} {Theoretical and Computational Fluid Dynamics}\ }\textbf
  {\bibinfo {volume} {34}},\ \bibinfo {pages} {497--519} (\bibinfo {year}
  {2020})}\BibitemShut {NoStop}%
\bibitem [{\citenamefont {Sun}\ and\ \citenamefont
  {Wang}(2020)}]{flow_reconstruction_5}%
  \BibitemOpen
  \bibfield  {author} {\bibinfo {author} {\bibfnamefont {L.}~\bibnamefont
  {Sun}}\ and\ \bibinfo {author} {\bibfnamefont {J.-X.}\ \bibnamefont {Wang}},\
  }\bibfield  {title} {\enquote {\bibinfo {title} {Physics-constrained bayesian
  neural network for fluid flow reconstruction with sparse and noisy data},}\
  }\href {\doibase https://doi.org/10.1016/j.taml.2020.01.031} {\bibfield
  {journal} {\bibinfo  {journal} {Theoretical and Applied Mechanics Letters}\
  }\textbf {\bibinfo {volume} {10}},\ \bibinfo {pages} {161--169} (\bibinfo
  {year} {2020})}\BibitemShut {NoStop}%
\bibitem [{\citenamefont {Crowdy}(2020)}]{scmap_book}%
  \BibitemOpen
  \bibfield  {author} {\bibinfo {author} {\bibfnamefont {D.}~\bibnamefont
  {Crowdy}},\ }\href@noop {} {\emph {\bibinfo {title} {Solving problems in
  multiply connected domains}}}\ (\bibinfo  {publisher} {SIAM},\ \bibinfo
  {year} {2020})\BibitemShut {NoStop}%
\bibitem [{\citenamefont {Ronneberger}, \citenamefont {Fischer},\ and\
  \citenamefont {Brox}(2015)}]{unet}%
  \BibitemOpen
  \bibfield  {author} {\bibinfo {author} {\bibfnamefont {O.}~\bibnamefont
  {Ronneberger}}, \bibinfo {author} {\bibfnamefont {P.}~\bibnamefont
  {Fischer}}, \ and\ \bibinfo {author} {\bibfnamefont {T.}~\bibnamefont
  {Brox}},\ }\bibfield  {title} {\enquote {\bibinfo {title} {U-net:
  Convolutional networks for biomedical image segmentation},}\ }in\ \href@noop
  {} {\emph {\bibinfo {booktitle} {International Conference on Medical image
  computing and computer-assisted intervention}}}\ (\bibinfo {organization}
  {Springer},\ \bibinfo {year} {2015})\ pp.\ \bibinfo {pages}
  {234--241}\BibitemShut {NoStop}%
\bibitem [{\citenamefont {Li}\ \emph {et~al.}(2020)\citenamefont {Li},
  \citenamefont {Kovachki}, \citenamefont {Azizzadenesheli}, \citenamefont
  {Liu}, \citenamefont {Bhattacharya}, \citenamefont {Stuart},\ and\
  \citenamefont {Anandkumar}}]{fourier_neural_operator}%
  \BibitemOpen
  \bibfield  {author} {\bibinfo {author} {\bibfnamefont {Z.}~\bibnamefont
  {Li}}, \bibinfo {author} {\bibfnamefont {N.}~\bibnamefont {Kovachki}},
  \bibinfo {author} {\bibfnamefont {K.}~\bibnamefont {Azizzadenesheli}},
  \bibinfo {author} {\bibfnamefont {B.}~\bibnamefont {Liu}}, \bibinfo {author}
  {\bibfnamefont {K.}~\bibnamefont {Bhattacharya}}, \bibinfo {author}
  {\bibfnamefont {A.}~\bibnamefont {Stuart}}, \ and\ \bibinfo {author}
  {\bibfnamefont {A.}~\bibnamefont {Anandkumar}},\ }\bibfield  {title}
  {\enquote {\bibinfo {title} {Fourier neural operator for parametric partial
  differential equations},}\ }\href@noop {} {\bibfield  {journal} {\bibinfo
  {journal} {arXiv preprint arXiv:2010.08895}\ } (\bibinfo {year}
  {2020})}\BibitemShut {NoStop}%
\bibitem [{\citenamefont {De~Vito}\ \emph {et~al.}(2005)\citenamefont
  {De~Vito}, \citenamefont {Rosasco}, \citenamefont {Caponnetto}, \citenamefont
  {De~Giovannini}, \citenamefont {Odone},\ and\ \citenamefont
  {Bartlett}}]{de2005learning}%
  \BibitemOpen
  \bibfield  {author} {\bibinfo {author} {\bibfnamefont {E.}~\bibnamefont
  {De~Vito}}, \bibinfo {author} {\bibfnamefont {L.}~\bibnamefont {Rosasco}},
  \bibinfo {author} {\bibfnamefont {A.}~\bibnamefont {Caponnetto}}, \bibinfo
  {author} {\bibfnamefont {U.}~\bibnamefont {De~Giovannini}}, \bibinfo {author}
  {\bibfnamefont {F.}~\bibnamefont {Odone}}, \ and\ \bibinfo {author}
  {\bibfnamefont {P.}~\bibnamefont {Bartlett}},\ }\bibfield  {title} {\enquote
  {\bibinfo {title} {Learning from examples as an inverse problem.}}\
  }\href@noop {} {\bibfield  {journal} {\bibinfo  {journal} {Journal of Machine
  Learning Research}\ }\textbf {\bibinfo {volume} {6}} (\bibinfo {year}
  {2005})}\BibitemShut {NoStop}%
\bibitem [{\citenamefont {Taira}\ \emph {et~al.}(2020)\citenamefont {Taira},
  \citenamefont {Hemati}, \citenamefont {Brunton}, \citenamefont {Sun},
  \citenamefont {Duraisamy}, \citenamefont {Bagheri}, \citenamefont {Dawson},\
  and\ \citenamefont {Yeh}}]{pod_review1}%
  \BibitemOpen
  \bibfield  {author} {\bibinfo {author} {\bibfnamefont {K.}~\bibnamefont
  {Taira}}, \bibinfo {author} {\bibfnamefont {M.~S.}\ \bibnamefont {Hemati}},
  \bibinfo {author} {\bibfnamefont {S.~L.}\ \bibnamefont {Brunton}}, \bibinfo
  {author} {\bibfnamefont {Y.}~\bibnamefont {Sun}}, \bibinfo {author}
  {\bibfnamefont {K.}~\bibnamefont {Duraisamy}}, \bibinfo {author}
  {\bibfnamefont {S.}~\bibnamefont {Bagheri}}, \bibinfo {author} {\bibfnamefont
  {S.~T.~M.}\ \bibnamefont {Dawson}}, \ and\ \bibinfo {author} {\bibfnamefont
  {C.-A.}\ \bibnamefont {Yeh}},\ }\bibfield  {title} {\enquote {\bibinfo
  {title} {Modal analysis of fluid flows: Applications and outlook},}\ }\href
  {\doibase 10.2514/1.J058462} {\bibfield  {journal} {\bibinfo  {journal} {AIAA
  Journal}\ }\textbf {\bibinfo {volume} {58}},\ \bibinfo {pages} {998--1022}
  (\bibinfo {year} {2020})},\ \Eprint
  {http://arxiv.org/abs/https://doi.org/10.2514/1.J058462}
  {https://doi.org/10.2514/1.J058462} \BibitemShut {NoStop}%
\bibitem [{\citenamefont {Colburn}, \citenamefont {Cessna},\ and\ \citenamefont
  {Bewley}(2011)}]{kalman_filter}%
  \BibitemOpen
  \bibfield  {author} {\bibinfo {author} {\bibfnamefont {C.~H.}\ \bibnamefont
  {Colburn}}, \bibinfo {author} {\bibfnamefont {J.~B.}\ \bibnamefont {Cessna}},
  \ and\ \bibinfo {author} {\bibfnamefont {T.~R.}\ \bibnamefont {Bewley}},\
  }\bibfield  {title} {\enquote {\bibinfo {title} {State estimation in
  wall-bounded flow systems. part 3. the ensemble kalman filter},}\ }\href
  {\doibase 10.1017/jfm.2011.222} {\bibfield  {journal} {\bibinfo  {journal}
  {Journal of Fluid Mechanics}\ }\textbf {\bibinfo {volume} {682}},\ \bibinfo
  {pages} {289–303} (\bibinfo {year} {2011})}\BibitemShut {NoStop}%
\bibitem [{\citenamefont {Gustafsson}\ \emph {et~al.}(2018)\citenamefont
  {Gustafsson}, \citenamefont {Janjić}, \citenamefont {Schraff}, \citenamefont
  {Leuenberger}, \citenamefont {Weissmann}, \citenamefont {Reich},
  \citenamefont {Brousseau}, \citenamefont {Montmerle}, \citenamefont
  {Wattrelot}, \citenamefont {Bučánek}, \citenamefont {Mile}, \citenamefont
  {Hamdi}, \citenamefont {Lindskog}, \citenamefont {Barkmeijer}, \citenamefont
  {Dahlbom}, \citenamefont {Macpherson}, \citenamefont {Ballard}, \citenamefont
  {Inverarity}, \citenamefont {Carley}, \citenamefont {Alexander},
  \citenamefont {Dowell}, \citenamefont {Liu}, \citenamefont {Ikuta},\ and\
  \citenamefont {Fujita}}]{meteorology_fr_variational_3}%
  \BibitemOpen
  \bibfield  {author} {\bibinfo {author} {\bibfnamefont {N.}~\bibnamefont
  {Gustafsson}}, \bibinfo {author} {\bibfnamefont {T.}~\bibnamefont {Janjić}},
  \bibinfo {author} {\bibfnamefont {C.}~\bibnamefont {Schraff}}, \bibinfo
  {author} {\bibfnamefont {D.}~\bibnamefont {Leuenberger}}, \bibinfo {author}
  {\bibfnamefont {M.}~\bibnamefont {Weissmann}}, \bibinfo {author}
  {\bibfnamefont {H.}~\bibnamefont {Reich}}, \bibinfo {author} {\bibfnamefont
  {P.}~\bibnamefont {Brousseau}}, \bibinfo {author} {\bibfnamefont
  {T.}~\bibnamefont {Montmerle}}, \bibinfo {author} {\bibfnamefont
  {E.}~\bibnamefont {Wattrelot}}, \bibinfo {author} {\bibfnamefont
  {A.}~\bibnamefont {Bučánek}}, \bibinfo {author} {\bibfnamefont
  {M.}~\bibnamefont {Mile}}, \bibinfo {author} {\bibfnamefont {R.}~\bibnamefont
  {Hamdi}}, \bibinfo {author} {\bibfnamefont {M.}~\bibnamefont {Lindskog}},
  \bibinfo {author} {\bibfnamefont {J.}~\bibnamefont {Barkmeijer}}, \bibinfo
  {author} {\bibfnamefont {M.}~\bibnamefont {Dahlbom}}, \bibinfo {author}
  {\bibfnamefont {B.}~\bibnamefont {Macpherson}}, \bibinfo {author}
  {\bibfnamefont {S.}~\bibnamefont {Ballard}}, \bibinfo {author} {\bibfnamefont
  {G.}~\bibnamefont {Inverarity}}, \bibinfo {author} {\bibfnamefont
  {J.}~\bibnamefont {Carley}}, \bibinfo {author} {\bibfnamefont
  {C.}~\bibnamefont {Alexander}}, \bibinfo {author} {\bibfnamefont
  {D.}~\bibnamefont {Dowell}}, \bibinfo {author} {\bibfnamefont
  {S.}~\bibnamefont {Liu}}, \bibinfo {author} {\bibfnamefont {Y.}~\bibnamefont
  {Ikuta}}, \ and\ \bibinfo {author} {\bibfnamefont {T.}~\bibnamefont
  {Fujita}},\ }\bibfield  {title} {\enquote {\bibinfo {title} {Survey of data
  assimilation methods for convective-scale numerical weather prediction at
  operational centres},}\ }\href {\doibase https://doi.org/10.1002/qj.3179}
  {\bibfield  {journal} {\bibinfo  {journal} {Quarterly Journal of the Royal
  Meteorological Society}\ }\textbf {\bibinfo {volume} {144}},\ \bibinfo
  {pages} {1218--1256} (\bibinfo {year} {2018})},\ \Eprint
  {http://arxiv.org/abs/https://rmets.onlinelibrary.wiley.com/doi/pdf/10.1002/qj.3179}
  {https://rmets.onlinelibrary.wiley.com/doi/pdf/10.1002/qj.3179} \BibitemShut
  {NoStop}%
\bibitem [{\citenamefont {Dimet}\ and\ \citenamefont
  {Talagrand}(1986)}]{meteorology_fr_variational}%
  \BibitemOpen
  \bibfield  {author} {\bibinfo {author} {\bibfnamefont {F.-X.~L.}\
  \bibnamefont {Dimet}}\ and\ \bibinfo {author} {\bibfnamefont
  {O.}~\bibnamefont {Talagrand}},\ }\bibfield  {title} {\enquote {\bibinfo
  {title} {Variational algorithms for analysis and assimilation of
  meteorological observations: theoretical aspects},}\ }\href {\doibase
  10.3402/tellusa.v38i2.11706} {\bibfield  {journal} {\bibinfo  {journal}
  {Tellus A: Dynamic Meteorology and Oceanography}\ }\textbf {\bibinfo {volume}
  {38}},\ \bibinfo {pages} {97--110} (\bibinfo {year} {1986})},\ \Eprint
  {http://arxiv.org/abs/https://doi.org/10.3402/tellusa.v38i2.11706}
  {https://doi.org/10.3402/tellusa.v38i2.11706} \BibitemShut {NoStop}%
\bibitem [{\citenamefont {Courtier}(1997)}]{meteorology_fr_variational_2}%
  \BibitemOpen
  \bibfield  {author} {\bibinfo {author} {\bibfnamefont {P.}~\bibnamefont
  {Courtier}},\ }\bibfield  {title} {\enquote {\bibinfo {title} {Dual
  formulation of four-dimensional variational assimilation},}\ }\href {\doibase
  https://doi.org/10.1002/qj.49712354414} {\bibfield  {journal} {\bibinfo
  {journal} {Quarterly Journal of the Royal Meteorological Society}\ }\textbf
  {\bibinfo {volume} {123}},\ \bibinfo {pages} {2449--2461} (\bibinfo {year}
  {1997})},\ \Eprint
  {http://arxiv.org/abs/https://rmets.onlinelibrary.wiley.com/doi/pdf/10.1002/qj.49712354414}
  {https://rmets.onlinelibrary.wiley.com/doi/pdf/10.1002/qj.49712354414}
  \BibitemShut {NoStop}%
\bibitem [{\citenamefont {Foures}\ \emph {et~al.}(2014)\citenamefont {Foures},
  \citenamefont {Dovetta}, \citenamefont {Sipp},\ and\ \citenamefont
  {Schmid}}]{fr_history_variational}%
  \BibitemOpen
  \bibfield  {author} {\bibinfo {author} {\bibfnamefont {D.~P.~G.}\
  \bibnamefont {Foures}}, \bibinfo {author} {\bibfnamefont {N.}~\bibnamefont
  {Dovetta}}, \bibinfo {author} {\bibfnamefont {D.}~\bibnamefont {Sipp}}, \
  and\ \bibinfo {author} {\bibfnamefont {P.~J.}\ \bibnamefont {Schmid}},\
  }\bibfield  {title} {\enquote {\bibinfo {title} {A data-assimilation method
  for reynolds-averaged navier–stokes-driven mean flow reconstruction},}\
  }\href {\doibase 10.1017/jfm.2014.566} {\bibfield  {journal} {\bibinfo
  {journal} {Journal of Fluid Mechanics}\ }\textbf {\bibinfo {volume} {759}},\
  \bibinfo {pages} {404–431} (\bibinfo {year} {2014})}\BibitemShut {NoStop}%
\bibitem [{\citenamefont {He}\ and\ \citenamefont
  {Liu}(2020)}]{fr_history_lse_4}%
  \BibitemOpen
  \bibfield  {author} {\bibinfo {author} {\bibfnamefont {C.}~\bibnamefont
  {He}}\ and\ \bibinfo {author} {\bibfnamefont {Y.}~\bibnamefont {Liu}},\
  }\bibfield  {title} {\enquote {\bibinfo {title} {Time-resolved reconstruction
  of turbulent flows using linear stochastic estimation and sequential data
  assimilation},}\ }\href {\doibase 10.1063/5.0014249} {\bibfield  {journal}
  {\bibinfo  {journal} {Physics of Fluids}\ }\textbf {\bibinfo {volume} {32}},\
  \bibinfo {pages} {075106} (\bibinfo {year} {2020})},\ \Eprint
  {http://arxiv.org/abs/https://doi.org/10.1063/5.0014249}
  {https://doi.org/10.1063/5.0014249} \BibitemShut {NoStop}%
\bibitem [{\citenamefont {Butcher}\ and\ \citenamefont
  {Spencer}(2020)}]{fr_history_lse_5}%
  \BibitemOpen
  \bibfield  {author} {\bibinfo {author} {\bibfnamefont {D.}~\bibnamefont
  {Butcher}}\ and\ \bibinfo {author} {\bibfnamefont {A.}~\bibnamefont
  {Spencer}},\ }\bibfield  {title} {\enquote {\bibinfo {title} {Linear
  stochastic estimation of the coherent structures in internal combustion
  engine flow},}\ }\href {\doibase 10.1177/1468087418824896} {\bibfield
  {journal} {\bibinfo  {journal} {International Journal of Engine Research}\
  }\textbf {\bibinfo {volume} {21}},\ \bibinfo {pages} {1738--1749} (\bibinfo
  {year} {2020})},\ \Eprint
  {http://arxiv.org/abs/https://doi.org/10.1177/1468087418824896}
  {https://doi.org/10.1177/1468087418824896} \BibitemShut {NoStop}%
\bibitem [{\citenamefont {Podvin}\ \emph {et~al.}(2018)\citenamefont {Podvin},
  \citenamefont {Nguimatsia}, \citenamefont {Foucaut}, \citenamefont {Cuvier},\
  and\ \citenamefont {Fraigneau}}]{fr_history_lse_3}%
  \BibitemOpen
  \bibfield  {author} {\bibinfo {author} {\bibfnamefont {B.}~\bibnamefont
  {Podvin}}, \bibinfo {author} {\bibfnamefont {S.}~\bibnamefont {Nguimatsia}},
  \bibinfo {author} {\bibfnamefont {J.-M.}\ \bibnamefont {Foucaut}}, \bibinfo
  {author} {\bibfnamefont {C.}~\bibnamefont {Cuvier}}, \ and\ \bibinfo {author}
  {\bibfnamefont {Y.}~\bibnamefont {Fraigneau}},\ }\bibfield  {title} {\enquote
  {\bibinfo {title} {On combining linear stochastic estimation and proper
  orthogonal decomposition for flow reconstruction},}\ }\href@noop {}
  {\bibfield  {journal} {\bibinfo  {journal} {Experiments in Fluids}\ }\textbf
  {\bibinfo {volume} {59}},\ \bibinfo {pages} {1--12} (\bibinfo {year}
  {2018})}\BibitemShut {NoStop}%
\bibitem [{\citenamefont {Willcox}(2006)}]{gappy_pod_ex2}%
  \BibitemOpen
  \bibfield  {author} {\bibinfo {author} {\bibfnamefont {K.}~\bibnamefont
  {Willcox}},\ }\bibfield  {title} {\enquote {\bibinfo {title} {Unsteady flow
  sensing and estimation via the gappy proper orthogonal decomposition},}\
  }\href@noop {} {\bibfield  {journal} {\bibinfo  {journal} {Computers \&
  fluids}\ }\textbf {\bibinfo {volume} {35}},\ \bibinfo {pages} {208--226}
  (\bibinfo {year} {2006})}\BibitemShut {NoStop}%
\bibitem [{\citenamefont {Saini}, \citenamefont {Arndt},\ and\ \citenamefont
  {Steinberg}(2016)}]{gappy_pod_ex4}%
  \BibitemOpen
  \bibfield  {author} {\bibinfo {author} {\bibfnamefont {P.}~\bibnamefont
  {Saini}}, \bibinfo {author} {\bibfnamefont {C.~M.}\ \bibnamefont {Arndt}}, \
  and\ \bibinfo {author} {\bibfnamefont {A.~M.}\ \bibnamefont {Steinberg}},\
  }\bibfield  {title} {\enquote {\bibinfo {title} {Development and evaluation
  of gappy-pod as a data reconstruction technique for noisy piv measurements in
  gas turbine combustors},}\ }\href@noop {} {\bibfield  {journal} {\bibinfo
  {journal} {Experiments in Fluids}\ }\textbf {\bibinfo {volume} {57}},\
  \bibinfo {pages} {1--15} (\bibinfo {year} {2016})}\BibitemShut {NoStop}%
\bibitem [{\citenamefont {Wright}\ \emph {et~al.}(2009)\citenamefont {Wright},
  \citenamefont {Yang}, \citenamefont {Ganesh}, \citenamefont {Sastry},\ and\
  \citenamefont {Ma}}]{sparse_representation}%
  \BibitemOpen
  \bibfield  {author} {\bibinfo {author} {\bibfnamefont {J.}~\bibnamefont
  {Wright}}, \bibinfo {author} {\bibfnamefont {A.~Y.}\ \bibnamefont {Yang}},
  \bibinfo {author} {\bibfnamefont {A.}~\bibnamefont {Ganesh}}, \bibinfo
  {author} {\bibfnamefont {S.~S.}\ \bibnamefont {Sastry}}, \ and\ \bibinfo
  {author} {\bibfnamefont {Y.}~\bibnamefont {Ma}},\ }\bibfield  {title}
  {\enquote {\bibinfo {title} {Robust face recognition via sparse
  representation},}\ }\href {\doibase 10.1109/TPAMI.2008.79} {\bibfield
  {journal} {\bibinfo  {journal} {IEEE Transactions on Pattern Analysis and
  Machine Intelligence}\ }\textbf {\bibinfo {volume} {31}},\ \bibinfo {pages}
  {210--227} (\bibinfo {year} {2009})}\BibitemShut {NoStop}%
\bibitem [{\citenamefont {Callaham}, \citenamefont {Maeda},\ and\ \citenamefont
  {Brunton}(2019)}]{fr_history_sparse_reconstruction}%
  \BibitemOpen
  \bibfield  {author} {\bibinfo {author} {\bibfnamefont {J.~L.}\ \bibnamefont
  {Callaham}}, \bibinfo {author} {\bibfnamefont {K.}~\bibnamefont {Maeda}}, \
  and\ \bibinfo {author} {\bibfnamefont {S.~L.}\ \bibnamefont {Brunton}},\
  }\bibfield  {title} {\enquote {\bibinfo {title} {Robust flow reconstruction
  from limited measurements via sparse representation},}\ }\href {\doibase
  10.1103/PhysRevFluids.4.103907} {\bibfield  {journal} {\bibinfo  {journal}
  {Phys. Rev. Fluids}\ }\textbf {\bibinfo {volume} {4}},\ \bibinfo {pages}
  {103907} (\bibinfo {year} {2019})}\BibitemShut {NoStop}%
\bibitem [{\citenamefont {Espath}\ \emph {et~al.}(2021)\citenamefont {Espath},
  \citenamefont {Kabanov}, \citenamefont {Kiessling},\ and\ \citenamefont
  {Tempone}}]{fr_history_sfdf}%
  \BibitemOpen
  \bibfield  {author} {\bibinfo {author} {\bibfnamefont {L.}~\bibnamefont
  {Espath}}, \bibinfo {author} {\bibfnamefont {D.}~\bibnamefont {Kabanov}},
  \bibinfo {author} {\bibfnamefont {J.}~\bibnamefont {Kiessling}}, \ and\
  \bibinfo {author} {\bibfnamefont {R.}~\bibnamefont {Tempone}},\ }\bibfield
  {title} {\enquote {\bibinfo {title} {Statistical learning for fluid flows:
  Sparse fourier divergence-free approximations},}\ }\href {\doibase
  10.1063/5.0064862} {\bibfield  {journal} {\bibinfo  {journal} {Physics of
  Fluids}\ }\textbf {\bibinfo {volume} {33}},\ \bibinfo {pages} {097108}
  (\bibinfo {year} {2021})},\ \Eprint
  {http://arxiv.org/abs/https://doi.org/10.1063/5.0064862}
  {https://doi.org/10.1063/5.0064862} \BibitemShut {NoStop}%
\bibitem [{\citenamefont {Sonoda}\ and\ \citenamefont
  {Murata}(2017)}]{nn_universal_approx}%
  \BibitemOpen
  \bibfield  {author} {\bibinfo {author} {\bibfnamefont {S.}~\bibnamefont
  {Sonoda}}\ and\ \bibinfo {author} {\bibfnamefont {N.}~\bibnamefont
  {Murata}},\ }\bibfield  {title} {\enquote {\bibinfo {title} {Neural network
  with unbounded activation functions is universal approximator},}\ }\href
  {\doibase https://doi.org/10.1016/j.acha.2015.12.005} {\bibfield  {journal}
  {\bibinfo  {journal} {Applied and Computational Harmonic Analysis}\ }\textbf
  {\bibinfo {volume} {43}},\ \bibinfo {pages} {233--268} (\bibinfo {year}
  {2017})}\BibitemShut {NoStop}%
\bibitem [{\citenamefont {Brown}\ \emph {et~al.}(2020)\citenamefont {Brown},
  \citenamefont {Mann}, \citenamefont {Ryder}, \citenamefont {Subbiah},
  \citenamefont {Kaplan}, \citenamefont {Dhariwal}, \citenamefont
  {Neelakantan}, \citenamefont {Shyam}, \citenamefont {Sastry}, \citenamefont
  {Askell} \emph {et~al.}}]{gpt3}%
  \BibitemOpen
  \bibfield  {author} {\bibinfo {author} {\bibfnamefont {T.~B.}\ \bibnamefont
  {Brown}}, \bibinfo {author} {\bibfnamefont {B.}~\bibnamefont {Mann}},
  \bibinfo {author} {\bibfnamefont {N.}~\bibnamefont {Ryder}}, \bibinfo
  {author} {\bibfnamefont {M.}~\bibnamefont {Subbiah}}, \bibinfo {author}
  {\bibfnamefont {J.}~\bibnamefont {Kaplan}}, \bibinfo {author} {\bibfnamefont
  {P.}~\bibnamefont {Dhariwal}}, \bibinfo {author} {\bibfnamefont
  {A.}~\bibnamefont {Neelakantan}}, \bibinfo {author} {\bibfnamefont
  {P.}~\bibnamefont {Shyam}}, \bibinfo {author} {\bibfnamefont
  {G.}~\bibnamefont {Sastry}}, \bibinfo {author} {\bibfnamefont
  {A.}~\bibnamefont {Askell}},  \emph {et~al.},\ }\bibfield  {title} {\enquote
  {\bibinfo {title} {Language models are few-shot learners},}\ }\href@noop {}
  {\bibfield  {journal} {\bibinfo  {journal} {arXiv preprint arXiv:2005.14165}\
  } (\bibinfo {year} {2020})}\BibitemShut {NoStop}%
\bibitem [{\citenamefont {Karras}\ \emph {et~al.}(2020)\citenamefont {Karras},
  \citenamefont {Laine}, \citenamefont {Aittala}, \citenamefont {Hellsten},
  \citenamefont {Lehtinen},\ and\ \citenamefont {Aila}}]{stylegan2}%
  \BibitemOpen
  \bibfield  {author} {\bibinfo {author} {\bibfnamefont {T.}~\bibnamefont
  {Karras}}, \bibinfo {author} {\bibfnamefont {S.}~\bibnamefont {Laine}},
  \bibinfo {author} {\bibfnamefont {M.}~\bibnamefont {Aittala}}, \bibinfo
  {author} {\bibfnamefont {J.}~\bibnamefont {Hellsten}}, \bibinfo {author}
  {\bibfnamefont {J.}~\bibnamefont {Lehtinen}}, \ and\ \bibinfo {author}
  {\bibfnamefont {T.}~\bibnamefont {Aila}},\ }\href@noop {} {\enquote {\bibinfo
  {title} {Analyzing and improving the image quality of stylegan},}\ }
  (\bibinfo {year} {2020}),\ \Eprint {http://arxiv.org/abs/1912.04958}
  {arXiv:1912.04958 [cs.CV]} \BibitemShut {NoStop}%
\bibitem [{\citenamefont {Kumar}, \citenamefont {Bahl},\ and\ \citenamefont
  {Chakraborty}(2021)}]{flow_reconstruction_9}%
  \BibitemOpen
  \bibfield  {author} {\bibinfo {author} {\bibfnamefont {Y.}~\bibnamefont
  {Kumar}}, \bibinfo {author} {\bibfnamefont {P.}~\bibnamefont {Bahl}}, \ and\
  \bibinfo {author} {\bibfnamefont {S.}~\bibnamefont {Chakraborty}},\
  }\bibfield  {title} {\enquote {\bibinfo {title} {State estimation with
  limited sensors--a deep learning based approach},}\ }\href@noop {} {\bibfield
   {journal} {\bibinfo  {journal} {arXiv preprint arXiv:2101.11513}\ }
  (\bibinfo {year} {2021})}\BibitemShut {NoStop}%
\bibitem [{\citenamefont {Carter}\ \emph {et~al.}(2021)\citenamefont {Carter},
  \citenamefont {De~Voogt}, \citenamefont {Soares},\ and\ \citenamefont
  {Ganapathisubramani}}]{flow_reconstruction_7}%
  \BibitemOpen
  \bibfield  {author} {\bibinfo {author} {\bibfnamefont {D.~W.}\ \bibnamefont
  {Carter}}, \bibinfo {author} {\bibfnamefont {F.}~\bibnamefont {De~Voogt}},
  \bibinfo {author} {\bibfnamefont {R.}~\bibnamefont {Soares}}, \ and\ \bibinfo
  {author} {\bibfnamefont {B.}~\bibnamefont {Ganapathisubramani}},\ }\bibfield
  {title} {\enquote {\bibinfo {title} {Data-driven sparse reconstruction of
  flow over a stalled aerofoil using experimental data},}\ }\href {\doibase
  10.1017/dce.2021.5} {\bibfield  {journal} {\bibinfo  {journal} {Data-Centric
  Engineering}\ }\textbf {\bibinfo {volume} {2}},\ \bibinfo {pages} {e5}
  (\bibinfo {year} {2021})}\BibitemShut {NoStop}%
\bibitem [{\citenamefont {Chen}, \citenamefont {Hachem},\ and\ \citenamefont
  {Viquerat}(2021)}]{2d_flowpred_random_1}%
  \BibitemOpen
  \bibfield  {author} {\bibinfo {author} {\bibfnamefont {J.}~\bibnamefont
  {Chen}}, \bibinfo {author} {\bibfnamefont {E.}~\bibnamefont {Hachem}}, \ and\
  \bibinfo {author} {\bibfnamefont {J.}~\bibnamefont {Viquerat}},\ }\bibfield
  {title} {\enquote {\bibinfo {title} {Graph neural networks for laminar flow
  prediction around random two-dimensional shapes},}\ }\href {\doibase
  10.1063/5.0064108} {\bibfield  {journal} {\bibinfo  {journal} {Physics of
  Fluids}\ }\textbf {\bibinfo {volume} {33}},\ \bibinfo {pages} {123607}
  (\bibinfo {year} {2021})},\ \Eprint
  {http://arxiv.org/abs/https://doi.org/10.1063/5.0064108}
  {https://doi.org/10.1063/5.0064108} \BibitemShut {NoStop}%
\bibitem [{\citenamefont {Blom}(1983)}]{karmantrefftz}%
  \BibitemOpen
  \bibfield  {author} {\bibinfo {author} {\bibfnamefont {J.}~\bibnamefont
  {Blom}},\ }\href@noop {} {\enquote {\bibinfo {title} {Some characteristic
  quantities of karman-trefftz profiles},}\ }\bibinfo {type} {Tech. Rep.}\
  \bibinfo {number} {NASA-TM-77013}\ (\bibinfo {year} {1983})\BibitemShut
  {NoStop}%
\bibitem [{\citenamefont {Hu}(1998)}]{dscpack}%
  \BibitemOpen
  \bibfield  {author} {\bibinfo {author} {\bibfnamefont {C.}~\bibnamefont
  {Hu}},\ }\bibfield  {title} {\enquote {\bibinfo {title} {Algorithm 785: A
  software package for computing schwarz-christoffel conformal transformation
  for doubly connected polygonal regions},}\ }\href {\doibase
  10.1145/292395.291204} {\bibfield  {journal} {\bibinfo  {journal} {ACM Trans.
  Math. Softw.}\ }\textbf {\bibinfo {volume} {24}},\ \bibinfo {pages}
  {317–333} (\bibinfo {year} {1998})}\BibitemShut {NoStop}%
\bibitem [{\citenamefont {Driscoll}\ and\ \citenamefont
  {Trefethen}(2002)}]{scmap_book2}%
  \BibitemOpen
  \bibfield  {author} {\bibinfo {author} {\bibfnamefont {T.~A.}\ \bibnamefont
  {Driscoll}}\ and\ \bibinfo {author} {\bibfnamefont {L.~N.}\ \bibnamefont
  {Trefethen}},\ }\href {\doibase 10.1017/CBO9780511546808} {\emph {\bibinfo
  {title} {Schwarz-Christoffel Mapping}}},\ Cambridge Monographs on Applied and
  Computational Mathematics\ (\bibinfo  {publisher} {Cambridge University
  Press},\ \bibinfo {year} {2002})\BibitemShut {NoStop}%
\bibitem [{\citenamefont {Trefethen}(1980)}]{scmap_background1}%
  \BibitemOpen
  \bibfield  {author} {\bibinfo {author} {\bibfnamefont {L.~N.}\ \bibnamefont
  {Trefethen}},\ }\bibfield  {title} {\enquote {\bibinfo {title} {Numerical
  computation of the schwarz--christoffel transformation},}\ }\href@noop {}
  {\bibfield  {journal} {\bibinfo  {journal} {SIAM Journal on Scientific and
  Statistical Computing}\ }\textbf {\bibinfo {volume} {1}},\ \bibinfo {pages}
  {82--102} (\bibinfo {year} {1980})}\BibitemShut {NoStop}%
\bibitem [{Note1()}]{Note1}%
  \BibitemOpen
  \bibinfo {note} {Vertex coordinates in the $w$-domain}\BibitemShut {NoStop}%
\bibitem [{\citenamefont {Viquerat}\ \emph {et~al.}(2021)\citenamefont
  {Viquerat}, \citenamefont {Rabault}, \citenamefont {Kuhnle}, \citenamefont
  {Ghraieb}, \citenamefont {Larcher},\ and\ \citenamefont
  {Hachem}}]{bezier_code}%
  \BibitemOpen
  \bibfield  {author} {\bibinfo {author} {\bibfnamefont {J.}~\bibnamefont
  {Viquerat}}, \bibinfo {author} {\bibfnamefont {J.}~\bibnamefont {Rabault}},
  \bibinfo {author} {\bibfnamefont {A.}~\bibnamefont {Kuhnle}}, \bibinfo
  {author} {\bibfnamefont {H.}~\bibnamefont {Ghraieb}}, \bibinfo {author}
  {\bibfnamefont {A.}~\bibnamefont {Larcher}}, \ and\ \bibinfo {author}
  {\bibfnamefont {E.}~\bibnamefont {Hachem}},\ }\bibfield  {title} {\enquote
  {\bibinfo {title} {Direct shape optimization through deep reinforcement
  learning},}\ }\href {\doibase https://doi.org/10.1016/j.jcp.2020.110080}
  {\bibfield  {journal} {\bibinfo  {journal} {Journal of Computational
  Physics}\ }\textbf {\bibinfo {volume} {428}},\ \bibinfo {pages} {110080}
  (\bibinfo {year} {2021})}\BibitemShut {NoStop}%
\bibitem [{\citenamefont {Witherden}, \citenamefont {Farrington},\ and\
  \citenamefont {Vincent}(2014)}]{pyfr}%
  \BibitemOpen
  \bibfield  {author} {\bibinfo {author} {\bibfnamefont {F.}~\bibnamefont
  {Witherden}}, \bibinfo {author} {\bibfnamefont {A.}~\bibnamefont
  {Farrington}}, \ and\ \bibinfo {author} {\bibfnamefont {P.}~\bibnamefont
  {Vincent}},\ }\bibfield  {title} {\enquote {\bibinfo {title} {Pyfr: An open
  source framework for solving advection–diffusion type problems on streaming
  architectures using the flux reconstruction approach},}\ }\href {\doibase
  https://doi.org/10.1016/j.cpc.2014.07.011} {\bibfield  {journal} {\bibinfo
  {journal} {Computer Physics Communications}\ }\textbf {\bibinfo {volume}
  {185}},\ \bibinfo {pages} {3028--3040} (\bibinfo {year} {2014})}\BibitemShut
  {NoStop}%
\bibitem [{\citenamefont {Huynh}(2007)}]{flux_reconstruction}%
  \BibitemOpen
  \bibfield  {author} {\bibinfo {author} {\bibfnamefont {H.~T.}\ \bibnamefont
  {Huynh}},\ }\bibfield  {title} {\enquote {\bibinfo {title} {A flux
  reconstruction approach to high-order schemes including discontinuous
  galerkin methods},}\ }in\ \href@noop {} {\emph {\bibinfo {booktitle} {18th
  AIAA computational fluid dynamics conference}}}\ (\bibinfo {year} {2007})\
  p.\ \bibinfo {pages} {4079}\BibitemShut {NoStop}%
\bibitem [{\citenamefont {Huang}\ \emph {et~al.}(2020)\citenamefont {Huang},
  \citenamefont {Qin}, \citenamefont {Zhou}, \citenamefont {Zhu}, \citenamefont
  {Liu},\ and\ \citenamefont {Shao}}]{dnn_normalization_general_survey}%
  \BibitemOpen
  \bibfield  {author} {\bibinfo {author} {\bibfnamefont {L.}~\bibnamefont
  {Huang}}, \bibinfo {author} {\bibfnamefont {J.}~\bibnamefont {Qin}}, \bibinfo
  {author} {\bibfnamefont {Y.}~\bibnamefont {Zhou}}, \bibinfo {author}
  {\bibfnamefont {F.}~\bibnamefont {Zhu}}, \bibinfo {author} {\bibfnamefont
  {L.}~\bibnamefont {Liu}}, \ and\ \bibinfo {author} {\bibfnamefont
  {L.}~\bibnamefont {Shao}},\ }\href@noop {} {\enquote {\bibinfo {title}
  {Normalization techniques in training dnns: Methodology, analysis and
  application},}\ } (\bibinfo {year} {2020}),\ \Eprint
  {http://arxiv.org/abs/2009.12836} {arXiv:2009.12836 [cs.LG]} \BibitemShut
  {NoStop}%
\bibitem [{\citenamefont {Gundersen}\ \emph {et~al.}(2021)\citenamefont
  {Gundersen}, \citenamefont {Oleynik}, \citenamefont {Blaser},\ and\
  \citenamefont {Alendal}}]{flow_reconstruction_8}%
  \BibitemOpen
  \bibfield  {author} {\bibinfo {author} {\bibfnamefont {K.}~\bibnamefont
  {Gundersen}}, \bibinfo {author} {\bibfnamefont {A.}~\bibnamefont {Oleynik}},
  \bibinfo {author} {\bibfnamefont {N.}~\bibnamefont {Blaser}}, \ and\ \bibinfo
  {author} {\bibfnamefont {G.}~\bibnamefont {Alendal}},\ }\bibfield  {title}
  {\enquote {\bibinfo {title} {Semi-conditional variational auto-encoder for
  flow reconstruction and uncertainty quantification from limited
  observations},}\ }\href {\doibase 10.1063/5.0025779} {\bibfield  {journal}
  {\bibinfo  {journal} {Physics of Fluids}\ }\textbf {\bibinfo {volume} {33}},\
  \bibinfo {pages} {017119} (\bibinfo {year} {2021})},\ \Eprint
  {http://arxiv.org/abs/https://doi.org/10.1063/5.0025779}
  {https://doi.org/10.1063/5.0025779} \BibitemShut {NoStop}%
\bibitem [{\citenamefont {Kingma}\ and\ \citenamefont
  {Ba}(2017)}]{adam_optimizer}%
  \BibitemOpen
  \bibfield  {author} {\bibinfo {author} {\bibfnamefont {D.~P.}\ \bibnamefont
  {Kingma}}\ and\ \bibinfo {author} {\bibfnamefont {J.}~\bibnamefont {Ba}},\
  }\href@noop {} {\enquote {\bibinfo {title} {Adam: A method for stochastic
  optimization},}\ } (\bibinfo {year} {2017}),\ \Eprint
  {http://arxiv.org/abs/1412.6980} {arXiv:1412.6980 [cs.LG]} \BibitemShut
  {NoStop}%
\bibitem [{\citenamefont {Abadi}\ \emph {et~al.}(2016)\citenamefont {Abadi},
  \citenamefont {Barham}, \citenamefont {Chen}, \citenamefont {Chen},
  \citenamefont {Davis}, \citenamefont {Dean}, \citenamefont {Devin},
  \citenamefont {Ghemawat}, \citenamefont {Irving}, \citenamefont {Isard} \emph
  {et~al.}}]{tensorflow}%
  \BibitemOpen
  \bibfield  {author} {\bibinfo {author} {\bibfnamefont {M.}~\bibnamefont
  {Abadi}}, \bibinfo {author} {\bibfnamefont {P.}~\bibnamefont {Barham}},
  \bibinfo {author} {\bibfnamefont {J.}~\bibnamefont {Chen}}, \bibinfo {author}
  {\bibfnamefont {Z.}~\bibnamefont {Chen}}, \bibinfo {author} {\bibfnamefont
  {A.}~\bibnamefont {Davis}}, \bibinfo {author} {\bibfnamefont
  {J.}~\bibnamefont {Dean}}, \bibinfo {author} {\bibfnamefont {M.}~\bibnamefont
  {Devin}}, \bibinfo {author} {\bibfnamefont {S.}~\bibnamefont {Ghemawat}},
  \bibinfo {author} {\bibfnamefont {G.}~\bibnamefont {Irving}}, \bibinfo
  {author} {\bibfnamefont {M.}~\bibnamefont {Isard}},  \emph {et~al.},\
  }\bibfield  {title} {\enquote {\bibinfo {title} {Tensorflow: A system for
  large-scale machine learning},}\ }in\ \href@noop {} {\emph {\bibinfo
  {booktitle} {12th $\{$USENIX$\}$ symposium on operating systems design and
  implementation ($\{$OSDI$\}$ 16)}}}\ (\bibinfo {year} {2016})\ pp.\ \bibinfo
  {pages} {265--283}\BibitemShut {NoStop}%
\bibitem [{\citenamefont {Ioffe}\ and\ \citenamefont
  {Szegedy}(2015)}]{batchnorm}%
  \BibitemOpen
  \bibfield  {author} {\bibinfo {author} {\bibfnamefont {S.}~\bibnamefont
  {Ioffe}}\ and\ \bibinfo {author} {\bibfnamefont {C.}~\bibnamefont
  {Szegedy}},\ }\bibfield  {title} {\enquote {\bibinfo {title} {Batch
  normalization: Accelerating deep network training by reducing internal
  covariate shift},}\ }in\ \href@noop {} {\emph {\bibinfo {booktitle}
  {International conference on machine learning}}}\ (\bibinfo {organization}
  {PMLR},\ \bibinfo {year} {2015})\ pp.\ \bibinfo {pages}
  {448--456}\BibitemShut {NoStop}%
\bibitem [{\citenamefont {Agarap}(2019)}]{relu_good}%
  \BibitemOpen
  \bibfield  {author} {\bibinfo {author} {\bibfnamefont {A.~F.}\ \bibnamefont
  {Agarap}},\ }\href@noop {} {\enquote {\bibinfo {title} {Deep learning using
  rectified linear units (relu)},}\ } (\bibinfo {year} {2019}),\ \Eprint
  {http://arxiv.org/abs/1803.08375} {arXiv:1803.08375 [cs.NE]} \BibitemShut
  {NoStop}%
\bibitem [{spe(2008)}]{spearman_rcc}%
  \BibitemOpen
  \enquote {\bibinfo {title} {Spearman rank correlation coefficient},}\ in\
  \href {\doibase 10.1007/978-0-387-32833-1_379} {\emph {\bibinfo {booktitle}
  {The Concise Encyclopedia of Statistics}}}\ (\bibinfo  {publisher} {Springer
  New York},\ \bibinfo {address} {New York, NY},\ \bibinfo {year} {2008})\ pp.\
  \bibinfo {pages} {502--505}\BibitemShut {NoStop}%
\bibitem [{\citenamefont {Cover}, \citenamefont {Thomas}\ \emph
  {et~al.}(1991)\citenamefont {Cover}, \citenamefont {Thomas} \emph
  {et~al.}}]{mutual_information}%
  \BibitemOpen
  \bibfield  {author} {\bibinfo {author} {\bibfnamefont {T.~M.}\ \bibnamefont
  {Cover}}, \bibinfo {author} {\bibfnamefont {J.~A.}\ \bibnamefont {Thomas}},
  \emph {et~al.},\ }\bibfield  {title} {\enquote {\bibinfo {title} {Entropy,
  relative entropy and mutual information},}\ }\href@noop {} {\bibfield
  {journal} {\bibinfo  {journal} {Elements of information theory}\ }\textbf
  {\bibinfo {volume} {2}},\ \bibinfo {pages} {12--13} (\bibinfo {year}
  {1991})}\BibitemShut {NoStop}%
\bibitem [{\citenamefont {Pedregosa}\ \emph {et~al.}(2011)\citenamefont
  {Pedregosa}, \citenamefont {Varoquaux}, \citenamefont {Gramfort},
  \citenamefont {Michel}, \citenamefont {Thirion}, \citenamefont {Grisel},
  \citenamefont {Blondel}, \citenamefont {Prettenhofer}, \citenamefont {Weiss},
  \citenamefont {Dubourg}, \citenamefont {Vanderplas}, \citenamefont {Passos},
  \citenamefont {Cournapeau}, \citenamefont {Brucher}, \citenamefont {Perrot},\
  and\ \citenamefont {Duchesnay}}]{scikit-learn}%
  \BibitemOpen
  \bibfield  {author} {\bibinfo {author} {\bibfnamefont {F.}~\bibnamefont
  {Pedregosa}}, \bibinfo {author} {\bibfnamefont {G.}~\bibnamefont
  {Varoquaux}}, \bibinfo {author} {\bibfnamefont {A.}~\bibnamefont {Gramfort}},
  \bibinfo {author} {\bibfnamefont {V.}~\bibnamefont {Michel}}, \bibinfo
  {author} {\bibfnamefont {B.}~\bibnamefont {Thirion}}, \bibinfo {author}
  {\bibfnamefont {O.}~\bibnamefont {Grisel}}, \bibinfo {author} {\bibfnamefont
  {M.}~\bibnamefont {Blondel}}, \bibinfo {author} {\bibfnamefont
  {P.}~\bibnamefont {Prettenhofer}}, \bibinfo {author} {\bibfnamefont
  {R.}~\bibnamefont {Weiss}}, \bibinfo {author} {\bibfnamefont
  {V.}~\bibnamefont {Dubourg}}, \bibinfo {author} {\bibfnamefont
  {J.}~\bibnamefont {Vanderplas}}, \bibinfo {author} {\bibfnamefont
  {A.}~\bibnamefont {Passos}}, \bibinfo {author} {\bibfnamefont
  {D.}~\bibnamefont {Cournapeau}}, \bibinfo {author} {\bibfnamefont
  {M.}~\bibnamefont {Brucher}}, \bibinfo {author} {\bibfnamefont
  {M.}~\bibnamefont {Perrot}}, \ and\ \bibinfo {author} {\bibfnamefont
  {E.}~\bibnamefont {Duchesnay}},\ }\bibfield  {title} {\enquote {\bibinfo
  {title} {Scikit-learn: Machine learning in {P}ython},}\ }\href@noop {}
  {\bibfield  {journal} {\bibinfo  {journal} {Journal of Machine Learning
  Research}\ }\textbf {\bibinfo {volume} {12}},\ \bibinfo {pages} {2825--2830}
  (\bibinfo {year} {2011})}\BibitemShut {NoStop}%
\bibitem [{\citenamefont {Chen}\ \emph {et~al.}(2018)\citenamefont {Chen},
  \citenamefont {Papandreou}, \citenamefont {Kokkinos}, \citenamefont
  {Murphy},\ and\ \citenamefont {Yuille}}]{conv_good}%
  \BibitemOpen
  \bibfield  {author} {\bibinfo {author} {\bibfnamefont {L.-C.}\ \bibnamefont
  {Chen}}, \bibinfo {author} {\bibfnamefont {G.}~\bibnamefont {Papandreou}},
  \bibinfo {author} {\bibfnamefont {I.}~\bibnamefont {Kokkinos}}, \bibinfo
  {author} {\bibfnamefont {K.}~\bibnamefont {Murphy}}, \ and\ \bibinfo {author}
  {\bibfnamefont {A.~L.}\ \bibnamefont {Yuille}},\ }\bibfield  {title}
  {\enquote {\bibinfo {title} {Deeplab: Semantic image segmentation with deep
  convolutional nets, atrous convolution, and fully connected crfs},}\ }\href
  {\doibase 10.1109/TPAMI.2017.2699184} {\bibfield  {journal} {\bibinfo
  {journal} {IEEE Transactions on Pattern Analysis and Machine Intelligence}\
  }\textbf {\bibinfo {volume} {40}},\ \bibinfo {pages} {834--848} (\bibinfo
  {year} {2018})}\BibitemShut {NoStop}%
\bibitem [{\citenamefont {Goodfellow}, \citenamefont {Bengio},\ and\
  \citenamefont {Courville}(2016)}]{denoising_autoencoder}%
  \BibitemOpen
  \bibfield  {author} {\bibinfo {author} {\bibfnamefont {I.}~\bibnamefont
  {Goodfellow}}, \bibinfo {author} {\bibfnamefont {Y.}~\bibnamefont {Bengio}},
  \ and\ \bibinfo {author} {\bibfnamefont {A.}~\bibnamefont {Courville}},\
  }\href@noop {} {\emph {\bibinfo {title} {Deep Learning}}}\ (\bibinfo
  {publisher} {MIT Press},\ \bibinfo {year} {2016})\ pp.\ \bibinfo {pages}
  {504--505},\ \bibinfo {note}
  {\url{http://www.deeplearningbook.org}}\BibitemShut {NoStop}%
\bibitem [{\citenamefont {Kim}\ \emph {et~al.}(2021)\citenamefont {Kim},
  \citenamefont {Kim}, \citenamefont {Won},\ and\ \citenamefont
  {Lee}}]{gan_flow_superresolution}%
  \BibitemOpen
  \bibfield  {author} {\bibinfo {author} {\bibfnamefont {H.}~\bibnamefont
  {Kim}}, \bibinfo {author} {\bibfnamefont {J.}~\bibnamefont {Kim}}, \bibinfo
  {author} {\bibfnamefont {S.}~\bibnamefont {Won}}, \ and\ \bibinfo {author}
  {\bibfnamefont {C.}~\bibnamefont {Lee}},\ }\bibfield  {title} {\enquote
  {\bibinfo {title} {Unsupervised deep learning for super-resolution
  reconstruction of turbulence},}\ }\href {\doibase 10.1017/jfm.2020.1028}
  {\bibfield  {journal} {\bibinfo  {journal} {Journal of Fluid Mechanics}\
  }\textbf {\bibinfo {volume} {910}},\ \bibinfo {pages} {A29} (\bibinfo {year}
  {2021})}\BibitemShut {NoStop}%
\end{thebibliography}%

\clearpage
\appendix

\clearpage
\section{Velocity and pressure predictions using the spatio-temporal model with a zero temporal gap ($k=0$)}
\label{sec:app:velocity-pressure-pred}

\begin{figure}[h!]
    \raggedleft
    \includegraphics[width=0.06\columnwidth]{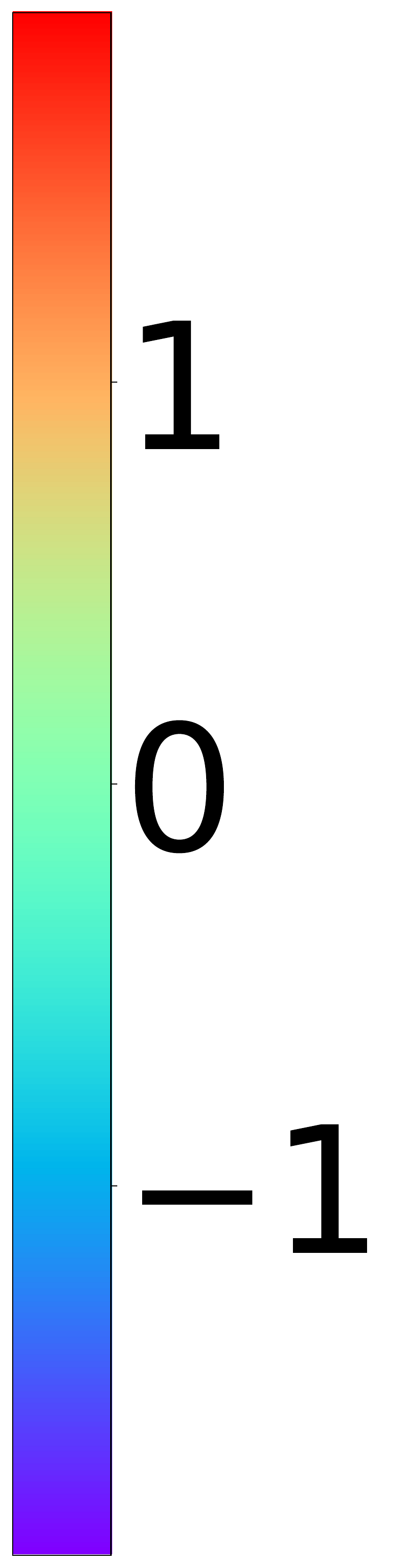}
    \includegraphics[width=0.30\columnwidth]{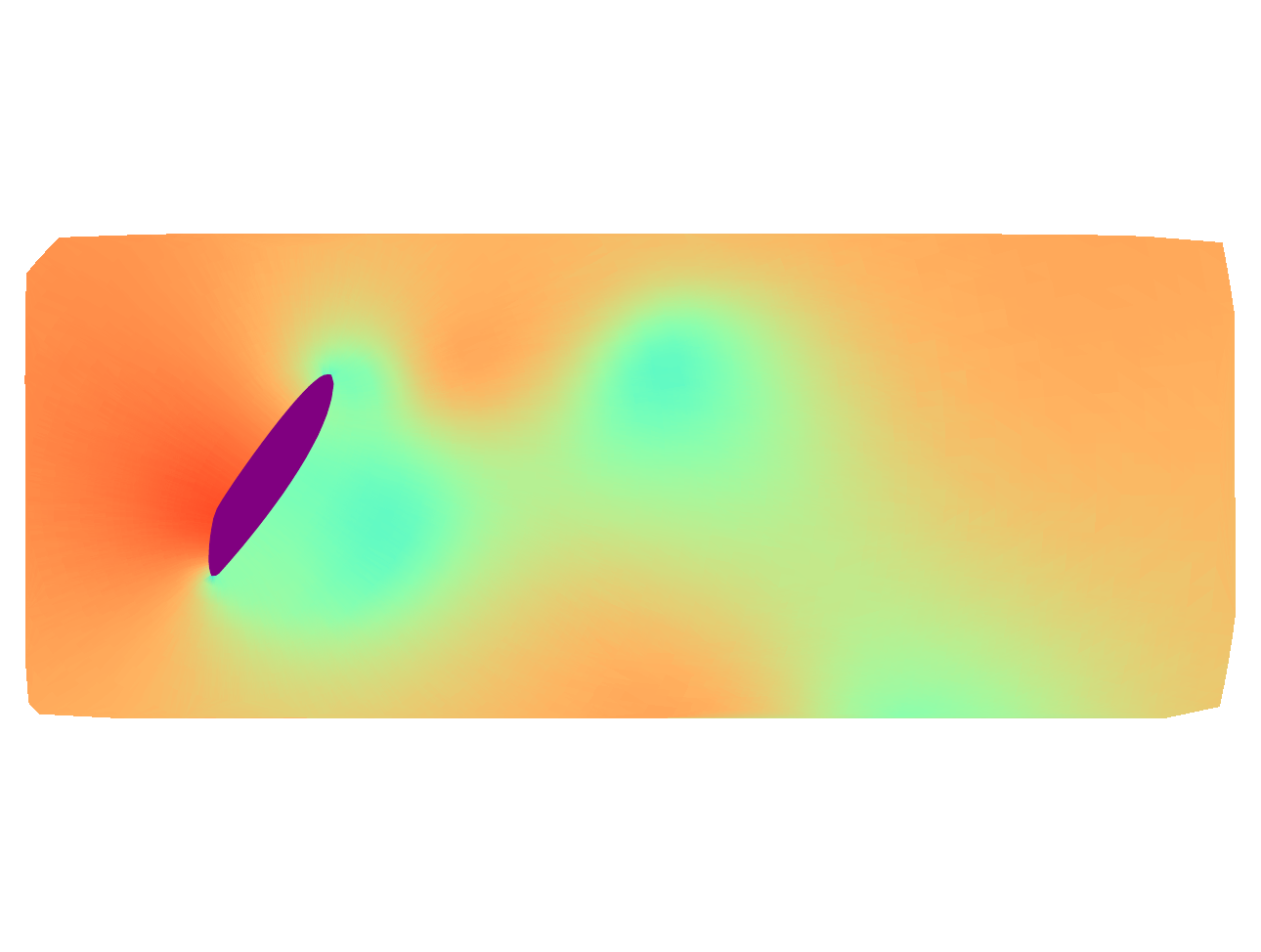}
    \includegraphics[width=0.30\columnwidth]{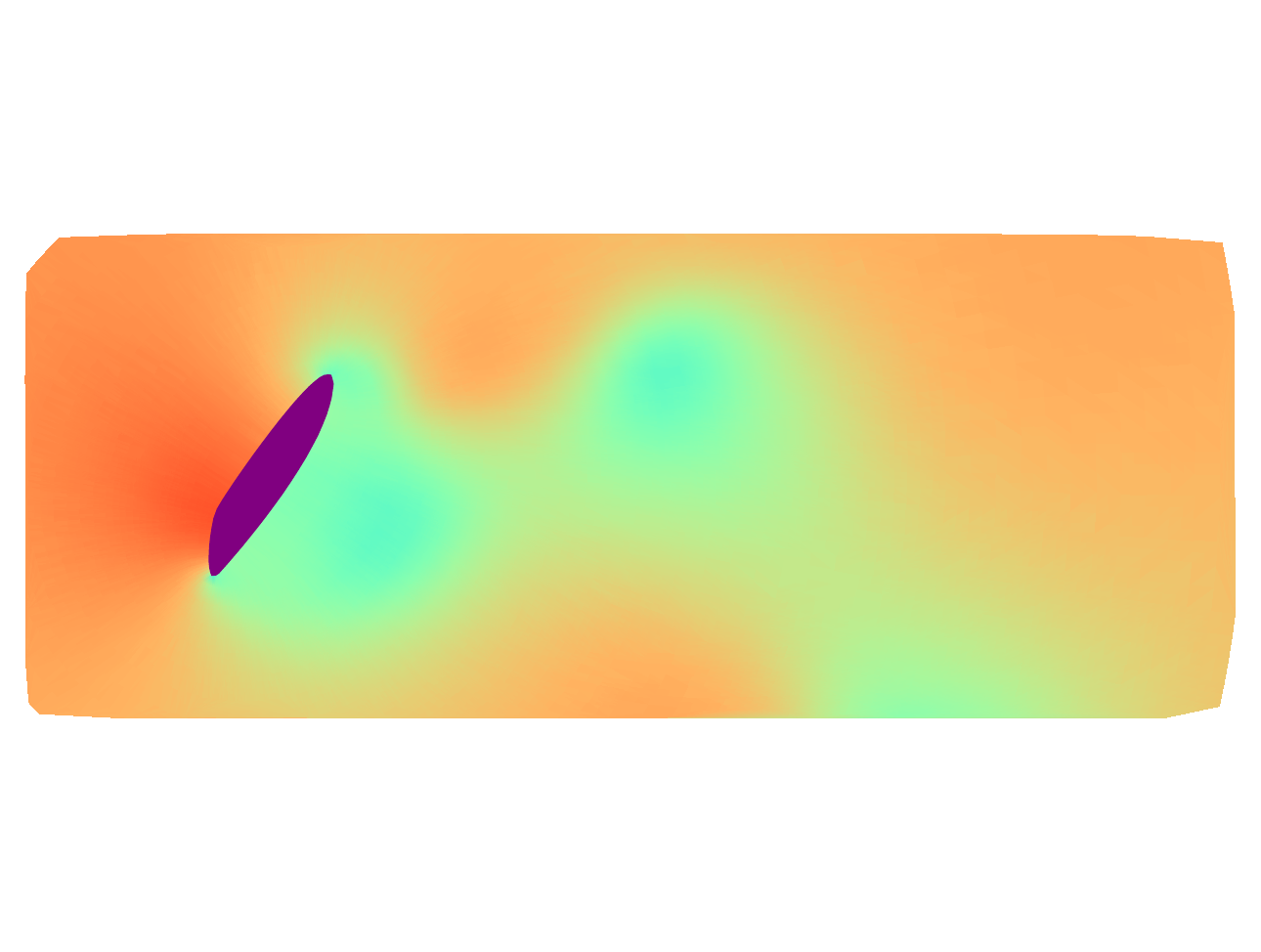}
    \includegraphics[width=0.30\columnwidth]{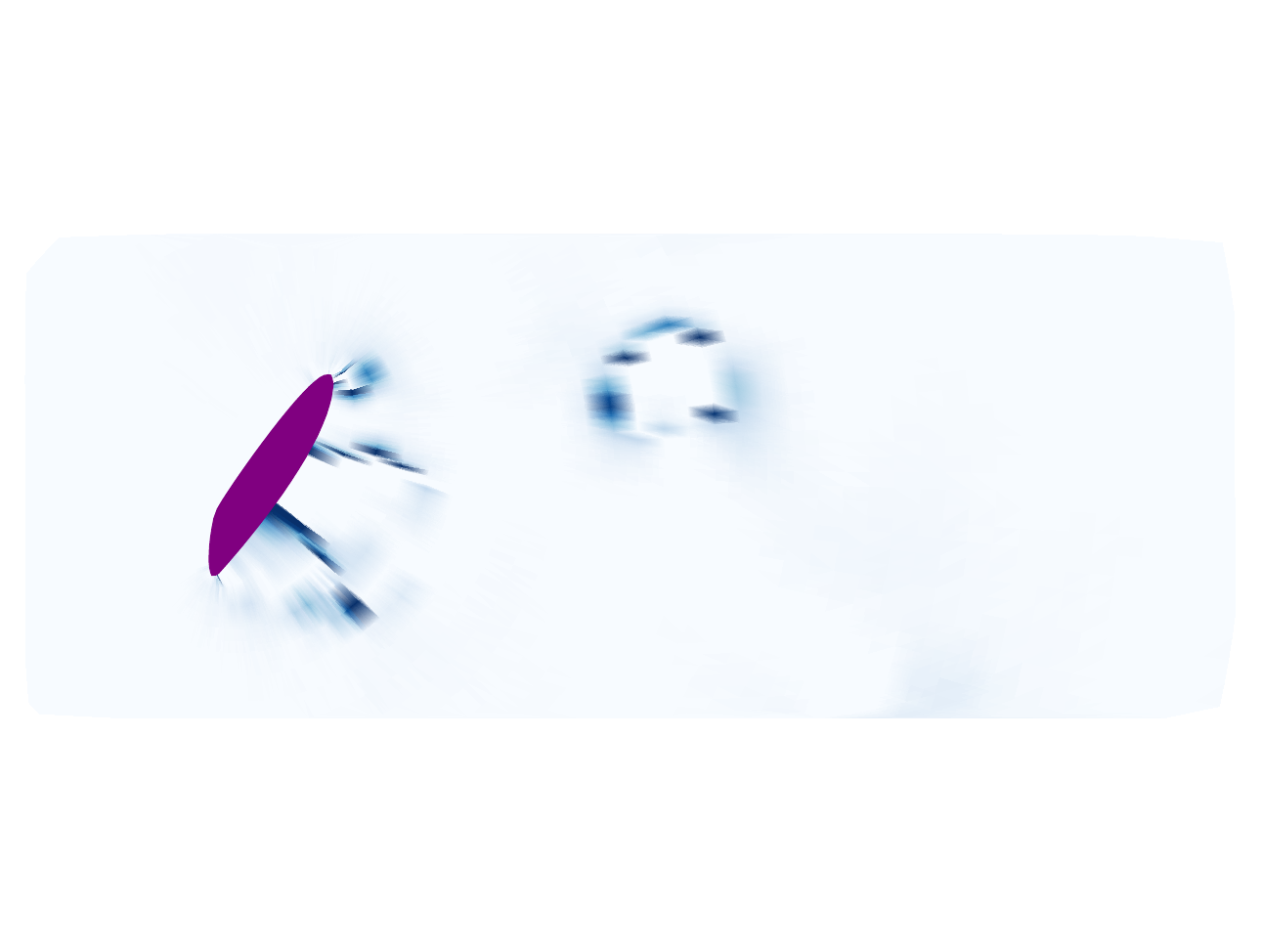} \\
    \includegraphics[width=0.06\columnwidth]{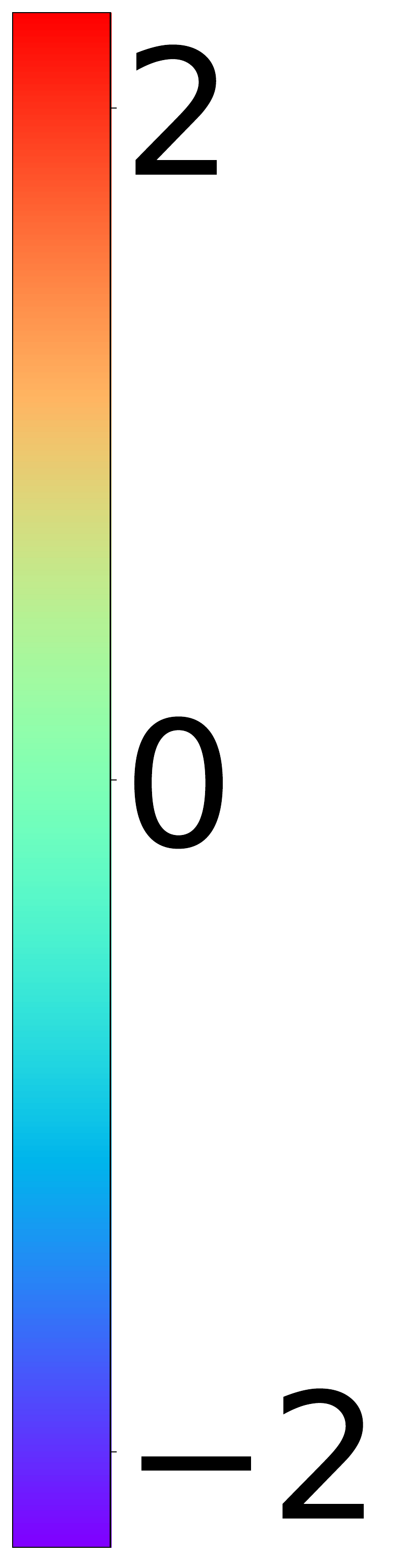}
    \includegraphics[width=0.30\columnwidth]{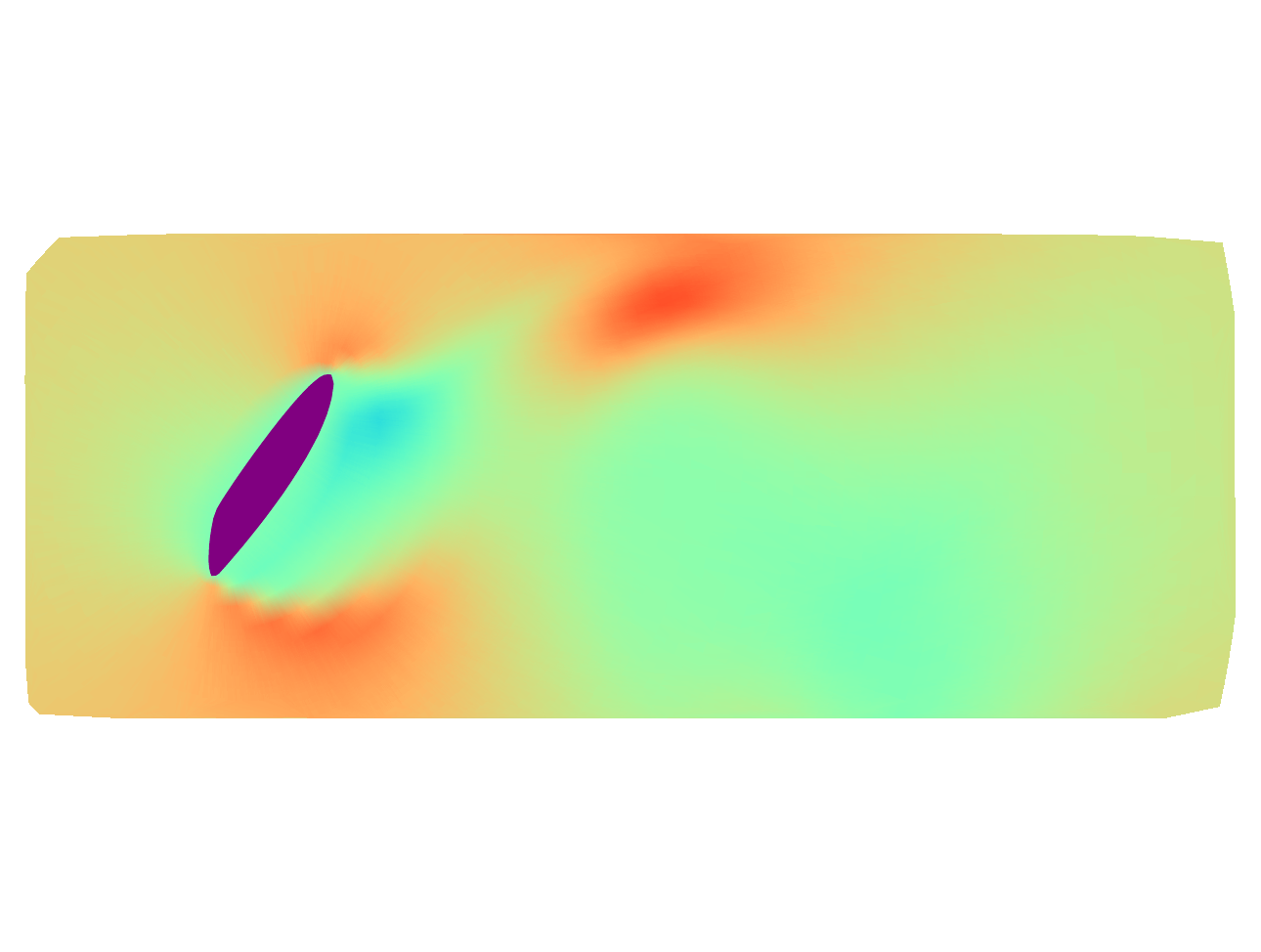}
    \includegraphics[width=0.30\columnwidth]{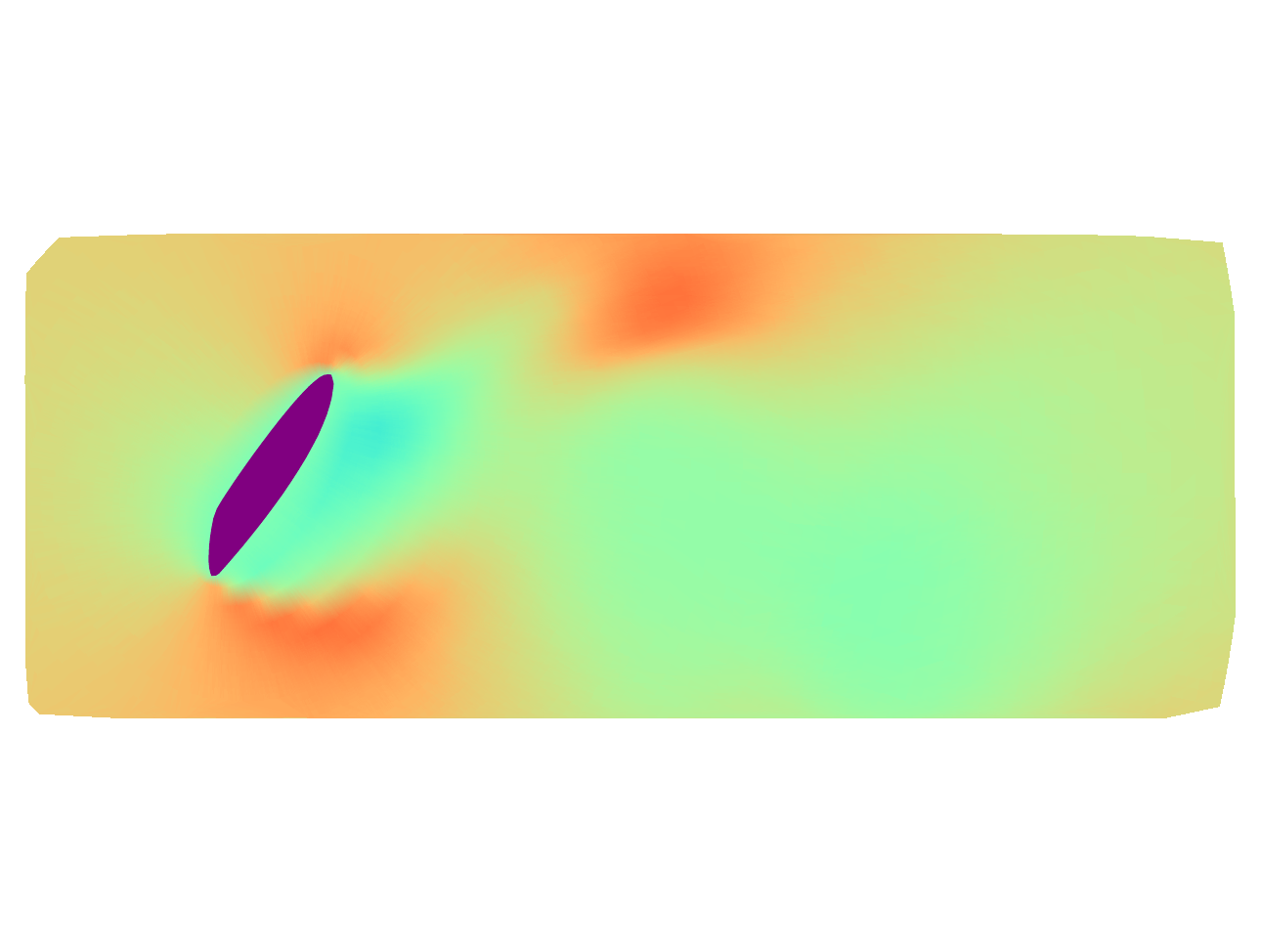}
    \includegraphics[width=0.30\columnwidth]{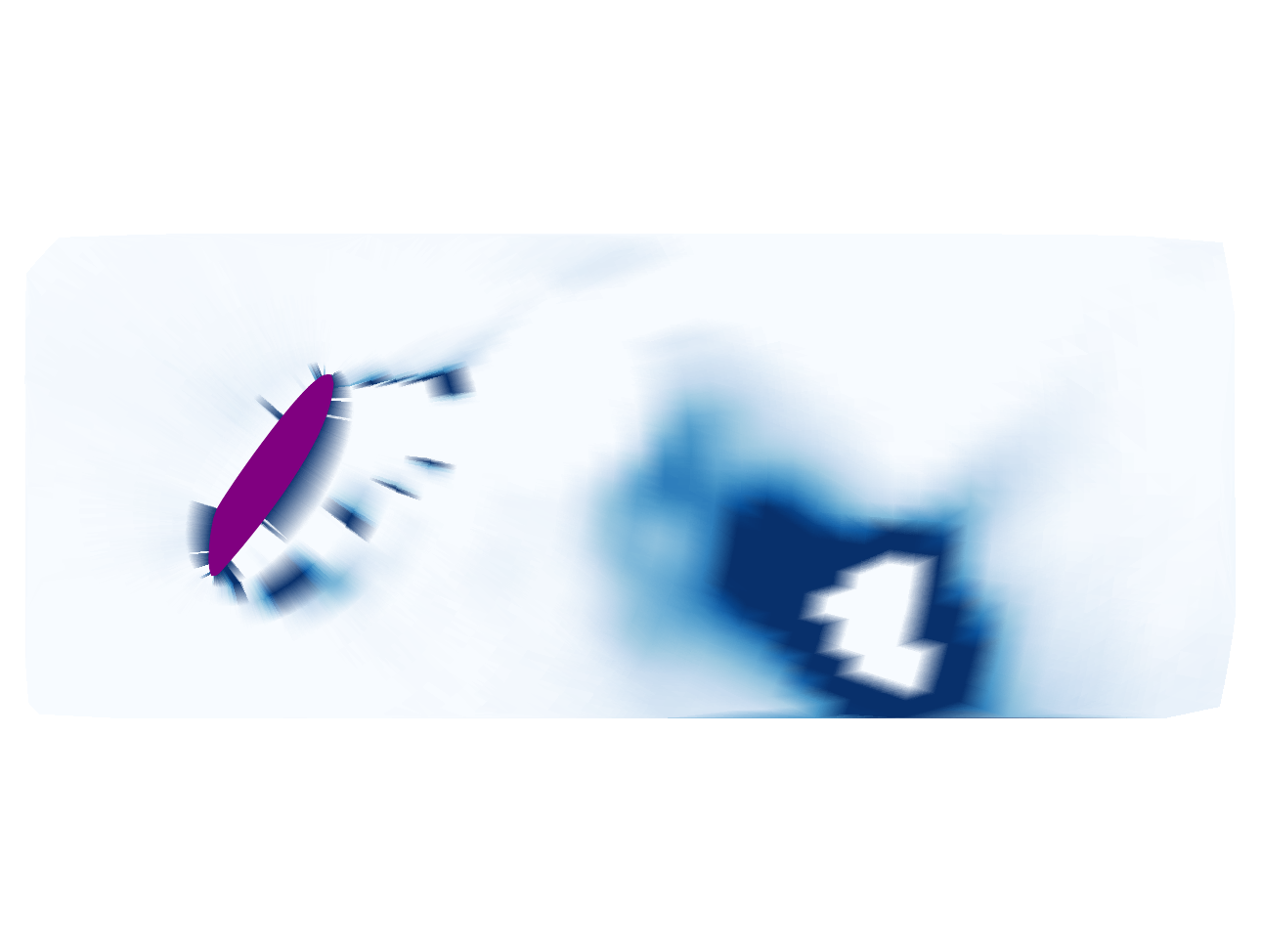} \\
    \includegraphics[width=0.06\columnwidth]{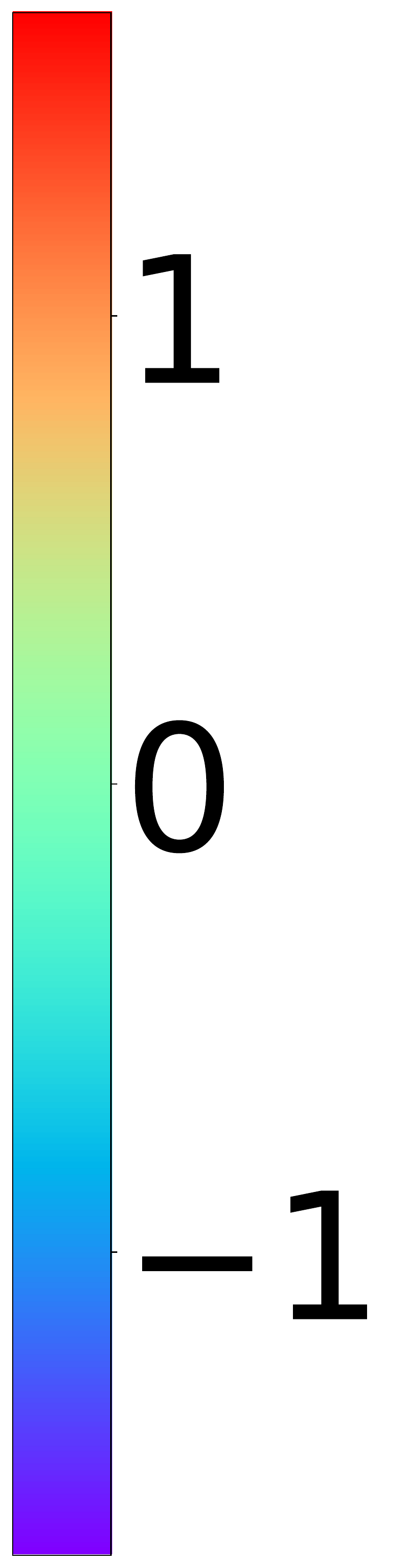}
    \includegraphics[width=0.30\columnwidth]{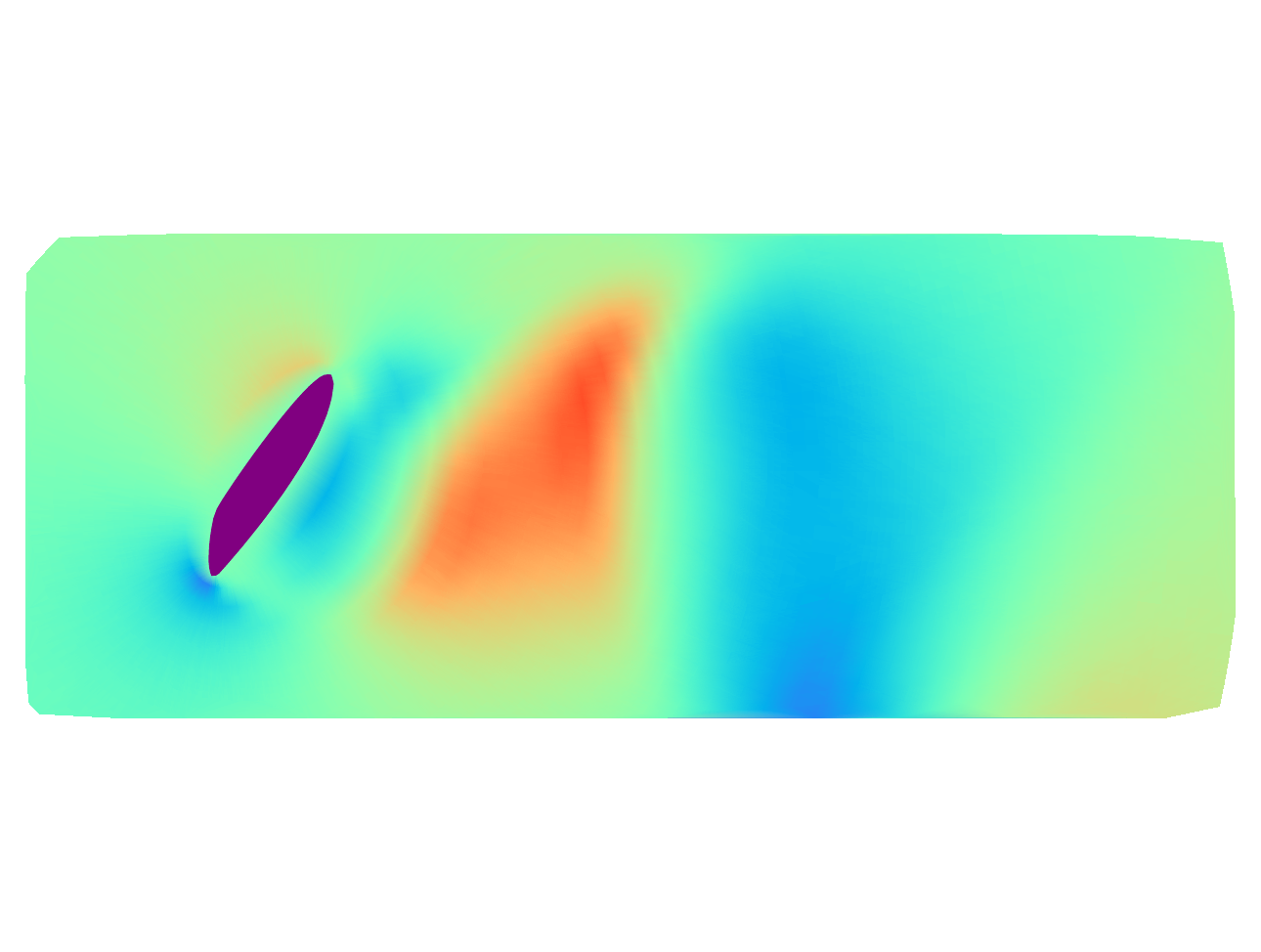}
    \includegraphics[width=0.30\columnwidth]{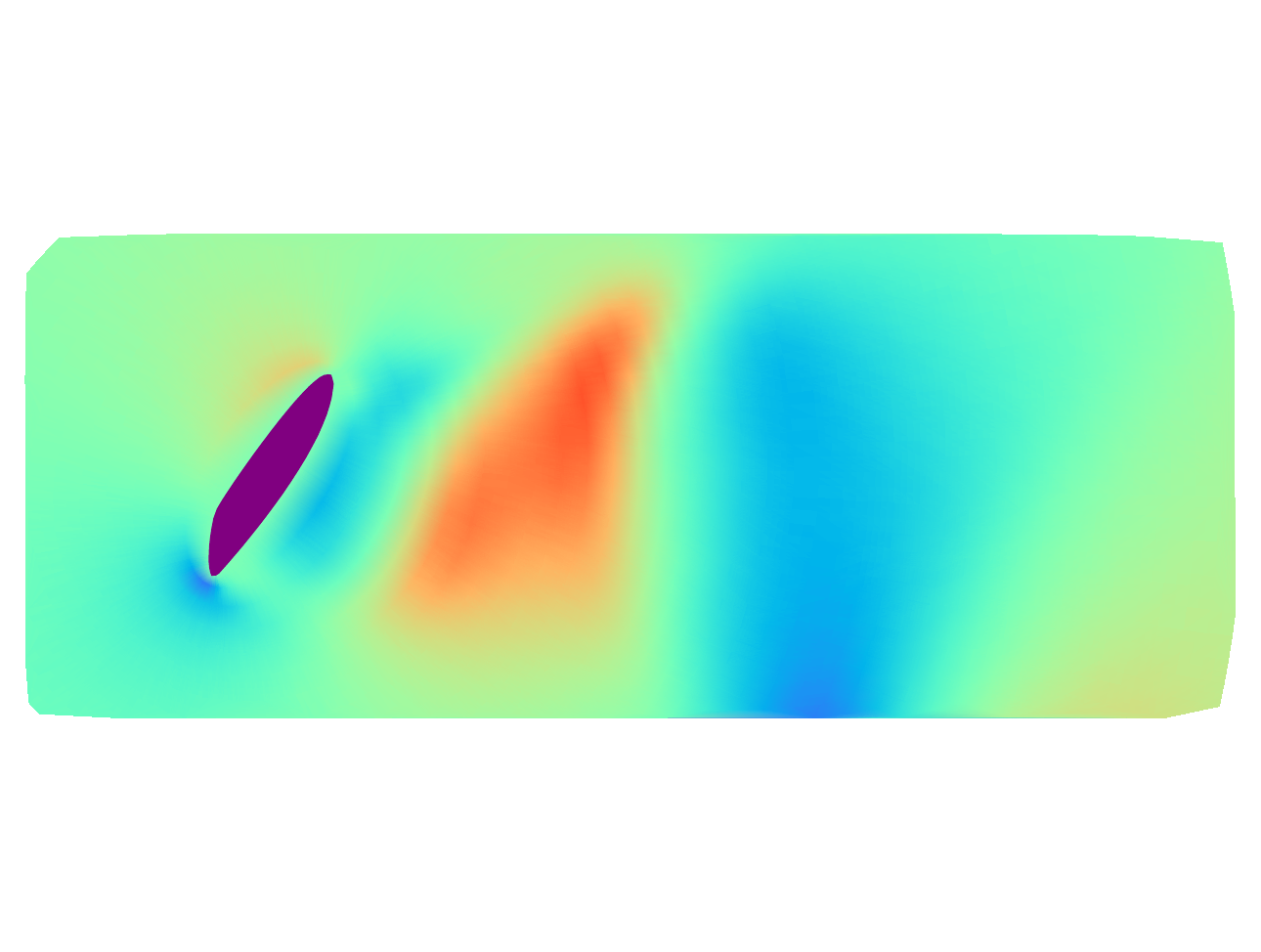}
    \includegraphics[width=0.30\columnwidth]{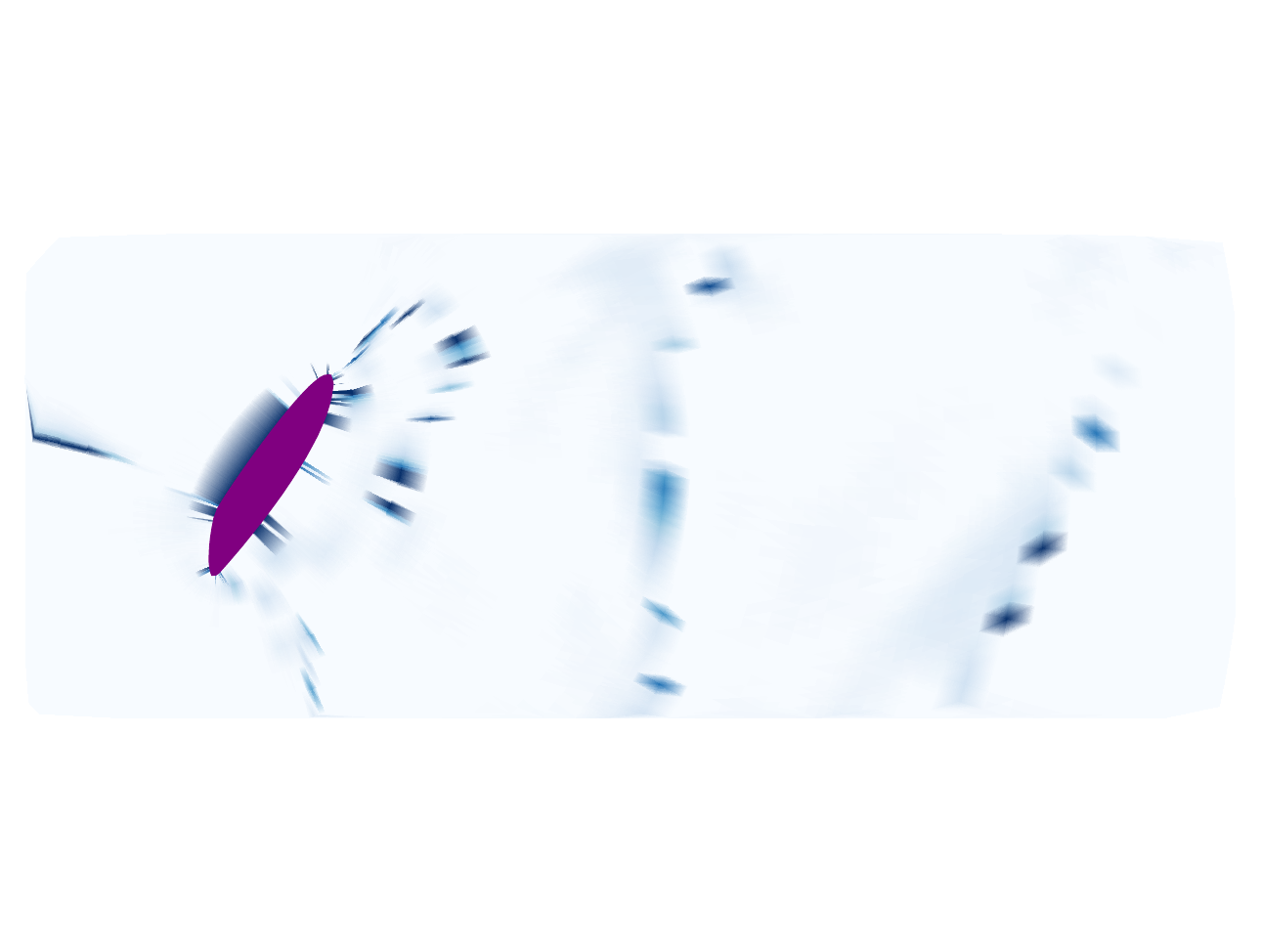} \\
    \includegraphics[width=0.22\textwidth]{images/spatiotemporal/pcterrorcbarr0.pdf}
    \caption{Ground truth (left), predicted (middle) and percentage error (right) fields for pressure (top), $u$-velocity (middle) and $v$-velocity (bottom) for a validation snapshot.}
    \label{fig:my_label}
\end{figure}

\clearpage
\section{Vorticity predictions using the spatio-temporal model with a zero temporal gap ($k=0$)}
\label{sec:app:spatio-temporalpred-k0}

\begin{figure}[h!]
    \centering
    \includegraphics[width=0.30\columnwidth]{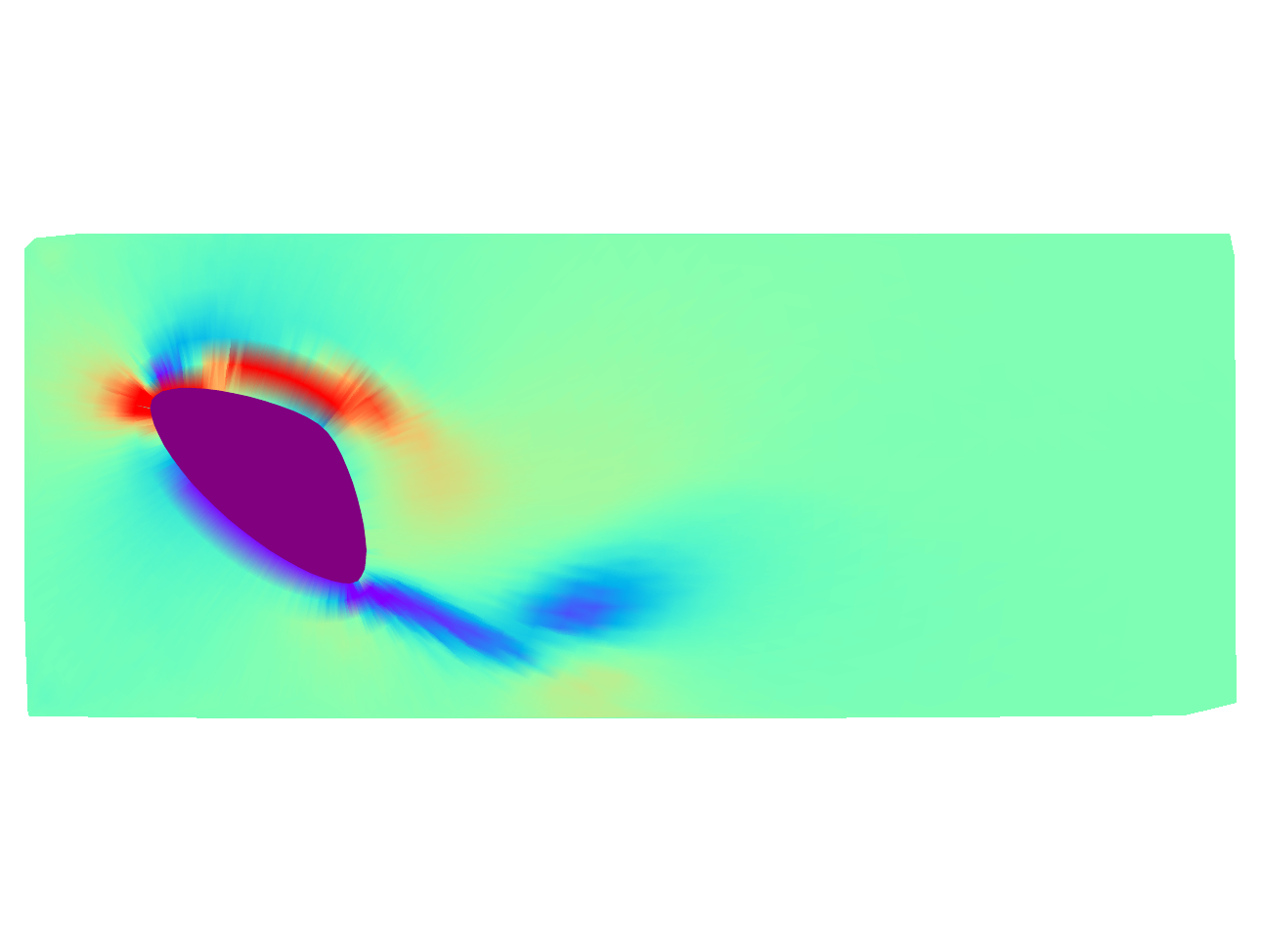}
    \includegraphics[width=0.30\columnwidth]{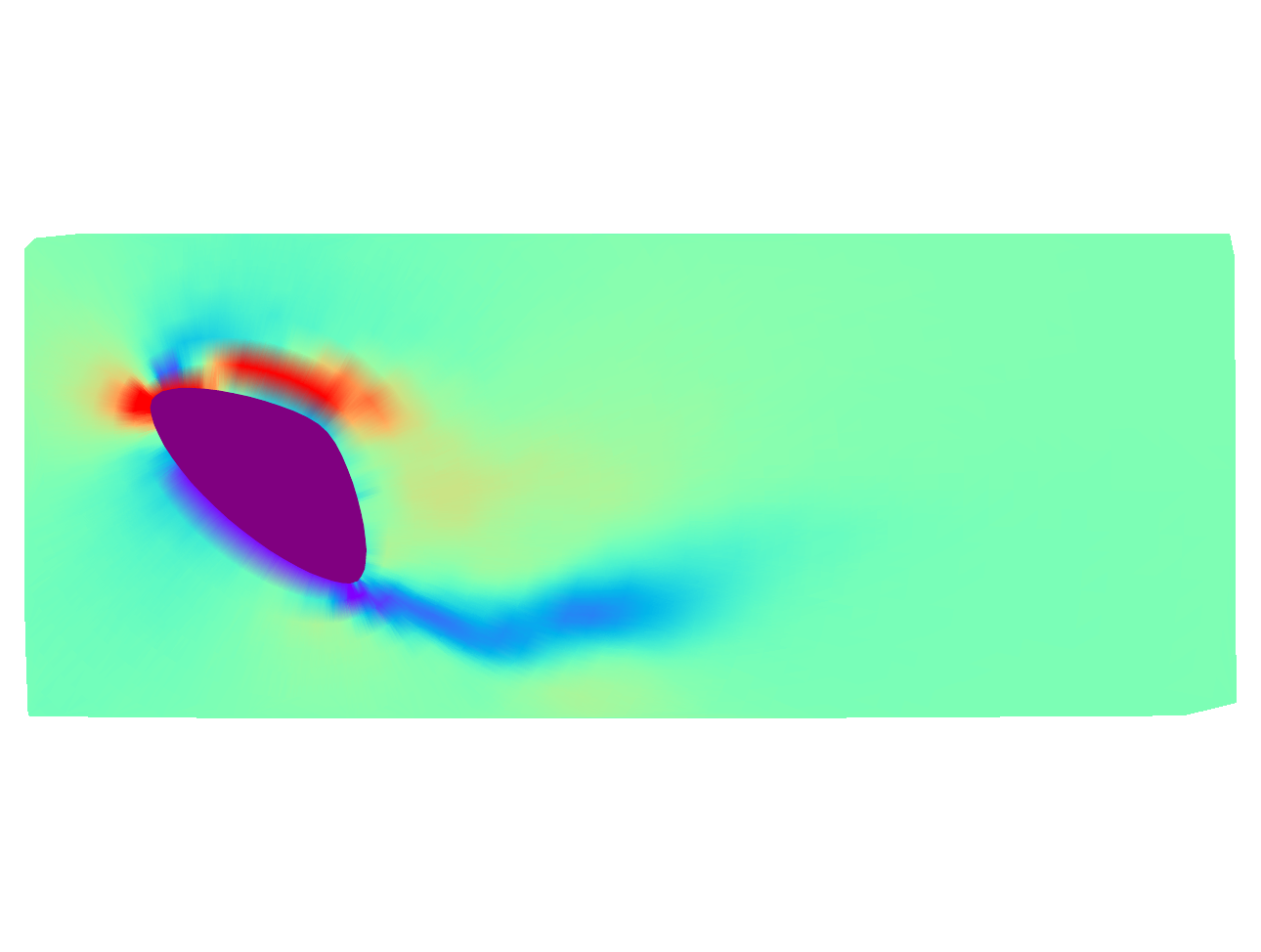}
    \includegraphics[width=0.30\columnwidth]{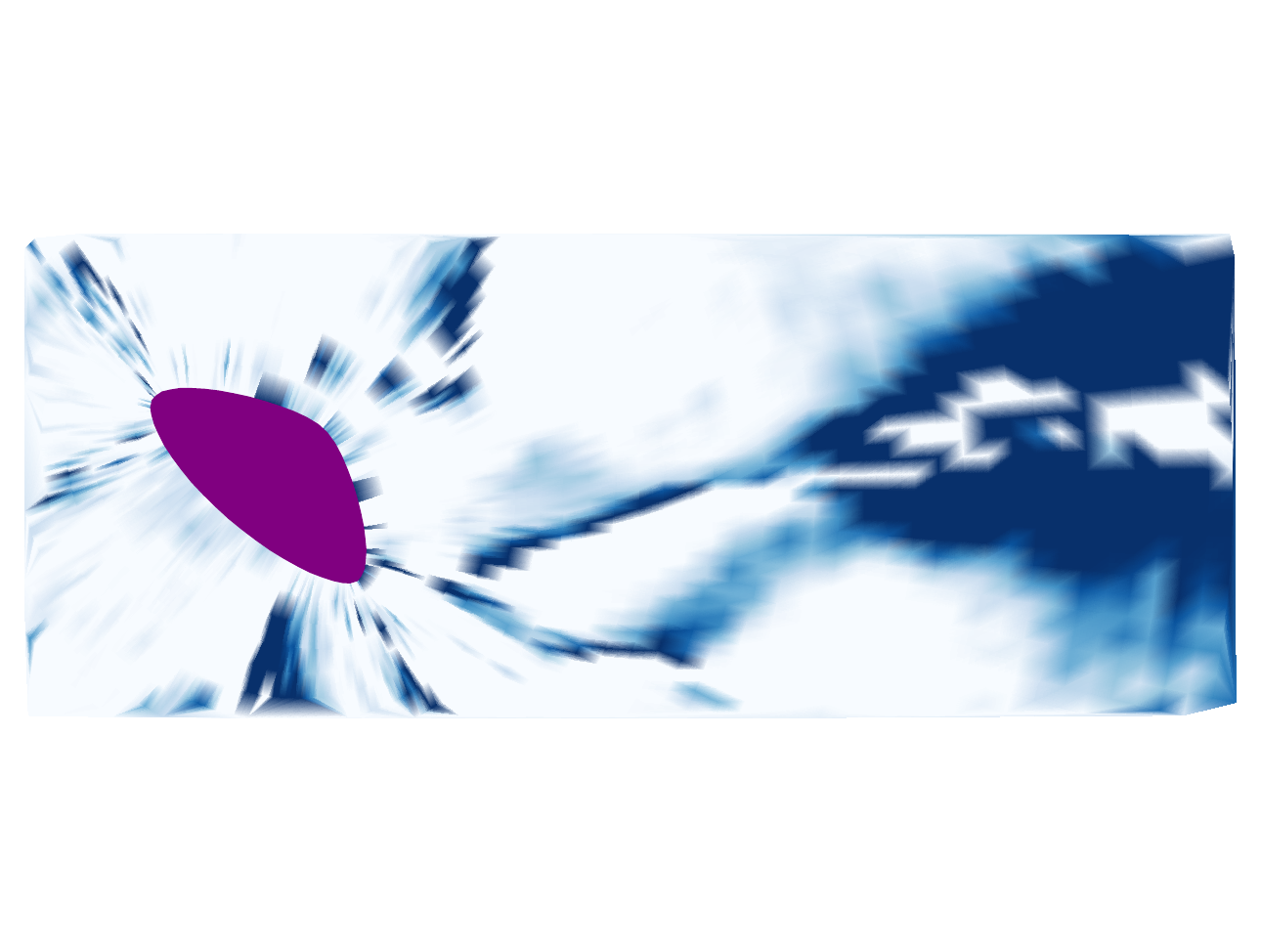}
    \includegraphics[width=0.30\columnwidth]{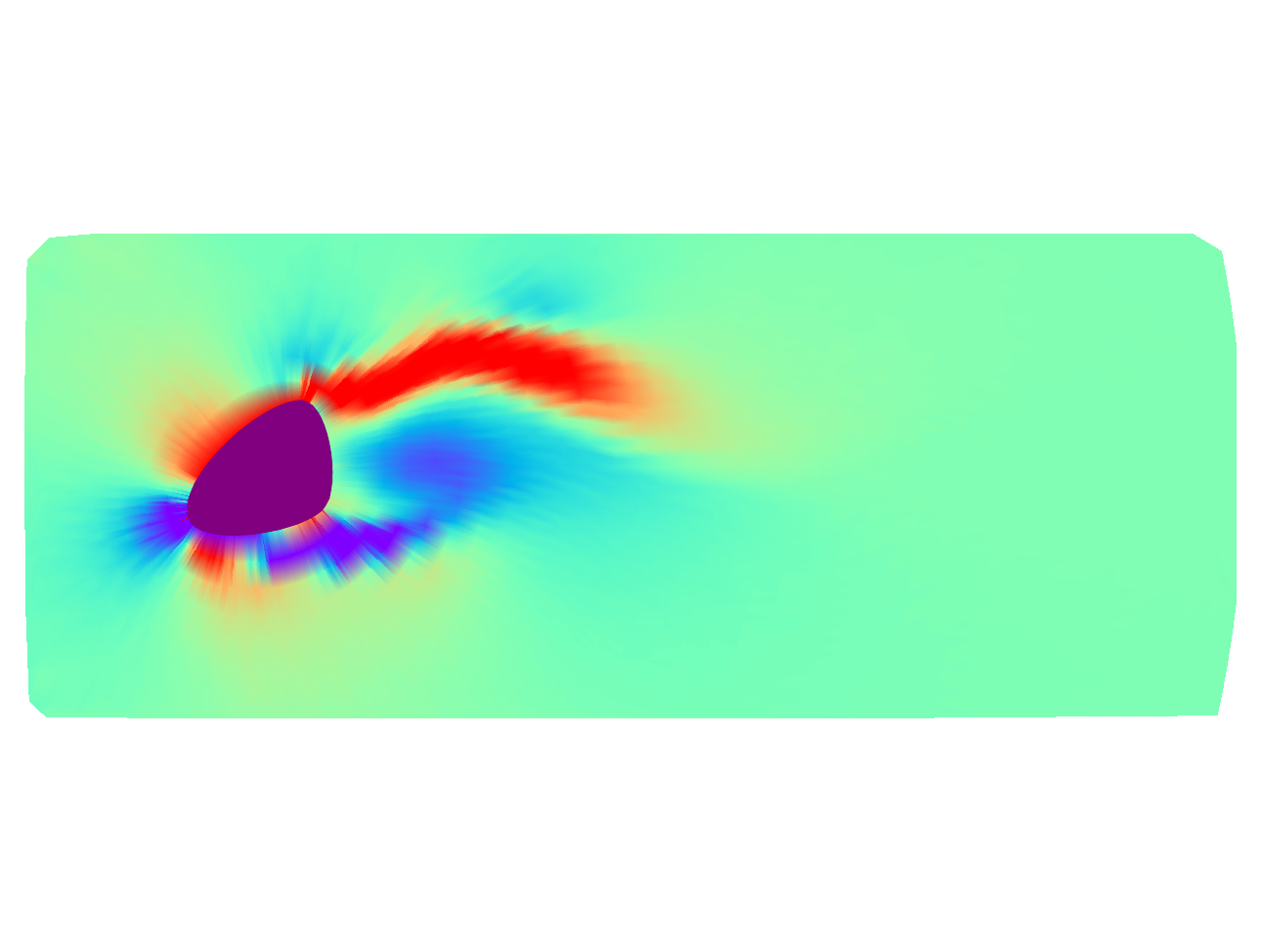}
    \includegraphics[width=0.30\columnwidth]{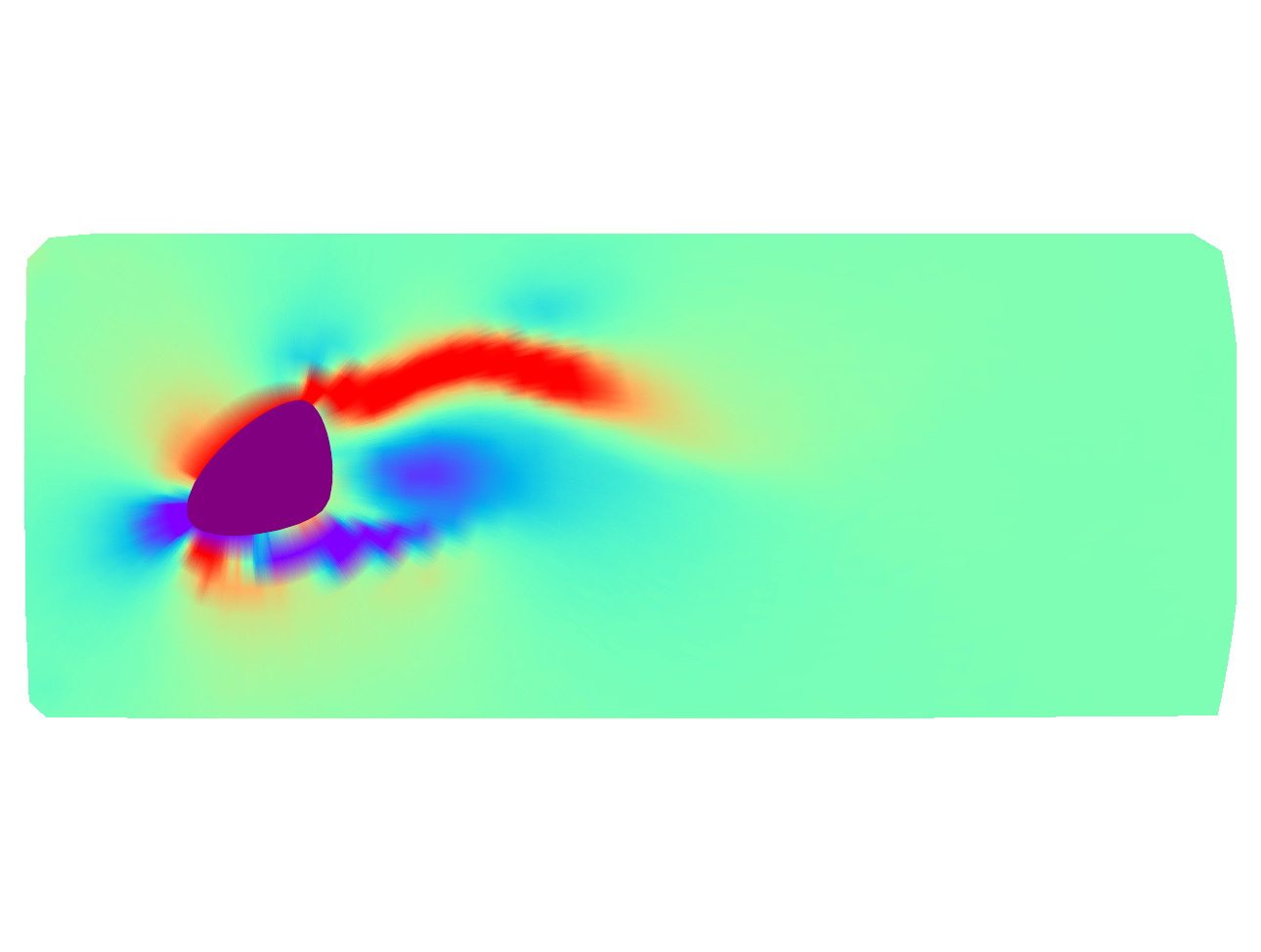}
    \includegraphics[width=0.30\columnwidth]{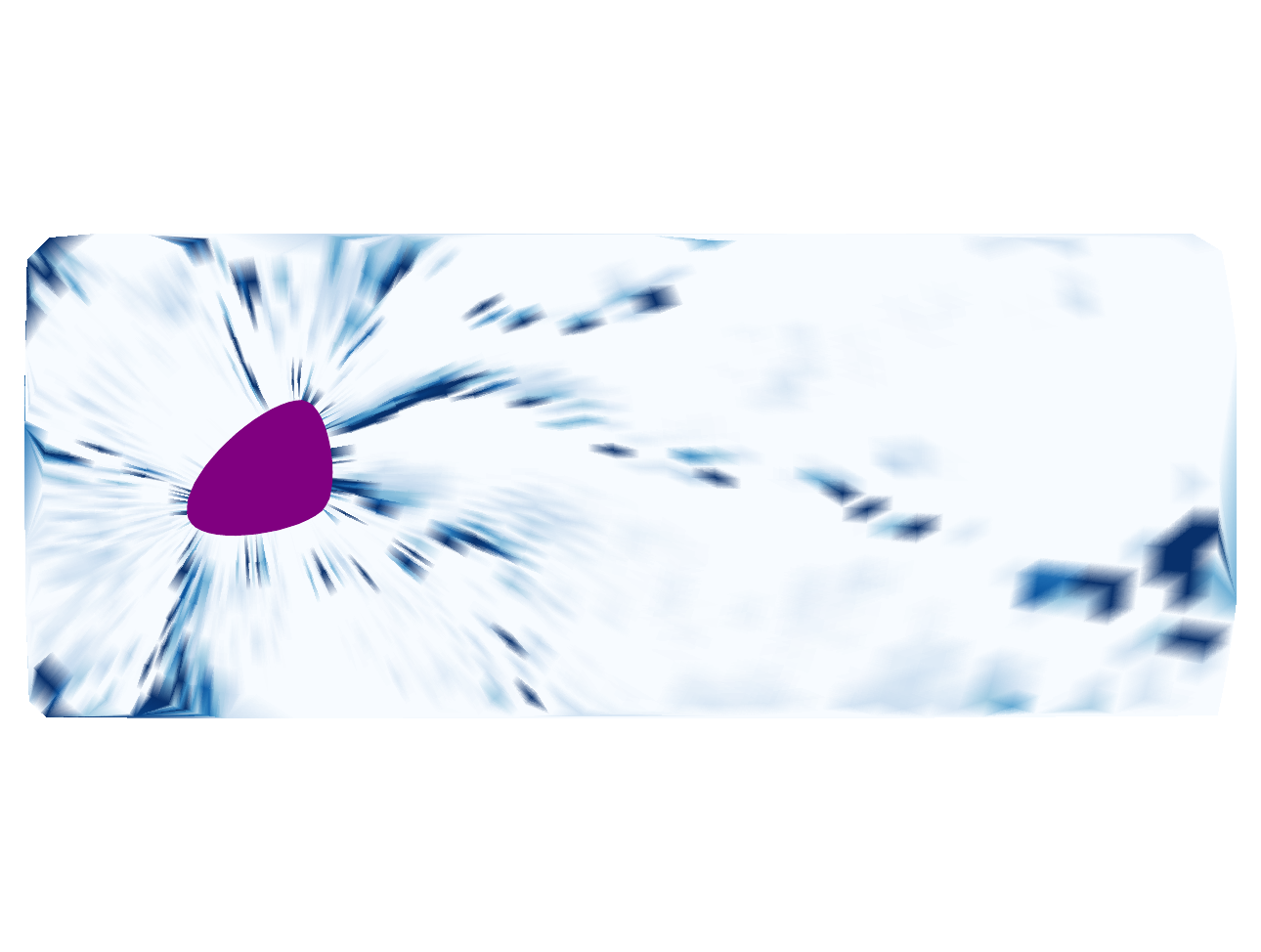}
    \includegraphics[width=0.30\columnwidth]{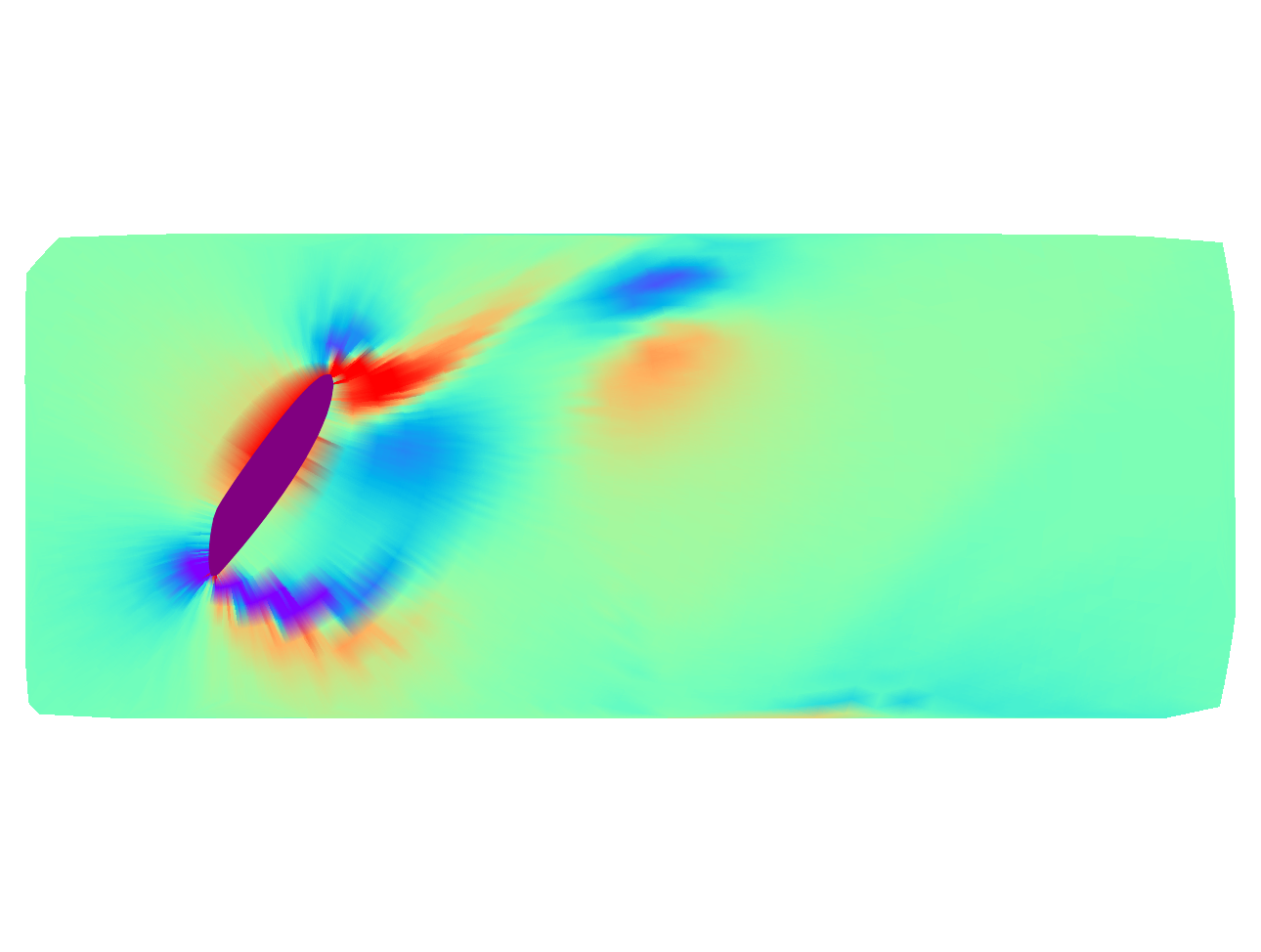}
    \includegraphics[width=0.30\columnwidth]{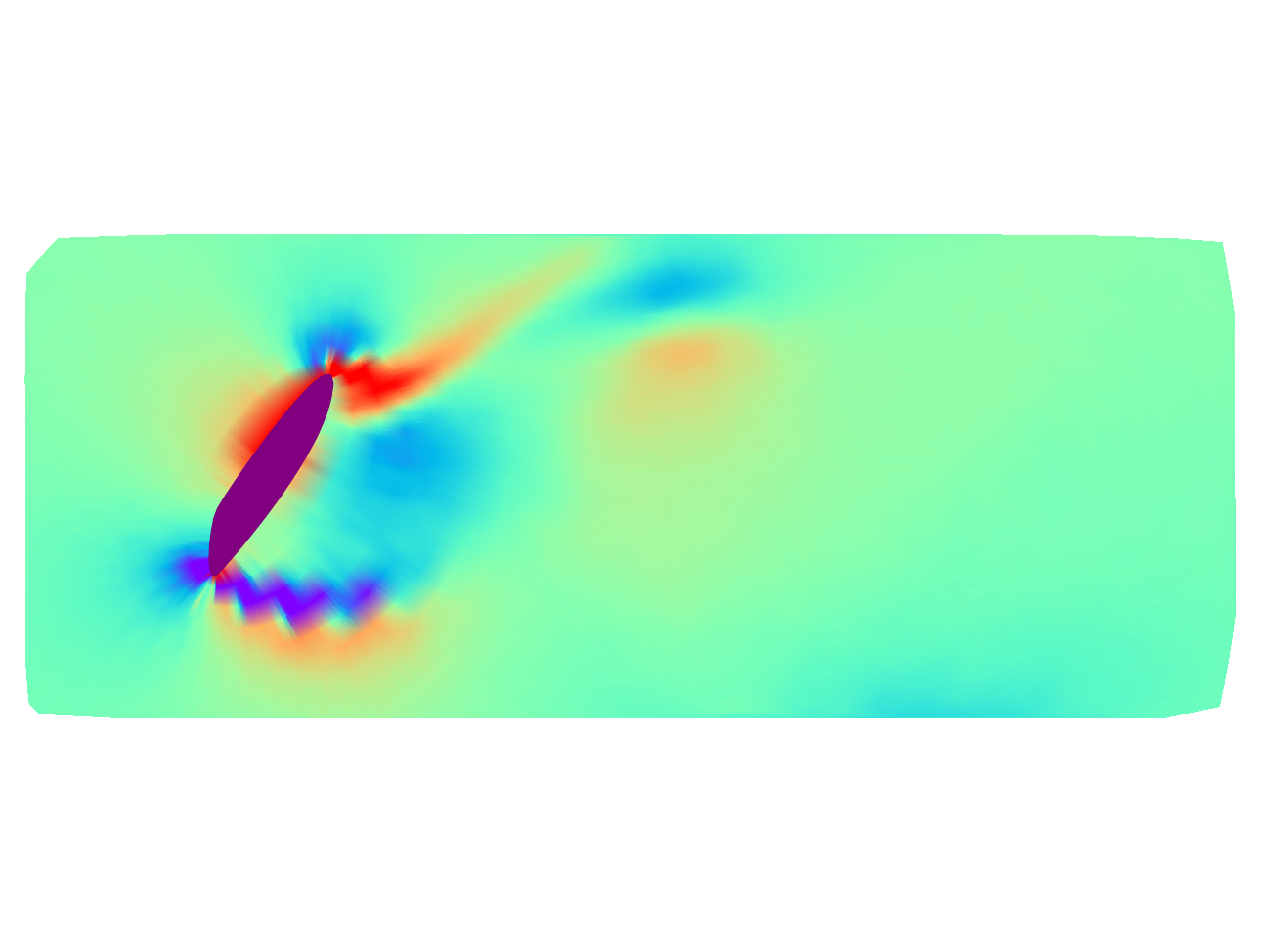}
    \includegraphics[width=0.30\columnwidth]{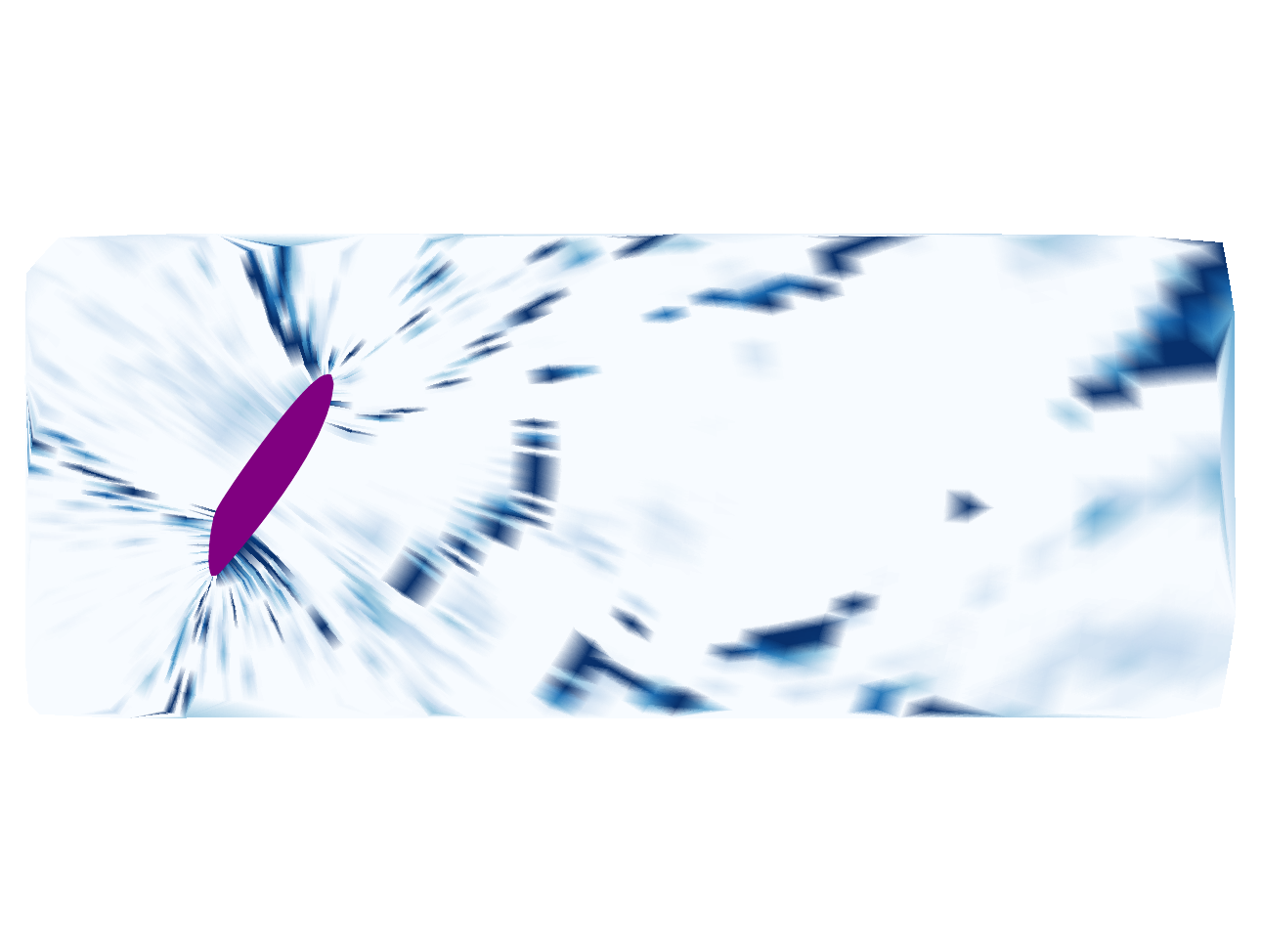}
    \includegraphics[width=0.20\textwidth]{images/spatiotemporal/cbar5530.pdf}
    \includegraphics[width=0.22\textwidth]{images/spatiotemporal/pcterrorcbarr0.pdf}
    \caption{Examples using the spatial (SD-UNet) plus temporal (FNO) model architecture with $k=0$ from \secref{sec:model_spatio-temporal_reconstruction}. Left column contains the ground truth snapshots, middle column contains the predictions and the right column contains the percentage error maps.}
    \label{fig:sdunet_fno_appendix}
\end{figure}

\end{document}